\documentclass[11pt]{report}

\usepackage[online]{suthesis-2e} 
\usepackage{suthesis-abstract}
\usepackage{graphicx}

\usepackage[cmex10]{amsmath}

\usepackage{amssymb}
\usepackage{multirow}
\usepackage{tikz}
\usetikzlibrary{shapes}
\usepackage{pgfplots}

\usepackage{color}   
\usepackage{hyperref}
\hypersetup{
    colorlinks=true, 
    linktoc=all,     
    linkcolor=black,  
    citecolor=black,
    urlcolor=black
}

\usepackage{algorithm}
\usepackage{algorithm,eurosym}
\usepackage{algorithmic}

\usepackage{amsthm}

\usepackage{url}

\usepackage{ascii}
\usepackage[cmex10]{amsmath}
\usepackage{amssymb}
\usepackage{amssymb,amsmath}
\newtheorem{definition}{Definition}
\newtheorem{lemma}{Lemma}
\newtheorem{scenario}{Scenario}
\newtheorem{example}{Example}

\newtheorem{theorem}{Theorem}
\newtheorem{proposition}{Proposition}
\newtheorem{observation}{Observation}

\begin{document}

    \title{Simultaneous Discrimination Prevention and Privacy Protection\\
            in Data Publishing and Mining}
   \author{Sara Hajian}
    \dept{Department of Computer Engineering and Mathematics}
    \principaladviser{Josep Domingo-Ferrer}
    \firstreader{Dino Pedreschi}
%

    \beforepreface

\thesiscopyrightpage

\signature{Prof. Dr. Josep Domingo-Ferrer (Advisor)}
\vspace{\signaturespace}
\signature{Prof. Dr. Dino Pedreschi (Co-advisor)}
\vspace{\signaturespace}
\ucgssignature
\vspace{\signaturespace}

\prefacesection{Abstract} 

Data mining is an increasingly important technology for extracting useful knowledge hidden in large collections of
data. There are, however, negative social perceptions about data mining, among which potential privacy violation and potential
discrimination. The former is an unintentional or deliberate disclosure of a user profile or activity data as part of the
output of a data mining algorithm or as a result of data sharing. For this reason, privacy preserving data mining has been introduced to trade off the utility of the resulting data/models for protecting individual privacy. The latter consists of treating people unfairly on the basis of their belonging to a specific group. Automated data
collection and data mining techniques such as classification have paved the way to making automated decisions,
like loan granting/denial, insurance premium computation, etc. If the training datasets are biased in what regards discriminatory attributes like gender, race, religion, etc., discriminatory decisions may ensue. For this reason, anti-discrimination
techniques including discrimination discovery and prevention have been introduced in data mining. Discrimination can be either
direct or indirect. Direct discrimination occurs when decisions are made based on discriminatory attributes. Indirect discrimination
occurs when decisions are made based on non-discriminatory attributes which are strongly correlated with biased discriminatory ones.

In the first part of this thesis, we tackle discrimination prevention in data mining and propose new techniques applicable for direct or indirect
discrimination prevention individually or both at the same time. We discuss how to clean training datasets and outsourced
datasets in such a way that direct and/or indirect discriminatory decision rules are converted to legitimate (non-discriminatory)
classification rules. The experimental evaluations demonstrate that the proposed techniques are effective at removing direct and/or
indirect discrimination biases in the original dataset while preserving data quality.

In the second part of this thesis, by presenting samples of privacy violation and potential discrimination in data mining, we argue that privacy and discrimination risks should be tackled together. We explore the relationship between privacy preserving data mining and discrimination prevention in data mining to design holistic approaches capable of addressing both threats simultaneously during 
the knowledge discovery process. As part of this effort, we have investigated for the first time the problem of discrimination and privacy aware frequent pattern discovery, {\em i.e.} the sanitization of the collection of patterns mined from a transaction database in such a way that neither privacy-violating nor discriminatory inferences can be inferred on the released patterns. Moreover, we investigate  the problem of discrimination and privacy aware data publishing, {\em i.e.} transforming the data, instead of patterns, in order to simultaneously
fulfill privacy preservation and discrimination prevention. In the above cases, it turns out that the impact of our transformation on the quality of data or patterns
is the same or only slightly higher than the impact of achieving just privacy preservation.


\newpage

\vspace*{\fill}
\begingroup
\centering
{\em To my family, advisors and friends}

\endgroup
\vspace*{\fill}

%

\newpage




    \afterpreface
 
    \chapter{Introduction}
\label{chap:intro}


Data mining is an increasingly important technology for extracting useful
knowledge hidden in large collections of data, especially human and social data sensed by the ubiquitous technologies that support most human activities in our age.
As a matter of fact, the new opportunities to extract knowledge and understand human and social complex phenomena
increase hand in hand with the risks of violation of fundamental human
rights, such as privacy and non-discrimination.
\emph{Privacy} refers to the individual right to choose freely what to do with
one's own personal information, while \emph{discrimination} refers to unfair or unequal treatment of people based on membership to a category, group or minority, without regard to individual merit.
Human rights laws not only have concern about data protection \cite{eurpprivacy} but also prohibit discrimination \cite{AustralianLeg,Europeandisc} against protected groups on the grounds of race, color, religion, nationality, sex, marital status, age and pregnancy; and in a number of settings, like credit and insurance,  personnel selection and wages, and access to public services.
Clearly, preserving the great benefits of data mining within a privacy-aware
and discrimination-aware technical ecosystem would lead to a wider
social acceptance of a multitude of new services
and applications based on the knowledge discovery process.

\section{Privacy Challenges of Data Publishing and Mining}

We live in times of unprecedented opportunities of sensing, storing and analyzing micro-data on human activities at extreme detail and resolution, at society level \cite{pederFair}. Wireless networks and mobile devices record the traces of our movements. Search engines record the logs of our queries for finding information on the web. Automated payment systems record the tracks of our purchases. Social networking services record our connections to friends, colleagues, collaborators.

Ultimately, these big data of human activity are at the heart of the very idea of a knowledge society \cite{pederFair}: a society where small or big decisions made by businesses or policy makers or ordinary citizens – can be informed by reliable knowledge, distilled from the ubiquitous digital traces generated as a side effect of our living. Although increasingly sophisticated data analysis and data mining techniques support knowledge discovery from human activity data to improve the quality of on-line and off-line services for users, they are increasingly raising user privacy concerns on the other side. 

From the users' perspective, insufficient privacy protections on the part of a service they use
and entrust with their activity, personal or sensitive information could lead to significant emotional,
financial, and physical harm. An unintentional or deliberate disclosure of a user profile or activity data as part of the
output of an internal data mining algorithm or as a result of data sharing may potentially lead to embarrassment, identity theft and discrimination \cite{Korolova}. It is hard to foresee all the privacy risks that a digital dossier consisting of detailed
profile and activity data could pose in the future, but it is not inconceivable that it could harm users' lives. In general, during knowledge discovery, privacy violation is an unintentional or deliberate intrusion into the personal data of the data subjects, namely, of the (possibly unaware) people whose data are being collected, analyzed and mined \cite{pederFair}.

Thus, although users appreciate the continual innovation and improvement in the quality of
on-line and off-line services using sophisticated data analysis and mining techniques, they are also becoming increasingly concerned about their privacy, and about the
ways their personal and activity data are compiled, mined, and shared \cite{Purcell}.  On the other hand, for institutions and companies such as banks, insurance companies and search engines that offer different kinds of online and/or off-line services, the trust of users in their privacy practices is a strategic product and business advantage. Therefore, it is in the interest of 
these organizations to strike a balance between mining and sharing user data in order to improve their
services and  protecting user privacy to retain the trust of users~\cite{Vascellaro}.

In order to respond to the above challenges, data protection
technology needs to be developed in tandem with data mining and publishing techniques \cite{FuturICT}. The framework to advance is thinking of \emph{privacy by design}\footnote{The European Data Protection Supervisor Peter Hustinx is a staunch defender of the Privacy by Design approach and has recommended it as the standard approach to data protection for the EU, see 
\url{http://www.edps.europa.eu/EDPSWEB/webdav/site/mySite/shared/Documents/Consultation/Opinions/2010/10-03-19_Trust_Information_Society_EN.pdf}
Ann Cavoukian – Privacy Commissioner of Ontario, Canada – has been one of the early
defenders of the Privacy by Design approach. See the principles that have been formulated
\url{http://www.privacybydesign.ca/index.php/about-pbd/7-foundational-principles/}}. The basic idea is to inscribe
privacy protection into the analytical technology by design and construction, so that the analysis takes the privacy requirements in consideration from the very start.
Privacy by design, in the research field of \emph{privacy preserving data
mining} (PPDM), is a recent paradigm that promises a quality leap in the conflict between data protection and data utility. 
PPDM has become increasingly popular because it allows sharing and using sensitive data for analysis purposes. Different PPDM methods have been developed for different purposes, such as data hiding, knowledge (rule) hiding, distributed PPDM and privacy-aware knowledge sharing in different data mining tasks.

%


\section{Discrimination Challenges of Data Publishing and Mining}
Discrimination refers to an unjustified difference in treatment on the basis of any physical or cultural trait, such as sex, ethnic origin, religion or political opinions. From the legal perspective, privacy violation is not the only risk which threatens fundamental human rights; discrimination risks are also concerned when mining and sharing personal data. 
In most European and North-American countries, it is forbidden by law to discriminate against certain protected groups \cite{ChaldersChap}. The European Union has one of the strongest anti-discrimination legislations (See, {\em e.g.}, Directive 2000/43/EC, Directive 2000/78/EC/ Directive 2002/73/EC, Article 21 of the Charter of Fundamental Rights and Protocol 12/Article 14 of the European Convention on Human Rights), describing discrimination on the basis of race, ethnicity, religion, nationality, gender, sexuality, disability, marital status, genetic features, language and age. It does so in a number of settings, such as employment and training, access to housing, public services, education and health care; credit and insurance; and adoption. European efforts on the non-discrimination front make clear the fundamental importance of the effective implementation and enforcement of non-discrimination norms \cite{ChaldersChap} for 
Europe's citizens.

From the user's perspective, many people may not mind other people knowing about their ethnic origins, but they would strenuously object to be denied a credit or a grant if their ethnicities were part of that decision.
As mentioned above, nowadays socially sensitive decisions may be taken by automatic systems, {\em e.g.}, for screening or ranking applicants to a job position, to a loan, to school admission and so on. For instance, data mining and machine learning classification models are constructed on the basis of historical data exactly with the purpose of learning the distinctive elements of different classes or profiles, such as good/bad debtor in credit/insurance scoring systems.
Automatically generated decision support models may exhibit discriminatory behavior toward certain groups based upon, {\em e.g.} gender or ethnicity. In general, during knowledge discovery, discrimination risk is the unfair use of the discovered knowledge in making discriminatory decisions about the (possibly unaware) people who are classified, or profiled \cite{pederFair}.
Therefore, it is in the interest
of banks, insurance companies, employment agencies, the police and other institutions that employ data mining models for decision making upon individuals, to ensure that these computational models are free from discrimination \cite{ChaldersChap}.


Then, data mining and data analytics on data about people need to  incorporate many ethical values by design, not only data protection but also non-discrimination. We need novel, disruptive technologies
for the construction of human knowledge discovery systems that, by design, offer native technological safeguards against discrimination. 
\emph{Anti-discrimin\allowbreak ation by design}, in the research field of \emph{discrimination prevention in data mining} (DPDM), is a more recent paradigm that promises a quality leap in the conflict between non-discrimination and data/model utility. 
More specifically, discrimination has been recently considered from a data mining perspective. Some proposals are oriented to the discovery and measurement of discrimination, while others deal with preventing data mining from becoming itself a source of discrimination, due to automated decision making based on discriminatory models extracted from biased datasets. 
In fact, DPDM consists of extracting models (typically, classifiers) that trade off utility of the resulting data/model with non-discrimination.

\section{Simultaneous Discrimination Prevention and Privacy Protection 
            in Data Publishing and Mining}

In this thesis, by exploring samples of privacy violation and discrimination risks in contexts of data and knowledge publishing, we realize that privacy and anti-discrimination are two intimately intertwined concepts: they share common challenges, common methodological problems to be solved and, in certain contexts, directly interact with each other. Despite this striking commonality, there is an evident gap between the large body of research in data privacy technologies and the recent early results in anti-discrimination technologies. This thesis aims to answer the following research questions:

\begin{itemize}

\item What is the relationship between PPDM and DPDM? Can privacy protection achieve anti-discrimination (or the other way round)?

\item Can we adapt and use some of the existing approaches from the PPDM literature for DPDM?

\item Is it enough to tackle only privacy or discrimination risks to make a truly trustworthy technology for knowledge discovery? If not, how can we design a holistic method capable of addressing both threats together in significant data mining processes?

\end{itemize}

This thesis aims to find ways to mine and share personal data while protecting users' privacy and preventing discrimination against protected groups of users, and to motivate companies, institutions and the research community to consider a need for \emph{simultaneous} privacy and anti-discrimination \emph{by design}. 
This thesis is the first work inscribing simultaneously privacy and anti-discrimination with a  by design approach in data publishing and mining. It is a first example of a more comprehensive set of ethical values that is inscribed into the analytical process. 
   
\subsection{Contributions}

Specifically, this thesis makes the following concrete contributions to answer the above questions:

\begin{enumerate}

\item {\bf A methodology for direct and indirect discrimination prevention in data mining}. We tackle discrimination prevention in data mining and propose new techniques applicable for direct or indirect
discrimination prevention individually or both at the same time. Inspired by the data transformation methods for knowledge (rule) hiding in PPDM, we devise new data transformation methods ({\em i.e.} direct and indirect rule protection, rule generalization) for converting direct and/or indirect discriminatory decision rules to legitimate (non-discriminatory) classification rules. We also propose new metrics to evaluate the utility of the proposed approaches and we compare these
approaches. The experimental evaluations demonstrate that the proposed techniques are effective at removing direct and/or
indirect discrimination biases in the original dataset while preserving data quality.

\item {\bf Discrimination- and privacy-aware frequent pattern discovery}. Consider the case when a set of patterns extracted from the personal data of a population of individual persons is released for a subsequent use into a decision making process, such as granting or denying credit. First, the set of patterns may reveal sensitive information about individual persons in the training population and, second, decision rules based on such patterns may lead to unfair discrimination, depending on what is represented in the training cases. We argue that privacy and discrimination risks should be tackled {\em together}, and we present a methodology for doing so while publishing frequent pattern mining results. We describe a set of pattern sanitization methods, one for each discrimination measure used in the legal literature, to achieve a fair publishing of frequent patterns in combination with a privacy transformation based on $k$-anonymity. Our proposed pattern sanitization methods yield both privacy- and discrimination-protected patterns, while introducing reasonable (controlled) pattern distortion. We also explore the possibility to combine anti-discrimination with differential privacy.

\item {\bf Generalization-based privacy preservation and discrimination
prevention in data publishing}. We investigate the problem of discrimination and privacy aware data publishing, {\em i.e.} transforming the data, instead of patterns, in order to simultaneously
fulfill privacy preservation and discrimination prevention. Our approach falls into the pre-processing category: it sanitizes the data before they are used in data mining tasks rather than sanitizing the knowledge patterns extracted by data mining tasks (post-processing).
Very often, knowledge publishing (publishing the sanitized patterns) is not enough for the users or researchers, who want to be able to mine the data themselves. This gives researchers greater flexibility in performing the required data analyses. We observe that published data must be {\em both} privacy-preserving and unbiased regarding discrimination. We present the first generalization-based approach to simultaneously offer privacy preservation and discrimination prevention. We formally define the problem, give an optimal algorithm to tackle it and evaluate the algorithm in terms of both general and specific data analysis metrics. It turns out that the impact of our transformation on the quality of data is the same or only slightly higher than the impact of achieving just privacy preservation. In addition, we show how to extend our approach to different privacy models and anti-discrimination
legal concepts.

\end{enumerate}

The presentation of this thesis is divided into eight chapters according to the contributions.

\subsection{Structure}

This thesis is organized as follows.
In Chapter \ref{chap:discrelatedwork}, we review the related works on discrimination discovery and prevention in data mining (Section \ref{sec:RWDISC}). Moreover, we introduce some basic definitions and concepts that are used throughout
this thesis related to data mining (Section \ref{sec:BDDISC}) and measures of discrimination (Section \ref{sec:discmeasures}). In Chapter \ref{chap:privrelatedwork}, we first present a brief review on PPDM in Section~\ref{sec:BRPRIV}. After that, we introduce some basic definitions and concepts that are used throughout
the thesis related to data privacy (Section \ref{sec:BDPRIV}), and we elaborate on models (measures) of privacy (Section \ref{sec:privmeasures}) and samples of anonymization techniques (Section \ref{sec:samples}). Finally, 
we review the approaches and algorithms of PPDM in Section~\ref{sec:ALGSPRIV}.

In Chapter \ref{chap:methodology}, we propose a methodology for direct and indirect discrimination prevention in data mining. Our contributions on discrimination prevention are presented in Section~\ref{sec:CONTRIMETHO}. 
Section~\ref{sec:DIDMMETH} introduces direct and indirect discrimination measurement. Section~\ref{sec:prevention}
describes our proposal for direct and indirect discrimination prevention.
The proposed data transformation methods for direct and indirect discrimination prevention are presented in Section~\ref{sec:DTDDP} and Section~\ref{sec:DTIDP}, respectively. Section~\ref{DRP-IRP} introduces the proposed data transformation methods for simultaneous direct and indirect discrimination prevention. We describe our algorithms and their computational cost based on the proposed direct and indirect discrimination prevention methods in Section~\ref{sec:ALGSMETHOD} and Section~\ref{sec:time}, respectively.
Section~\ref{sec:results} shows the tests
we have performed to assess the validity and quality of our
proposal and we compare different methods.
Finally, Section~\ref{sec:conclusions} summarizes conclusions. 



In Chapter ~\ref{chap:DPAFPD}, we propose the first approach for achieving simultaneous discrimination and privacy awareness in frequent pattern discovery. 
Section~\ref{sec:INTROFPD} presents the motivation
example (Section~\ref{sec:MOTIVEXAMPLE1}) and our contributions (Section~\ref{sec:CONSFPD}) in frequent pattern discovery.
Section \ref{sec:privacy} presents the method that we use for obtaining
an anonymous version of an original pattern set. Section \ref{sec:dispatterns} describes the notion of discrimination protected (Section~\ref{sec:DPFP}) and unexplainable discrimination protected (Section~\ref{sec:UNECPDISPRTOCFPS}) frequent patterns. Then, our proposed methods and algorithms to obtain these pattern sets are presented in Sections~\ref{sec:nDG} and \ref{sec:AUNEXPDISCPROTPS},  respectively. In Section \ref{sec:problemsol}, we formally define the problem of simultaneous privacy and anti-discrimination
pattern protection, and we introduce our solution. Section \ref{sec:experiments} reports the evaluation of our sanitization methods.
In Section \ref{sec:DPdiscussion}, we study and discuss the use of a privacy protection method based on differential privacy and its implications. Finally, Section \ref{sec:conclusionfpd} concludes the chapter.

In Chapter \ref{chap:study}, we present a study on the impact of well-known  data anonymization techniques on anti-discrimination. 
Our proposal for releasing discrimination-free version of original data is presented in Section \ref{sec:NDDT}. In Section \ref{sec:privacydisc}, we study the impact of different generalization and suppression schemes on discrimination prevention.
Finally, Section \ref{sec:conclusionstudy} summarizes conclusions.
In Chapter \ref{chap:GBPPDPDM}, we present a generalization-based approach for privacy preservation and discrimination prevention in data publishing. 
Privacy and anti-discrimination models are presented in Section \ref{sec:privacymodel} and \ref{discriminationmodel}, respectively. In Section \ref{sec:problemdefinition}, we formally
define the problem of simultaneous privacy and anti-discrimination
data protection. Our proposed approach and an algorithm for discrimination-
and privacy-aware data publishing are presented in Sections \ref{sec:approach} and \ref{sec:algorithm}. Section \ref{sec:expri} reports experimental work. An extension of the approach to alternative privacy-preserving requirements and anti-discrimination legal constraints is presented in Section \ref{sec:alternative}. Finally, Section \ref{sec:conclusiongen} summarizes conclusions.



Finally, Chapter~\ref{conclusions} is the closing chapter.
It briefly recaps the thesis contributions, it lists the publications
that have resulted from our work, it states some conclusions
and it identifies open issues for future work.


\chapter{Background on Discrimination-aware Data Mining}
\label{chap:discrelatedwork}


In sociology, discrimination is the prejudicial treatment
of an individual based on their membership in a certain group or category.
It involves denying to members of one group
opportunities that are available to other groups. There is a list of
anti-discrimination acts, which are laws designed to prevent
discrimination on the basis of a number of attributes ({\em e.g.} race,
religion, gender, nationality, disability, marital status and age)
in various settings ({\em e.g.} employment and training, access to public
services, credit and insurance, etc.).
For example, the European Union implements 
in~\cite{eudisc2004}
the principle of
equal treatment between men and women in the access to and supply
of goods and services; also, it implements equal 
treatment in matters of
employment and occupation in~\cite{eudisc2006}.
Although there are some laws against discrimination, all of them
are reactive, not proactive. Technology can add proactivity
to legislation by contributing
discrimination discovery and prevention techniques.


\section{Related Work}
\label{sec:RWDISC}
The collection and analysis of observational and experimental data are the main tools for assessing
the presence, the extent, the nature, and the trend of discrimination phenomena. Data analysis techniques
have been proposed in the last fifty years in the economic, legal, statistical, and, recently, in data mining literature. This is not surprising, since discrimination analysis is a multi-disciplinary problem, involving
sociological causes, legal argumentations, economic models, statistical techniques, and computational issues.
For a multidisciplinary survey on discrimination analysis see \cite{Romeimulti}. In this thesis, we focus on a 
knowledge discovery (or data mining) perspective of discrimination analysis.

Recently, the issue of anti-discrimination has 
been considered from
a data mining perspective \cite{peder2008}, under the name of discrimination-aware data analysis.
A substantial part
of the existing literature on anti-discrimination in data mining is oriented
to {\em discovering} and {\em measuring}
discrimination.
Other contributions deal
with {\em preventing} discrimination. Summaries of contributions in discrimination-aware data analysis are collected in \cite{CaldersBook}.

\subsection{Discrimination Discovery from Data}

Unfortunately, the actual discovery of discriminatory situations and practices, hidden in a dataset of historical decision records, is an extremely difficult task. The reason is twofold:

\begin{itemize}

\item First, personal data in decision records are typically highly dimensional: as a consequence, a huge number of possible contexts may, or may not, be the theater for discrimination. To see this point, consider the case of gender discrimination in credit approval: although an analyst may observe that no discrimination occurs in general, it may turn out that older women obtain car loans only rarely. Many small or large niches  that conceal discrimination may exist, and therefore all possible specific situations should be considered as candidates, consisting of all possible combinations of variables and variable values: personal data, demographics, social, economic and cultural indicators, etc. The anti-discrimination analyst is thus faced with a combinatorial explosion of possibilities, which make her work hard: albeit the task of checking some known suspicious situations can be conducted using available statistical methods and known stigmatized groups, the task of discovering niches of discrimination in the data is unsupported.

\item The second source of complexity is \emph{indirect discrimination}: the feature that may be the object of discrimination, {\em e.g.}, the race or ethnicity, is not directly recorded in the data. Nevertheless, racial discrimination may be hidden in the data, for instance in the case where a \emph{redlining practice} is adopted: people living in a certain neighborhood are frequently denied credit, but from demographic data we can learn that most people living in that neighborhood belong to the same ethnic minority. Once again, the anti-discrimination analyst is faced with a large space of possibly discriminatory situations: all interesting discriminatory situations that emerge from the data, both directly and in combination with further background knowledge need to be discovered ({\em e.g.}, census data).

\end{itemize}

Pedreschi {\em et al.} \cite{peder2008,pederruggi2009,ruggi2010,pedreschiICAIL} have introduced the first data mining approaches for discrimination discovery. The approaches have followed the legal principle
of {\em under-representation} to unveil
contexts of possible discrimination against {\em protected-by-law}
groups ({\em e.g.}, women). This is done by
extracting classification rules from a dataset of historical
decision records (inductive part); then, rules are ranked according
to some {\em legally grounded} measures of discrimination (deductive part). The approach has been implemented on top of an Oracle database \cite{rugipedre2010} by relying on
tools for frequent itemset mining. A GUI for visual exploratory analysis has been developed by Gao and Berendt in \cite{Gao2011}.

This discrimination discovery approach opens a promising avenue for research, based on an apparently paradoxical idea: data mining, that has a clear potential to create discriminatory profiles and classifications, can also be used the other way round, as a powerful aid to the anti-discrimination analyst, capable of automatically discovering the patterns of discrimination that emerge from the available data with strongest evidence.

The result of the above knowledge
discovery process is a (possibly large) set of classification rules, which provide local and overlapping
niches of possible discrimination: a global description is lacking of who is and is not discriminated against. Luong et al. \cite{Ruggieri2011} exploit the idea of \emph{situation-testing}. For each member of the protected group with a
negative decision outcome, testers with similar characteristics are searched for in a dataset of historical decision records. If there are significantly different decision outcomes between the testers of the protected group and the testers of the unprotected group, the negative decision can be ascribed to a bias against the 
protected group, thus labeling the individual as discriminated against. Similarity is modeled via a distance
function. Testers are searched for among the $k$-nearest neighbors, and the difference is measured by some legally grounded measures of discrimination calculated over the two sets of testers. After this kind of labeling, a global description of those labeled as discriminated against can be extracted as a standard classification task. A
real case study in the context of the evaluation of scientific projects for funding is presented by Romei et al. \cite{Romei2012}.

The approaches described so far assume that the dataset under analysis contains attributes that denote
protected groups ({\em i.e.}, case of \emph{direct discrimination}). This may not be the case when such attributes are not available, or not even collectable
at a micro-data level ({\em i.e.}, case of indirect discrimination), as in the case of the loan applicant's race. Ruggieri et al. \cite{pedreschiICAIL,ruggi2010} adopt a form of
rule inference to cope with the indirect discovery of discrimination. 
The correlation information is called background knowledge, and is itself coded as an
{\em association rule}.

The above results do not yet explain how to build a discrimination-free knowledge discovery and deployment (KDD) technology for decision making. This is crucial as we are increasingly surrounded by 
automatic decision making software that makes decisions on people based on profiles and categorizations. We need novel, disruptive technologies for the construction of human knowledge discovery systems that, {\em by design}, offer native technological safeguards against discrimination. Here we evoke the concept of “Privacy by Design” coined in the ‟90s by Ann Cavoukian, the Information and Privacy Commissioner of Ontario, Canada. In brief, Privacy by Design refers to the philosophy and approach of embedding privacy into the design, operation and management of information processing technologies and systems.

In different contexts, adopting different techniques for inscribing discrimination protection within the KDD process will be needed, in order to go beyond the discovery of unfair discrimination, and achieve the much more challenging goal of {\em preventing} discrimination, {\em before} it takes place.

\subsection{Discrimination Prevention in Data Mining}

With the advent of data mining, decision support systems become increasingly intelligent and versatile, since effective decision models can be constructed on the basis of historical decision records by means of machine learning and data mining methods, up to the point that decision making is sometimes fully automated, {\em e.g.}, in credit scoring procedures and in credit card fraud detection systems. However, there is no guarantee that the deployment of the extracted knowledge does not incur discrimination against minorities and disadvantaged groups, {\em e.g.}, because the data from which the knowledge is extracted contain patterns with implicit discriminatory bias.
Such patterns will then be replicated in the decision rules derived from the data by mining and learning algorithms. Hence, learning from historical data may
lead to the discovery of traditional prejudices that are endemic in reality, and to assigning the status of general rules to such practices (maybe unconsciously, as these rules can end up deeply hidden within a piece of software).

%
%
%
%
%

Thus, beyond discrimination
discovery, preventing knowledge-based decision support systems
from making discriminatory decisions
is a more challenging issue. A straightforward approach to avoid that the classifier's prediction be based on the discriminatory attribute would be to remove that attribute from the training dataset. This approach, however, does not work \cite{KamiKAIS,ChaldersChap}. The reason is that there may be other attributes that are highly correlated with the discriminatory one. In such a situation the classifier will use these correlated attributes to indirectly discriminate. 
 In the banking example, {\em e.g.}, postal code may be highly correlated with ethnicity. Removing ethnicity would not solve much, as postal code is an excellent predictor for this attribute. Obviously, one could decide to also remove the highly correlated attributes from the dataset as well. Although this would resolve the discrimination problem, in this process much useful information will get lost, leading to suboptimal predictors \cite{KamiKAIS,ChaldersChap}. Hence, there are two important challenges regarding discrimination prevention: one challenge is to consider
both direct and indirect discrimination instead of only
direct discrimination; the other challenge is to find
a good trade off between discrimination removal and the
utility of the data/models for data mining.
In such a context, the challenging problem of discrimination prevention consists of re-designing existing data publishing and knowledge discovery techniques in order to incorporate a legally grounded notion of non-discrimination in the extracted knowledge, with the objective that the deployment phase leads to non-discriminatory decisions. Up to now, four non mutually-exclusive strategies have been proposed to prevent discrimination in the data mining and knowledge discovery process.

The first strategy consists of a controlled distortion of the training set (a pre-processing approach). Kamiran and Calders \cite{KamiKAIS} compare sanitization techniques such as changing class labels based on prediction confidence, instance re-weighting, and sampling. Zliobaitye et al. \cite{Calders2011} prevent excessive sanitization by taking into account legitimate explanatory variables that are correlated with grounds of
discrimination, {\em i.e.}, \emph{genuine occupational requirement}. The approach of Luong et al. \cite{Ruggieri2011} extends to
discrimination prevention by changing the class label of individuals that are labeled as discriminated. The advantage of the pre-processing
approach is that it does not require changing the standard data mining algorithms, unlike the in-processing approach, and it allows 
data publishing (rather than just knowledge publishing), unlike 
the post-processing approach.

The second strategy is to modify the classification learning algorithm (an in-processing approach), by integrating it with anti-discrimination criteria. Calders and Verwer \cite{Calders2010} consider three approaches to
deal with naive Bayes models, two of which consist in modifying the learning algorithm: training a separate model for each protected group; and adding a latent variable to model the class value in the
absence of discrimination. Kamiran et al. \cite{CaldersICDM} modify the entropy-based splitting criterion in decision tree induction to account for attributes denoting protected groups. Kamishima et al. \cite{Kamishima2012} measure the indirect causal effect of variables modeling grounds of discrimination on the independent variable in a classification model by their mutual information. Then, 
they apply a regularization ({\em i.e.}, a change in the
objective minimization function) to probabilistic discriminative models, such as logistic regression.

The third strategy is to post-process the classification model once it has been extracted. Pedreschi et al. \cite{pederruggi2009} alter the confidence of classification rules inferred by the CPAR algorithm. Calders and Verwer
\cite{Calders2010} act on the probabilities of a naive Bayes model. Kamiran et al. \cite{CaldersICDM} re-label the class predicted
at the leaves of a decision tree induced by C4.5.
Finally, the fourth strategy assumes no change in the construction of a classifier. At the time of application, instead, predictions are corrected to keep proportionality of decisions among protected and
unprotected groups. Kamiran et al. \cite{Kamiran2012} propose correcting predictions of probabilistic classifiers that are close to the decision boundary, given that (statistical) discrimination may occur when there is no clear feature supporting a positive or a negative decision.

Moreover, on the relationship between privacy and anti-discrimination from legal perspective, the chapter by Gellert et al. \cite{GellertChap}
reports a comparative analysis of data protection and anti-discrimination legislations. And from the technology perspective, Dwork et al. \cite{Dwork2012}
propose a model of fairness of classifiers and relate it to differential privacy in databases. The model
imposes that the predictions over two similar cases be also similar. The similarity of cases is formalized
by a distance measure between tuples. The similarity of predictions is formalized by the distance between
the distributions of probability assigned to class values.

\section{Preliminaries}
\label{sec:preliminaries}

In this section, we briefly review the background knowledge
required in the remainder of this thesis. First, we recall some
basic definitions related to data mining \cite{Tan2006}. After that, we
elaborate on measuring and discovering discrimination.

\subsection{Basic Definitions}
\label{sec:BDDISC}

Let $\mathcal{I} = \{i_1, \ldots, i_n \}$ be a set of items,
where each {\em item} $i_j$ has the form {\em attribute=value}
({\em e.g.}, \emph{Sex=female}). An {\em itemset} $X \subseteq  \mathcal{I}$ is a collection of one or more items,
{\em e.g.} \emph{\{Sex=female, Credit\_history=no-taken\}}.
A {\em database} is a collection of data objects (records)
and their attributes; more formally, a
(transaction) database $\mathcal{D}= \{r_1, \ldots, r_m\}$ is a set of
data records or transactions where each $r_i \subseteq \mathcal{I}$.
Civil rights laws \cite{AustralianLeg,Europeandisc} explicitly identify the groups to be protected against discrimination, such as
minorities and disadvantaged people, {\em e.g.}, women. In our context, these groups can be represented as items, {\em e.g.}, \emph{Sex=female}, which we call potentially discriminatory (PD) items; a collection of PD items can be represented as an itemset, {\em e.g.}, \emph{\{Sex=female, Foreign\_worker=yes\}}, which we call PD itemset or protected-by-law (or protected for short)
groups, denoted by $DI_b$. An itemset $X$ is potentially non-discriminatory (PND) if $X \cap DI_b = \emptyset$, {\em e.g.}, \emph{\{credit\_history=no-taken\}} is a PND itemset where $DI_b$:\emph{\{Sex=female\}}.
{\em PD attributes} are those that can take PD items as values; for instance,
\emph{Race} and \emph{Gender} where $DI_b$:\emph{\{Sex=female, Race=black\}}. 
A {\em decision (class) attribute} is one
taking as values {\em yes} or {\em no} to report
		the outcome of a decision made on an individual;
		an example is the attribute {\em credit\_approved}, which can be
		{\em yes} or {\em no}. A {\em class item} is an item of class attribute, {\em e.g.}, \emph{Credit\_approved=no}.
The {\em support} of an itemset $X$ in a database $\mathcal{D}$
is the number of records that contain $X$, {\em i.e.}
$supp_\mathcal{D}(X) = |\{r_i \in \mathcal{D} | X \subseteq r_i\}|$, where $|\ .\ |$ is the cardinality operator. 
From patterns, it is possible to derive association rules. An {\em association rule} is an expression $X\rightarrow Y$, where X and Y are itemsets.  We say that $X\rightarrow Y$ is a {\em classification rule} if   $Y$ is a class item and $X$ is an itemset containing no class item, {\em e.g.}
{\em Sex=female,
Cedit\_history=no-taken $\rightarrow$  Credit\_approved=no}.  The itemset $X$ is
called the premise of the rule. We say that a rule $X \rightarrow C$
is {\em completely supported} by a record if both $X$ and $C$ appear in
the record.
The {\em confidence} of a classification rule, $conf_\mathcal{D}(X\rightarrow C)$, measures how often the class item $C$ appears in records that contain $X$. Hence, if $supp_\mathcal{D}(X)> 0$ then
\begin{equation}
\label{eq:conf}
conf_\mathcal{D}(X\rightarrow C)=\frac{supp_\mathcal{D}(X, C)}{supp_\mathcal{D}(X)}
\end{equation}
Confidence ranges over $[0, 1]$. We omit the subscripts in $supp_\mathcal{D}(\cdot)$	 and $conf_\mathcal{D}(\cdot)$ when there is no ambiguity.
A {\em frequent classification rule} is a classification rule with
support and confidence greater than respective specified
lower bounds. 
The {\em negated itemset, i.e. $\neg X$} is an itemset with
the same attributes as $X$, but the attributes in $\neg X$
take any value except those taken by attributes in $X$.
In this chapter, we use the $\neg$ notation for itemsets with binary or
non-binary categorical attributes. For a binary attribute, {\em e.g.}
\{Foreign worker=Yes/No\}, if $X$ is \{Foreign worker=Yes\},
then $\neg X$ is \{Foreign worker=No\}. If $X$ is binary,
it can be converted to $\neg X$ and vice versa, that is,
the negation works in both senses.
In the previous example,
we can select the records in $\mathcal{DB}$ so that the
value of the Foreign worker attribute is ``Yes'' and change that
attribute's value to ``No'', and conversely.
However, for a non-binary categorical attribute,
{\em e.g.} \{Race=Black/White/Indian\}, if
$X$ is \{Race=Black\}, then
$\neg X$ is \{Race=White\} or \{Race=Indian\}.
In this case, $\neg X$ can be converted to $X$ without ambiguity,
but the conversion of $X$ into $\neg X$ is not uniquely defined.
In the previous example,
we can select the records in $\mathcal{DB}$ such that the
Race attribute is ``White'' or ``Indian'' and change that attribute's value
to ``Black'';  but if we want to negate \{Race=Black\}, we do not know
whether to change it to \{Race=White\} or \{Race=Indian\}.
In this thesis, we use only non-ambiguous negations.
    
\subsection{Measures of Discrimination}
\label{sec:discmeasures}
The legal principle of under-representation has inspired existing approaches for discrimination discovery based on rule/pattern mining.

Given $DI_b$ and starting from a dataset $\mathcal{D}$ of historical decision records, the authors of \cite{peder2008} propose to extract frequent classification rules of the form $A,B\rightarrow C$, called PD rules, to unveil contexts $B$ of possible discrimination, where the non-empty protected group $A \subseteq DI_b$ suffers from
over-representation with respect to the {\em negative} decision $C$
($C$ is a class item reporting a negative decision, such as
credit denial, application rejection,
job firing, and so on). In other words, $A$
is under-represented w.r.t. the corresponding positive decision $\neg C$.
As an example, rule {\em Sex=female, Job=veterinarian $\rightarrow$ Credit\_approved=no} is a PD rule about denying credit (the decision $C$) to women (the protected group $A$) among those who are veterinarians (the context B), where $DI_b$:\emph{\{Sex=female\}}. And a classification rule of the form $X \rightarrow C$ is called PND rule if $X$ is a PND itemset. As an example, rule {\em Credit\_history=paid-delay, Job=veterinarian $\rightarrow$ Credit\_approved=no} is a PND rule, where $DI_b$:\emph{\{Sex=female\}}.

Then, the degree of under-representation should be measured over each PD rule by one of the {\em legally grounded} measures
introduced in Pedreschi {\em et al.} \cite{pederruggi2009}.
\begin{definition}
\label{def:slift}
Let $A,B\rightarrow C$ be a PD classification rule extracted from $\mathcal{D}$ with $conf(\neg A, B\rightarrow C)>0$. The selection lift\footnote{Discrimination on the basis of an
attribute happens
if a person with an attribute is treated
less favorably than a person without the attribute.} (slift) of the rule is
\begin{equation}
\label{eq:slift}
slift(A,B\rightarrow C)=\frac{conf(A,B\rightarrow C)}{conf(\neg A, B\rightarrow C)}
\end{equation}
\end{definition}
In fact, $slift$ is the ratio
of the proportions of benefit denial, {\em e.g.}, credit denial, between the protected and unprotected groups, {\em e.g.} women and men resp., in the given context, {\em e.g.} new applicants.
A special case of  {\em slift} occurs when we deal with non-binary
attributes, for instance, when comparing the credit denial
ratio of blacks with the ratio for other groups of the population.
This yields a third measure called {\em contrasted lift (clift)} which,
given $A$ as a single item $a = v_1$ ({\em e.g.} black race), compares
it with the most favored item $a = v_2$ ({\em e.g.} white race).
\begin{definition}
\label{def:clift}
Let $a=v_1,B\rightarrow C$ be a PD classification rule extracted from $\mathcal{D}$, and $v_2 \in dom(a)$ with
$conf(a=v_2, B\rightarrow C)$ minimal and non-zero. The {\em contrasted lift  (clift)} of the
rule is
\begin{equation}
\label{eq:clift}
clift(a=v_1, B \rightarrow C)=\frac{conf(a=v_1,B\rightarrow C)}{conf(a=v_2, B\rightarrow C)}
\end{equation}
\end{definition}
%
%
%
%
%
%
%
\begin{definition}
\label{def:elift}
Let $A,B\rightarrow C$ be a PD classification rule extracted from $\mathcal{D}$ with
$conf(B\rightarrow C)>0$. The {\em extended lift \footnote{Discrimination occurs when a higher proportion of people not in the group is able to comply.} (elift)} of the
rule is
\begin{equation}
\label{eq:elift}
elift(A, B \rightarrow C)=\frac{conf(A,B\rightarrow C)}{conf(B\rightarrow C)}
\end{equation}
\end{definition}
In fact, {\em elift}
is the ratio
of the proportions of benefit denial, {\em e.g.} credit denial, between the protected groups and all people who were not granted the benefit in the given context, {\em e.g.}
women versus all men and women who
were denied credit, in the given context, {\em e.g.} those who live in NYC.

The last ratio measure is the {\em odds lift (olift)}, the ratio between the odds
of the proportions of benefit denial between the protected and
unprotected groups.

\begin{definition}
\label{def:olift}

Let $A,B\rightarrow C$ be a PD classification rule extracted from $\mathcal{D}$ with
$conf(\neg A, B\rightarrow C)>0$ and $conf(A, B \rightarrow C) <1$. The {\em odds lift  (olift)} of the
rule is
\begin{equation}
olift(A, B \rightarrow C)=\frac{odds(A,B\rightarrow C)}{odds(\neg A, B\rightarrow C)}
\end{equation}

where

\begin{equation}
odds(A, B \rightarrow C)=\frac{conf(A,B\rightarrow C)}{conf(A, B\rightarrow \neg C)}
\end{equation}

\end{definition}

Although the measures
introduced so far are defined in terms of ratios, measures based on the difference of confidences have been
considered on the legal side as well.
\begin{definition}
\label{def:dlift}
Let $A,B\rightarrow C$ be a PD classification rule extracted from $\mathcal{D}$.
The {\em difference measures} are defined as
\begin{equation}
\label{eq:sliftd}
slift_d(A, B \rightarrow C)=conf(A,B\rightarrow C) - conf(\neg A, B \rightarrow C)
\end{equation}
\begin{equation}
elift_d(A, B \rightarrow C)=conf(A,B\rightarrow C) - conf(B \rightarrow C)
\end{equation}
\end{definition}
Difference-based measures range over $[-1, 1]$.
Lastly, the following measures are also defined in terms of ratios and known as chance measures.
\begin{definition}
\label{def:chancelift}
Let $A,B\rightarrow C$ be a PD classification rule extracted from $\mathcal{D}$.The {\em chance measures} are defined as
\begin{equation}
\label{eq:sliftc}
slift_c(A, B \rightarrow C)=\frac{1-conf(A,B\rightarrow C) }{1-conf(\neg A, B \rightarrow C)}
\end{equation}
\begin{equation}
elift_c(A, B \rightarrow C)=\frac{1-conf(A,B\rightarrow C)}{1-conf(B \rightarrow C)}
\end{equation}
\end{definition}
For $slift$, $elift$ and $olift$, the values of interest
(potentially indicating discrimination)
are those greater than 1; for $slift_d$ and $elift_d$, they are those greater than 0; and for $slift_c$ and $elift_c$, they are those less than 1.
On the legal side, different measures are adopted worldwide.
For example, UK law mentions mostly $slift_d$.
The EU court of justice has made more emphasis in $slift$, and
US laws courts mainly refer to $slift_c$.

\begin{figure}[ht]
\centering
Classification rule: $c=A,B\rightarrow C$
\begin{displaymath}
\begin{array}{r|c|c|c}
B & C & \neg C \\
\hline
A & a_1 & n_1-a_1 & n_1\\
\neg A & a_2 & n_2 - a_2 & n_2
\end{array}
\end{displaymath}
\[p_1=a_1/n_1 \quad p_2=a_2/n_2 \quad p=(a_1+a_2)/(n_1+n_2)\]
\[elift(c)=\frac{p_1}{p},\quad slift(c)=\frac{p_1}{p_2},\quad olift(c)=\frac{p_1(1-p_2)}{p_2(1-p_1)}\]
\[elift_d(c)=p_1-p, \quad  slift_d(c)=p_1-p_2\]
\[elift_c(c)=\frac{1-p_1}{1-p}, \quad  slift_c(c)=\frac{1-p_1}{1-p_2}\]
\caption{Discrimination measures}
\label{fig:Measures}
\end{figure}

An alternative view of the measures introduced so far can be given starting from the contingency table of $c: A, B \rightarrow C$ shown in Fig. \ref{fig:Measures}. Each cell in the table is filled
in with the number of records in the data table $\mathcal{D}$ satisfying B and the coordinates ({\em i.e.}, their absolute support). Using the notation of the figure, confidence of $c: A, B \rightarrow C$ is $p_1=a_1/n_1$. Similarly, other measures can be
defined as shown in Fig. \ref{fig:Measures}. Confidence intervals and tests of statistical significant of the above measures are discussed in \cite{pederruggi2009}. Here, we only mention that statistical tests will rank the rules according to how unlikely it is that they would be observed if there was equal treatment, not according to the severity of discrimination. The rankings imposed
by the discrimination measures in Fig. \ref{fig:Measures} are investigated by Pedreschi et al. \cite{pedreSAC}: the choice of the
reference measure critically affects the rankings of PD rules, with the $slift_c$ and the $slift$ measures exhibiting the largest differences.

 \chapter{Background on Privacy-aware Data Mining}
\label{chap:privrelatedwork}


Privacy protection is a basic right, stated
in Article 12 of the Universal Declaration of Human Rights. 
It is also an important concern
in today's digital world. Data security and privacy are two concepts that  are often used in
conjunction; however, they represent two different facets of data protection and various techniques have
been developed for them \cite{Giannotti2006}. Privacy is not just a goal or service like security, but it is the people's
expectation to reach a protected and controllable situation, possibly without having to actively look for it by
themselves. Therefore, privacy is defined as "the rights of individuals to determine for themselves when,
how, and what information about them is used for different purposes" \cite{Agrawal2002}. In information technology, the protection of sensitive data is a crucial issue,
which has attracted many researchers. In knowledge discovery, efforts at guaranteeing privacy when mining and sharing personal data have led to developing privacy preserving data mining (PPDM)  techniques. PPDM have become
increasingly popular because they allow publishing and sharing sensitive data for secondary analysis. Different PPDM methods and models (measures) have been proposed to trade off the
utility of the resulting data/models for protecting individual privacy against different kinds of privacy
attacks. 

\section{Brief Review}
\label{sec:BRPRIV}

The problem of protecting privacy within data mining
has been extensively studied
since the 1970s, when Dalenius was the first to formulate
the statistical disclosure control problem~\cite{Dale74}.
Research on data anonymization has carried on ever since
in the official statistics community,
and several computational
procedures were proposed during the 1980s and 1990s,
based on random noise addition,
generalization, suppression, microaggregation, bucketization,
etc. (see~\cite{Hund12,PPDPBOOK2010} for a compendium).
In that literature, the approach was first to anonymize
and then measure how much anonymity had been achieved,
by either computing the probability of re-identification or
performing record linkage experiments.
In the late 1990s, researchers in the database community
stated the $k$-anonymity model~\cite{Samarati,Sweeney}:
a data set is $k$-anonymous if its records are indistinguishable
by an intruder within groups of $k$. The novelty of this approach
was that the anonymity target was established {\em ex ante}
and then computational
procedures were used to reach that target.
The computational procedures initially proposed for {\em $k$-anonymity}
were generalization and suppression; microaggregation was proposed
later as a natural alternative~\cite{Domi05}.
In 2000, the database community re-discovered anonymization via
random noise addition, proposed in the statistical community as far
back as 1986~\cite{kim86}, and coined the new term
privacy-preserving data mining (PPDM,\cite{agra00,lind02}).
{\em Differential privacy} \cite{Dwork2006} is a more recent
anonymity model that holds much promise: it seeks to render
the influence of the presence/absence of any individual on
the released outcome negligible. The computational
approach initially proposed to achieve differential privacy
was Laplace noise addition, although other approaches have
recently been proposed~\cite{Soria12}. SDC and PPDM have become increasingly popular because they allow
publishing and sharing
sensitive data for secondary analysis. Detailed descriptions of different PPDM
models and methods can be found in \cite{PPDMBOOK2008,PPDPBOOK2010,Giannotti2006}.

\section{Preliminaries}

In this section, we briefly review the background knowledge
required in the remainder of this thesis from data privacy technologies. First, we recall some
basic definitions. After that, we
elaborate on privacy measures (models) and samples of anonymization techniques.

\subsection{Basic Definitions}
\label{sec:BDPRIV}

Given the data table $\mathcal{D} (A_1,\cdots, A_n)$, a set of attributes $\mathcal{A}=\{A_1,\cdots, A_n\}$, and a record/tuple $t \in \mathcal{D}$, $t[A_i,\cdots,A_j]$ denotes the sequence of the values of $A_i,\cdots,A_j$ in $t$, where $\{A_i,\cdots, A_j\} \subseteq \{A_1,\cdots, A_n\}$. Let $\mathcal{D}[A_i,\cdots,A_j]$ be the projection, maintaining duplicate records, of attributes $A_i,\cdots,A_j$ in $\mathcal{D}$. Let $|\mathcal{D}|$ be the cardinality of $\mathcal{D}$,
that is, the number of records it contains.
The attributes $\mathcal{A}$ in a database $\mathcal{D}$ can be classified
into several categories. {\em Identifiers}
are attributes that uniquely identify individuals in the database,
like {\em Passport number}.
A {\em quasi-identifier} (QI)
is a set of attributes
that, in combination, can be linked to external
		identified information
		for re-identifying  						
		an individual;
		for example, \emph{Zipcode}, \emph{Birthdate} and
		\emph{Gender}.
{\em Sensitive attributes} (S) are those that
		contain sensitive information, such as \emph{Disease}
		or \emph{Salary}. Let $S$ be a set of sensitive attributes in $\mathcal{D}$.

\subsection{Models of Privacy}
\label{sec:privmeasures}
 
As mentioned in the beginning of this chapter, in the last
fifteen years plenty of privacy models 
have been proposed  to trade off the utility of the resulting data/models for protecting individual privacy against different kinds of privacy attacks. Defining privacy is a difficult task. One of the key challenges is how to model the background knowledge of an adversary.
Simply removing explicit identifiers ({\em e.g.}, name, passport number)
does not preserve privacy, given that the adversary has some
background knowledge about the victim. Sweeney \cite{Sweeney} illustrates that 87\% of the U.S. population can be uniquely identified
based on 5-digit zip code, gender, and date of birth.
These attributes are QI and the adversary may know these values from publicly available sources such as a voter list. An individual can be identified from published data by simply joining the QI attributes
with an external data source ({\em i.e.}, record linkage). 

In order to prevent record linkage attacks between the released data and external identified data sources through quasi-identifiers,
Samarati and
Sweeney \cite{Samarati1998,SWEENEY1998} proposed the
notion of $k$-anonymity.

\begin{definition}[$k$-anonymity]
Let $\mathcal{D}(A_1, \cdots, A_n)$ be a data table and
$QI=\{Q_1,\cdots,$\allowbreak$Q_m\} \subseteq \{A_1,\cdots, A_n\}$
be a quasi-identifier.
$\mathcal{D}$ is said to satisfy $k$-anonymity w.r.t. $QI$
if each combination of values of attributes in $QI$ is shared by at least $k$
tuples (records) in $\mathcal{D}$.
\end{definition}
   
Consequently, the probability of
linking a victim to a specific record through QI is at most $1/k$.
A data table satisfying this requirement is called $k$-anonymous.
Other privacy measures to prevent record linkage include $(X,Y)$-anonymity \cite{WangFung} and multi-relational $k$-anonymity \cite{Nergiz2007}.

$k$-Anonymity can protect the original data against record linkage attacks, but it cannot protect the data
against attribute linkage (disclosure).  In the attack of attribute linkage, the attacker may not precisely identify the record of
the specific individual, but could infer his/her sensitive values ({\em e.g.}, salary, disease) from the published data table $\mathcal{D}$. 
Some models have been proposed to address this type of threat. The most popular ones are $l$-diversity \cite{Machanavajjhala2006} and $t$-closeness \cite{Li2007}. The general idea of these models is to diminish the correlation between QI and sensitive attributes.

$l$-Diversity requires at least $l$ distinct values for the
sensitive attribute in each group of QI. Let $q^*$-block be the set of records in $\mathcal{D}$ whose QI attribute values generalize to $q$.

\begin{definition}[$l$-diversity]
\label{def:ldiversity}
A $q^*$-block is $l$-diverse 
if it contains at least $l$ “well-represented” values for the sensitive attribute $S$. A data table $\mathcal{D}$ is $l$-diverse if every $q^*$-block is $l$-diverse.
\end{definition}

$t$-Closeness requires the distribution of a sensitive
attribute in any group on QI to be close to the distribution of
the attribute in the overall table.

\begin{definition}[$t$-closeness]
\label{def:tcloseness}
A $q^*$-block is said to have $t$-closeness if the distance between the distribution of a sensitive attribute in
this $q^*$-block and the distribution of the attribute in the whole table is no more than a threshold $t$. A data table $\mathcal{D}$ is said to have $t$-closeness
if all $q^*$-blocks have $t$-closeness.
\end{definition}

Other privacy models for attribute disclosure protection include  $(\alpha, k)$-anonymity \cite{Wong2006}, $(k, e)$-anonymity \cite{Koudas2007}, $(c, k)$-safety \cite{Martin2007}, privacy skyline \cite{Chen2007}, $m$-confidentiality \cite{Wong2007} and $(\epsilon,m)$-anonymity \cite{JLi2008}. 


Differential privacy is a privacy model that provides
a worst-case privacy guarantee in the presence of arbitrary external information.
It protects against any privacy breaches resulting from joining
different databases.
It guarantees that an adversary learns nothing
about an individual, regardless of whether
the individual's record
is present or absent in the data.
Informally, differential privacy \cite{Dwork2006} requires that the output of
a data analysis mechanism be approximately the same, even
if any single record in the input database is arbitrarily added
or removed.

\begin{definition}[Differential privacy]
\label{def:diffpriv}
A randomized algorithm $\mathcal{ALG}$ is $\epsilon$-differentially private if for all datasets $\mathcal{D}$, $\mathcal{D}'$
that differ in one individual ({\em i.e.} data of
one person), and for all $S \subseteq Range(\mathcal{ALG})$, it
holds that
$Pr[\mathcal{ALG}(\mathcal{D})\in S] \leq e^\epsilon  Pr[\mathcal{ALG}(\mathcal{D}') \in S]$.
\end{definition}

There are alternative privacy measures which are extensions of $\epsilon$-differentially privacy including $(\epsilon, \delta)$-differentially
privacy \cite{DworkEURO} and the crowd-blending privacy model \cite{Gehrke2012}.  
\subsection{Sanitization Mechanisms}
\label{sec:samples}


Typically, the original data or data mining results do not satisfy a specified privacy
model and,
before being published, they must be modified through an anonymization method,
also called {\em sanitization mechanism} in the literature.

\subsubsection{Generalization and Suppression}

Samarati and Sweeney \cite{Samarati1998,Samarati,Sweeney} gave  methods for $k$-anonymization based on {\em generalization} and {\em suppression}.
Computational procedures alternative to generalization have
thereafter been proposed to attain $k$-anonymity, like
microaggregation~\cite{Domi05}. Nonetheless, generalization
remains not only the
main method
for $k$-anonymity, but it can also be used to satisfy other privacy models
({\em e.g.}, $l$-diversity in \cite{Machanavajjhala2006}, $t$-closeness in \cite{Li2007} and differential privacy in \cite{FungDiff}).

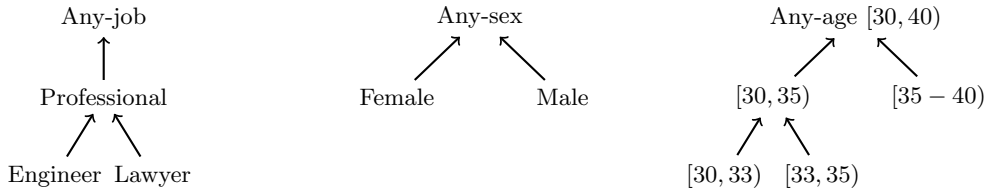
\begin{figure*}[t]
\begin{tikzpicture}
\footnotesize
\tikzstyle{level 1}=[sibling distance=22mm]
\tikzstyle{level 2}=[sibling distance=13mm]
\tikzset{level distance=30pt}

\node (1ar) at (6,0) {Any-job}
	child { node (1w) {Professional} edge from parent [thick,<-]
		child { node (1ww) {Engineer} edge from parent [thick,<-]}
            child { node (2ww) {Lawyer} edge from parent [thick,<-]}
};

	child { node (1c) {Artist} edge from parent [thick,<-]
		child { node (1b) {Dancer} edge from parent [thick,<-]}
		child { node (1a) {Writer} edge from parent [thick,<-]}
		};

\node (2as) at (11,0) {Any-sex}
	child { node (1w) {Female} edge from parent [thick,<-]}
	child { node (1c) {Male} edge from parent [thick,<-]
		};

\node (3ah) at (16,0) {Any-age $[30,40)$}
	child { node (3a) {$[30,35)$} edge from parent [thick,<-]
		child { node (3e){$[30,33)$} edge from parent [thick,<-]}
		child { node (3f) {$[33,35)$} edge from parent [thick,<-]}
		}
	child { node (3b) {$[35-40)$} edge from parent [thick,<-]
		};
\end{tikzpicture}

%
%
\caption{Generalization taxonomy tree for Sex, Job and Age attributes}
\label{fig:taxonomy}
\end{figure*}

A generalization replaces QI attribute values with a generalized version of them using the generalization taxonomy tree of QI attributes, {\em e.g.} Figure \ref{fig:taxonomy}. Five possible generalization schemes \cite{PPDPBOOK2010} are summarized below.


In {\em full-domain generalization}, all values in an attribute are generalized to the same level of the taxonomy tree. For example, consider Figure \ref{fig:taxonomy}; if {\em Lawyer} and {\em Engineer} are generalized to {\em Professional}, then it also requires generalizing {\em Dancer} and {\em Writer} to {\em Artist}.
In {\em subtree generalization}, at a nonleaf node, either all child values or none are generalized. For example, consider Figure \ref{fig:taxonomy}; if {\em Engineer} is generalized to {\em Professional}, it also requires generalizing {\em Lawyer} to {\em Professional}, but {\em Dancer} and {\em Writer} can remain ungeneralized.

{\em Sibling generalization} is similar to the subtree generalization, except for the fact that some siblings may remain ungeneralized. For example, consider Figure \ref{fig:taxonomy}; if {\em Engineer} is generalized to {\em Professional}, {\em Lawyer} can remain ungeneralized.
In all of the above schemes, if a value is generalized, all its instances are generalized. Such schemes are called {\em global recoding}. In {\em cell generalization}, also known as {\em local recoding}, some instances of a value may remain ungeneralized while other instances are generalized. For example, consider Figure \ref{fig:taxonomy}; {\em Female} in one record in a data table is generalized to {\em Any-sex}, while {\em Female} in another record can remain ungeneralized.
{\em Multidimensional generalization} flexibly allows two QI groups, even having the same value, to be independently generalized into different parent groups. For example, consider Figure \ref{fig:taxonomy}; $\langle${\em Engineer, Male}$\rangle$ can be generalized to $\langle${\em Engineer, Any-sex}$\rangle$ while $\langle${\em Engineer, Female}$\rangle$ can be generalized to $\langle${\em Professional, Female}$\rangle$.
Although algorithms using multi-dimensional or cell generalizations  cause less information loss than algorithms using full-domain generalization, the former suffer from the problem of data exploration \cite{PPDPBOOK2010}. This problem is caused
by the co-existence of specific and generalized values in
the generalized data set, which make data exploration and interpretation
difficult for the data analyst.


A suppression consists in suppressing some values of the QI attributes for some (or all) records. Three possible suppression schemes are {\em record suppression}, {\em value suppression} and {\em cell suppression}.
Record suppression refers to suppressing an entire record.
Value suppression refers to suppressing every instance of a given value in a table.
Cell suppression refers to suppressing some instances of a given value in a table.

\subsubsection{Laplacian and Exponential Mechanisms}

To satisfy $\epsilon$-differential privacy, several randomized mechanisms have been proposed. Here, we mention the ones which are mostly used in literature. The first approach is the {\em Laplacian mechanism}. It
computes a function $F$ on the dataset $\mathcal{D}$ in a differentially 
private way, by adding to $F(\mathcal{D})$ Laplace-distributed random noise.
The magnitude of the noise depends on the {\em sensitivity} $S_F$ of $F$:
$$S_F = \max_{(\mathcal{D}, \mathcal{D}')}|F(\mathcal{D}) - F(\mathcal{D}')|$$
where $(\mathcal{D}, \mathcal{D}')$ is any pair of datasets that
differ in one individual and belong to the domain of $F$.
Formally, the Laplacian mechanism
$\mathcal{ALG}_F$ can be written as
$$\mathcal{ALG}_F(\mathcal{D}) = F(\mathcal{D}) + Lap(\frac{S_F}{\epsilon})$$
where
$Lap(\beta)$ denotes a random variable sampled
from the Laplace distribution with scale parameter $\beta$.
The second approach is the {\em exponential mechanism} proposed by McSherry and Talwar in \cite{McSherry2007}, which can work on any kind of data.
It computes a function $F$ on a dataset $\mathcal{D}$ by sampling from the set of all possible outputs in the
range of $F$ according to an exponential distribution, with outputs
that are ``more accurate'' being sampled with higher
probability. This approach requires specifying
a utility function $u : \mathcal{D} \times \mathcal{R} \rightarrow \mathbb{R}$, where the real valued
score $u(\mathcal{D}, t)$ indicates how accurate it is to return $t$ when
the input dataset is $\mathcal{D}$. Higher scores mean better utility
outputs which should be returned with higher probabilities. For any function $u$, an algorithm $\mathcal{ALG}_F$ that chooses an output $t$ with probability proportional to $exp(\frac{\epsilon u(\mathcal{D},t)}{2S_u})$ satisfies $\epsilon$-differential privacy, where
$S_u$ is the sensitivity of utility function.

\section{Approaches and Algorithms}
\label{sec:ALGSPRIV}

As mentioned in Chapter~\ref{chap:discrelatedwork}, privacy by design refers to the philosophy and approach of embedding privacy into the design, operation and management of information processing technologies and systems. Privacy by design, in the research field of privacy preserving data mining and publishing, is a recent paradigm that promises a quality leap in the conflict between data protection and data utility. 

The development of a theory for privacy by design will be needed to adopt different techniques for inscribing privacy protection within the KDD process. As in the case of anti-discrimination, there are three non mutually-exclusive strategies to protect privacy in the KDD process, according to the phase of the data mining process in which they operate. 

The first strategy is a pre-processing approach. It sanitizes the original data before they are used in data mining tasks. The goal of this strategy can be the safe disclosure of transformed data under suitable formal safeguards ({\em e.g.}, $k$-anonymity) to protect sensitive information in the original data set, or the safe disclosure of transformed data to protect certain specified secret patterns, hidden in the original data. The PPDM methods which aim to achieve the first goal are known as {\em privacy preserving data publishing} (PPDP) and the ones which aim to achieve the second goal are known as knowledge hiding \cite{Verykios2008}. It is clear that these methods allow the execution of data mining algorithms by parties other than the data holder. 

The second strategy is an in-processing approach. It modifies the data mining algorithms by integrating them with privacy protection constraints. The goal of this strategy is the safe disclosure of mined models under suitable formal safeguards ({\em e.g.}, $k$-anonymity). It is due to the fact that the disclosure of data mining results themselves can violate individual privacy \cite{KantarciogluJC04}. The PPDM methods which aim to achieve this goal are known as privacy preserving knowledge publishing or privacy-aware data analysis. It is clear that these methods require the execution of data mining algorithms only by the data holder.

The third strategy is a post-processing approach. It takes into consideration the privacy model constraints after data mining
is complete by sanitizing (transforming) data mining results. Similar to the second strategy, the goal is the disclosure of mined models while protecting individual privacy. These methods are also known as privacy preserving knowledge publishing methods and require the execution of data mining algorithms only by the data holder. The difference between post-processing and in-processing is that the former does not require any modification of the mining process and therefore it can use any data mining tool available. However, the latter requires instead to redesign data mining algorithms and tools to directly enforce privacy protection criteria.

In this thesis, we concentrate on 
privacy preserving data publishing 
and privacy preserving knowledge publishing.

\subsection{Privacy Preserving Data Publishing}
\label{sec:PPDRelatedwork}
There are many works available on this problem
that focus on the privacy of the individuals whose data is collected in the database. 
They study the problem of how to transform the original data while satisfying the requirement of a particular privacy model.
Almost all privacy models have been studied in PPDP. Note that measuring data quality loss as a side effect of data distortion can be general or tailored to specific data mining tasks. 

\subsubsection{$k$-Anonymity and its Extensions}

The following algorithms adopt $k$-anonymity or its extensions as the underlying privacy principle to prevent record linkage attacks. These algorithms are minimal or optimal \cite{PPDPBOOK2010}. A data table is optimally anonymous if it satisfies the given privacy requirement and contains most information according to the chosen information metric among all satisfying tables. Sweeney's \cite{Sweeney} {\em MinGen} algorithm exhaustively examines all potential full-domain generalizations to identify the optimal generalization measured by a general information metric ({\em e.g.}, similarity between the original data and the anonymous data).
Samarati \cite{Samarati} proposed a {\em binary search} algorithm that first identifies all minimal generalizations, and then finds the optimal generalization measured by a general metric. Enumerating all minimal generalizations is an expensive operation, and hence not scalable for large data sets. LeFevre et al. \cite{Lefevre2005} presented a suite of optimal bottom-up generalization algorithms, called {\em Incognito}, to generate all possible $k$-anonymous full-domain generalizations.  Another algorithm called {\em $K$-Optimize} \cite{BAYARDO2005} effectively prunes non-optimal anonymous tables by modeling the search space using a set enumeration tree.  

The second family of algorithms produces a
minimal $k$-anonymous table by employing a greedy search guided by a search metric. A data table is minimally anonymous if it satisfies
the given privacy requirement and its sequence of anonymization 
operations cannot be reduced without violating the requirement.
Being heuristic in nature, these algorithms find a minimally anonymous solution, but are more scalable than optimal algorithms. The {\em $\mu$-Argus} algorithm \cite{Hund1996} computes the frequency of all 3-value combinations of domain values, then greedily applies subtree generalizations and cell suppressions to achieve $k$-anonymity. Since the method limits the size of attribute combination, the resulting data may not be $k$-anonymous when
more than 3 attributes are considered. Sweeney's \cite{SWEENEY1998} {\em Datafly} system was the first $k$-anonymization algorithm
scalable to handle real-life large data sets. It achieves $k$-anonymization by generating an array of QI group sizes and greedily generalizing those combinations with less than $k$ occurrences based on a heuristic search metric  that selects the attribute with the largest number of distinct values. Datafly employs full-domain generalization and
record suppression schemes. 
{\em Iyengar}~\cite{IYENGAR2002} was among 
the first to aim at preserving classification information
in $k$-anonymous data by employing a genetic algorithm with an incomplete stochastic search based on a classification metric and a subtree generalization scheme. To address the efficiency issue in $k$-anonymization, a {\em Bottom-Up Generalization} algorithm was proposed in Wang et al. \cite{Wang2004} to find
a minimal $k$-anonymization for classification. The algorithm starts from the original data that violate $k$-anonymity and greedily selects a generalization operation at each step according to a search metric. The generalization process is terminated as soon as all groups have the minimum size $k$. Wang et al. \cite{Wang2004} showed that this heuristic significantly reduces the search space. Instead of bottom-up, the {\em Top-Down Specialization} (TDS) method \cite{Fung2005} generalizes a table by specializing it from the most general state in which all values are generalized to the most general values of their taxonomy trees. At each step, TDS selects the specialization according to a search
metric. The specialization process terminates if no specialization can
be performed without violating $k$-anonymity. LeFevre et al. \cite{LEFEVRE2006} presented a greedy top-down
specialization algorithm {\em Mondrian} for finding a minimal $k$-anonymization in the case of the multidimensional generalization scheme. This algorithm is very similar to TDS. Xu et al. \cite{Xu} showed that employing cell generalization could further improve the data quality.
Although the multidimensional and cell generalization schemes cause less information loss, they suffer from the data exploration problem discussed in Section \ref{sec:samples}. 

\subsubsection{$l$-Diversity and its Extensions}

The following algorithms adopt $l$-diversity or its extensions  as the underlying privacy principle to prevent attribute disclosure.
 Though their privacy models are different from those of record linkage, many algorithms for attribute linkage are simple extensions
from algorithms for record linkage.
Machanavajjhala et al. \cite{Machanavajjhala2006} modified the bottom-up Incognito \cite{Lefevre2005} to identify optimal $l$-diverse full-domain generalizations with which original 
data are $l$-diverse; the modified algorithm is 
called {\em $l$-Diversity Incognito}. In other words, generalizations
help to achieve $l$-diversity, just as generalizations help to achieve $k$-anonymity. Therefore, $k$-anonymization algorithms that employ full-domain and subtree generalization
can also be extended into $l$-diversity algorithms.
LeFevre et al. \cite{LEFEVRE2006b} proposed a suite of greedy algorithms {\em InfoGain Mondrian} to identify a minimally anonymous table satisfying $k$-anonymity and/or entropy $l$-diversity with the consideration of a specific data analysis task such as classification modeling multiple target attributes and query answering with minimal imprecision. 

\subsubsection{$\epsilon$-Differential privacy and its Extensions}

Differential privacy has recently received much attention in data privacy, especially for interactive databases \cite{Dwork2011}. There are also some works available in literature studying the problem of differentially private data release. Rastogi et al. \cite{Rastogi} design the
$\alpha\beta$ algorithm for data perturbation that satisfies differential
privacy. Machanavajjhala et al. \cite{Machanavajjhala} apply the notion of differential privacy for synthetic data generation.
Barak et al. \cite{Barak2007} address the problem of releasing a set of 
consistent marginals of a contingency table. Their method
ensures that each count of the marginals is non-negative
and their sum is consistent for a set of marginals. Xiao et
al. \cite{Xiao} propose Privelet, a wavelet-transformation-based approach
that lowers the magnitude of noise needed to ensure
differential privacy to publish a multidimensional frequency
matrix. Hay et al. \cite{hay2010} propose a method to publish differentially private histograms for a one-dimensional data set.
Although Privelet and Hay et al.'s approach can achieve differential
privacy by adding polylogarithmic noise variance,
the latter is only limited to a one-dimensional data set. In~\cite{FungDiff}, a generalization-based algorithm
for differentially private data release is presented.
They show that differentially private data can be released by adding uncertainty in the generalization procedure.
It first probabilistically generates a generalized contingency table and then adds noise to the counts.

\subsection{Privacy Preserving Knowledge Publishing}
\label{sec:difffreq}
One key challenge in PPDM originates from the following
privacy question \cite{KantarciogluJC04}: do the data mining results themselves violate privacy? In other words, may
the disclosure of extracted patterns reveal sensitive information?
Some works on this problem are available, under the name of privacy-aware data analysis,
that focus on the privacy of the individuals whose data is collected in the database \cite{Friedman2008,AtzoriBGP08,Bhaskar2010,Friedman2010,Li2012,lee2012}. They study the problem of how to run a particular data mining algorithm on databases while satisfying the requirement of a particular privacy model.
Among the above-mentioned privacy models, $k$-anonymity and differential privacy have been studied in privacy-aware data analysis.
In \cite{Friedman2008} the problem of sanitizing decision trees is studied and
a method is given for directly building a $k$-anonymous decision
tree from a private data set. The proposed algorithm
is basically an improvement of the classical decision tree building algorithm,
combining mining and anonymization in a single process (in-processing approach). In \cite{AtzoriBGP08}
the anonymity problem is addressed
in the setting of frequent patterns. The authors define
the notion of $k$-anonymous patterns and propose a methodology to
guarantee the $k$-anonymity property in a collection
of  published frequent patterns (post-processing approach).
In \cite{Friedman2010}, an algorithm is proposed for building a classifier while guaranteeing differential privacy (in-processing approach).
In \cite{Bhaskar2010} and~\cite{Li2012}, post-processing and
in-processing approaches, respectively, are
used to perform frequent pattern mining in a transactional database while
satisfying differential privacy.


\chapter[A Methodology for Direct and Indirect Discrimination
Prevention in Data Mining]{A Methodology for Direct and Indirect Discrimination
Prevention in Data Mining}
\chaptermark{A Methodology for Direct and Indirect DPDM}
\label{chap:methodology}


Automated data collection and data mining techniques
such as classification rule mining have paved the way to making
automated decisions, like loan granting/denial,
insurance premium computation, etc. If the training datasets
are biased in what regards discriminatory attributes
like gender, race, religion, etc., discriminatory decisions may ensue.
Discrimination can be either direct or indirect.
Direct discrimination occurs when decisions are made based on discriminatory attributes. Indirect discrimination occurs when decisions are made
based on non-discriminatory attributes which are strongly correlated
with biased sensitive ones. In this chapter, we tackle discrimination
prevention in data mining and propose new techniques applicable
for direct or indirect discrimination prevention individually or
both at the same time. We discuss how to clean training datasets
and outsourced datasets in such a way that direct and/or indirect
discriminatory decision rules are converted to legitimate
(non-discriminatory) classification rules. We also propose
new metrics to evaluate the utility of the proposed approaches and we
compare these approaches.
The experimental evaluations demonstrate that the proposed
techniques are effective at removing direct and/or
indirect discrimination biases in the original dataset while
preserving data quality.

\section{Contributions}
\label{sec:CONTRIMETHO}

Discrimination prevention methods based on pre-processing
published so far \cite{KamiKAIS}
present some limitations, which we next highlight:
\begin{itemize}
\item They attempt to detect discrimination in the original data
only for one discriminatory item and based on a single measure.
This approach
cannot guarantee that the transformed dataset is really discrimination-free,
because it is known that discriminatory behaviors can often be hidden
behind several discriminatory items, and even behind combinations of them.
\item They only consider direct discrimination.
\item They do not include any measure to evaluate how much discrimination
has been removed and how much information loss has been incurred.
\end{itemize}

In this chapter, we propose pre-processing
methods which overcome the above limitations.
Our new data transformation methods ({\em i.e.} rule protection and
rule generalization) are based on measures for both direct and indirect
discrimination and can deal with several discriminatory items.
Also, we provide utility measures. Hence, our approach to discrimination
prevention is broader than in previous work.

In our earlier work~\cite{hajian2010}, we introduced the initial
idea of using rule protection and rule generalization for
direct discrimination prevention, but we gave no experimental results.
In ~\cite{hajian2011LNC}, we introduced the use of rule protection
in a different way for indirect discrimination prevention and
we gave some preliminary experimental results. In this thesis,
we present a {\em unified approach to direct and indirect
discrimination prevention}, with finalized algorithms and all possible data
transformation methods based on rule protection
and/or rule generalization
that could be applied for direct or indirect discrimination
prevention. We specify the different features of each method.

As part of this effort, we have developed metrics that
specify which records should be changed, how many
records should be changed and how those records should be changed during data transformation.
In addition, we propose new utility measures to evaluate the
different proposed discrimination prevention methods in terms of data
quality and discrimination removal for both direct and indirect
discrimination. Based on the proposed measures, we present
extensive experimental results for two well-known datasets and
compare the different possible methods for direct or indirect
discrimination prevention to find out which methods could be more
successful in terms of low information
loss and high discrimination removal.


\section{Direct and Indirect Discrimination Measurement}
\label{sec:DIDMMETH}
Let $\mathcal{FR}$ be the database of frequent classification
rules extracted from $\mathcal{D}$. Whether a PD rule in $\mathcal{FR}$ has to be considered discriminatory or not can be
assessed by thresholding one of the measures in Fig. \ref{fig:Measures}.

\begin{definition}
\label{label:disc}
Let $f$ be one of the measures in Fig. \ref{fig:Measures}.
Given protected groups $DI_b$ and $\alpha\in R$, a fixed threshold\footnote{$\alpha$ states an acceptable level of discrimination according to laws and
regulations. For example, the U.S. Equal Pay Act \cite{uspay} states that "a selection rate for any race, sex, or ethnic group which is less than four-fifths of the rate for the group with the highest rate will generally be regarded as evidence of adverse impact". This amounts to using $clift$ with $\alpha=1.25$.}, a PD
classification rule $r: A,B\rightarrow C$, where $C$ denies some benefit and $A \subseteq DI_b$, is $\alpha$-protective
w.r.t. $f$ if $f(r)<\alpha$. Otherwise,
$c$ is $\alpha$-discriminatory.
\end{definition}

The purpose of direct discrimination discovery is to identify
$\alpha$-discriminatory rules. In fact, $\alpha$-discriminatory rules
indicate biased rules that are
directly inferred from discriminatory items ({\em e.g.} Foreign worker = Yes).
We call these rules direct $\alpha$-discriminatory rules. 

The purpose of indirect discrimination discovery is to
identify {\em redlining rules}. In fact, redlining rules indicate biased rules
that are indirectly inferred
from non-discriminatory items ({\em e.g.} Zip = 10451) because
of their correlation with discriminatory ones. To determine the redlining rules, Pedreschi {\em et al.} in \cite{peder2008} stated the theorem below which
gives a lower bound for $\alpha$-discrimination of PD
classification rules, given information available
in PND rules ($\gamma$, $\delta$)
and information available from background rules ($\beta_1$, $\beta_2$).
They assume that background knowledge takes the form of association rules
relating a PND itemset $D$ to a PD
itemset $A$ within the context $B$.

\begin{theorem}
\label{def:elb}
Let $r: D,B \rightarrow C$ be a PND classification rule, and let
$$ \gamma = conf(r: D, B \rightarrow C) \hspace{5 mm} \delta
= conf(B \rightarrow C) > 0.$$
Let $A$ be a PD itemset,  and let $\beta_1$, $\beta_2$ such that
$$conf(r_{b1}: A, B \rightarrow D) \geq \beta_1$$
$$conf(r_{b2}: D, B \rightarrow A) \geq \beta_2 > 0.$$
Call
$$f(x) = \frac{\beta_1}{\beta_2}(\beta_2 + x -1)$$

\[ elb(x, y) = \left\{ \begin{array}{l}f(x)/y\;\;\mbox{if f(x) \textgreater 0}\\
0 \;\; \mbox{otherwise} \end{array} \right. \]

It holds that, for $\alpha \geq 0$, if $elb(\gamma, \delta) \geq \alpha$, the PD classification rule $r': A, B \rightarrow C$ is $\alpha$-discriminatory.
\end{theorem}

Based on the above theorem, the following formal definitions of redlining and non-redlining rules are presented:

\begin{definition}
\label{def:redr}

A PND classification rule $r: D, B\rightarrow C$ is a {\em redlining rule}
if it could yield an $\alpha$-discriminatory rule $r': A, B \rightarrow C$
in combination with currently available background knowledge rules of
the form $r_{b1}:A,B \rightarrow D$ and $r_{b2}:D,B \rightarrow A$,
where $A$ is a PD itemset. For example
$\{\mbox{Zip=10451}, \mbox{City=NYC}\} \rightarrow \mbox{Hire=No}$.
\end{definition}

\begin{definition}
\label{def:nredr}

A PND classification rule $r: D, B\rightarrow C$ is a
{\em non-redlining or legitimate rule}
if it cannot yield any $\alpha$-discriminatory rule
$r': A, B \rightarrow C$ in combination with
currently available background knowledge rules
of the form $r_{b1}: A,B \rightarrow D$ and $r_{b2}: D,B \rightarrow A$,
where $A$ is a PD itemset. For
example $\{\mbox{Experience=Low}, \mbox{City=NYC}\} \rightarrow \mbox{Hire=No}$.
\end{definition}

We call $\alpha$-discriminatory rules that ensue
from redlining rules {\em indirect $\alpha$-discriminatory rules}.

\section{The Approach}
\label{sec:prevention}

In this section, we present our approach,
including the 
data transformation methods that can be used for 
direct and/or indirect discrimination prevention.
For each method, its algorithm and its computational
cost are specified.
Our approach for direct
and indirect discrimination prevention can
be described in terms of two phases:
\begin{itemize}
\item {\bf Discrimination Measurement}.

Direct and indirect discrimination discovery includes
identifying $\alpha$-discriminatory rules and redlining rules. To this end,
first, based on predetermined discriminatory items in $\mathcal{D}$,
frequent classification rules in $\mathcal{FR}$ are divided in two groups:
PD and PND rules. Second, direct discrimination is
measured by identifying $\alpha$-discriminatory rules among the PD rules
using a direct discrimination measure ({\em e.g.}, $elift$) and a
discriminatory threshold ($\alpha$). Third,
indirect discrimination is measured by identifying redlining rules among
the PND rules combined with background knowledge, using
an indirect discriminatory measure ($elb$) and a discriminatory
threshold ($\alpha$).
Let $\mathcal{MR}$ be the database of direct $\alpha$-discriminatory
rules  obtained with the above process. In addition, let
$\mathcal{RR}$ be the database of redlining rules and their
respective indirect $\alpha$-discriminatory rules obtained with
the above process.
\item {\bf Data Transformation}. Transform the original data
$\mathcal{D}$ in such a way to remove direct and/or indirect
discriminatory biases, with minimum impact on the data and
on legitimate decision rules, so that no unfair decision rule
can be mined from the transformed data. In the following subsections,
we present the data transformation methods that can be used for this purpose.
\end{itemize}

\begin{figure*}
\centering
\includegraphics{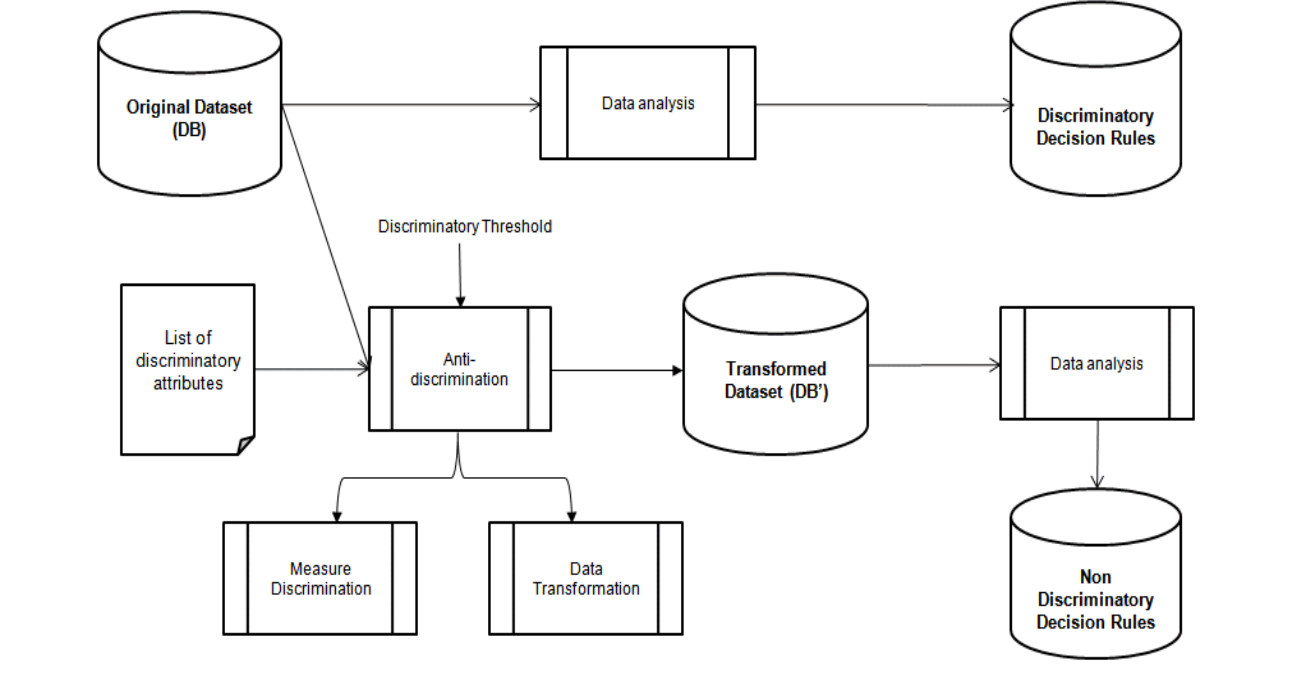}
\caption{The process of extracting biased and unbiased decision rules}
\label{fig:framework}
\end{figure*}

Figure~\ref{fig:framework} illustrates that if the original
biased dataset $\mathcal{D}$
goes through an anti-discrimination process including discrimination
measurement and data transformation, the rules extracted from
transformed dataset $\mathcal{D'}$ could lead to automated unfair decisions.

As mentioned before,
background knowledge
might be obtained from the original dataset itself because of
the existence of PND attributes that are highly
correlated with the PD ones in the original dataset. Let $BK$
be a database of background rules that is defined as
\begin{eqnarray*}
 \mathcal{BK} = \{r_{b2}: D, B \rightarrow A|A \mbox{ discriminatory itemset and }    
 supp(D, B \rightarrow A) \geq ms\}
\end{eqnarray*}
In fact, $\mathcal{BK}$ is the set of classification rules $D, B \rightarrow A$ with a given minimum support $ms$ that shows the correlation between the PD itemset $A$ and the PND itemset $D$ with context $B$. Although rules of the form $r_{b1} :A,B \rightarrow D$ (in
Theorem~\ref{def:elb}) are not
included in $\mathcal{BK}$, $conf(r_{b1}:A,B \rightarrow D)$ could be obtained as $supp(r_{b2}:D,B \rightarrow A)/supp(B \rightarrow A)$.

\section{Data Transformation for Direct Discrimination}
\label{sec:DTDDP}

The proposed solution to prevent direct discrimination
is based on the fact that the
dataset of decision rules would be free of direct
discrimination if it only contained PD rules that
are $\alpha$-protective or are instances of at least one non-redlining
PND rule. Therefore, a suitable data transformation
with minimum information loss should be applied
in such a way that each $\alpha$-discriminatory rule
either becomes $\alpha$-protective or an instance of a non-redlining
PND rule. We call the first procedure {\em direct rule protection} and
the second one {\em rule generalization}.

\subsection{Direct Rule Protection (DRP)}
\label{sec:DRP}

In order to convert each $\alpha$-discriminatory rule into
an $\alpha$-protective rule, based on the direct discriminatory
measure $elift$ ({\em i.e.} Definition~\ref{def:elift}), we should enforce
the following inequality for each $\alpha$-discriminatory
rule $r': A, B \rightarrow C$
in $\mathcal{MR}$, where $A$ is a PD itemset:
\begin{equation}
\label{ineq1}
elift(r') < \alpha
\end{equation}
By using the statement of the $elift$ Definition,
Inequality (\ref{ineq1}) can be rewritten as

\begin{equation}
\label{ineq2}
\frac{conf(r': A, B \rightarrow C)}{conf(B \rightarrow C)} < \alpha
\end{equation}
Let us rewrite Inequality (\ref{ineq2}) in the following way
\begin{equation}
\label{ineq3}
conf(r': A, B\rightarrow C) < \alpha \cdot conf(B\rightarrow C)
\end{equation}
So, it is clear that Inequality (\ref{ineq1}) can be satisfied by
decreasing the confidence of the $\alpha$-discriminatory rule
$r': A, B \rightarrow C$
to a value less than the right-hand side of Inequality (\ref{ineq3}),
without affecting the confidence of its base rule $B\rightarrow C$.
A possible solution for decreasing
\begin{equation}
\label{ineq4}
conf (r': A, B\rightarrow C) = \frac{supp(A,B,C)}{supp(A,B)}
\end{equation}
is to perturb the discriminatory itemset from $\neg A$
to $A$ in the subset $\mathcal{D}_c$
of all records of the original dataset which completely support
the rule $\neg A, B\rightarrow \neg C$ and have minimum impact on other rules;
doing so increases the denominator of Expression (\ref{ineq4})
while keeping the numerator and $conf(B\rightarrow C$) unaltered.

There is also another way to provide direct rule protection.
Let us rewrite Inequality (\ref{ineq2}) in the following different way

\begin{equation}
\label{ineq5}
conf(B \rightarrow C) > \frac{conf(r': A, B \rightarrow C)}{\alpha}
\end{equation}
It is clear that Inequality (\ref{ineq1}) can be satisfied by
increasing the confidence of the base rule
($B \rightarrow C$) of the $\alpha$-discriminatory rule
$r': A, B \rightarrow C$
to a value higher than the right-hand side of Inequality (\ref{ineq5}),
without affecting the value of $conf(r': A, B \rightarrow C)$.
A possible solution for increasing Expression
\begin{equation}
\label{ineq6}
conf (B\rightarrow C) = \frac{supp(B,C)}{supp(B)}
\end{equation}
is to perturb the class item from $\neg C$ to $C$ in
the subset $\mathcal{DB}_c$
of all records of the original dataset which completely support
the rule $\neg A, B\rightarrow \neg C$ and have minimum impact on other rules;
doing so increases the numerator of Expression (\ref{ineq6})
while keeping the denominator and $conf(r': A, B\rightarrow C)$
unaltered.

Therefore, there are two methods that could be applied for direct
rule protection. One method (Method 1) changes the
PD itemset in some records
({\em e.g.} gender changed
from male to female in the records with granted credits)
and the other method (Method 2)
changes the class item in some records
({\em e.g.} from grant credit to deny credit in the records with male gender).
Thus, a suitable data transformation with minimum information loss should be applied in such a way that each $\alpha$-discriminatory rule $r$ becomes $\alpha$-protective by ensuring that $f(r) < \alpha$, 
where $f(r)$ could be one of the measures in Fig.~\ref{fig:Measures}. 
Then the general idea of our proposed method DRP for all these measures is the same. However, in the details ({\em i.e.} which records should be changed, how many records should be changed and how those records should be changed during data transformation), there could be some differences because of the different definition of these measures. For instance, in Table~\ref{tab:DTM} we present all the possible 
data transformation methods for measures $elift$, $slift$ and $olift$ based 
on our proposed direct rule protection method.

\begin{table*}[t]
\caption{Data transformation methods (DTMs) for different measures}
\label{tab:DTM}
\centering
\begin{tabular}{|c|c|c|}
\hline
{\bf Measures} & {\bf DTMs for DRP} & {\bf Transformation requirement} \\

\hline
\multirow{2}{*}{$elift$} & 	$\neg A, B \rightarrow \neg C \Rightarrow A, B \rightarrow \neg C$ & \multirow{2}{*}{$elift(A, B\rightarrow C) < \alpha$} \\
 & $\neg A, B \rightarrow \neg C \Rightarrow \neg A, B \rightarrow C$ & \\

\hline
\multirow{2}{*}{$slift$} & 	$A, B \rightarrow C \Rightarrow A, B \rightarrow \neg C$ & \multirow{2}{*}{$slift(A, B \rightarrow C) < \alpha$} \\
 & $\neg A, B \rightarrow \neg C \Rightarrow \neg A, B \rightarrow C$ & \\

\hline

\multirow{4}{*}{$olift$} & 	$\neg A, B \rightarrow \neg C \Rightarrow A, B \rightarrow \neg C$ & \multirow{4}{*}{$olift(A, B\rightarrow C) < \alpha$} \\
 & $\neg A, B \rightarrow \neg C \Rightarrow \neg A, B \rightarrow C$ & \\
&  $A, B \rightarrow C \Rightarrow A, B \rightarrow \neg C$ & \\
&  $A, B \rightarrow C \Rightarrow \neg A, B \rightarrow C$ & \\

\hline

\end{tabular}
\end{table*}

As shown in Table~\ref{tab:DTM}, two possible data transformation methods for $elift$ are the same as two possible ones for $olift$ and one of them is the same as the
possible ones for $slift$. Another point is that there is a method 
(the second one for each measure) that is the same for all three measures. 
Then we can conclude that for DRP not only the general idea is the same 
for different measures, but also in the details, there is a data transformation method applicable 
for all the measures.

\subsection{Rule Generalization (RG)}

Rule generalization is another data transformation method for direct
discrimination prevention. It is based on the fact
that if each $\alpha$-discriminatory rule $r': A, B \rightarrow C$ in the
database of decision rules was an instance of at least
one non-redlining (legitimate) PND rule $r: D, B \rightarrow C$,
the dataset would be free of direct discrimination.

In rule generalization, we consider the relation between rules
instead of discrimination measures. The following example illustrates
this principle. Assume that a complainant claims
discrimination against foreign workers among applicants for a job position.
A classification rule
$\{\mbox{Foreign worker=Yes}, \mbox{City=NYC}\} \rightarrow \mbox{Hire=No}$
with high $elift$ supports the complainant's claim.
However, the decision maker could argue that this rule is
an instance of a more general rule
$\{\mbox{Experience=Low}, \mbox{City=NYC}\} \rightarrow \mbox{Hire=No}$.
In other words, foreign workers are rejected because
of their low experience, not just because they are foreign.
The general rule rejecting low-experienced applicants
is a legitimate one, because experience can be considered a genuine/legitimate
requirement for some jobs. To formalize this dependency
among rules ({\em i.e.} $r'$ is an instance of $r$), Pedreschi {\em et al.}
in \cite{pedreschiICAIL} say that a PD classification rule
$r': A,B\rightarrow C$  is an instance of a PND
rule $r: D,B \rightarrow C$ if rule $r$ holds with the same or higher
confidence, namely $conf(r: D,B \rightarrow C) \geq
conf(r': A,B\rightarrow C)$, and a case (record) satisfying
discriminatory itemset $A$ in context $B$ satisfies
legitimate itemset $D$ as well, namely $conf(A, B \rightarrow D) = 1$.
The two conditions can be relaxed as follows:

\begin{definition}
\label{def:dinstance}
Let $d \in [0, 1]$. A classification rule $r': A,B\rightarrow C$
is a $d$-instance of $r: D,B \rightarrow C$ if
both conditions below are true:
\begin{itemize}
\item{\bf Condition 1:} $conf(r) \geq d\cdot conf(r')$
\item{\bf Condition 2:} $conf (r'': A, B\rightarrow D) \geq d$.
\end {itemize}
\end{definition}

Then, if $r'$ is a $d$-instance of $r$ (where $d$ is 1 or a value near 1),
$r'$ is free of direct discrimination. 
Based on this concept, we propose a data transformation method
({\em i.e.} rule generalization)
to transform each $\alpha$-discriminatory $r'$ in $\mathcal{MR}$ into
a $d$-instance of a legitimate rule.
An important issue to perform rule generalization is to find
a suitable PND rule ($r:D, B \rightarrow C$) or, equivalently,
to find a suitable $D$ ({\em e.g.} Experience=Low).
Although choosing non-redlining rules, as done in this chapter,
is a way to obtain legitimate PND rules, sometimes it is not enough
and a semantic hierarchy is needed to find the most suitable
legitimate itemset.

At any rate, rule generalization can
be attempted for $\alpha$-discriminatory rules $r'$ for which
there is at least one non-redlining PND rule $r$ satisfying at least one
of the two conditions of Definition~\ref{def:dinstance}.
If any of the two conditions does not hold,
the original data should be transformed for it to hold.
Let us assume that Condition (2) is satisfied but Condition (1)
is not. Based on the definition of $d$-instance,
to satisfy the first condition of
Definition~\ref{def:dinstance}, we should enforce
for each $\alpha$-discriminatory rule
$r': A, B \rightarrow C$ in $\mathcal{MR}$
the following inequality, with respect to its PND
rule $r: D, B \rightarrow C$:
\begin{equation}
\label{ineq7}
conf(r: D, B \rightarrow C) \geq d\cdot conf(r': A, B \rightarrow C)
\end{equation}
Let us rewrite Inequality (\ref{ineq7}) in the following way
\begin{equation}
\label{ineq8}
conf (r': A, B\rightarrow C) \leq \frac{conf(r: D, B\rightarrow C)}{d}
\end{equation}
So, it is clear that Inequality (\ref{ineq7}) can be satisfied by
decreasing the confidence of the $\alpha$-discriminatory
rule ($r': A, B \rightarrow C$)
to values less than the right-hand side of Inequality (\ref{ineq8}),
without affecting the confidence of rule $r: D, B\rightarrow C$
or the satisfaction of Condition (2) of Definition~\ref{def:dinstance}.
The confidence of $r'$ was previously specified in
Expression (\ref{ineq4}). A possible solution to decrease
this confidence is
to perturb the class item from $C$ to $\neg C$ in
the subset $\mathcal{D}_c$ of all records in the original dataset which
completely support the
rule $A, B, \neg D \rightarrow C$ and have minimum impact on other rules;
doing so decreases the
numerator of Expression (\ref{ineq4}) while keeping its denominator,
$conf (r: D, B \rightarrow C)$ and also $conf(r": A, B \rightarrow D)$
(Condition (2) for rule generalization) unaltered.

Let us see what happens if
if Condition (1) of Definition~\ref{def:dinstance}
is satisfied but Condition (2) is not. In this case,
based on the definition of $d$-instance, to satisfy Condition (2)
we should enforce the following inequality for each
$\alpha$-discriminatory rule $r': A, B \rightarrow C$
in $\mathcal{MR}$ with respect to its PND rule $r: D, B \rightarrow C$:
\begin{equation}
\label{ineq9}
conf(r": A, B \rightarrow D) \geq d
\end{equation}
Inequality (\ref{ineq9}) must be satisfied by
increasing the confidence of rule $r": A, B \rightarrow D$
to a value higher than $d$, without affecting the satisfaction
of Condition (1). However, any effort at increasing the confidence
of $r"$ impacts on the confidence of the $r$ or $r'$ rules and
might threaten the satisfaction of Condition (1) of
Definition~\ref{def:dinstance};
indeed, in order to increase the confidence of $r"$ we must
either decrease $supp(A,B)$ (which increases $conf(r')$)
or change $\neg D$ to $D$ for those records
satisfying $A$ and $B$ (which decreases $conf(r)$).
Hence, rule generalization
can only be applied if Condition (2) is satisfied
without any data transformation.

To recap, we see that rule generalization can be achieved
provided that Condition (2) is satisfied,
because Condition (1) can be reached by changing the
class item in some records ({\em e.g.} from ``Hire no'' to ``Hire yes''
in the records of foreign and high-experienced people in NYC city).

\subsection{Direct Rule Protection and Rule Generalization}
\label{sec:DRPRG}

Since rule generalization might not be applicable
for all $\alpha$-discriminatory rules in $\mathcal{MR}$,
rule generalization cannot be used alone
for direct discrimination prevention and must be combined
with direct rule protection. When applying both rule generalization
and direct rule protection,
$\alpha$-discriminatory rules are divided into two groups:

\begin{itemize}
\item $\alpha$-discriminatory rules $r'$ for which
there is at least one non-redlining PND rule $r$ satisfying
Condition (2) of Definition~\ref{def:dinstance}.
For these rules, rule generalization is performed
unless direct rule protection requires less data transformation
(in which case direct rule protection is used).
\item $\alpha$-discriminatory rules such that there is no such PND
rule. For these rules, direct rule protection is performed.
\end{itemize}

We propose Algorithm~\ref{alg31} to select
the most appropriate discrimination prevention approach for
each $\alpha$-discriminatory rule.
First, for each $\alpha$-discriminatory rule in $\mathcal{MR}$
of type $r': A, B\rightarrow C$, a collection $D_{pn}$ of
non-redlining PND rules of type $r: D, B\rightarrow C$ is found (Step 2).
Then, the conditions
of Definition~\ref{def:dinstance} are checked
for each rule in $D_{pn}$, for $d\geq 0.8$ (Steps 4-18). Three cases
arise depending on whether
Conditions (1) and (2) hold:

\begin{itemize}
\item {\bf Case 1: There is at least one rule $r \in D_{pn}$ such that both
Conditions (1) and (2) of Definition~\ref{def:dinstance} hold.}
In this case $r'$ is a $d$-instance of $r$
for $d \geq 0.8$ and no transformation is required (Steps 19-20).

\item {\bf Case 2: There is no rule in $D_{pn}$ satisfying both
Conditions (1) and (2) of Definition~\ref{def:dinstance}, but there
is at least one rule satisfying Condition (2).} In this case (Step 23),
the PND rule $r_b$ in $D_{pn}$ should be
selected (Step 24) which requires the minimum data transformation to fulfill
Condition (1). A smaller difference between the values of the
two sides of Condition
(1) for each $r$ in $D_{pn}$ indicates a smaller
required data transformation. In this case
the $\alpha$-discriminatory rule is transformed by rule
generalization (Step 25).

\item {\bf Case 3: No rule in $D_{pn}$ satisfies Condition (2)
of Definition~\ref{def:dinstance}.}
In this case (Step 21), rule generalization is not possible and
direct rule protection should be performed (Step 22).

\end{itemize}

\begin{algorithm}[h]
\caption{{\sc Determining Discrimination Prevention Approaches}}
\label{alg31}
{\scriptsize  \begin{algorithmic}[1]
\REQUIRE $\mathcal{DB}$, $\mathcal{MR}$, $\mathcal{FR}$, $d\geq 0.8$, $\alpha$
\FOR{each $r': A, B \rightarrow C \in \mathcal{MR}$}
\STATE $D_{pn}\gets \mbox {Collection of non-redlining PND rules } r: D, B \rightarrow C$ from $FR$
\IF{$|D_{pn}| \neq 0$}
\FOR{each $r \in D_{pn}$}
\STATE // Assess conditions of $p$-instance
\STATE Compute $conf(r") \mbox{, where } r": A, B \in D$
\IF{$conf(r) \geq d\cdot conf(r')$}
\STATE $C1_r  \gets true$
\ELSE
\STATE $C1_r \gets false$
\STATE $dif1_r\gets d\cdot conf(r') - conf(r)$
\ENDIF
\IF{$conf (r") \geq d$}
\STATE $C2_r \gets true$
\ELSE
\STATE $C2_r \gets false$
\ENDIF
\ENDFOR
\IF{$\exists r \in D_{pn}\mbox{ s.t. } C1_r = true \wedge  C2_r = true$}
\STATE $TR_{r'}\gets Nt$ // No transformation needed
\ELSIF {for all $r \in D_{pn}\mbox{, }C2_r = false$}
\STATE $TR_{r'}\gets DRP$ // Direct Rule Protection
\ELSIF{$\exists r \in D_{pn}\mbox{ s.t. }C2_r = true$}
\STATE $r_b\gets r \in D_{pn}\mbox{ with minimum }dif1_r$
\STATE $TR_{r'}\gets RG$ // Rule Generalization
\ENDIF
\ELSE
\STATE // $|D_{pn}|=0$
\STATE $TR_{r'} \gets DRP$ // Direct Rule Protection
\ENDIF
\IF{$TR_{r'} = RG$}
\STATE $dif_r' \gets conf (r')-\alpha \cdot conf(B \rightarrow C$)
\IF{$dif_r' < dif1_r$}
\STATE $TR_{r'} \gets DRP$
\ENDIF
\ENDIF
\ENDFOR
\ENSURE $\mathcal{TR}$ containing all $r' \in \mathcal{MR}$ and their respective $TR_{r'}$ and $r_b$

\end{algorithmic}}
\end{algorithm}

For the $\alpha$-discriminatory rules to which
rule generalization can be applied,
it is possible that direct rule protection can be achieved with
a smaller data transformation.
For these rules the algorithm
{\em should select the approach
with minimum transformation} (Steps 31-36).
The algorithm yields as output a database
$\mathcal{TR}$ with all $r'\in \mathcal{MR}$, their respective
rule $r_b$ and their respective discrimination
prevention approaches ($TR_{r'}$).

\section{Data Transformation for Indirect Discrimination}
\label{sec:DTIDP}

The proposed solution to prevent indirect discrimination
is based on the fact that the dataset of decision rules would
be free of indirect discrimination
if it contained no redlining rules. To achieve this, a suitable data
transformation with minimum information loss should be applied
in such a way that
redlining rules are converted to non-redlining rules. We call
this procedure {\em indirect rule protection}.

\subsection{Indirect Rule Protection (IRP)}
\label{sec:IRP}

In order to turn a redlining rule into
an non-redlining rule, based on the indirect discriminatory measure ({\em i.e.} $elb$ in
Theorem~\ref{def:elb}), we should enforce the
following inequality for each redlining rule $r: D, B \rightarrow C$
in $\mathcal{RR}$:
\begin{equation}
\label{ineq10}
elb(\gamma, \delta) < \alpha
\end{equation}
By using the definitions stated when introducing $elb$
in Theorem~\ref{def:elb}\footnote{It is worth noting that
$\beta_1$ and $\beta_2$ are lower bounds for $conf(r_{b1})$ and $conf(r_{b2})$,
respectively, so it is correct if we replace $\beta_1$ and $\beta_2$
with $conf(r_{b1})$ and $conf(r_{b2})$ in the $elb$ formulation.},
Inequality (\ref{ineq10}) can be rewritten as

\begin{equation}
\label{ineq11}
\frac{\frac{conf(r_{b1})}{conf(r_{b2})}(conf(r_{b2}) + conf(r: D, B \rightarrow C) -1)}{conf(B \rightarrow C)} < \alpha
\end{equation}

Note
that the discriminatory itemset (\textit{i.e.} $A$) is not
removed from the original database $\mathcal{D}$ and
the rules $r_{b1}: A, B \rightarrow D$ and $r_{b2}: D, B \rightarrow A$
are obtained from $\mathcal{D}$,
so that their confidences might change as a result of data transformation
for indirect discrimination prevention.
Let us rewrite Inequality (\ref{ineq11}) in the following way
\begin{equation}
\label{ineq13}
conf(r_{b1}: A, B \rightarrow D) <
\frac{\alpha \cdot conf(B \rightarrow C) \cdot conf(r_{b2})}{conf(r_{b2}) + conf(r: D, B \rightarrow C) -1}
\end{equation}
Clearly, in this case Inequality (\ref{ineq10}) can
be satisfied by decreasing the confidence of rule $r_{b1}: A, B \rightarrow D$
to values
less than the right-hand side of Inequality (\ref{ineq13})
without affecting either the confidence of the redlining rule
or the confidence of the $B \rightarrow C$ and $r_{b2}$ rules.
Since the values of both inequality
sides are dependent,
a transformation is required that decreases the left-hand side of
the inequality without any impact on the right-hand side.
A possible solution for decreasing
\begin{equation}
\label{ineq14}
conf(A, B\rightarrow D) =\frac{supp(A, B, D)}{supp(A, B)}
\end{equation}
in Inequality (\ref{ineq13}) to the target value is to
perturb the discriminatory
itemset from $\neg A$ to $A$ in the subset $\mathcal{D}_c$
of all records of the original dataset which completely support
the rule $\neg A, B, \neg D \rightarrow \neg C$ and have
minimum impact on other rules; this
increases the denominator of Expression (\ref{ineq14})
while keeping the numerator and $conf(B \rightarrow C)$,
$conf(r_{b2}: D, B\rightarrow A$), and $conf(r: D, B\rightarrow C$)
unaltered.

There is another way to provide indirect rule protection.
Let us rewrite Inequality (\ref{ineq11}) as
Inequality (\ref{ineq12}), where the confidences
of $r_{b1}$ and $r_{b2}$ rules are not constant.

\begin{equation}
\label{ineq12}
conf(B \rightarrow C) >
\frac{\frac{conf(r_{b1})}{conf(r_{b2})}(conf(r_{b2}) + conf(r: D, B \rightarrow C) -1)}{\alpha}
\end{equation}
Clearly, in this case Inequality (\ref{ineq10}) can
be satisfied by increasing the confidence of the base rule ($B \rightarrow C$)
of the redlining rule $r: D, B \rightarrow C$ to values
greater than the right-hand side of Inequality (\ref{ineq12})
without affecting either the confidence of the redlining rule
or the confidence of the $r_{b1}$ and $r_{b2}$ rules.
A possible solution for increasing Expression
(\ref{ineq6}) in Inequality (\ref{ineq12}) to the target value is to
perturb the class item from $\neg C$ to $C$ in the subset $\mathcal{D}_c$
of all records of the original dataset which completely support
the rule $\neg A, B, \neg D \rightarrow \neg C$ and have
minimum impact on other rules; this
increases the numerator of Expression (\ref{ineq6})
while keeping the denominator and $conf(r_{b1}: A, B\rightarrow D$),
$conf(r_{b2}: D, B\rightarrow A$), and $conf(r:D, B\rightarrow C$)
unaltered.

Hence, like in direct rule protection,
there are also two methods that could be applied for
indirect rule protection. One method (Method 1)
changes the  discriminatory itemset in some
records ({\em e.g.} from non-foreign worker to foreign
worker in the records of hired people in NYC city with Zip$\neq$10451)
and the other method (Method 2) changes the class item in some
records ({\em e.g.} from ``Hire yes'' to ``Hire no''
in the records of  non-foreign worker of people in NYC
city with Zip$\neq$10451).
\section{Data Transformation for Both Direct and Indirect Discrimination}
\label{DRP-IRP}

We deal here with the key problem of transforming data
with minimum information loss to prevent {\em at the same time}
both direct and indirect
discrimination. We will give a pre-processing
solution to {\em simultaneous direct and indirect discrimination prevention}.
First, we explain when direct and indirect discrimination could
simultaneously occur. This depends on whether the original dataset
($\mathcal{D}$) contains discriminatory itemsets or not. Two cases arise:

\begin{itemize}
\item PD itemsets (\textit{i.e.} $A$) did not exist in the original database $\mathcal{D}$ or have previously been removed from it due to privacy constraints or for preventing discrimination. However, if background knowledge from publicly available data
({\em e.g.} census data) is available, indirect
discrimination remains possible. In fact, in this case,
only PND rules are extracted from $\mathcal{D}$ so
only indirect discrimination could happen.

\item At least one PD itemset (\textit{i.e.} $A$) is not
removed from the original database ($\mathcal{D}$). So it is clear
that PD rules could be extracted from $\mathcal{D}$ and
direct discrimination could happen. However,
in addition to direct discrimination, indirect discrimination might
occur because of background knowledge obtained from $\mathcal{D}$
itself due to the existence of PND items
that are highly correlated with the
sensitive (discriminatory) ones. Hence,
in this case both direct and indirect discrimination could happen.
\end{itemize}


To provide both direct rule protection (DRP) and
indirect rule protection (IRP) at the same time,
an important point is the relation between the data transformation
methods. Any
data transformation to eliminate direct $\alpha$-discriminatory
rules should not
produce new redlining rules or prevent the existing ones from
being removed. Also
\begin{table*}[t]
\caption{Direct and indirect rule protection methods}
\label{tab:RPM}
\centering
\scalebox{0.78}{
\begin{tabular}{|c|c|c|}
\hline
{\bf } & {\bf Method 1} & {\bf Method 2} \\

\hline
Direct Rule Protection & $\neg A, B \rightarrow \neg C \Rightarrow A, B \rightarrow \neg C$ & $\neg A, B \rightarrow \neg C \Rightarrow \neg A, B \rightarrow C$ \\

\hline
Indirect Rule Protection & $\neg A, B, \neg D \rightarrow \neg C \Rightarrow A, B, \neg D \rightarrow \neg C$  &  $\neg A, B, \neg D \rightarrow \neg C \Rightarrow \neg A, B, \neg D \rightarrow C$ \\

\hline

\end{tabular}
}
\end{table*}
any data transformation to eliminate redlining rules should
not produce new direct $\alpha$-discriminatory rules or prevent
the existing ones from being removed.

For subsequent use in this section,
we summarize in Table~\ref{tab:RPM} the methods for DRP and IRP
described in Sections~\ref{sec:DRP} and~\ref{sec:IRP} above.
We can see in Table~\ref{tab:RPM} that DRP and IRP
operate the same kind of data transformation: in both
cases Method 1 changes the PD itemset,
whereas Method 2 changes the class item.
Therefore, {\em in principle} any data transformation for DRP
(resp. IRP) not only does not need to have
a negative impact on IRP (resp. DRP),
but both kinds of protection could even be beneficial to each other.

However, there is a difference between DRP and IRP:
the set of records chosen for transformation.
As shown in Table~\ref{tab:RPM}, in IRP
the chosen records should not satisfy the $D$ itemset
(chosen records are those with
$\neg A, B, \neg D \rightarrow \neg C$), whereas
DRP does not care about $D$ at all
(chosen records are those with  $\neg A, B \rightarrow \neg C$).
The following interactions
between direct and indirect rule protection become apparent.

\begin{lemma}
Method 1 for DRP cannot be used if
simultaneous DRP and IRP are desired.
\end{lemma}

{\bf Proof:}
Method 1 for DRP might undo the protection provided by Method 1 for IRP, as we next justify. Method 1 for DRP decreases $conf(A,B \rightarrow C)$ until the direct rule protection requirement (Inequality (\ref{ineq3})) is met and Method 1 for IRP needs to decrease $conf(A,B \rightarrow D)$ until the indirect rule protection requirement is met (Inequality (\ref{ineq13})). Assume that decreasing $conf(A,B \rightarrow C)$ to meet the direct rule protection requirement is achieved by changing $y$ (how $y$ is obtained will be discussed in Section~\ref{sec:time}) number of records with $\neg A, B, \neg C$ to records with $A, B, \neg C$ (as done by Method 1 for DRP). This actually could increase $conf(A,B \rightarrow D)$ if  $z$ among the changed records, with $z \leq y$,
turn out to satisfy $D$. This increase can undo the protection provided by Method 1 for IRP  (i.e. $conf(A, B \rightarrow D) < IRP_{req1}$, where $IRP_{req1}= \frac{\alpha \cdot conf(B \rightarrow C) \cdot conf(r_{b2})}{conf(r_{b2}) + conf(r: D, B \rightarrow C) -1}$) if the new value $conf(A, B \rightarrow D)=\frac{supp(A, B, D)+z}{supp(A, B)+y}$  is greater than or equal to $IRP_{req1}$, which
happens if $z \geq IRP_{req1}\cdot (supp(A, B)+Y)- supp(A, B, D)$. 
\hfill $\Box$
\begin{lemma}
Method 2 for IRP is beneficial for Method 2 for DRP.
On the other hand, Method 2 for DRP is at worst neutral
for Method 2 for IRP.
\end{lemma}

{\bf Proof:} Method 2 for DRP and Method 2 for IRP
are both aimed at increasing $conf(B \rightarrow C)$.
In fact, Method 2 for IRP
changes a subset of the records changed by Method 2 for DRP.
This proves that Method 2 for IRP is beneficial for Method 2 for DRP.
On the other hand, let us check that, in
the worst case, Method 2 for DRP is neutral
for Method 2 for IRP: such a worst case is the one in which all changed
records satisfy $D$, which could result in increasing {\em both} sides
of Inequality (\ref{ineq12}) by an equal amount
(due to increasing $conf(B \rightarrow C)$
and $conf(D,B \rightarrow C)$); even in this case, there is no change in
whatever protection is achieved by Method 2 for IRP. 
\hfill $\Box$

Thus, we conclude that Method 2 for DRP and Method 2 for IRP
are the only methods among those described that can be applied to achieve
simultaneous direct and indirect discrimination prevention.
In addition, in the cases where either only direct or only
indirect discrimination
exist, there is no interference
between the described methods: Method 1 for DRP,
Method 2 for DRP and Rule Generalization can be used to
prevent direct discrimination;
Method 1 for IRP and Method 2 for IRP can be used
to prevent indirect discrimination.
In what follows,
we propose algorithms based on the described
methods that cover direct and/or indirect discrimination prevention.

\section{The Algorithms}
\label{sec:ALGSMETHOD}

We describe in this section our algorithms based on the
direct and indirect discrimination prevention methods
proposed in Sections~\ref{sec:DTDDP}, ~\ref{sec:DTIDP}
and~\ref{DRP-IRP}. 
There are some assumptions common to all algorithms in this section.
First, we assume the class attribute
in the original dataset $\mathcal{D}$
to be binary ({\em e.g.} denying or granting credit). Second,
we consider classification rules with negative decision
({\em e.g.} denying credit) to be in $\mathcal{FR}$.
Third, we assume the PD itemsets ({\em i.e.} $A$)
and the PND itemsets ({\em i.e.} $D$) to be
binary or non-binary categorical.

\subsection{Direct Discrimination Prevention Algorithms}
\label{sec:DDPAs}

We start with direct rule protection. Algorithm~\ref{alg32} details Method 1
for DRP. For each direct $\alpha$-discriminatory rule $r'$
in $\mathcal{MR}$ (Step 3), after finding the subset $\mathcal{D}_c$
(Step 5), records in $\mathcal{D}_c$ should be changed until the direct
rule protection requirement (Step 10) is met for
each respective rule (Steps 10-14).

Among the records of $\mathcal{D}_c$, one should change
those with lowest impact on the other ($\alpha$-protective or non-redlining)
rules. Hence, for each record $db_c \in \mathcal{D}_c$,
the number of rules whose premise is supported by
$db_c$ is taken as the impact of $db_c$ (Step 7), that is $impact(db_c)$;
the rationale is that changing $db_c$ impacts on the confidence
of those rules. Then the records $db_c$ with minimum $impact(db_c)$ are
selected for change (Step 9), with the aim of scoring well in terms
of the utility measures proposed in the next section. We call this procedure
(Steps 6-9) {\em impact minimization} and we
re-use it in the pseudocodes of the rest of algorithms specified
in this chapter.

\begin{algorithm}[h]
\caption{{\sc Direct Rule Protection (Method 1)}}
\label{alg32}
{\small \begin{algorithmic}[1]
\STATE Inputs: $\mathcal{D}$, $\mathcal{FR}$, $\mathcal{MR}$, $\alpha$, $DI_b$
\STATE Output: $\mathcal{D'}$ (transformed dataset)
\FOR{each $r': A, B\rightarrow C \in \mathcal{MR}$}
\STATE $\mathcal{FR} \gets \mathcal{FR} -\{ r'\}$
\STATE $\mathcal{D}_c\gets \mbox{All records completely supporting }\neg A, B\rightarrow \neg C$
\FOR {each $db_c \in \mathcal{D}_c$}
\STATE Compute $impact(db_c)=|\{r_a\in \mathcal{FR} | db_c$ supports
the premise of $r_a\}|$
\ENDFOR
\STATE Sort $\mathcal {D}_c$ by ascending impact
\WHILE{$conf(r') \geq \alpha \cdot conf(B\rightarrow C)$}
\STATE Select first record in $\mathcal{D}_c$
\STATE Modify PD itemset of $db_c$ from $\neg A$ to $A$ in $\mathcal{D}$
\STATE Recompute $conf(r')$
\ENDWHILE
\ENDFOR
\STATE Output: $\mathcal{D'} = \mathcal{D}$
\end{algorithmic}}
\end{algorithm}

Algorithm~\ref{alg33} details Method 2 for DRP.
The parts of Algorithm~\ref{alg33} to find subset $\mathcal{D}_c$
and perform {\em impact minimization} (Step 4) are the same
as in Algorithm~\ref{alg32}. However, the transformation
requirement that should be met for each
$\alpha$-discriminatory rule in $\mathcal{MR}$ (Step 5) and the
kind of data transformation are different (Steps 5-9).

\begin{algorithm}[h]
\caption{{\sc Direct Rule Protection (Method 2)}}
\label{alg33}
{\small \begin{algorithmic}[1]
\STATE Inputs: $\mathcal{D}$, $\mathcal{FR}$, $\mathcal{MR}$, $\alpha$, $DI_b$
\STATE Output: $\mathcal{D'}$ (transformed dataset)
\FOR{each $r': A, B\rightarrow C \in \mathcal{MR}$}
\STATE Steps 4-9 Algorithm~\ref{alg32}
\WHILE{$conf(B \rightarrow C) \leq \frac{conf(r')}{\alpha}$}
\STATE Select first record in $\mathcal{D}_c$
\STATE Modify the class item of $db_c$ from $\neg C$ to $C$ in $\mathcal{D}$
\STATE Recompute $conf(B \rightarrow C)$
\ENDWHILE
\ENDFOR
\STATE Output: $\mathcal{D'} = \mathcal{D}$
\end{algorithmic}}
\end{algorithm}

As mentioned in Section~\ref{sec:DRPRG}, rule generalization cannot
be applied alone for solving direct discrimination prevention, but it can
be used in combination with Method 1 or Method 2 for DRP.
In this case, after specifying the discrimination prevention method
({\em i.e.} direct rule protection or rule generalization) to
be applied for each $\alpha$-discriminatory rule based on the
algorithm in Section~\ref{sec:DRPRG},
Algorithm~\ref{alg34} should be run to combine
rule generalization and one of the two direct rule protection methods.

\begin{algorithm}[h]
\caption{{\sc Direct Rule Protection and Rule Generalization}}
\label{alg34}
{\small \begin{algorithmic}[1]
\STATE Inputs: $\mathcal{D}$, $\mathcal{FR}$, $\mathcal{TR}$, $p \geq 0.8$, $\alpha$, $DI_b$
\STATE Output: $\mathcal{D'}$ (transformed dataset)
\FOR{each $r': A, B\rightarrow C \in \mathcal{TR}$}
\STATE $\mathcal{FR} \gets \mathcal{FR} -\{ r'\}$
\IF{$TR_{r'}$ = RG} \STATE \mbox{// Rule Generalization}
 \STATE $\mathcal{D}_c\gets \mbox{All records completely supporting }A, B, \neg D \rightarrow C$
\STATE Steps 6-9 Algorithm~\ref{alg32}
\WHILE{$conf(r') > \frac{conf(r_b: D, B \rightarrow C)}{p}$}
\STATE Select first record in $\mathcal{D}_c$
\STATE Modify class item of $db_c$ from $C$ to $\neg C$ in $\mathcal{D}$
\STATE Recompute $conf(r')$
\ENDWHILE
\ENDIF
\IF{$TR_{r'}$ = DRP}
\STATE \mbox{// Direct Rule Protection}
\STATE Steps 5-14 Algorithm~\ref{alg32} or Steps 4-9 Algorithm~\ref{alg33}
\ENDIF
\ENDFOR
\STATE Output: $\mathcal{D'} = \mathcal{D}$
\end{algorithmic}}
\end{algorithm}

Algorithm~\ref{alg34} takes as input $\mathcal{TR}$,
which is the output of the algorithm in Section~\ref{sec:DRPRG}, containing
all $r' \in \mathcal{MR}$ and their respective $TR_{r'}$ and $r_b$.
For each $\alpha$-discriminatory rule $r'$ in $\mathcal{TR}$,
if $TR_{r'}$ shows that rule generalization should be performed (Step 5),
after determining the records that should be changed for
{\em impact minimization} (Steps 7-8), these records should be changed
until the rule generalization requirement is met (Steps 9-13).
Also, if $TR_{r'}$ shows that direct rule protection should be performed
(Step 15), based on either Method 1 or Method 2, the relevant
sections of Algorithm~\ref{alg32} or ~\ref{alg33} are called,
respectively (Step 17).

\subsection{Indirect Discrimination Prevention Algorithms}

A detailed algorithm implementing Method 2 for IRP
is provided in~\cite{hajian2011LNC}, from which an
algorithm implementing Method 1 for IRP can be easily derived.
Due to similarity with the previous
algorithms, we do not recall those two algorithms for IRP here.

\subsection{Direct and Indirect Discrimination Prevention Algorithms}

Algorithm~\ref{alg35} details our proposed data
transformation method for simultaneous
direct and indirect discrimination prevention. The algorithm
starts with redlining rules. From each redlining
rule ($r: X \rightarrow C$), more than one indirect $\alpha$-discriminatory
rule ($r': A, B \rightarrow C$) might be generated because of two reasons:
1) existence of different ways to group the items in $X$ into
a context itemset $B$ and a PND itemset
$D$ correlated to some PD itemset $A$;
and 2) existence of more than one item in $DI_b$.
Hence, as shown in Algorithm~\ref{alg35} (Step 5),
given a redlining rule $r$, proper data transformation
should be conducted for all indirect $\alpha$-discriminatory rules
$r': (A \subseteq DI_b), (B \subseteq X) \rightarrow C$ ensuing from $r$.

\begin{algorithm}[h]

\caption{{\sc Direct and Indirect Discrimination Prevention}}
\label{alg35}
{\scriptsize \begin{algorithmic}[1]

\STATE Inputs: $\mathcal{D}$, $\mathcal{FR}$, $\mathcal{RR}$,
$\mathcal{MR}$, $\alpha$, $DI_b$

 \STATE Output: $\mathcal{D'}$ (transformed dataset)

\FOR{each $r: X\rightarrow C \in \mathcal{RR}$, where $D, B \subseteq X$}

\STATE $\gamma = conf(r)$

\FOR{each $r':(A \subseteq DI_b), (B \subseteq X) \rightarrow C \in \mathcal{RR}$}
\STATE $\beta_2=conf(r_{b2}:X\rightarrow A)$
\STATE $\Delta_1=supp(r_{b2}:X\rightarrow A)$
\STATE $\delta = conf(B \rightarrow C)$
\STATE $\Delta_2= supp(B \rightarrow A)$
\STATE $\beta_1= \frac{\Delta_1}{\Delta_2}$ \hspace{10 mm}//$conf(r_{b1}: A, B \rightarrow D)$
\STATE Find $\mathcal{D}_c$: all records in $\mathcal{D}$ that completely
support $\neg A, B, \neg D \rightarrow \neg C$

\STATE Steps 6-9 Algorithm~\ref{alg32}

\IF{$r' \in \mathcal{MR}$}
\WHILE{$(\delta \leq \frac{\beta_1(\beta_2+\gamma-1)}{\beta_2 \cdot \alpha})$ and $(\delta \leq \frac{conf(r')}{\alpha})$}
\STATE Select first record $db_c$ in $\mathcal{D}_c$
\STATE Modify the class item of $db_c$ from $\neg C$ to $C$ in $\mathcal{D}$
\STATE Recompute $\delta = conf(B \rightarrow C)$
\ENDWHILE
\ELSE
\WHILE{$\delta \leq \frac{\beta_1(\beta_2+\gamma-1)}{\beta_2 \cdot \alpha}$}
\STATE Steps 15-17 Algorithm~\ref{alg35}
\ENDWHILE

\ENDIF
\ENDFOR

\ENDFOR
\FOR{each $r':(A, B \rightarrow C) \in \mathcal{MR} \setminus \mathcal{RR}$}
\STATE $\delta = conf(B \rightarrow C)$
\STATE Find $\mathcal{D}_c$: all records in $\mathcal{D}$ that completely
support $\neg A, B \rightarrow \neg C$

\STATE Step 12

\WHILE{$(\delta \leq \frac{conf(r')}{\alpha})$}
\STATE Steps 15-17 Algorithm~\ref{alg35}
\ENDWHILE

\ENDFOR

\STATE Output: $\mathcal{D'} = \mathcal{D}$

\end{algorithmic}}

\end{algorithm}

If some rules can be extracted from $\mathcal{D}$ as both
direct and indirect $\alpha$-discriminatory rules,
it means that there is overlap between
$\mathcal{MR}$ and $\mathcal{RR}$; in such case, data
transformation is performed until both the direct and the
indirect rule protection requirements are satisfied (Steps 13-18).
This is possible because, as we showed in Section~\ref{DRP-IRP},
the same data transformation method (Method 2 consisting
of changing the class item) can provide both DRP and IRP.
However, if there is no overlap between $\mathcal{MR}$ and $\mathcal{RR}$,
the data transformation is performed according to Method 2 for IRP,
until the indirect discrimination prevention requirement
is satisfied (Steps 19-23) for each indirect
$\alpha$-discriminatory rule ensuing from
each redlining rule in $\mathcal{RR}$;
this can be done without any negative impact on direct discrimination
prevention, as justified in Section~\ref{DRP-IRP}.
Then, for each direct $\alpha$-discriminatory rule
$r' \in \mathcal{MR} \setminus \mathcal{RR}$
(that is only directly extracted from $\mathcal{D}$),
data transformation for satisfying the direct
discrimination prevention requirement is performed (Steps 26-33),
based on Method 2 for DRP;
this can be done without any negative
impact on indirect discrimination prevention, as justified in
Section~\ref{DRP-IRP}.
Performing rule protection or generalization for each rule in $\mathcal{MR}$ by each of Algorithms~\ref{alg32}-~\ref{alg35} has no adverse effect on protection for other rules ({\em i.e.} rule protection at Step $i+x$ to make $r'$ protective
cannot turn into discriminatory a rule $r$ made 
protective at Step $i$) because of the two following reasons: the kind of data transformation for each rule is the same (change the PD itemset or the class item of records) and there are no two $\alpha$-discriminatory rules $r$ and $r'$ in $\mathcal{MR}$ such that $r=r'$. 

\section{Computational Cost}
\label{sec:time}

The computational cost of Algorithm~\ref{alg32}
can be broken down as follows:

\begin{itemize}
\item Let $m$ be the number of records in $\mathcal{D}$.
The cost of finding subset $\mathcal{D}_c$ (Step 5) is $O(m)$.

\item Let $k$ be the number of rules in $\mathcal{FR}$ and $h$ the
number of records in subset $\mathcal{D}_c$. The cost of computing
$impact(db_c)$ for all records in $\mathcal{D}_c$ (Steps 6-8) is $O(hk)$.

\item The cost of sorting $\mathcal{D}_c$ by ascending impact (Step 9) is $O(h \log h)$. Then, the cost of the {\em impact minimization}
procedure (Steps 6-9) in all algorithms is $O(hk+h \log h)$.
\item During each iteration of the inner loop (Step 10),
the number of records supporting the premise of rule
$r': A, B \rightarrow C$ is increased by one. After $i$ iterations,
the confidence of $r': A, B \rightarrow C$ will
be $conf(r':A, B \rightarrow C)^{(i)}=\frac{N_{ABC}}{N_{AB}+i}$,
where $N_{ABC}$ is the number of records supporting rule $r'$
and $N_{AB}$ is the number of records supporting the premise
of rule $r'$. If we let
$DRP_{req1}=\alpha \cdot conf(B \rightarrow C)$, the inner loop (Step 10)
is iterated until $conf(r': A, B \rightarrow C)^{(i)} < DRP_{req1}$ or
equivalently $\frac{N_{ABC}}{N_{AB}+i} < DRP_{req1}$. This inequality
can be rewritten as $i > (\frac{N_{ABC}}{DRP_{req1}}-N_{BC})$. From this
last inequality we can derive that
$i=\lceil \frac{N_{ABC}}{DRP_{req1}}-N_{BC} \rceil$. Hence,
iterations in the inner loop (Step 10) will stop as soon as the first
integer value greater than (or equal) $\frac{N_{ABC}}{DRP_{req1}}-N_{BC}$
is reached. Then, the cost spent on the inner loop to satisfy the
direct rule protection requirement (Steps 10-14)
will be $O(m*\lceil \frac{N_{ABC}}{DRP_{req1}}-N_{BC} \rceil)$.
\end{itemize}

Therefore, assuming $n$ is the number of $\alpha$-discriminatory
rules in $\mathcal{MR}$ (Step 3), the total computational
time of Algorithm~\ref{alg32} is bounded by $O(n*\{m+hk+h \log h+im\})$,
where $i=\lceil \frac{N_{ABC}}{DRP_{req1}}-N_{BC} \rceil$.

The {\em impact minimization} procedure substantially
increases the complexity. Without computing the impact,
the time complexity of Algorithm~\ref{alg32}
decreases to $O(n*\{m+im\})$. In addition,
it is clear that the execution time of Algorithm~\ref{alg32}
increases linearly with the number $m$ of original data records
as well as the number $k$ of frequent classification rules and
the number $n$ of $\alpha$-discriminatory rules.

The computational cost of the other algorithms can be computed
similarly, with some small differences. In summary,
the total computational time of Algorithm~\ref{alg33} is
also bounded by $O(n*\{m+hk+h\log h+im\})$, where
$i=\lceil ({N_B}*DRP_{req2})-N_{BC} \rceil$, $N_{BC}$ is the number
of records supporting rule $B \rightarrow C$, $N_B$ is the number of
records supporting itemset $B$ and $DRP_{req2}=\frac{conf(r')}{\alpha}$. The computational cost of Algorithm~\ref{alg34} is the same as the last ones
with the difference that $i=\lceil {N_{ABC}-(RG_{req}*N_{AB})}\rceil$,
where $RG_{req}=\frac{conf(r_b)}{d}$, or
$i=\lceil ({N_B}*DRP_{req2})-N_{BC} \rceil$, depending
on whether rule generalization or direct rule protection is performed.

Finally, assuming $f$ is the number
of indirect $\alpha$-discriminatory rules in $\mathcal{RR}$ and $n$ is
the number of direct $\alpha$-discriminatory rules in $\mathcal{MR}$ that
no longer exist in $\mathcal{RR}$, the total computational time of
Algorithm~\ref{alg35} is bounded by $O((f+n)*\{m+hk+h \log h+im\})$, where
$i=\lceil ({N_B}*max_{req})-N_{BC} \rceil$ and
$max_{req} = \max( \frac{\beta_1(\beta_2+\gamma-1)}{\beta_2 \cdot \alpha}, \frac{conf(r')}{\alpha})$.

\section{Experiments}
\label{sec:results}

This section presents the experimental evaluation of the proposed
direct and/or indirect discrimination prevention approaches and algorithms.
To obtain $\mathcal{FR}$ and $\mathcal{BK}$ we used the
Apriori algorithm~\cite{Apriori},
which is a common algorithm to extract frequent rules.
All algorithms and utility measures were implemented using
the C\# programming language. The tests were performed on an 2.27 GHz
Intel\textregistered~Core\texttrademark i3 machine, equipped
with 4 GB of RAM, and running under Windows 7 Professional.

First, we describe the datasets used in our experiments.
Then, we introduce the new utility measures we propose
to evaluate direct and indirect discrimination prevention methods
in terms of their success at discrimination removal and impact on
data quality. Finally, we present the evaluation results of the
different methods and also the comparison between them.

\subsection{Datasets}
\label{sec:datasets}

{\em Adult dataset:} We used the Adult dataset from ~\cite{UCI}, also
known as Census Income, in our experiments. This dataset consists of
48,842 records, split into a ``train'' part with 32,561 records
and a ``test'' part with 16,281 records. The dataset has
14 attributes (without class attribute). We used the ``train''
part in our experiments. The prediction task associated to the
Adult dataset is to determine whether a person makes more than
50K\$ a year 
based on census and demographic information about people. The dataset
contains both categorical and numerical attributes.

For our experiments with the Adult dataset, we set
$DI_b$=\{Sex=Female, Age=Young\}. Although the Age attribute
in the Adult dataset is numerical, we converted it to categorical
by partitioning its domain into two fixed
intervals: Age $\leq$ 30 was renamed as Young and
Age $>$ 30 was renamed as old.

{\em German Credit dataset:} We also used the
German Credit dataset from~\cite{UCI}. This dataset
consists of 1,000 records and 20 attributes (without class attribute)
of bank account holders. This is a well-known real-life dataset,
containing both numerical and categorical attributes.
It has been frequently used in the anti-discrimination
literature ~\cite{peder2008,KamiKAIS}.
The class attribute in the German Credit dataset
takes values representing good or bad classification of the bank
account holders. For our experiments with this dataset,
we set $DI_b$= \{Foreign worker=Yes, Personal Status=Female and not Single,
Age=Old\} (cut-off for Age=Old: 50 years old).

\subsection{Utility Measures}
\label{sec:utility}

Our proposed techniques should be evaluated based on two aspects.
On the one hand, we need to measure the success of the method
in removing all evidence of direct and/or indirect discrimination
from the original dataset; on
the other hand, we need to measure the impact of the method
in terms of information loss ({\em i.e.} data quality loss).
To measure discrimination removal, four metrics were used:
\begin{itemize}

\item {\bf Direct Discrimination Prevention Degree (DDPD)}. This measure
quantifies the percentage of $\alpha$-discriminatory
rules that are no longer $\alpha$-discriminatory in the transformed dataset. We define DDPD as
\[ DDPD=\frac{|\mathcal{MR}|-|\mathcal{MR}'|}{|\mathcal{MR}|} \]
where $\mathcal{MR}$ is the database of $\alpha$-discriminatory rules
extracted from $\mathcal{D}$ and
$\mathcal{MR}'$ is the
database of $\alpha$-discriminatory rules extracted
from the transformed dataset $\mathcal{D}'$. Note
that  $|\cdot|$ is the cardinality operator.

\item {\bf Direct Discrimination Protection Preservation (DDPP)}.
This measure quantifies the percentage of the $\alpha$-protective
rules in the original dataset that remain
$\alpha$-protective in the transformed dataset. It is defined as
\[ DDPP=\frac{|\mathcal{PR} \bigcap \mathcal{PR}'|}{|\mathcal{PR}|} \]



where $\mathcal{PR}$ is the database of $\alpha$-protective rules
extracted from the original dataset $\mathcal{D}$ and
$\mathcal{PR}'$ is the database of $\alpha$-protective rules extracted from the transformed dataset $\mathcal{D}'$.

\item {\bf Indirect Discrimination Prevention Degree (IDPD)}
This measure quantifies the percentage of redlining rules that are no
longer redlining in the transformed dataset. It is defined like DDPD but 
substituting $\mathcal{MR}$ and $\mathcal{MR}'$ with the database of redlining rules
extracted from $\mathcal{D}$ and $\mathcal{D}'$, respectively.

\item {\bf Indirect Discrimination Protection Preservation (IDPP)}
This measure quantifies the percentage of non-redlining
rules in the original dataset that remain non-redlining in the
transformed dataset. It is defined like DDPP but substituting $\mathcal{PR}$ and $\mathcal{PR}'$ with the database of non-redlining extracted from $\mathcal{D}$ and $\mathcal{D}'$, respectively.

\end{itemize}

Since the above measures are used to evaluate the success of the
proposed method in direct and indirect discrimination prevention,
ideally their value should be 100\%. To measure data quality,
we use two metrics proposed in the literature as
information loss measures in the context of
rule hiding for privacy-preserving data mining (PPDM)~\cite{Verykios2008}.
\begin{itemize}
\item {\bf Misses Cost} (MC). This measure quantifies the percentage of
rules among those extractable from the original dataset
that cannot be extracted from the transformed dataset
(side-effect of the transformation process). 

\item {\bf Ghost Cost} (GC). This measure quantifies the
percentage of the rules among those extractable
from the transformed dataset that
were not extractable from the original dataset
(side-effect of the transformation process). 
\end{itemize}

MC and GC should ideally be 0\%. However, MC and GC may not be 0\% as a
side-effect of the transformation process.

\subsection{Empirical Evaluation}
\label{sec:Results}

We implemented
the algorithms for all proposed methods
for direct and/or indirect discrimination prevention,
and we evaluated them in terms of the proposed utility
measures. We report the performance results in this section.

Tables~\ref{tab:utility2} and ~\ref{tab:utility1} show the
utility scores obtained by our methods
on the Adult dataset and the German Credit dataset, respectively.
Within each table, the first row relates 
to the simple approach of deleting discriminatory attributes,  
the next four rows relate
to direct discrimination prevention methods, the next two ones
relate to indirect discrimination prevention methods and the last one
relates to the combination of direct and indirect discrimination.

\begin{table*}[t]\footnotesize
\caption{Adult dataset: Utility measures for minimum support 2\% and confidence 10\% for all the methods. Value ``n.a.'' denotes that the
respective measure is not applicable.}
\label{tab:utility2}
\centering
\scalebox{0.72}{
\begin{tabular}{|c|c|c|c|c|c|c|c|c|c|c|c|}
\hline
\multicolumn{ 1}{|c|}{Methods} & \multicolumn{ 1}{|c|}{$\alpha$} &  \multicolumn{ 1}{|c|}{$p$} &   No. & No. Indirect & No. Direct &      \multicolumn{ 4}{|c|}{Disc. Removal} & \multicolumn{ 2}{|c|}{Data Quality} \\

\multicolumn{ 1}{|c|}{} & \multicolumn{ 1}{|c|}{} & \multicolumn{ 1}{|c|}{} & Redlining  & $\alpha$-Disc.  & $\alpha$-Disc.  & \multicolumn{ 2}{|c|}{Direct} & \multicolumn{ 2}{|c|}{Indirect} &  \multicolumn{ 2}{|c|}{} \\

\multicolumn{ 1}{|c|}{} & \multicolumn{ 1}{|c|}{}  & \multicolumn{ 1}{|c|}{} &    Rules &      Rules &      Rules &       DDPD &       DDPP &       IDPD &       IDPP &         MC &         GC \\
\hline
Removing. Disc. Attributes   &  n.a. &  n.a  &     n.a. &    n.a  &        n.a. &     n.a. &       n.a. &      n.a. &      n.a. &         66.08  &      0  \\
\hline
DRP (Method 1)   &  1.2 &  n.a  &     n.a. &    n.a  &        274 &      100 &        100 &      n.a. &      n.a. &          4.16  &       4.13  \\
\hline
DRP (Method 2)  &  1.2 &  n.a.  &    n.a. &         n.a. &        274 &        100  &        100 &        n.a. &        n.a. &          0 &          0  \\
\hline
DRP (Method 1) + RG &  1.2 &   0.9    &   n.a.  &         n.a. &        274 &        100 &        100 &        n.a. &        n.a. &   4.1  &          4.1 \\
\hline
DRP (Method 2) + RG &  1.2 & 0.9  &    n.a. &       n.a. &         274 &        91.58 &        100 &       n.a. &       n.a. &          0 &       0  \\
\hline
IRP (Method 1) & 1.1 &  n.a.   & 21  &      30  &         n.a. &        n.a. &        n.a. &       100 &   100  & 0.54          &   0.38          \\
\hline
IRP (Method 2) & 1.1 &  n.a.   &  21 &       30  &         n.a. &        n.a. &        n.a. &       100 &       100 &       0   &      0      \\
\hline
DRP(Method 2) + IRP(Method 2) & 1.1 &  n.a.   &  21 &       30  &        280 &        100 &        100 &      100 &       100 &          0 &          0   \\
\hline
\multicolumn{ 6}{|c|}{No of Freq. Class. Rules: 5,092} &                                     \multicolumn{ 6}{|c|}{No. of Back. Know. Rules: 2089} \\
\hline
\end{tabular}
}
\end{table*}

\begin{table*}[t]\footnotesize
\caption{German Credit dataset: Utility measures for
minimum support 5\% and confidence 10\% for all methods.
Value ``n.a.'' denotes that the respective measure is not applicable.}
\label{tab:utility1}
\centering
\scalebox{0.72}{
\begin{tabular}{|c|c|c|c|c|c|c|c|c|c|c|c|}
\hline
\multicolumn{ 1}{|c|}{Methods} & \multicolumn{ 1}{|c|}{$\alpha$} &  \multicolumn{ 1}{|c|}{$p$} &   No. & No. Indirect & No. Direct &      \multicolumn{ 4}{|c|}{Discrimination Removal} & \multicolumn{ 2}{|c|}{Data Quality} \\

\multicolumn{ 1}{|c|}{} & \multicolumn{ 1}{|c|}{} & \multicolumn{ 1}{|c|}{} & Redlining  & $\alpha$-Disc.  & $\alpha$-Disc.  & \multicolumn{ 2}{|c|}{Direct} & \multicolumn{ 2}{|c|}{Indirect} &  \multicolumn{ 2}{|c|}{} \\

\multicolumn{ 1}{|c|}{} & \multicolumn{ 1}{|c|}{}  & \multicolumn{ 1}{|c|}{} &    Rules &      Rules &      Rules &       DDPD &       DDPP &       IDPD &       IDPP &         MC &         GC \\
\hline
Removing. Disc. Attributes   &  n.a. &  n.a  &     n.a. &    n.a  &        n.a. &     n.a. &       n.a. &      n.a. &      n.a. &         64.35  &      0  \\
\hline
DRP (Method 1)   &  1.2 &  n.a  &     n.a. &    n.a  &        991 &      100 &        100 &      n.a. &      n.a. &          15.44  &       13.52  \\
\hline
DRP (Method 2)  &  1.2 &  n.a.  &    n.a. &         n.a. &        991 &        100  &        100 &        n.a. &        n.a. &          0 &          4.06  \\
\hline
DRP (Method 1) + RG &  1.2 &   0.9    &   n.a.  &         n.a. &        991 &        100 &        100 &        n.a. &        n.a. &   13.34  &          12.01 \\
\hline
DRP (Method 2) + RG  & 1.2 & 0.9  &    n.a. &       n.a. &         991 &        100 &        100 &       n.a. &       n.a. &          0.01 &       4.06  \\
\hline
IRP (Method 1) & 1 &  n.a.   &  37 &      42  &        n.a. &   n.a.      &        n.a. &   100 & 100 &  1.62     &  1.47                     \\
\hline
IRP (Method 2) & 1 &  n.a.   &  37 &       42  &         n.a. &        n.a. &        n.a. &       100 &       100 &       0   &      0.96      \\
\hline
DRP(Method 2) + IRP(Method 2) & 1 &  n.a.   &  37 &       42  &        499 &        99.97 &        100 &      100 &       100 &      0 &          2.07   \\
\hline
\multicolumn{ 6}{|c|}{No of Freq. Class. Rules: 32,340} &                                     \multicolumn{ 6}{|c|}{No. of Back. Know. Rules: 22,763} \\
\hline
\end{tabular}
}
\end{table*}

Table~\ref{tab:utility2} shows
the results for minimum support 2\% and minimum confidence 10\%.
Table~\ref{tab:utility1} shows the results for minimum support 5\%
and minimum confidence 10\%.
In Tables~\ref{tab:utility2} and ~\ref{tab:utility1}, the results
of direct discrimination prevention methods are
reported for discriminatory threshold $\alpha=1.2$ and, in the cases
where direct rule protection is applied in combination with rule
generalization, we used $d=0.9$, and
$DI_b$=\{Sex=Female, Age=Young\} in the Adult dataset,
and $DI_b$= \{Foreign worker=Yes, Personal Status=Female and not Single,
Age=Old\} in the German Credit dataset.
In addition,
in Table~\ref{tab:utility2}, the results of the indirect discrimination
prevention methods and both direct and indirect discrimination
prevention are reported for discriminatory
threshold $\alpha=1.1$ and  $DI_b$=\{Sex=Female, Age=Young\};
in Table~\ref{tab:utility1},
these results are reported for $\alpha=1$ and
$DI_b$= \{Foreign worker=Yes\}).

We selected the discriminatory threshold values and $DI_
b$
for each dataset in such a way that the number of redlining
rules and $\alpha$-discriminatory rules extracted from $\mathcal{D}$ could be suitable to test all our methods. In addition to the scores of
utility measures, the number of redlining rules,
the number of indirect $\alpha$-discriminatory rules and the
number of direct $\alpha$-discriminatory rules are also reported in
Tables~\ref{tab:utility2} and ~\ref{tab:utility1}. These tables
also show the number of frequent classification rules found, as well
as the number of background knowledge rules related to this experiment.

As shown in Tables~\ref{tab:utility2} and ~\ref{tab:utility1}, we get
very good results for all methods in terms of discrimination removal:
DDPD, DDPP, IDPD, IDPP are near 100\% for both datasets.
In terms of data quality, the best results for direct discrimination
prevention are obtained with Method 2 for DRP or Method 2 for DRP
combined with Rule Generalization. The best results for indirect
discrimination prevention are obtained with Method 2 for IRP.
This shows that lower information loss is obtained with the methods
changing the class item ({\em i.e.} Method 2) than with those
changing the discriminatory itemset ({\em i.e.} Method 1).
As mentioned above, in direct discrimination prevention,
rule generalization cannot be applied alone and must
be applied in combination with direct rule protection; however, direct
rule protection can be applied alone. The results in the last row of the above tables
({\em i.e.} Method 2 for DRP +
Method 2 for IRP) based on Algorithm~\ref{alg35} for the case
of simultaneous direct and indirect
discrimination demonstrate that the proposed solution achieves
a high degree of simultaneous direct and indirect discrimination removal
with very little information loss.

For all methods, Tables~\ref{tab:utility2} and~\ref{tab:utility1}
show that we obtained lower information loss in terms of MC and GC in the
Adult dataset than in the German Credit dataset.
In terms of discrimination removal, results on both datasets were
almost the same. 
In addition, the highest value of information loss is obtained by the simple approach of removing discriminatory attributes (first row of each table): 
as it could be expected, entirely suppressing the PD attributes 
is much more information-damaging than modifying 
the values of these attributes in a few records. 

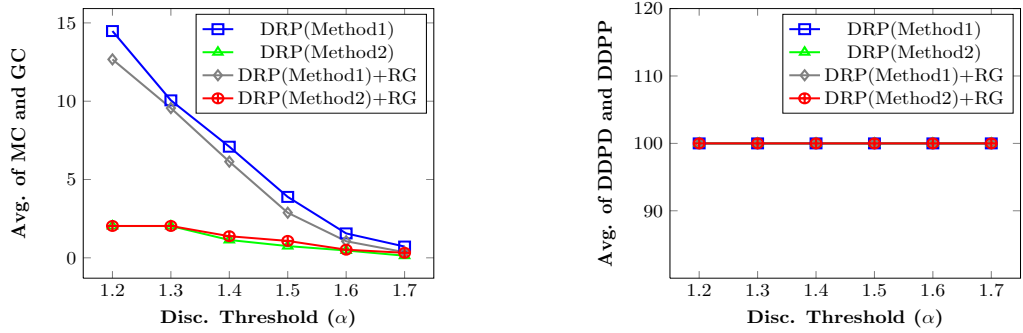
\begin{figure*}
\centering
\begin{tikzpicture}[thick, scale=1]
\def \plotdist {7.8 cm} 
\def \xsc {0.68} 
\def \ysc {0.63} 
\begin{scope}[xscale=\xsc , yscale=\ysc]
\begin{axis}[
        xlabel={\bf Disc. Threshold
 ($\alpha$)},
        ylabel={\bf Avg. of MC and GC
} ,
        ]
    \addplot[mark=square,blue,mark size=3pt,very thick] plot coordinates {
        (1.2,14.48)
        (1.3,10.055)
        (1.4,7.095)
        (1.5,3.89)
        (1.6,1.56)
        (1.7,0.72)
         
    };
    \addlegendentry{DRP(Method1)}
    \addplot[mark=triangle,green,mark size=3pt,very thick]
        plot coordinates {
          (1.2,2.03)
        (1.3,2.03)
        (1.4,1.14)
        (1.5,0.75)
        (1.6,0.47)
        (1.7,0.135)
        };
    \addlegendentry{DRP(Method2)}
    \addplot[mark=diamond,gray,mark size=3pt,very thick] plot coordinates {
        (1.2,12.67)
        (1.3,9.56)
        (1.4,6.14)
        (1.5,2.88)
        (1.6,1.06)
        (1.7,0.36)
    };
    \addlegendentry{DRP(Method1)+RG}
 \addplot[mark=oplus,red,mark size=3pt,very thick]
        plot coordinates {
        (1.2,2.035)
        (1.3,2.035)
        (1.4,1.38)
        (1.5,1.08)
        (1.6,0.52)
        (1.7,0.32)
        };
    \addlegendentry{DRP(Method2)+RG}
    \end{axis}
    \end{scope}
\begin{scope}[xshift=\plotdist, xscale=\xsc , yscale=\ysc]
\begin{axis}[
        xlabel={\bf Disc. Threshold ($\alpha$)
 },
        ylabel={\bf Avg. of DDPD and DDPP},
        ]
    \addplot[mark=square,blue,mark size=3pt,very thick] plot coordinates {
        (1.2,100)
        (1.3,100)
        (1.4,100)
        (1.5,100)
        (1.6,100)
        (1.7,100)
    };
    \addlegendentry{DRP(Method1)}
  \addplot[mark=triangle,green,mark size=3pt,very thick] plot coordinates {
       (1.2,100)
        (1.3,100)
        (1.4,100)
        (1.5,100)
        (1.6,100)
        (1.7,100)
      };
    \addlegendentry{DRP(Method2)}
    \addplot[mark=diamond,gray,mark size=3pt,very thick]
        plot coordinates {
        (1.2,100)
        (1.3,100)
        (1.4,100)
        (1.5,100)
        (1.6,100)
        (1.7,100)
        };
    \addlegendentry{DRP(Method1)+RG}
  \addplot[mark=oplus,red,mark size=3pt,very thick]
        plot coordinates {
        (1.2,100)
        (1.3,100)
        (1.4,100)
        (1.5,100)
        (1.6,100)
        (1.7,100)
        };
    \addlegendentry{DRP(Method2)+RG}
    \end{axis}
    \end{scope}
\end{tikzpicture}
\caption{Information loss (left) and discrimination removal degree (right)
for direct discrimination prevention methods for $\alpha \in [1.2, 1.7]$. DRP(Method $i$): Method $i$ for DRP; RG: Rule Generalization.
}
\label{fig:results1}
\end{figure*}

After the above general results and comparison between methods, we now present more specific
results on each method for different parameters $\alpha$ and $d$.
Figure~\ref{fig:results1} shows on the left the degree of information
loss (as average of MC and GC) and on the right
the degree of discrimination removal (as average of DDPD and DDPP)
of direct discrimination prevention methods for the German Credit
dataset when the value of the discriminatory threshold $\alpha$
varies from 1.2 to 1.7, $d$ is 0.9, the minimum support is 5\%
and the minimum confidence is 10\%. The number of direct
$\alpha$-discriminatory rules extracted from the dataset is 991
for $\alpha=1.2$, 415 for $\alpha=1.3$, 207 for
$\alpha=1.4$, 120 for $\alpha=1.5$, 63 for $\alpha=1.6$ and 30 for
$\alpha=1.7$, respectively. As shown in Figure~\ref{fig:results1}, the degree of discrimination removal provided by all
methods for different values of $\alpha$ is also 100\%. However,
the degree of information loss decreases substantially as $\alpha$ increases;
the reason is that, as $\alpha$ increases, the number of
$\alpha$-discriminatory rules to be dealt with decreases.
In addition, as shown in Figure ~\ref{fig:results1}, the lowest
information loss for most values of $\alpha$ is obtained by Method 2 for
DRP.

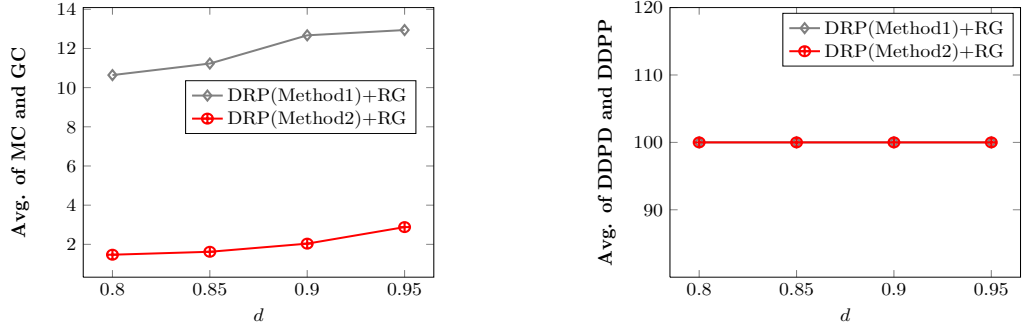
\begin{figure*}
\centering
\begin{tikzpicture}[thick, scale=1]
\def \plotdist {7.8 cm} 
\def \xsc {0.68} 
\def \ysc {0.63} 
\begin{scope}[xscale=\xsc , yscale=\ysc]
\begin{axis}[
legend style={
at={(0.95,0.73)},
},
        xlabel={\bf $d$},
        ylabel={\bf Avg. of MC and GC
},
        ]
%
    \addplot[mark=diamond,gray,mark size=3pt,very thick] plot coordinates         {
        (0.8,10.64)
        (0.85,11.23)
        (0.9,12.67)
        (0.95,12.94)
        
    };
    \addlegendentry{DRP(Method1)+RG}
 \addplot[mark=oplus,red,mark size=3pt,very thick]
        plot coordinates {
         (0.8,1.47)
        (0.85,1.62)
        (0.9,2.035)
        (0.95,2.88)

        };
    \addlegendentry{DRP(Method2)+RG}
    \end{axis}
    \end{scope}
\begin{scope}[xshift=\plotdist, xscale=\xsc , yscale=\ysc]
\begin{axis}[
        xlabel={\bf $d$
 },
        ylabel={\bf Avg. of DDPD and DDPP},
        ]
    \addplot[mark=diamond,gray,mark size=3pt,very thick]
        plot coordinates {
       (0.8,100)
        (0.85,100)
        (0.9,100)
        (0.95,100)
        };
    \addlegendentry{DRP(Method1)+RG}
  \addplot[mark=oplus,red,mark size=3pt,very thick]
        plot coordinates {
        (0.8,100)
        (0.85,100)
        (0.9,100)
        (0.95,100)
        };
    \addlegendentry{DRP(Method2)+RG}
    \end{axis}
    \end{scope}
\end{tikzpicture}
\caption{Information loss (left) and discrimination removal (right) degree for direct discrimination prevention
methods for $d \in [0.8, 0.95]$.
DRP(Method$i$): Method $i$ for DRP; RG: Rule Generalization.
}
\label{fig:results3}
\end{figure*}

In addition, to demonstrate the impact of varying $d$ on
the utility measures in the methods
using Rule Generalization, Figure~\ref{fig:results3} (left) shows the
degree of information loss and Figure~\ref{fig:results3} (right) shows the degree of discrimination removal for different values of $d$ (0.8, 0.85, 0.9, 0.95)
and $\alpha$=1.2 for the German Credit dataset. Although the values
of DDPD and DDPP achieved for different values of $p$ remain almost the same,
increasing the value of $d$ leads to an increase of MC and GC because,
to cope with the rule generalization requirements, more data records must
be changed.


\begin{table*}[t]\footnotesize
\caption{Adult dataset: Utility measures for minimum support 2\% and
confidence 10\% for direct and indirect rule protection;
columns show the results for different values of $\alpha$.
Value ``n.a.'' denotes that the respective measure is not applicable.}
\label{tab:utility3}
\centering
\scalebox{0.9}{
\begin{tabular}{|c|c|c|c|c|c|c|c|c|c|}
\hline
\multicolumn{ 1}{|c|}{$\alpha$} &     No. of & No. of Indirect & No. of Direct &      \multicolumn{ 4}{|c|}{Discrimination Removal} & \multicolumn{ 2}{|c|}{Data Quality}  \\

\multicolumn{ 1}{|c|}{} & redlining  & $\alpha$-Disc.  & $\alpha$-Disc.  & \multicolumn{ 2}{|c|}{Direct} & \multicolumn{ 2}{|c|}{Indirect} &  \multicolumn{ 2}{|c|}{} \\

\multicolumn{ 1}{|c|}{} &      rules &      Rules &      rules &       DDPD &       DDPP &       IDPD &       IDPP &         MC &         GC         \\
\hline
       $\alpha$=1 &         43 &         71 &        804 &      89.45 &        100 &      95.35 &      100 &          0 &       0.03  \\
\hline
     $\alpha$=1.1 &         21 &         30 &        280 &        100 &        100 &        100 &        100 &          0 &          0  \\
\hline
     $\alpha$=1.2 &          9 &         14 &        140 &        100 &        100 &        100 &        100 &          0 &          0 \\
\hline
     $\alpha$=1.3 &       0 &       0 &         67 &        100 &        100 &       n.a. &       100 &          0 &       0.01 \\
\hline
     $\alpha$=1.4 &       0 &       0 &         32 &        100 &        100 &       n.a. &       100 &          0 &          0  \\
\hline
     $\alpha$=1.5 &       0 &       0 &          7 &        100 &        100 &       n.a. &       100 &          0 &          0 \\
\hline
\multicolumn{ 4}{|c|}{No of Freq. Class. Rules: 5,092} &                                     \multicolumn{ 6}{|c|}{No. of Back. Know. Rules: 2,089} \\
\hline
\end{tabular}
}
\end{table*}

\begin{table*}[t]\footnotesize
\caption{German Credit dataset: Utility measures for minimum support 5\% and confidence 10\% for direct and indirect rule protection; columns show the results for different values of $\alpha$. Value ``n.a.'' denotes that the respective measure is not applicable.}
\label{tab:utility4}
\centering
\scalebox{0.9}{
\begin{tabular}{|c|c|c|c|c|c|c|c|c|c|}
\hline
\multicolumn{ 1}{|c|}{$\alpha$} &     No. of & No. of Indirect & No. of Direct &      \multicolumn{ 4}{|c|}{Discrimination Removal} & \multicolumn{ 2}{|c|}{Data Quality} \\

\multicolumn{ 1}{|c|}{} & redlining  & $\alpha$-Disc.  & $\alpha$-Disc.  & \multicolumn{ 2}{|c|}{Direct} & \multicolumn{ 2}{|c|}{Indirect} &  \multicolumn{ 2}{|c|}{} \\

\multicolumn{ 1}{|c|}{} &      rules &      Rules &      rules &       DDPD &       DDPP &       IDPD &       IDPP &         MC &         GC \\
\hline
       $\alpha$=1 &         37 &         42 &        499 &      99.97 &        100 &      100 &      100 &          0 &       2.07  \\
\hline
     $\alpha$=1.1 &         0 &         0 &        312 &        100  &        100 &        n.a. &        100 &          0 &          2.07  \\
\hline
     $\alpha$=1.2 &          0 &         0 &        26 &        100 &        100 &        n.a. &        100 &          0 &          1.01 \\
\hline
     $\alpha$=1.3 &      0 &       0 &         14 &        100 &        100 &       n.a. &       100 &          0 &       1.01  \\
\hline
     $\alpha$=1.4 &       0 &       0  &         9 &        100 &        100 &       n.a. &       100 &          0 &          0.69   \\
\hline
\multicolumn{ 4}{|c|}{No of Freq. Class. Rules: 32,340} &                                     \multicolumn{ 6}{|c|}{No. of Back. Know. Rules: 22,763} \\
\hline
\end{tabular}
}
\end{table*}

Tables~\ref{tab:utility3} and ~\ref{tab:utility4} show the utility measures
obtained by running Algorithm~\ref{alg35} to achieve 
simultaneous direct and indirect discrimination prevention
({\em i.e.} Method 2 for DRP+ Method 2 for IRP) on the Adult and German credit
datasets, respectively.
In Table~\ref{tab:utility3} the results are reported for
different values of $\alpha \in [1, 1.5]$; in Table~\ref{tab:utility4}
different values of $\alpha \in [1, 1.4]$ are considered.
We selected these $\alpha$ intervals in such a way that, with respect to
the predetermined discriminatory items in this experiment for the Adult dataset
({\em i.e.} $DI_b$=\{Sex=Female, Age=Young\}) and the German Credit dataset ({\em i.e.} $DI_b$= \{Foreign worker=Yes\}),
both direct $\alpha$-discriminatory and redlining rules could be extracted.
The reason is that we need to detect some cases with
both direct and indirect discrimination to be able to test our method.
Moreover, we restricted the lower bound to limit the number
of direct $\alpha$-discriminatory and redlining rules. In addition
to utility measures, the number of redlining rules, the number of
indirect $\alpha$-discriminatory rules and the number
of direct $\alpha$-discriminatory rules are also reported for
different values of $\alpha$. 

The values of both direct discrimination removal measures
({\em i.e.} DDPD and DDPP) and indirect discrimination removal
measures ({\em i.e.} IDPD and IDPP) shown in Tables~\ref{tab:utility3}
and ~\ref{tab:utility4} demonstrate that the proposed solution achieves
a high degree of both direct and indirect discrimination prevention for
different values of the discriminatory threshold. The important point is that,
by applying the proposed method, we get good results for both
direct and indirect discrimination prevention at the same time.
In addition, the values of MC and GC demonstrate that
the proposed solution incurs low information loss. 

Tables~\ref{tab:utility3} and ~\ref{tab:utility4} show that
we obtained lower information loss in terms of the GC measure in the
Adult dataset than in the German Credit dataset.
Another remark on these tables is that,
although no redlining rules are detected
in the Adult dataset for $\alpha \geq 1.3$ and in the
German Credit dataset for $\alpha \geq 1.1$,
the IDPP measure is computed and reported to show that in the
cases where only direct discrimination exists, the elimination
of direct discrimination by Algorithm~\ref{alg35}
does not have a negative impact on indirect discrimination
({\em i.e.} non-redlining rules do not become redlining rules).


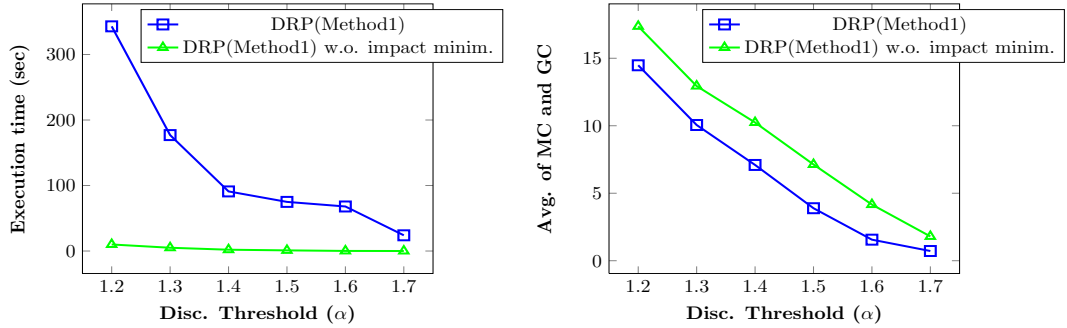
\begin{figure*}
\centering
\begin{tikzpicture}[thick, scale=1]
\def \plotdist {7 cm} 
\def \xsc {0.68} 
\def \ysc {0.63} 
\begin{scope}[xscale=\xsc , yscale=\ysc]
\begin{axis}[
legend style={
at={(1.2,0.98)},
},
        xlabel={\bf Disc. Threshold
 ($\alpha$)},
        ylabel={\bf Execution time (sec)
} ,
        ]
    \addplot[mark=square,blue,mark size=3pt,very thick] plot coordinates {
        (1.2,343)
        (1.3,177)
        (1.4,91)
        (1.5,75)
        (1.6,68)
        (1.7,24)
         
    };
    \addlegendentry{DRP(Method1)}
    \addplot[mark=triangle,green,mark size=3pt,very thick]
        plot coordinates {
        (1.2,10)
        (1.3,5)
        (1.4,2)
        (1.5,1)
        (1.6,0.15)
        (1.7,0.027)
        };
    \addlegendentry{DRP(Method1) w.o. impact minim.}
    \end{axis}
    \end{scope}
\begin{scope}[xshift=\plotdist, xscale=\xsc , yscale=\ysc]
\begin{axis}[
legend style={
at={(1.3,0.98)},
},
        xlabel={\bf Disc. Threshold ($\alpha$)
 },
        ylabel={\bf Avg. of MC and GC},
        ]
    \addplot[mark=square,blue,mark size=3pt,very thick] plot coordinates {
        (1.2,14.48)
        (1.3,10.055)
        (1.4,7.095)
        (1.5,3.89)
        (1.6,1.56)
        (1.7,0.72)
    };
    \addlegendentry{DRP(Method1)}
  \addplot[mark=triangle,green,mark size=3pt,very thick] plot coordinates {
       (1.2,17.365)
        (1.3,12.94)
        (1.4,10.24)
        (1.5,7.13)
        (1.6,4.16)
        (1.7,1.8)
      };
    \addlegendentry{DRP(Method1) w.o. impact minim.}
    \end{axis}
    \end{scope}
\end{tikzpicture}
\caption{Execution times (left) and Information loss degree (right) of Method 1 for DRP for $\alpha \in [1.2, 1.7]$ with and without impact minimization.}
\label{fig:results4}
\end{figure*}

Figure~\ref{fig:results4} illustrates
the effect of the {\em impact minimization} procedure, described
in Section~\ref{sec:DDPAs}, on execution times and information loss of
Method 1 for DRP, respectively. As shown in this figure (right)
{\em impact minimization} has a noticeable
effect on information loss (decreasing MC and GC). However, as discussed
in Section~\ref{sec:time} and shown in Figure~\ref{fig:results4} (left),
{\em impact minimization} substantially increases the execution time
of the algorithm. For other methods, the same happens.
Figure~\ref{fig:results4} (left) also shows that,
by increasing $\alpha$, the number of $\alpha$-discriminatory rules
and hence the execution time are decreased. 
Additional experiments are presented in 
the Appendix to show the effect of varying the minimum support 
and the minimum confidence on the proposed techniques. 

\begin{table*}[t]\footnotesize
\renewcommand{\arraystretch}{1.2}
\caption{Adult dataset: number of frequent classification rules and $\alpha-$discriminatory rules found during the tests, for minimum confidence 10\% and different values of minimum support (2\%, 5\% and 10\%)}
\label{tab:utility5}
\centering
\begin{tabular}{|c|ccc|}
\hline
           & \multicolumn{ 3}{|c|}{{\bf No. of $\alpha$-disc. rules}} \\

 {\bf $\alpha$} &       2\% &       5\% &     10\%\\

\hline
     
       1.2 &        274 &       46 &       27 \\
\hline
No. of Freq. Class. Rules &      5,092 &      1,646 &     545\\
\hline
\end{tabular}  

\end{table*}

\begin{table*}[t]\footnotesize
\renewcommand{\arraystretch}{1.2}
\caption{Adult dataset: utility measures for minimum confidence 10\%, $\alpha$=1.2 and $d=0.9$; columns show the results for different values of minimum 
support (2\%, 5\% and 10\%) and different methods.}
\label{tab:utility6}
\centering
\scalebox{0.87}{
\begin{tabular}{|c|ccc|ccc|ccc|ccc|}
\hline
           &                    \multicolumn{ 3}{|c}{{\bf MC}} &                    \multicolumn{ 3}{|c}{{\bf GC}} &                   \multicolumn{ 3}{|c}{{\bf DDPD}} &                   \multicolumn{ 3}{|c|}{{\bf DDPP}} \\

Methods &    2\% &    5\% &    10\%  &   2\% &    5\% &    10\% &        2\% &    5\% &    10\%  &    2\% &    5\% &    10\% \\
\hline
\hline

DRP (Method 1) &    4.16    &   2.91  &   1.61    &   4.13    &  3.21     &       0.39  & 100    & 100     & 100     &  100    &    100   &   100    \\

DRP (Method 2)  &     0   &   0   &    0   & 0  &   0     &  0      &    100   &  100    & 100      &    100  &   100    & 100  \\

DRP (Method 1) + RG &       4.1 &     2.67 &     1.61 &      4.1 &       3.26 &      0.39 &    100 &      100 &      100 &      100 &      100 &      100\\

DRP (Method 2) + RG &       0 &       0 &      0&       0 &     0 &      0 &            91.58  &       100 &       100 &      100 &      100 &      100       \\
\hline
\end{tabular}  
}
\end{table*}

As shown in Table~\ref{tab:utility5}, different values of the minimum 
support have an impact on the number of frequent rules and hence on 
the number of $\alpha$-discriminatory rules. As shown in
Table~\ref{tab:utility6}, by increasing the value of the
minimum support the information loss degrees (MC and GC) achieved 
by the different techniques decrease. Meanwhile, as shown 
in Table~\ref{tab:utility6}, the discrimination removal degrees 
(DDPD and DDPP) achieved by the different techniques remain 
the same (discrimination removal is maximum) 
for different values of the minimum support.


\begin{figure*}
\centering
\begin{tikzpicture}[thick, scale=1]
\def \plotdist {7.8 cm} 
\def \xsc {0.68} 
\def \ysc {0.63} 
\begin{scope}[xscale=\xsc , yscale=\ysc]
\begin{axis}[
legend style={
at={(0.99,0.4)},
},
        xlabel={\bf Minimum confidence (\%)
},
        ylabel={\bf Avg. of MC and GC
} ,
        ]
    \addplot[mark=square,blue,mark size=3pt,very thick] plot coordinates {
        (10,4.145)
        (20,3.685)
        (30,4.625)
        (40,4.135)
        (50,4.795)
        (60,5.36)
        (70,7.39)
        (80,3.665)
        (90,4.685)
         
    };
    \addlegendentry{DRP(Method1)}
    \addplot[mark=triangle,green,mark size=3pt,very thick]
        plot coordinates {
        (10,0)
        (20,0.225)
        (30,1.79)
        (40,3.51)
        (50,5.02)
        (60,6.075)
        (70,6.605)
        (80,5.135)
        (90,4.545)

        };
    \addlegendentry{DRP(Method2)}
    \addplot[mark=diamond,gray,mark size=3pt,very thick] plot coordinates {
         (10,4.1)
        (20,3.615)
        (30,4.56)
        (40,3.975)
        (50,4.735)
        (60,5.3)
        (70,6.625)
        (80,3.63)
        (90,4.545)
    };
    \addlegendentry{DRP(Method1)+RG}
 \addplot[mark=oplus,red,mark size=3pt,very thick]
        plot coordinates {
        (10,0)
        (20,0.22)
        (30,1.65)
        (40,3.37)
        (50,4.92)
        (60,6)
        (70,6.58)
        (80,5.1)
        (90,4.5)
        };
    \addlegendentry{DRP(Method2)+RG}
    \end{axis}
    \end{scope}
\begin{scope}[xshift=\plotdist, xscale=\xsc , yscale=\ysc]
\begin{axis}[
        xlabel={\bf Minimum confidence (\%)
 },
        ylabel={\bf Avg. of DDPD and DDPP},
        ]
    \addplot[mark=square,blue,mark size=3pt,very thick] plot coordinates {
        (10,100)
        (20,100)
        (30,100)
        (40,100)
        (50,100)
        (60,100)
        (70,100)
        (80,100)
        (90,100)
    };
    \addlegendentry{DRP(Method1)}
  \addplot[mark=triangle,green,mark size=3pt,very thick] plot coordinates {
       (10,100)
        (20,100)
        (30,100)
        (40,100)
        (50,100)
        (60,100)
        (70,100)
        (80,100)
        (90,100)
      };
    \addlegendentry{DRP(Method2)}
    \addplot[mark=diamond,gray,mark size=3pt,very thick]
        plot coordinates {
        (10,100)
        (20,100)
        (30,100)
        (40,100)
        (50,100)
        (60,100)
        (70,100)
        (80,100)
        (90,100)
        };
    \addlegendentry{DRP(Method1)+RG}
  \addplot[mark=oplus,red,mark size=3pt,very thick]
        plot coordinates {
       (10,100)
        (20,100)
        (30,100)
        (40,100)
        (50,100)
        (60,100)
        (70,100)
        (80,100)
        (90,100)
        };
    \addlegendentry{DRP(Method2)+RG}
    \end{axis}
    \end{scope}
\end{tikzpicture}
\caption{Adult dataset: Information loss (left) and discrimination removal degree (right) for discrimination prevention methods for minimum support=2\%, 
$\alpha=1.2$, $p=0.9$ and minimum confidence in $[10, 90]$. DRP(Method $i$): Method $i$ for DRP; RG: Rule Generalization.}
\label{fig:resultsapp}
\end{figure*}
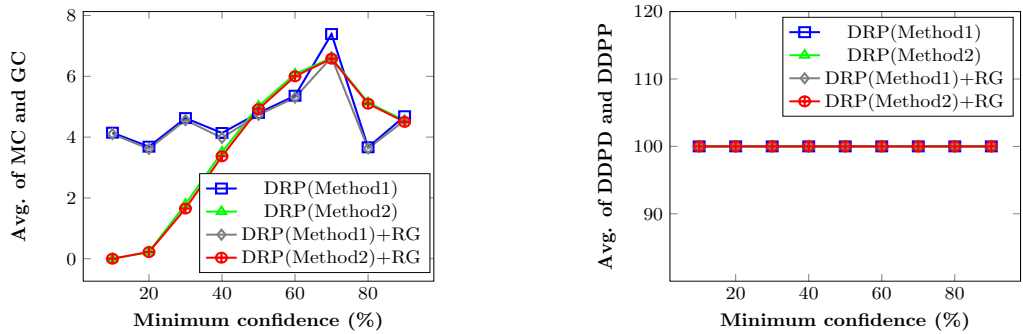

As shown in the left-hand side graph of 
Figure~\ref{fig:resultsapp}, different values of minimum confidence 
have a non-uniform impact on the information loss degree 
(average of MC and MC): sometimes increasing the minimum confidence
can decrease the information loss degree and sometimes it 
can increase the information loss degree.
On the other hand, the right-hand side graph of Figure~\ref{fig:resultsapp}
shows that the average of the discrimination removal degrees DDPD and DDPP 
achieved by different techniques remains the same 
(discrimination removal is maximum) for different values of the
minimum confidence.

\section{Conclusions}
\label{sec:conclusions}

Along with privacy, discrimination is a very important issue when
considering the legal and ethical aspects of data mining.
It is more than obvious that most people do not want to be discriminated
because of their gender, religion, nationality, age and so on,
especially when those attributes are used for making decisions about
them like giving them a job, loan, insurance, etc. The perception of discrimination,
just like the perception of privacy, strongly depends on the legal
and cultural conventions of a society.

The purpose of this chapter was to develop a new pre-processing
discrimination prevention methodology including different data
transformation methods that can prevent direct discrimination,
indirect discrimination or both of them at the same time.
To attain this objective, the first step is to measure
discrimination and identify categories and groups of individuals that
have been directly and/or indirectly discriminated in the decision-making
processes; the second step is to transform data in the proper way to
remove all those discriminatory biases. Finally, discrimination-free
data models can be produced from the transformed dataset without seriously
damaging data quality. The experimental results
reported demonstrate that the proposed techniques are quite successful
in both goals of removing discrimination and preserving data quality. We show that indirect
discrimination removal can help direct discrimination removal.

\chapter[Discrimination- and Privacy-aware Frequent Pattern
Discovery]{Discrimination- and Privacy-aware Frequent Pattern
Discovery}
\chaptermark{Discrimination- and Privacy-aware FP Discovery}
\label{chap:DPAFPD}


Although methods independently addressing privacy
or discrimination in data mining have been proposed in the literature,
in this thesis we argue that privacy and discrimination risks should be tackled {\em together}, and we present a methodology for
doing so while publishing frequent pattern mining results.
We describe a set of pattern sanitization methods, one for each discrimination measure used in the legal literature, to achieve a fair publishing of frequent patterns in combination with a privacy transformation based on $k$-anonymity.
Our proposed pattern sanitization methods yield both privacy- and discrimination-protected patterns, while introducing reasonable (controlled) pattern distortion. We also explore the possibility to combine anti-discrimination with differential privacy instead of $k$-anonymity. Finally, the effectiveness of our proposals is assessed by extensive experiments.

\section{Introduction}
\label{sec:INTROFPD}

Up to now, PPDM and DPDM have been studied in isolation. We argue in this thesis that, in significant data mining processes, privacy and anti-discrimination protection should be addressed {\em together}. Consider the case in which a set of patterns extracted (mined) from the personal data of a population of individual persons is released for subsequent use in a decision making process, such as, {\em e.g.}, granting or denying credit. First, the set of patterns may reveal sensitive information about individual persons in the training population. Second, decision rules based on such patterns may lead to unfair discrimination, depending on what is represented in the training cases. The following example illustrates this point.

\subsection{Motivating Example}
\label{sec:MOTIVEXAMPLE1}

Assume a credit institution, {\em e.g.}, a bank, wants to release among its employees the rules to grant/deny credit, for the purpose of supporting future decision making. Assume that such rules have been mined from decision records accumulated during the past year in a certain city, such as those illustrated in Table \ref{tab:51}.  Consider two options:
\begin{itemize}
\item{\em Protection against the privacy threat only}. Only rules used in at least $k$ different
credit applications are published, in order to protect applicants' privacy according to $k$-anonymity.
This would allow releasing a
rule such that
{\em Sex = female $\rightarrow$ Credit\_approved = no}
if $k$ or more female applicants have been denied credit.
Clearly, using such a rule for credit scoring
is discriminatory against women.
\item {\em Protection against the discrimination threat only}. Discriminatory rules are sanitized, in order to prevent discrimination against female applicants. However, one could publish high-support high-confidence rules such as
{\em Job = veterinarian, salary $> 15000$ $\rightarrow$  Credit\_approved = yes} and {\em Job = veterinarian $\rightarrow$ Credit\_approved = yes}.
Assuming that the first rule holds for 40 people and
the second one for 41, their release would reveal
that there is only one veterinarian in the city that
has been granted credit even if s/he makes no more
than \euro 15000 a year. This is a potential privacy violation,
that is, a probable disclosure of the applicant's identity, and therefore of his/her income level.
\end{itemize}
\begin{table}[ht]
\caption{A data table of personal decision records}
\label{tab:51}
\centering
\begin{tabular}{|c|c|c|c|c|}
\hline
{\bf Sex} & {\bf Job} & {\bf Credit\_history} & {\bf Salary} & {\bf Credit\_approved} \\
\hline
Male & Writer & No-taken &  ... \euro & Yes  \\
\hline
Female & Lawyer & Paid-duly & ... \euro & No \\
\hline
Male & Veterinary & Paid-delay & ...\euro &  Yes  \\
\hline
... & ...& ...& ...&...\\
\hline
\end{tabular}
\end{table}
This simple example shows that protecting both privacy and non-discrimination
is needed when disclosing a set of patterns. The next question is: why not simply apply known techniques for PPDM and DPDM one after the other? We show in this chapter that this straightforward sequential approach does not work
in general: we have no guarantee that applying a
DPDM method after a PPDM one
preserves the desired privacy guarantee, because the DPDM sanitization of a set of patterns may destroy the effect of the earlier PPDM sanitization (and the other way round).
We therefore need a combined, holistic method capable of addressing the two goals together, so that we can safely publish the patterns resulting from a data mining process over a dataset of personal information, while keeping the distortion of the extracted patterns as low as possible. A truly trustworthy technology for knowledge discovery should face both privacy and discrimination threats as two sides of the same coin.
This line of reasoning also permeates the comprehensive reform of
the data protection law proposed in 2012 by the European Commission, currently under approval by the European Parliament, which introduces measures based on
profiling and discrimination within a broader concept of privacy and personal data\footnote{\url{http://ec.europa.eu/justice/newsroom/data-protection/news/120125_en.htm}}.

\subsection{Contributions}
\label{sec:CONSFPD}

The contributions of this chapter, towards the above stated aim, are summarized as follows. First, we define a natural scenario of pattern mining from personal data records containing sensitive attributes, potentially discriminatory attributes and decision attributes, and characterize the problem statement of publishing a collection of patterns which is at the same time both privacy-protected and discrimination-free. Second, we propose
new pattern sanitization methods for discrimination prevention when publishing frequent patterns. Moreover, we take into consideration the so-called genuine occupational requirement, {\em i.e.}, the fact that some apparent discriminatory rule may be actually explained by other admissible factors that are not specified explicitly in the rule. We propose an algorithm to make the frequent pattern protected against unexplainable discrimination only. 
Third, we propose the notion of $\alpha$-protective $k$-anonymous
({\em i.e.}, discrimination- and privacy-protected) patterns
to thwart both privacy and discrimination threats in a collection
of published frequent patterns. Fourth, we present a combined
pattern sanitization algorithm
to obtain an $\alpha$-protective $k$-anonymous version of the original pattern sets. In addition, we show how simultaneous anti-discrimination and privacy can be achieved in frequent pattern discovery while satisfying differential privacy, as an alternative. Fifth, we theoretically and empirically show that the proposed algorithm is effective at protecting against
both privacy and discrimination threats while introducing reasonable
(controlled) pattern distortion.

\section{Privacy-aware Frequent Pattern Discovery} 
 \label{sec:privacy}

In this section, we first describe the notion of $k$-anonymous frequent patterns and then we present a method to obtain a $k$-anonymous version of an
original pattern set.
\subsection{Anonymous Frequent Pattern Set}
\label{sec:anonymouspatt}

Given a support threshold $\sigma$, an itemset $X$ is called {\em $\sigma$-frequent} in a database $\mathcal{D}$ if $supp_\mathcal{D}(X) \geq \sigma$. A $\sigma$-frequent itemset is also called {\em  $\sigma$-frequent pattern}. The collection of all $\sigma$-frequent patterns in $\mathcal{D}$ is denoted by $\mathcal{F}(\mathcal{D},\sigma)$. The frequent pattern mining problem is formulated as follows: given a database $\mathcal{D}$ and a support threshold $\sigma$, find all $\sigma$-frequent patterns, {\em i.e.} the collection $\mathcal{F}(\mathcal{D},\sigma)$. 
Several algorithms have been proposed for finding $\mathcal{F}(\mathcal{D},\sigma)$. In this chapter we use the Apriori algorithm \cite{Apriori}, which is
a very common choice.

In \cite{AtzoriBGP08}, the notion of $k$-anonymous patterns is defined
as follows: a collection of patterns is
$k$-anonymous if each pattern $p$ in it is $k$-anonymous ({\em i.e.}  $supp(p)=0$ or $supp(p) \geq k$)
as well as any further pattern whose support can be
inferred from the collection.
The authors introduce a possible attack that exploits
non $k$-anonymous patterns whose support can be inferred from the collection.
Then they propose a framework for sanitizing 
patterns
and block this kind
of attacks.
\begin{example}
\label{exp:51}
Consider again the motivating example and take $k=8$.
The two patterns $p_1$:{\em \{Job=veterinarian, Credit\_approved=yes\}}
and $p_2$: {\em \{Job=veterinarian, Salary $>$ 15000, Credit\_approved=yes\}} are $8$-anonymous because $supp(p_2)=40 > 8$ and $supp(p_1)=41 > 8$.
However, an attacker can exploit a non-8-anonymous pattern {\em \{Job = veterinarian, $\neg$ (Salary $>$ 15000), Credit\_approved=yes\}}, whose support he
infers from $supp(p_1)-supp(p_2)=41-40=1$.
\hfill $\Box$
\end{example}
In order to check whether a collection of patterns is $k$-anonymous,
in \cite{AtzoriBGP08} the {\em inference channel} concept is introduced.
Informally, an inference channel is any collection of patterns (with their
respective supports) from which it is possible to infer non-$k$-anonymous patterns.
\begin{definition}
\label{def:inferencechannel}
Given a database $\mathcal{D}$ and two patterns $I$ and $J$,
with $I=\{i_1, \ldots,i_m\}$ and $J= I \cup \{a_1,\ldots,a_n\}$,
the set
$C^J_I$$=$$\{\langle X, supp_\mathcal{D}(X)\rangle | I \subseteq X \subseteq J\}$
constitutes an inference channel for the non $k$-anonymous pattern $p= I \cup \{\neg a_1, \dots, \neg a_n\}$ if $0 < supp_\mathcal{D}(C^J_I) < k$
where
\begin{equation}
\label{supchan}
supp_\mathcal{D}(C^J_I) = \sum_{I \subseteq X \subseteq J} (-1)^{|X\backslash I|}supp_\mathcal{D}(X).
\end{equation}
\end{definition}
See \cite{AtzoriBGP08} for details. An example of inference channel is given by any pattern such as $p: \{b\}$ which has a superset $p_s: \{b,d,e\}$ such that $0 < C^{p_s}_{p} < k$. In this case the pair $\langle p, supp(p) \rangle, \langle p_s, supp(p_s) \rangle$ constitutes an inference channel for the non-$k$-anonymous pattern $\{a, \neg b, \neg c\}$, whose support is given by $supp(b) - supp(b,d) - supp(b,e) + supp(b,d,e)$.
Then, we can formally define the collection of $k$-anonymous pattern set
as follows.
\begin{definition}[$k$-Anonymous pattern set]
\label{def:anonymouspatternrs}
Given a collection of frequent patterns $\mathcal{F(D, \sigma)}$ and
an anonymity threshold $k$, $\mathcal{F(D, \sigma)}$ is $k$-anonymous if
(1) $\nexists p \in \mathcal{F(D, \sigma)}$ s.t. $0<supp(p)<k$, and (2) $\nexists p_1$ and $p_2 \in \mathcal{F(D, \sigma)}$  s.t. $0<supp_\mathcal{D}(C^{p_2}_{p_1})<k$, where $p_1 \subset p_2$.
 \end{definition}
\subsection{Achieving an Anonymous Frequent Pattern Set}
\label{sec:IPG}
To generate a $k$-anonymous version of $\mathcal{F}(\mathcal{D}, \sigma)$,
Atzori {\em et al.}
\cite{AtzoriBGP08} proposed to first detect inference channels
violating $k$-anonymity in $\mathcal{F}(\mathcal{D}, \sigma)$
(Definition~\ref{def:inferencechannel}) and then block them in a
second step.
The pattern sanitization method  
blocks an inference channel $C^J_I$ due to a pair of patterns
$I = \{i_{1}, \ldots, i_{m}\}$ and $J = \{i_{1}, \ldots, i_{m},$ $a_{1}, \ldots, a_{n}\}$ in $\mathcal{F}(\mathcal{D}, \sigma)$  by increasing
the support of $I$ by $k$ to achieve $supp(C^I_J) \geq k$. In addition, to avoid contradictions
among the released patterns, the support of all subsets of $I$ is also
increased by $k$.
\begin{example}
\label{exp:52}
Let us resume Example \ref{exp:51} and take $k=8$. An
inference channel due to patterns $p_1$ and $p_2$ can be blocked by increasing the support of pattern
$p_1$:{\em \{Job=veterinarian, Credit\_approved=yes\}} and all its subsets
by 8. 
In this way, the non-8-anonymous pattern {\em \{Job=veterinarian, $\neg$(Salary $>$ 15000), Credit\_approved=yes\}} is 8-anonymous.
\hfill $\Box$
\end{example}
The privacy pattern sanitization method can avoid generating
new inference channels as a result of its transformation. 
In this way, we can obtain a $k$-anonymous version of $\mathcal{F}(\mathcal{D}, \sigma)$.

     \section{Discrimination-aware Frequent Pattern Discovery}
    \label{sec:dispatterns}
In this section, we first describe the notion of discrimination protected and unexplainable discrimination protected frequent patterns. Then, we present our proposed methods and algorithms to obtain these pattern sets, respectively.

\subsection{Discrimination Protected Frequent Pattern Set}
\label{sec:DPFP}

Starting from the definition of PD and PND classification rule in Section \ref{sec:discmeasures}, we define when a frequent pattern is PD.
\begin{definition}
	\label{def:PDpattern}
Given protected groups $DI_b$, a frequent pattern $p \in \mathcal{F}(\mathcal{D},\sigma)$ is said to be a PD if: (1)
$p$ contains a class item $C$ denying some benefit, i.e., $C \subset p$, and (2) $\exists p' \subset p$ s.t. $\ p' \subseteq DI_b$.
\end{definition}
In other words, a frequent pattern $p:\{A, B, C\}$ is a PD pattern if a PD classification rule $A, B\rightarrow C$ can be derived from it. As an example, pattern {\em \{Sex=female, Job=veterinarian, Credit\_approved=no\}} is a PD pattern, where $DI_b:$\emph{\{Sex=female\}}.
\begin{example}
\label{exp:53}
Let $f=slift$, $\alpha=1.25$ and $DI_b:$\emph{\{Sex=female\}}.
Assume that, in the data set of Table \ref{tab:51},
the total number of veterinarian women applicants and
the number of veterinarian women applicants who are denied credit are 34 and 20,
respectively, and the total number of veterinarian men applicants and the number of veterinarian men
applicants who are denied credit are 47 and 19, respectively.
The PD classification rule $r:${\em Sex=female, Job=veterinarian $\rightarrow$ Credit\_approved=no} extracted from Table \ref{tab:51} is 1.25-discriminatory, because $slift(r)=\frac{20/34}{19/47}=1.45$. 
\hfill $\Box$
\end{example}
Based on Definitions \ref{def:PDpattern} and \ref{label:disc},
we introduce the notions of $\alpha$-protective and $\alpha$-discriminatory patterns.
\begin{definition}
\label{label:disc_patterns}
Let $f$ be one of the measures in Fig.~\ref{fig:Measures}. Given protected groups $DI_b$ and $\alpha\in R$ a fixed threshold, a PD pattern $p: \{A,B,C\}$, where $C$ denies some benefit and $A \subseteq DI_b$, is $\alpha$-protective w.r.t. $f$ if the classification rule $r: A,B\rightarrow C$ is $\alpha$-protective. Otherwise,
$p$ is $\alpha$-discriminatory.
\end{definition}
\begin{example}
\label{exp:54}
Continuing Example \ref{exp:53}, a PD pattern
$p:${\em \{Sex = female, Job = veterinarian, Credit\_approved = no\}} extracted from Table \ref{tab:51}, is 1.25-discriminatory because rule $r$ is 1.25-discriminatory, where $r$ is {\em Sex = female, Job = veterinarian $\rightarrow$ Credit\_approved = no}.
\hfill $\Box$
\end{example}
Based on Definition \ref{label:disc_patterns}, we introduce the notion of discrimination protected pattern set.
\begin{definition}[$\alpha$-protective pattern set]
\label{def:protectivepatternrs}
Given a collection of frequent patterns $\mathcal{F(D, \sigma)}$, discrimination measure $f$, a discrimination threshold $\alpha$, and protected groups $DI_b$,  $\mathcal{F(D, \sigma)}$ is $\alpha$-protective w.r.t. $DI_b$ and $f$ if
$\nexists p \in \mathcal{F(D, \sigma)}$ s.t. $p$
is an $\alpha$-discriminatory pattern.
 \end{definition}
\subsection{Unexplainable Discrimination Protected Frequent Pattern Set}
\label{sec:UNECPDISPRTOCFPS}
Another legal concept we take into account in this chapter is the \emph{genuine occupational requirement}.
This requirement refers to discrimination
that may be partly explained by attributes {\em not} in $DI_b$ (\cite{Calders2011}), {\em e.g.},
denying credit to women may be explainable by the fact that most of them
have low salary or delay in returning previous credits.
Whether low salary or delay in returning previous credits is
an acceptable (legitimate) argument to deny credit is for lawyers (law) to be determined. We define them as legally-grounded groups, denoted by $DI_e$. In our context, $DI_e$ is a PND itemset which is legally admissible in a discrimination litigation, {\em e.g.}, \emph{\{Credit\_history=paid-delay\}}. Given $DI_e$ and $DI_b$, discrimination against protected groups is explained if
there is a high correlation between $DI_b$ and $DI_e$ and also between $DI_e$ and class item $C$.
As an example, discrimination against women in a given context is explainable by their delay in paying previous credits, first, if the majority of women in the given context have delay in paying previous credits and, second, if the delay in paying previous credit gives some objective information about the credit rejection. To determine which $\alpha$-discriminatory patters are explainable and which ones are not we use the notion of $d$-instance (see Definition \ref{def:dinstance}).
For high values of $d$ ({\em i.e.} 1 or near 1) Condition 1 of Definition~\ref{def:dinstance} shows high correlation between the class item ({\em i.e.} credit denial) and the legally-grounded group ({\em i.e.} delay in paying previous credits) and Condition 2 of Definition~\ref{def:dinstance} shows high correlation between the legally-grounded group ({\em i.e.} delay in paying previous credits) and the protected group ({\em i.e.} women) in a given context.
\begin{example}
\label{exp:pinstance}
Continuing Examples \ref{exp:53} and \ref{exp:54}, let $DI_e:$ {\em \{ Credit\_history = paid-delay\}} and $d=0.9$. Assume that in the dataset of Table \ref{tab:51}, the total number of veterinarian applicants who are delayed in paying previous credits and the total number of veterinarian applicants who are delayed in paying previous credits and are denied credit are 64 and 52, respectively, and the total number of women applicants who are veterinarian and are delayed in paying previous credits is 31. A PD classification rule $r':${\em Sex=female, Job=veterinarian $\rightarrow$ Credit\_approved=no} extracted from Table \ref{tab:51} is 0.9-instance of a PND classification rule
$r:${\em Credit\_history=paid-delay, Job=veterinarian $\rightarrow$ Credit\_approved=no} because (1) $\frac{52}{64} \geq 0.9 \cdot \frac{20}{34}$ and (2) $\frac{31}{34} \geq 0.9$. Thus, both conditions of Definition \ref{def:dinstance} are satisfied.
\hfill $\Box$
\end{example}
Based on Definitions \ref{def:dinstance} and \ref{def:PDpattern}, we introduce the notions of $d$-explainable and $d$-unexplainable frequent patterns.
\begin{definition}[d-(un)explainable pattern]
\label{label:exp_patterns}
Let $d \in [0, 1]$. An $\alpha$-discriminatory pattern $p': \{A,B,C\}$, where $C$ denies some benefit and $A \subseteq DI_b$, is a $d$-explainable pattern if a PD classification rule $r': A,B\rightarrow C$ is a $d$-instance of a PND classification rule $r: D, B \rightarrow C$, where $D \subseteq DI_e$. Otherwise $p'$ is a $d$-unexplainable pattern.
\end{definition}
\begin{example}
\label{exp:pinstancep}
Continuing Examples \ref{exp:53}-\ref{exp:pinstance}, the 1.25-discriminatory pattern $p:$ {\em\{Sex = female, Job = veterinarian, Credit\_approved = no\}} extracted from Table \ref{tab:51} is a 0.9-explainable pattern because rule $r'$ is 0.9-instance of rule $r$ where $r$ is
{\em Credit\_history = paid-delay, Job = veterinarian $\rightarrow$ Credit\_approved = no} and $r'$ is {\em Sex = female, Job = veterinarian $\rightarrow$ Credit\_approved = no}.
\hfill $\Box$
\end{example}
From Definition \ref{label:exp_patterns}, we introduce the notion of unexplainable discrimination protected pattern set.
\begin{definition}[$d$-explainable $\alpha$-protective pattern set]
\label{def:explainablepatternrs}
Given a collection of frequent patterns $\mathcal{F(D, \sigma)}$, a
discrimination measure $f$, a discrimination threshold $\alpha$, an explainable discrimination threshold $d$, protected groups $DI_b$ and legally-grounded groups $DI_e$, $\mathcal{F(D, \sigma)}$ is $d$-explainable $\alpha$-protective w.r.t. $DI_b$, $DI_e$ and $f$ if there is no $\alpha$-discriminatory pattern
$p \in \mathcal{F(D, \sigma)}$ s.t. $p$
is a $d$-unexplainable pattern.
 \end{definition}
    \subsection{Achieving a Discrimination Protected Frequent Pattern Set}
\label{sec:nDG}
In order to generate a discrimination protected ({\em i.e.} an $\alpha$-protective) version of $\mathcal{F}(\mathcal{D}, \sigma)$, we propose an approach including two steps. First,
detecting $\alpha$-discriminatory patterns in $\mathcal{F}(\mathcal{D}, \sigma)$ w.r.t. discriminatory measure $f$, $DI_b$ and $\alpha$ as discussed in Section \ref{sec:DPFP}.
We propose Algorithm \ref{alg:DetDiscPatt} for detecting $\alpha$-discriminatory patterns in $\mathcal{F}(\mathcal{D}, \sigma)$.
The algorithm starts by obtaining the subset $\mathcal{D}_{PD}$ which contains
the PD patterns in $\mathcal{F}(\mathcal{D}, \sigma)$ found according to $C$ and
$DI_b$ (Line 4). For each pattern $p:\{A, B, C\}$
in $\mathcal{D}_{PD}$, where $A \subseteq DI_b$, the value of $f$ (one of
the measures in Definitions \ref{def:elift}-\ref{def:dlift}) regarding its PD rule $r: X \rightarrow C$, where $X=A, B$, is computed to determine the subset $\mathcal{D}_{D}$ which contains
the $\alpha$-discriminatory patterns in $\mathcal{F}(\mathcal{D}, \sigma)$ (Lines 5-13). After obtaining $\mathcal{D}_{D}$, the second step of our approach is
sanitization for each pattern in $\mathcal{D}_{D}$, in order to make it $\alpha$-protective.
\begin{algorithm}[!h]
\caption{{\sc Detecting $\alpha$-discriminatory patterns}}
\label{alg:DetDiscPatt}
{\small \begin{algorithmic}[1]
\STATE Inputs: Database $\mathcal{D}$, $\mathcal{FP}:=\mathcal{F}(D, \sigma)$,
$DI_b$, discrimination measure $f$, $\alpha$, $C=${class item with
a negative decision value}
\STATE Output: $\mathcal{D}_D$: $\alpha$-discriminatory patterns in $\mathcal{FP}$
\STATE {\bf Function} \textsc{DetDiscPatt}($\mathcal{FP}$, $\mathcal{D}$, $DI_b$, $f$, $\alpha$, $C$)
\STATE $\mathcal{D}_{PD} \leftarrow$ All patterns $\langle p: A, B, C, supp(p)\rangle \in \mathcal{FP}$ with $p \cap C \neq \emptyset$ and $p \cap DI_b \neq \emptyset$
\FORALL{$p \in \mathcal{D}_{PD}$}
\STATE $X=p \backslash C$
\STATE $r=X \rightarrow C$
\STATE Compute $f(r)$ using $\mathcal{FP}$ and $\mathcal{D}$ where $f$ is one of the measures from Fig. \ref{fig:Measures}
\IF{$f(r) \geq \alpha$}
\STATE Add $p$ in $\mathcal{D}_D$
\ENDIF
\ENDFOR
\RETURN $\mathcal{D}_D$
\STATE {\bf End Function}
\end{algorithmic}}
\end{algorithm}
In the sequel, we study and propose a pattern sanitization solution for each possible measure of discrimination $f$.
\subsubsection{Anti-discrimination Pattern Sanitization for $slift$ and its Variants}
According to Definition~\ref{label:disc_patterns},
to make an $\alpha$-discriminatory pattern $p:\{A, B, C \}$ $\alpha$-protective where $f=slift$, we should enforce the following inequality
\begin{equation}
\label{ineq:1}
slift(A, B\rightarrow C) < \alpha
\end{equation}
where $A \subseteq DI_b$ and $C$ denies some benefit.
By using the definitions of confidence and $slift$ (Expressions
(\ref{eq:conf}) and (\ref{eq:slift}), resp.), Inequality (\ref{ineq:1}) can be rewritten as
\begin{equation}
\label{ineq:2}
\frac{\frac{supp(A,B,C)}{supp(A,B)}}{\frac{supp(\neg A,B,C)}{supp(\neg A,B)}} < \alpha.
\end{equation}
Then, it is clear that Inequality (\ref{ineq:1}) can be satisfied by decreasing
the left-hand side of Inequality (\ref{ineq:2}) to a value less than
the discriminatory threshold $\alpha$, which can be done in the following way:
\begin{itemize}
\item{\em Anti-discrimination pattern sanitization where} $f=slift$.
Increase the support of the pattern $\{A, B\}$ and all its subsets by a specific value $\Delta_{slift}$ to satisfy Inequality (\ref{ineq:2}). This increment decreases the numerator of equation $\frac{\frac{supp(A,B,C)}{supp(A,B)}}{\frac{supp(\neg A,B,C)}{supp(\neg A,B)}}$ while keeping the denominator unaltered.
\end{itemize}
Modifying the support of the subsets of respective patterns accordingly
is needed to avoid contradictions (maintain compatibility) among the released patterns.
In fact, {\em anti-discrimination pattern sanitization} makes pattern $p$ $\alpha$-protective by decreasing the proportion of people in the protected group and given context who were not granted the benefit ({\em e.g.} decreasing the proportion of veterinarian women applicants who were denied credit).
Let us compute the value $\Delta_{slift}$ to be used in anti-discrimination pattern sanitization where $f=slift$.
The support of the pattern $\{A, B\}$ should
be increased to satisfy Inequality (\ref{ineq:2}):
 $$slift(A, B\rightarrow C)=\frac{\frac{supp(A,B,C)}{supp(A,B)+\Delta_{slift}}}{\frac{supp(\neg A,B,C)}{supp(\neg A,B)}} < \alpha.$$
The above equality can be rewritten as
\begin{equation}
\label{eqdelta}
\Delta_{slift} > \frac{supp(A,B,C) \times supp(\neg A,B)}{supp(\neg A,B,C) \times \alpha}-supp(A,B).\end{equation}
Hence, taking $\Delta_{slift}$ equal to the ceiling of the right-hand side
of Equation (\ref{eqdelta}) suffices to make $p: \{A,B,C\}$ $\alpha$-protective w.r.t. $f=slift$. Considering the definitions of $slift_d$ and $slift_c$ (Expressions
(\ref{eq:sliftd}) and (\ref{eq:sliftc}), resp.), a similar method can make pattern $p$ $\alpha$-protective w.r.t. $f=slift_d$ and $f=slift_c$. The value of $\Delta_{slift_d}$ and  $\Delta_{slift_c}$ can be computed in the same way as we compute $\Delta_{slift}$.
\begin{example}
\label{exp:55}
Continuing Examples \ref{exp:53} and \ref{exp:54}, pattern
$p:$ {\em \{Sex = female, Job = veterinarian, Credit\_approved = no\}} can
be made $1.25$-protective by increasing the support of pattern
{\em \{Sex=female, Job=veterinarian\}} and all its subsets
by $\Delta_{slift}=6$, which is the value resulting from Inequality (\ref{eqdelta}).
\hfill $\Box$
\end{example}
As we define in Section \ref{sec:DPFP}, $clift$ is a special case of $slift$ and it has the same formula (see Definitions \ref{def:slift} and \ref{def:clift}). Then, a similar anti-discrimination pattern sanitization can make an $\alpha$-discriminatory $p:\{a=v_1, B, C \}$ $\alpha$-protective where $f=clift$. The value of $\Delta_{clift}$ is computed in the following way
\begin{equation}
\label{eqdeltaclift}
\Delta_{clift} = \lceil \frac{supp(a=v_1,B,C) \times supp(a=v_2,B)}{supp(a=v_2,B,C) \times \alpha}-supp(a=v_1,B)\rceil.\end{equation}
\subsubsection{Anti-discrimination Pattern Sanitization for $elift$ and its Variants}
According to Definition~\ref{label:disc_patterns},
to make an $\alpha$-discriminatory pattern $p:\{A, B, C \}$ $\alpha$-protective where $f=elift$, we should enforce the following inequality
\begin{equation}
\label{ineq:a}
elift(A, B\rightarrow C) < \alpha
\end{equation}
where $A \subseteq DI_b$ and $C$ denies some benefit.
By using the definitions of confidence and $elift$ (Expressions
(\ref{eq:conf}) and (\ref{eq:elift}), resp.), Inequality (\ref{ineq:a}) can be rewritten as
\begin{equation}
\label{ineq:b}
\frac{\frac{supp(A,B,C)}{supp(A,B)}}{\frac{supp(B,C)}{supp(B)}} < \alpha.
\end{equation}
Then, it is clear that Inequality (\ref{ineq:a}) can be satisfied by decreasing
the left-hand side of Inequality (\ref{ineq:b}) to a value less than
the discriminatory threshold $\alpha$. A similar anti-discrimination pattern sanitization proposed for $f=slift$ cannot make pattern $p$ $\alpha$-protective w.r.t. $f=elift$ because increasing the support of pattern $\{A, B\}$ and all its subsets by a specific value can decrease the numerator of equation $\frac{\frac{supp(A,B,C)}{supp(A,B)}}{\frac{supp(B,C)}{supp(B)}}$ and decrease the denominator of it as well.
Then, making pattern $p:\{A, B, C \}$ $\alpha$-protective w.r.t. $f=elift$ is possible using an alternative pattern sanitization method:
\begin{itemize}
\item{\em Anti-discrimination pattern sanitization where} $f=elift$.
Increase the support of the pattern $\{B, C\}$ and all its subsets by a specific value $\Delta_{elift}$ to satisfy Inequality (\ref{ineq:b}). This increment increases the denominator of equation $\frac{\frac{supp(A,B,C)}{supp(A,B)}}{\frac{supp(B,C)}{supp(B)}}$ while keeping the numerator unaltered.
\end{itemize}
In fact, the above method makes pattern $p$ $\alpha$-protective w.r.t. $elift$ by increasing the proportion of people in the given context who were not granted the benefit ({\em e.g.} increasing the proportion of veterinarian applicants who were denied credit while the proportion of veterinarian women applicants who were denied credit is unaltered).
Let us compute the value $\Delta_{elift}$ to be used in anti-discrimination pattern sanitization where $f=elift$.
The support of the pattern $\{B, C\}$ should
be increased to satisfy Inequality (\ref{ineq:b}):
 $$elift(A, B\rightarrow C)=\frac{\frac{supp(A,B,C)}{supp(A,B)}}{\frac{supp(B,C)+\Delta_{elift}}{supp(B)+\Delta_{elift}}} < \alpha.$$
Since the value of $\alpha$ is higher than 1 and
$\frac{supp(A,B,C)}{supp(A,B)} \leq \alpha$, from the above equality we obtain
\begin{equation}
\label{eqdeltae}
\Delta_{elift} > \frac{\alpha \times supp(A,B) \times supp(B,C)- supp(A,B,C) \times supp(B)}{supp(A,B,C)- \alpha \times supp(A,B)}.\end{equation}
Hence, taking $\Delta_{elift}$ equal to the ceiling of the right-hand side
of Equation (\ref{eqdeltae}) suffices to make $p: \{A,B,C\}$ $\alpha$-protective w.r.t. $f=elift$. Considering the definitions of $elift_d$ and $elift_c$ (Expressions
(\ref{eq:sliftd}) and (\ref{eq:sliftc}), resp.),
a similar method can make pattern $p$ $\alpha$-protective w.r.t. $f=elift_d$ and $f=elift_c$. The values of $\Delta_{elift_d}$ and $\Delta_{elift_c}$ can be computed in the same way as $\Delta_{elift}$.
\subsubsection{Discrimination Analysis}
An essential property of a valid anti-discrimination pattern sanitization method is not to
produce new discrimination as a result of the transformations it performs.
The following theorem shows that all the methods described above
satisfy this property.
\begin{theorem}
\label{theoDisc}
Anti-discrimination pattern sanitization methods for making $\mathcal{F}(\mathcal{D}, \sigma)$ $\alpha$-protective w.r.t. $f$ do not generate new discrimination as a result of their transformations, where $f$ is one of the measures from Fig. \ref{fig:Measures}.
\end{theorem}
\begin{proof}
It is enough to show that anti-discrimination pattern sanitization methods to
make each $\alpha$-discriminatory pattern in $\mathcal{F}(\mathcal{D}, \sigma)$ $\alpha$-protective w.r.t. $f$ cannot make $\alpha$-protective patterns in $\mathcal{F}(\mathcal{D}, \sigma)$ $\alpha$-discriminatory. Consider two
PD patterns $p_1:\{A, B, C\} $ and $p_2: \{A', B', C\} $, where $A, A'
\subseteq DI_b$ and $p_1 \neq p_2$.
The following possible relations between $p_1$ and $p_2$ are
conceivable:
\begin{itemize}
\item $A = A'$ and $B \neq B'$, special case: $B' \subset B$
\item $A \neq A'$ and $B = B'$, special case: $A' \subset A$
\item $A \neq A'$ and $B \neq B'$, special case: $A' \subset A$ and $B'
\subset B$
\end{itemize}
In all the above special cases ({\em i.e.} $p_2 \subset p_1$), making $p_1$ $\alpha$-protective w.r.t. $f$
involves increasing $supp(A',B')$ by $\Delta_{slift}$, $\Delta_{clift}$ or $\Delta_{slift_d}$ where $f=slift$, $f=clift$ or $f=slift_d$, resp., and involves increasing $supp(B',C)$ and $supp(B')$ by $\Delta_{elift}$,  $\Delta_{elift_d}$ where $f=elift$ or $f=elift_d$, respectively. This cannot make
$p_2$ less $\alpha$-protective w.r.t. $f$; actually, it can even make $p_2$
more protective because increasing $supp(A',B')$ can decrease $slift(A', B' \rightarrow C)$ and $slift_d(A', B' \rightarrow C)$ and increasing $supp(B',C)$ and $supp(B')$ can decrease $elift(A', B' \rightarrow C)$ and $elift_d(A', B' \rightarrow C)$.
On the other hand, making
$p_2$ $\alpha$-protective w.r.t. $f$ cannot make $p_1$ less or more protective since there is
no overlap between the modified patterns to make $p_2$ $\alpha$-protective
and the patterns whose changing support can change $f(A, B\rightarrow C)$.
Otherwise (no special cases), making $p_1$ (resp. $p_2$)
$\alpha$-protective w.r.t. $f$ cannot make $p_2$ (resp. $p_1$)
less or more protective since there is no
overlap between the modified patterns to make $p_1$ (resp. $p_2$) $\alpha$-protective w.r.t. $f$
and the patterns whose changing support can change $f(A', B' \rightarrow C)$ (resp. $f(A, B\rightarrow C)$).
Hence, the theorem holds. 
\end{proof}
Therefore, using the proposed anti-discrimination pattern sanitization methods, we can obtain an $\alpha$-protecti ve version of $\mathcal{F}(\mathcal{D}, \sigma)$ w.r.t. $f$. We propose Algorithm \ref{alg:AntiDiscPattSanit} for doing so.
The algorithm performs anti-discrimination pattern sanitization to make each $\alpha$-discriminatory pattern $p$ in $\mathcal{F}(D, \sigma)$ $\alpha$-protective w.r.t. $f$. The value of $\Delta_f$ is computed for each $\alpha$-discriminatory pattern w.r.t. the value of $f$.
\begin{algorithm}[h]
\caption{{\sc Anti-discrimination pattern sanitization}}
\label{alg:AntiDiscPattSanit}
{\small \begin{algorithmic}[1]
\STATE Inputs: Database $\mathcal{D}$, $\mathcal{FP}:=\mathcal{F}(D, \sigma)$, $\mathcal{D}_D$,
$DI_b$, discrimination measure $f$, $\alpha$, $C=$ class item with negative decision value
\STATE Output: $\mathcal{FP}^*$: $\alpha$-protective version of $\mathcal{FP}$
\STATE {\bf Function} \textsc{AntiDiscPattSanit}($\mathcal{FP}$, $\mathcal{D}$, $\mathcal{D}_D$, $DI_b$, $f$, $\alpha$, $C$)
\FORALL{$p: \{A, B, C\} \in \mathcal{D}_D$}
\STATE $X=p \backslash C$
\STATE Compute $\Delta_f$ for pattern $p$ w.r.t. the value of $f$ using $\mathcal{D}$, $\mathcal{FP}$ and $\alpha$
\IF{$\Delta_f \geq 1$}
\IF{$f=slift$ or $f=clift$ or $f=slift_d$}
\STATE $p_t=X$
\ELSIF{$f=elift$ or $elift_d$}
\STATE $Y= p \cap DI_b$
\STATE $p_t=p\backslash Y$
\ENDIF
\ENDIF
\STATE $\mathcal{D}_s= \{p_s \in  \mathcal{FP}| p_s \subseteq p_t\}$
\FORALL{$\langle p_s, supp(p_s)\rangle \in \mathcal{D}_s$}
\STATE $supp(p_s) = supp(p_s)+ \Delta_f$
\ENDFOR
\ENDFOR
\RETURN $\mathcal{FP}^* = \mathcal{FP}$
\STATE {\bf End Function}
\end{algorithmic}}
\end{algorithm}
\subsection{Achieving an Unexplainable Discrimination Protected Pattern Set}
\label{sec:AUNEXPDISCPROTPS}
In order to make $\mathcal{F}(D, \sigma)$ protected against only unexplainable discrimination ({\em i.e.} generating a $d$-explainable $\alpha$-protective  version of $\mathcal{F}(\mathcal{D}, \sigma)$), we propose an approach including three steps. First,
detecting $\alpha$-discriminatory patterns in $\mathcal{F}(\mathcal{D}, \sigma)$ w.r.t. discriminatory measure $f$, $DI_b$ and $\alpha$. Second, detecting $d$-unexplainable patterns among $\alpha$-discriminatory patterns obtained in the first step. Third,  sanitizing each $d$-unexplainable pattern to make it $\alpha$-protective.
We propose Algorithm \ref{alg:DetUnExplainPatt} for detecting $d$-unexplainable patterns in $\mathcal{F}(\mathcal{D}, \sigma)$.
The algorithm starts by obtaining the subset $\mathcal{D}_{PND}$ containing
the PND patterns in $\mathcal{F}(\mathcal{D}, \sigma)$ found according to $C$ and
$DI_b$ (Line 4). Then, the algorithm computes the subset $\mathcal{D}_{instance}$ containing the PND patterns which are legally-grounded according to
$DI_e$ (Line 5). Then, the algorithm uses the \textsc{DetDiscPatt} function in Algorithm \ref{alg:DetDiscPatt} to determine the subset $\mathcal{D}_{D}$ which contains $\alpha$-discriminatory patterns in $\mathcal{F}(\mathcal{D}, \sigma)$ (Line 6). Finally, the algorithm computes the subset $\mathcal{D}_{bad}$
containing patterns in $\mathcal{D}_{D}$ which are $d$-unexplainable according to $\mathcal{D}_{instance}$ and $d$ (Lines 7-19).
\begin{algorithm}[h]
\caption{{\sc Detecting $d$-unexplainable patterns}}
\label{alg:DetUnExplainPatt}
{\small \begin{algorithmic}[1]
\STATE Inputs: Database $\mathcal{D}$,$\mathcal{FP}:=\mathcal{F}(D, \sigma)$,
$DI_b$, $DI_e$, explainable discrimination threshold $d$, discrimination measure $f$, $\alpha$, $C=${class item with negative decision value}
\STATE Output: $\mathcal{D}_{bad}$: $d$-unexplainable patterns in $\mathcal{FP}$
\STATE {\bf Function} \textsc{DetUnExplainPatt}($\mathcal{FP}$, $\mathcal{D}$, $DI_e$, $DI_b$, $f$, $\alpha$, $d$, $C$)
\STATE $\mathcal{D}_{PND} \leftarrow$ All patterns $\langle p: A, B, C, supp(p)\rangle \in \mathcal{FP}$ with $p \cap C \neq \emptyset$ and $p \cap DI_b = \emptyset$
\STATE $\mathcal{D}_{instance} \leftarrow$ All patterns $\langle p: D, B, C, supp(p)\rangle \in \mathcal{D}_{PND}$ with $p \cap DI_e \neq \emptyset$
\STATE $\mathcal{D}_{D} \leftarrow$ {\bf Function} \textsc{DetDiscPatt}($\mathcal{FP}$, $\mathcal{D}$, $DI_b$, $f$, $\alpha$, $C$)
\FORALL{$p:\{A,B,C\} \in \mathcal{D}_{D}$}
\STATE $X=p \backslash C$
\STATE $r=X \rightarrow C$
\FOR{\textbf{each} $p_d \in \mathcal{D}_{instance}$}
\STATE $X_d=p_d \backslash C$
\STATE $r_d=X_d \rightarrow C$
\IF{$r$ is a $d$-instance of $r_d$}
\STATE Add $p$ in $\mathcal{D}_{legal}$
\ENDIF
\ENDFOR
\ENDFOR
\STATE $\mathcal{D}_{bad}=\mathcal{D}_{D} \backslash \mathcal{D}_{legal}$
\RETURN $\mathcal{D}_{bad}$
\STATE {\bf End Function}
\end{algorithmic}}
\end{algorithm}
After obtaining $\mathcal{D}_{bad}$, the third step is sanitizing each $d$-unexplainable pattern to make it $\alpha$-protective. In order to do this, we need to examine the impact of this transformation on $d$-explainable patterns in $\mathcal{F}(\mathcal{D}, \sigma)$.
\begin{theorem}
\label{theodinstanceelift}
Anti-discrimination pattern sanitization methods for making $\mathcal{F}(\mathcal{D}, \sigma)$ $d$-explainable $\alpha$-protective w.r.t. $f$ do not generate any new $d$-unexplainable pattern as a result of their transformations, where $f=elift$, $f=elift_d$, and $f=elift_c$.
\end{theorem}
\begin{proof}
It is enough to show that anti-discrimination pattern sanitization methods to
make each $d$-unexplainable pattern in $\mathcal{F}(\mathcal{D}, \sigma)$ $\alpha$-protective w.r.t. $f$ cannot make $d$-explainable patterns in $\mathcal{F}(\mathcal{D}, \sigma)$ $d$-unexplainable , where $f=elift$, $f=elift_d$ and $f=elift_c$. Consider two
PD patterns $p_1:\{A, B, C\} $ and $p_2: \{A', B', C\} $, where $A, A'
\subseteq DI_b$ and $p_1 \neq p_2$.
The following possible relations between $p_1$ and $p_2$ are
conceivable:
\begin{itemize}
\item $A = A'$ and $B \neq B'$, special case: $B' \subset B$
\item $A \neq A'$ and $B = B'$, special case: $A' \subset A$
\item $A \neq A'$ and $B \neq B'$, special case: $A' \subset A$ and $B'
\subset B$
\end{itemize}
In all the above cases ({\em i.e.}
special and non-special cases), making $d$-unexplainable pattern $p_1$ $\alpha$-protective w.r.t. $f$ involves increasing $supp(B',C)$ and $supp(B')$ by $\Delta_{elift}$, $\Delta_{elift_d}$, or $\Delta_{elift_c}$ where $f=elift$, $f=elift_d$ or $f=elift_c$, respectively. This cannot make
$d$-explainable pattern $p_2$ $d$-unexplainable w.r.t. $f$ because there is
no overlap between the modified patterns to make $p_1$ $\alpha$-protective and the patterns whose changing support can make $p_2$ $d$-unexplainable. On the other hand, making $d$-unexplainable pattern $p_2$ $\alpha$-protective cannot make
$d$-explainable pattern $p_1$ $d$-unexplainable for the same reason above.
Hence, the theorem holds. 
\end{proof}
\begin{theorem}
\label{theodinstanceslift}
Anti-discrimination pattern sanitization methods for making $\mathcal{F}(\mathcal{D}, \sigma)$ $d$-explainable $\alpha$-protective w.r.t. $f$ might generate new $d$-unexplainable patterns as a result of their transformations, where $f=slift$, $f=slift_d$, $f=clift$ and $f=slift_c$.
\end{theorem}
\begin{proof}
It is enough to show that anti-discrimination pattern sanitization methods to
make each $d$-unexplainable pattern in $\mathcal{F}(\mathcal{D}, \sigma)$ $\alpha$-protective w.r.t. $f$ can make a $d$-explainable in $\mathcal{F}(\mathcal{D}, \sigma)$ $d$-unexplainable, where $f=slift$, $f=slift_d$, $f=slift_c$ and $f=elift$. Consider two
PD patterns $p_1:\{A, B, C\} $ and $p_2: \{A', B', C\} $, where $A, A'
\subseteq DI_b$ and $p_1 \neq p_2$.
The following possible relations between $p_1$ and $p_2$ are
conceivable:
\begin{itemize}
\item $A = A'$ and $B \neq B'$, special case: $B' \subset B$
\item $A \neq A'$ and $B = B'$, special case: $A' \subset A$
\item $A \neq A'$ and $B \neq B'$, special case: $A' \subset A$ and $B'
\subset B$
\end{itemize}
In all the above special cases ({\em i.e.} $p_2 \subset p_1$), making $d$-unexplainable pattern $p_1$ $\alpha$-protective w.r.t. $f$ involves increasing the support of pattern $\{A',B'\}$ and all its subsets by $\Delta_{slift}$, $\Delta_{clift}$,  $\Delta_{slift_d}$, or $\Delta_{slift_c}$ where $f=slift$, $f=clift$ $f=slift_d$ or $f=clift_c$, respectively. This can make
$d$-explainable pattern $p_2$ $d$-unexplainable because this transformation can cause Condition 2 of Definition \ref{def:dinstance} to be
non-satisfied. This is because there is overlap between the modified patterns to make $p_1$ $\alpha$-protective and the patterns whose changing support can change the satisfaction of Condition 2 of Definition \ref{def:dinstance} w.r.t. pattern $p_2$. On the other hand, making $d$-unexplainable pattern $p_2$ $\alpha$-protective cannot make $d$-explainable pattern $p_1$ $d$-unexplainable since there is no overlap between the modified patterns to make $p_2$ $\alpha$-protective and the patterns whose changing support can change the satisfaction of Conditions 1 and 2 of Definition \ref{def:dinstance} w.r.t. pattern $p_1$.
Otherwise (no special cases), making $p_1$ (resp. $p_2$)
$\alpha$-protective w.r.t. $f$ cannot make $p_2$ (resp. $p_1$)
$d$-unexplainable since there is no
overlap between the modified patterns to make $p_1$ (resp. $p_2$) $\alpha$-protective w.r.t. $f$
and the patterns whose changing support can make $p_2$ (resp. $p_1$) $d$-unexplainable by changing the satisfaction of Conditions (1) and (2) in
Definition \ref{def:dinstance}.
Hence, the theorem holds. 
\end{proof}
Thus, according to the above theorems, Algorithm \ref{alg:AntiDiscPattSanit} cannot make $\mathcal{F}(\mathcal{D}, \sigma)$ $d$-explainable $\alpha$-protective w.r.t {\em all} possible values of $f$.
To attain
this desirable goal, we propose Algorithm \ref{alg:UnExplainAntiDiscPattSanit}.
Algorithm~\ref{alg:UnExplainAntiDiscPattSanit}
performs anti-discrimination pattern sanitization to make each $d$-unexplainable pattern $p$ in $\mathcal{D}_{bad}$ $\alpha$-protective by calling function \textsc{AntiDiscPattSanit} in Algorithm \ref{alg:AntiDiscPattSanit}, where $f=elift$, $f=elift_d$ or $f=elift_c$ (Line 6). As shown in Theorem \ref{theodinstanceslift}, given $d$-unexplainable pattern $p:\{A,B,C\}$ and $d$-explainable pattern $p_x:\{A',B',C\}$,
where $p_x \subset p$, making $p$ $\alpha$-protective w.r.t. $f$ might make $p_x$ $d$-unexplainable, where $f$ is $slift$, $slift_d$,
$f=clift$ or $f=slift_c$. For this reason and because pattern $p$
becomes $\alpha$-protective first
(see Algorithm~\ref{alg:AntiDiscAnonyPattSanit} in
Section~\ref{sec:PrivDisc}),
the algorithm checks whether making pattern $p$ $\alpha$-protective makes $p_x$ $d$-unexplainable. If yes, algorithm adds $p_x$ to $\mathcal{D}_{bad}$ (Lines 7-14).
%
\begin{algorithm}[h]
\caption{{\sc Unexplainable anti-discrimination pattern sanitization}}
\label{alg:UnExplainAntiDiscPattSanit}
{\small \begin{algorithmic}[1]
\STATE Inputs: Database $\mathcal{D}$, $\mathcal{FP}:=\mathcal{F}(D, \sigma)$, $\mathcal{D}_{bad}$,
$DI_b$, $d$, $\alpha$, discrimination measure $f$, $C=${class item with negative decision value}
\STATE Output: $\mathcal{FP}^*$: $d$-explainable $\alpha$-protective version of $\mathcal{FP}$
\STATE {\bf Function} \textsc{UnExplainAntiDiscPattSanit}($\mathcal{FP}$, $\mathcal{D}$, $\mathcal{D}_{bad}$, $DI_b$, $f$, $\alpha$, $C$)
\IF{$\mathcal{D}_{bad} \neq \emptyset$}
\IF{$f=elift$ or $f=elift_d$ or $f=elift_c$}
\STATE $\mathcal{FP}^* \leftarrow$ {\bf Function} \textsc{AntiDiscPattSanit}($\mathcal{FP}$, $\mathcal{D}$,  $\mathcal{D}_{bad}$, $DI_b$, $f$, $\alpha$, $C$)
\ELSIF{$f=slift$ or $f=slift_d$ or $f=clift$ or $f=slift_c$}
\FORALL{$p:\{A,B,C\} \in \mathcal{D}_{bad}$}
\STATE Lines 5-18 of Algorithm \ref{alg:AntiDiscPattSanit}
\IF{$\exists p_x \subset p$ in $\mathcal{FP}$ s.t. $p_x \notin \mathcal{D}_{bad}$ and $p_x$ is $d$-unexplainable}
\STATE Add $p_x$ in $\mathcal{D}_{bad}$
\ENDIF
\ENDFOR
\ENDIF
\ENDIF
\RETURN $\mathcal{FP}^* = \mathcal{FP}$
\STATE {\bf End Function}
\end{algorithmic}}
\end{algorithm}

      \section{Simultaneous Discrimination-Privacy Awareness in Frequent
Pattern Discovery}
        \label{sec:problemsol}
In this section, we present how simultaneous anti-discrimination and privacy can be achieved in frequent pattern discovery while satisfying $k$-anonymity and $\alpha$-protection. We first present our approach to obtain a discrimination-privacy protected version of an original pattern set. Then, we present our approach to obtain an unexplainable discrimination and privacy protected version of an original pattern set.
\subsection{Achieving a Discrimination and Privacy Protected Frequent Pattern Set}
\label{sec:PrivDisc}
In order to simultaneously achieve anti-discrimination and privacy in $\mathcal{F(D, \sigma)}$, we need to generate discrimination and privacy protected version of $\mathcal{F(D, \sigma)}$:
\begin{definition}[$\alpha$-protective $k$-anonymous pattern set]
\label{def:fair}
Given a collection of frequent patterns $\mathcal{F(D, \sigma)}$, anonymity threshold $k$, discrimination threshold $\alpha$, protected groups $DI_b$, and discrimination measure $f$, $\mathcal{F(D, \sigma)}$ is $\alpha$-protective $k$-anonymous if it is both $k$-anonymous and $\alpha$-protective w.r.t. $DI_b$ and $f$.
\end{definition}
We focus on the problem of producing a version of $\mathcal{F(D, \sigma)}$ that is $\alpha$-protective $k$-anonymous w.r.t. $DI_b$ and $f$. Like most works in $k$-anonymity~\cite{PPDPBOOK2010}, we
consider a single QI containing all attributes that can be potentially used in the
quasi-identifier. The more attributes included in QI, the more protection $k$-anonymity provides (and usually the more information loss).
Moreover, each QI attribute (unless it is the
class/decision attribute) can be a PD attribute or not depend on $DI_b$.
To obtain $\alpha$-protective $k$-anonymous version of $\mathcal{F(D, \sigma)}$, we should first examine the following issues: (1) how making
$\mathcal{F(D, \sigma)}$ $k$-anonymous impacts the
$\alpha$-protectiveness of $\mathcal{F(D, \sigma)}$ w.r.t. $f$; (2) how
making $\mathcal{F(D, \sigma)}$ $\alpha$-protective w.r.t $f$ impacts the
$k$-anonymity of $\mathcal{F(D, \sigma)}$.
Regarding the first issue, by presenting two scenarios we will
show that {\em privacy pattern sanitization} to achieve $k$-anonymity
in $\mathcal{F(D, \sigma)}$ can lead to different situations
regarding the $\alpha$-protectiveness of $\mathcal{F(D, \sigma)}$.
\begin{table}[ht]
\caption{Scenario \ref{scenario1}: Examples of frequent patterns extracted from Table \ref{tab:51}}
\label{tab:52}
\centering
\begin{tabular}{l c}
\hline
{\em Patterns} & {\em Support} \\
\hline
{\bf $p_s$}:\texttt{\{female, veterinarian\}}  & 45  \\
$p_2:$\texttt{\{female, veterinarian, salary $>$ 15000\}} & 42 \\
$p_1:$\texttt{\{female, veterinarian, No\}} & 32 \\
$p_n:$\texttt{\{male, veterinarian, No\}} & 16 \\
$p_{ns}:$\texttt{\{male, veterinarian\}} & 58 \\
\hline
\end{tabular}
\end{table}
\begin{scenario}
\label{scenario1}
{\em
Table \ref{tab:52} illustrates an example of frequent patterns
that come from
the data set in Table \ref{tab:51} with $\sigma=15$. Let $DI_b:$\emph{\{Sex=female\}}, $\alpha=1.25$, $k=8$ and $f=slift$. The PD pattern $p_1$ in Table \ref{tab:52} is 1.25-discriminatory since the value of $slift$ w.r.t. its PD rule is $\frac{32/45}{16/58}=2.58$ ({\em i.e.} this is inferred discrimination against veterinarian women applicants). On the other hand,
although the support of each pattern in the collection is higher than $k$,
there is an inference channel between patterns $p_s$ and $p_2$;
note that $supp(p_s)-supp(p_2)=3$ is smaller than 8
({\em i.e.} one can infer the
existence of only three veterinarian women in the city with salary no more than 15000 \euro). To block the inference channel between $p_s$ and $p_2$, the following privacy pattern sanitization is performed:
\begin{equation}
\label{eq:addsanti}
supp(p_s)+k, \forall p \subseteq p_s
\end{equation}
After this transformation, the new support of pattern $p_s$ is 53.
However, the supports of $p_{ns}$ and $p_n$ remain
unaltered since there is no inference channels between $p_{ns}$ and $p_n$ ($supp(p_{ns})-supp(p_n)=42$). Hence, the new value of $slift$ for the PD rule of pattern $p_1$ is $\frac{\frac{supp(p_1)}{supp(p_s)+k}}{\frac{supp(p_n)}{supp(p_{ns})}}$
which in this example is $\frac{32/(45+8)}{16/58}=2.19$. That is, the overall value of $slift$ is decreased.
Thus, in this scenario, making the collection of patterns in Table \ref{tab:52} $k$-anonymous can decrease discrimination; if the value of $slift$  became less than $\alpha$, pattern $p_1$ would even become
$\alpha$-protective.
}\hfill $\Box$
\end{scenario}
\begin{table}[ht]
\caption{Scenario \ref{scenario2}: Examples of frequent patterns extracted from Table \ref{tab:51}}
\label{tab:53}
\centering
\begin{tabular}{l c}
\hline
{\em Patterns} & {\em $Support$} \\
\hline
$p_s:$\texttt{\{male, veterinarian\}}  & 58  \\
$p_2:$\texttt{\{male, veterinarian, salary $>$ 15000\}} & 56 \\
$p_1:$\texttt{\{female, veterinarian, No\}} & 23 \\
$p_n:$\texttt{\{male, veterinarian, No\}} & 26 \\
$p_{ns}:$\texttt{\{female, veterinarian\}} & 45 \\
\hline
\end{tabular}
\end{table}
\begin{scenario}
\label{scenario2}
{\em
Table \ref{tab:53} illustrates an example of frequent patterns that
could come from the data set in Table \ref{tab:51} with $\sigma=20$.
Let $DI_s$:\emph{\{Sex=female\}}, $\alpha=1.25$, $k=8$, $f=slift$. A PD pattern $p_1$ in Table \ref{tab:53} is not 1.25-discriminatory since the value of $slift$ w.r.t. its PD rule is $\frac{23/45}{26/58}=1.14$. On the other hand, although the support of each pattern in the collection is higher than $k$, there is an inference channel between  $p_s$ and $p_2$;
note that $supp(p_s)-supp(p_2)=2$ is less than 8 ({\em i.e.} one can
infer the existence of only two veterinarian men in the city with salary no more than 15000 \euro). To block the inference channel between $p_s$ and $p_2$, pattern sanitization is performed according to Expression (\ref{eq:addsanti}).
After this transformation, the new support of pattern $p_s$ is 66 and the supports of $p_1$ and $p_{ns}$ stay unaltered  since there is no inference channel between $p_1$ and $p_{ns}$ ($supp(p_{ns})-supp(p_1)=22$). Hence, the new value of $slift$ for the PD rule of pattern $p_1$ is $\frac{\frac{supp(p_1)}{supp(p_{ns})}}{\frac{supp(p_n)}{supp(p_s)+k}}$
which is in this example $\frac{23/45}{26/(58+8)}=1.3$. That is, the overall value of $slift$ is increased.
Thus, in this scenario, making the collection of patterns in Table \ref{tab:53} $k$-anonymous can increase discrimination; in fact, with the numerical
values we have used, $p_1$ stops being $1.25$-protective
and becomes $1.25$-discriminatory.
}\hfill $\Box$
\end{scenario}
To summarize, using privacy pattern sanitization for making $\mathcal{F(D, \sigma)}$ $k$-anonymous can make $\mathcal{F(D, \sigma)}$ more or less $\alpha$-protective w.r.t. $slift$. We also observe a similar behavior for alternative discrimination measures. Then, achieving $k$-anonymity in frequent pattern discovery can achieve anti-discrimination or work against anti-discrimination.
Hence, detecting $\alpha$-discriminatory patterns in a $k$-anonymous version of $\mathcal{F(D, \sigma)}$ makes sense.
Regarding the second issue mentioned at the beginning of this section,
we will prove that if $\mathcal{F(D, \sigma)}$ is $k$-anonymous,
anti-discrimination pattern sanitization methods proposed in Section \ref{sec:nDG} to make it $\alpha$-protective w.r.t. $f$
cannot make $\mathcal{F(D, \sigma)}$ non-$k$-anonymous, {\em i.e.} they
cannot violate $k$-anonymity. Since the proposed anti-discrimination pattern sanitization methods are based
on adding support, they cannot make a $k$-anonymous pattern non-$k$-anonymous;
hence, we only need to prove that they cannot generate new inference channels.
\begin{theorem}
\label{theoprivacy}
Anti-discrimination pattern sanitization for making $\mathcal{F(D, \sigma)}$ $\alpha$-protective w.r.t. $f$ does not
generate inference channels, where $f$ is one of the measures from Fig. \ref{fig:Measures}.
\end{theorem}
\begin{proof}
For any $\alpha$-discriminatory pattern $p_a: \{A, B, C\}$, where $A \subseteq DI_b$ and $C$ denies some benefit,
anti-discrimination pattern sanitization is performed in one of the following ways w.r.t. the value of $f$:
\begin{equation}
\label{addnDGslift}
supp(p_s:\{A, B\})+\Delta_f, \forall p \subset p_s
\end{equation}
where $f=slift$, $f=clift$, $f=slift_d$ or $f=slift_c$.
\begin{equation}
\label{addnDGelift}
supp(p_s:\{B, C\})+\Delta_f, \forall p \subset p_s
\end{equation}
where $f=elift$, $f=elift_d$ or $f=elift_c$.
Inference channels could appear in two different cases:
(a) between the pattern $p_s$ or one of its supersets ($p_x$ s.t.
$p_s \subset p_x$), and (b) between the pattern $p_s$ and one
of its subsets ($p_x$ s.t. $p_x \subset p_s$).
{\em Case (a)}. Since $\mathcal{F(D, \sigma)}$ is $k$-anonymous, we have $supp(C^{p_x}_{p_s}) \geq k$.
Increasing the support of $p_s$ and its subsets by $\Delta_f$
as in Expressions (\ref{addnDGslift}-\ref{addnDGelift})
causes the value of $supp(C^{p_x}_{p_s})$ to
increase by $\Delta_f$, because only the first term of the sum
in Expression (\ref{supchan}) used to compute
$supp(C^{p_x}_{p_s})$ is increased (the support of $p_s$).
Hence, the support of $C^{p_x}_{p_s}$ stays above $k$.
{\em Case (b)}. Since $\mathcal{F(D, \sigma)}$ is $k$-anonymous, we have
$supp(C^{p_s}_{p_x})$ $\geq k$. Increasing the support
of $p_s$ and its subsets by $\Delta_f$ as in Expressions (\ref{addnDGslift}-\ref{addnDGelift})
means adding the same value $\Delta_f$ to each term of the sum in
Expression (\ref{supchan}) used to compute
$supp(C^{p_s}_{p_x})$. Hence, this support
of $C^{p_s}_{p_x}$ does not change.
Thus, the theorem holds. 
\end{proof}
Since using our anti-discrimination pattern sanitization methods
for making $\mathcal{F(D, \sigma)}$ $\alpha$-protective
w.r.t. $f$ cannot make $\mathcal{F(D, \sigma)}$ non-$k$-anonymous, a safe way to obtain
an $\alpha$-protective $k$-anonymous $\mathcal{F(D, \sigma)}$   w.r.t. $f$ is to apply
anti-discrimination pattern sanitization methods
to a $k$-anonymous version of $\mathcal{F(D, \sigma)}$, in order to turn
$\alpha$-discriminatory patterns detected in that $k$-anonymous version
into $\alpha$-protective patterns w.r.t $f$.
We propose
Algorithm \ref{alg:AntiDiscAnonyPattSanit} to obtain
an $\alpha$-protective $k$-anonymous version of an original frequent
pattern set w.r.t. to a discrimination measure $f$.
There are two assumptions in this algorithm: first,
the class attribute is binary; second, protected groups $DI_b$
correspond to nominal attributes.
Given an original pattern set $\mathcal{F}(\mathcal{D}, \sigma)$, denoted
by $\mathcal{FP}$ for short, a discriminatory threshold $\alpha$, an
anonymity threshold $k$, a discrimination measure $f$, protected groups $DI_b$ and a class item $C$ which denies some benefit, Algorithm \ref{alg:AntiDiscAnonyPattSanit} starts by obtaining $\mathcal{FP}'$, which is
a $k$-anonymous version of $FP$ (Line 3). It uses Algorithm 3 in
\cite{AtzoriBGP08} to do this.
Then, the algorithm uses the \textsc{DetDiscPatt} function in Algorithm \ref{alg:DetDiscPatt} to determine the subset $\mathcal{D}_{D}$ which contains $\alpha$-discriminatory patterns in $\mathcal{FP}'$ (Line 4).
\begin{algorithm}[h]
\caption{{\sc Anti-discrimination $k$-anonymous pattern sanitization}}
\label{alg:AntiDiscAnonyPattSanit}
{\small \begin{algorithmic}[1]
\STATE Inputs: Database $\mathcal{D}$, $\mathcal{FP}:=\mathcal{F}(D, \sigma)$, $k$,
$DI_b$, discrimination measure $f$, $\alpha$, $C=${class item with negative decision value}
\STATE Output: $\mathcal{FP}''$: $\alpha$-protective $k$-anonymous frequent pattern set
\STATE $\mathcal{FP}' \leftarrow PrivacyAdditiveSanitization(\mathcal{FP},k)$ \hspace{2mm}  //Algorithm 3 in \cite{AtzoriBGP08}
\STATE $\mathcal{D}_{D} \leftarrow$ {\bf Function} \textsc{DetDiscPatt}($\mathcal{FP}'$, $\mathcal{D}$, $DI_b$, $f$, $\alpha$, $C$)
\FORALL{$p \in \mathcal{D}_{D}$}
\STATE Compute $impact(p) = |\{p': A', B', C\} \in \mathcal{D}_{D}$ s.t. $p' \subset p|$
\ENDFOR
\STATE Sort $\mathcal{D}_{D}$ by descending $impact$
\STATE $\mathcal{FP}'' \leftarrow$ {\bf Function} \textsc{AntiDiscPattSanit}($\mathcal{FP}'$, $\mathcal{D}$, $\mathcal{D}_{D}$, $DI_b$, $f$, $\alpha$, $C$)
\STATE Output: $\mathcal{FP}''$
\end{algorithmic}}
\end{algorithm}
As we showed in Theorem \ref{theoDisc}, given two $\alpha$-discriminatory patterns $p_1:\{A, B, C\}$ and $p_2:\{A', B', C\}$, where $p_2 \subset p_1$, making $p_1$
$\alpha$-protective w.r.t. $f$ can make also $p_2$
less discriminatory or even $\alpha$-protective,
depending on the value of $\alpha$ and the support of patterns.
This justifies why, among the patterns in $\mathcal{D}_{D}$,
Algorithm~\ref{alg:AntiDiscAnonyPattSanit} transforms first those with maximum
impact on making other patterns $\alpha$-protective
w.r.t. $f$. For each pattern $p \in \mathcal{D}_{D}$, the number of patterns in $\mathcal{D}_{D}$ which are subsets of $p$ is taken
as the impact
of $p$ (Lines 4-7), that is $impact(p)$. Then,
the patterns in $\mathcal{D}_{D}$ will be made $\alpha$-protective w.r.t. $f$ by
descending order of $impact$ (Line 8). Thus, the patterns
with maximum $impact(p)$ will be made $\alpha$-protective first, with
the aim of minimizing the pattern distortion. 
Finally, the algorithm uses the function \textsc{AntiDiscPattSanit} in Algorithm \ref{alg:AntiDiscPattSanit} to make each pattern $p$ in $\mathcal{D}_{D}$ $\alpha$-protective using anti-discrmination pattern sanitization methods w.r.t. $f$ (Line 9).
\subsection{Achieving an Unexplainable Discrimination and Privacy Protected Pattern Set}
\label{sec:explainable}
In order to simultaneously achieve unexplainable discrimination and privacy awareness in $\mathcal{F(D, \sigma)}$, we need to generate an unexplainable discrimination and privacy protected version of $\mathcal{F(D, \sigma)}$:
\begin{definition}[$d$-explainable $\alpha$-protective $k$-anonymous pattern set]
\label{def:fairde}
Given a collection of frequent patterns $\mathcal{F(D, \sigma)}$,
an anonymity threshold $k$, an explainable discrimination threshold $d$, a discrimination threshold $\alpha$, protected groups $DI_b$, legally-grounded groups $DI_e$, and a discrimination measure $f$, $\mathcal{F(D, \sigma)}$ is $d$-explainable $\alpha$-protective $k$-anonymous if it is both $k$-anonymous and $d$-explainable $\alpha$-protective w.r.t. $DI_b$, $DI_e$ and $f$.
\end{definition}
In order to generate a $d$-explainable $\alpha$-protective $k$-anonymous version of $\mathcal{F}(D, \sigma)$, we need to examine the following issues: (1) how making $\mathcal{F}(D, \sigma)$ $k$-anonymous impacts $d$-explainable and $d$-unexplainable patterns in $\mathcal{F}(D, \sigma)$; (2) how making $\mathcal{F}(D, \sigma)$ $d$-explainable $\alpha$-protective w.r.t. $f$ impacts the $k$-anonymity of $\mathcal{F}(D, \sigma)$.
We study the first issue by presenting two scenarios.
\begin{table}[ht]
\caption{Scenario \ref{scenario3}: Examples of frequent patterns extracted from Table \ref{tab:51}}
\label{tab:54}
\centering
\begin{tabular}{l c}
\hline
{\em Patterns} & {\em Support} \\
\hline
{\bf $p_s$}:\texttt{\{female, veterinarian\}}  & 34  \\
$p_2:$\texttt{\{paid-delay, veterinarian, salary $>$ 15000\}} & 59 \\
$p_1:$\texttt{\{female, veterinarian, No\}} & 20 \\
$p_n:$\texttt{\{paid-delay, veterinarian, No\}} & 37 \\
$p_{ns}:$\texttt{\{paid-delay, veterinarian\}} &  64\\
$p_d:$\texttt{\{female, veterinarian, paid-delay\}} & 31 \\
\hline
\end{tabular}
\end{table}
\begin{scenario}
\label{scenario3}
{\em
Table \ref{tab:54} illustrates an example of frequent patterns
that come from
the data set in Table \ref{tab:51} with $\sigma=15$. Let $DI_b:$\emph{\{Sex = female\}},  $DI_e:$\emph{\{Credit\_history = paid-delay\}}, $f=slift$, $\alpha=1.25$, $k=8$ and $d=0.9$. Suppose the PD pattern $p_1$ in Table \ref{tab:54} is 1.25-discriminatory ({\em i.e.} there is inferred discrimination against veterinarian women applicants). However, pattern $p_1$ is a 0.9-explainable pattern  because both conditions of Definition \ref{def:dinstance} are satisfied ($\frac{supp(p_n)}{supp(p_{ns})}= \frac{37}{64}=0.57$ is higher than $0.9 \cdot \frac{supp(p_1)}{p_s}=0.9 \cdot \frac{20}{34}=0.53$ and $\frac{supp(p_d)}{supp(p_s)}=\frac{31}{34}=0.91$ is higher than 0.9). Then, $p_1$ is a $d$-explainable pattern w.r.t. pattern $p_n$ ({\em i.e.} the inferred discrimination against veterinarian women applicants is explainable by their delay in returning previous credits). On the other hand,
although the support of each pattern in the collection is higher than $k$,
there is an inference channel between patterns $p_{ns}$ and $p_2$;
note that $supp(p_{ns})-supp(p_2)=64-59=5$ is smaller than 8
({\em i.e.} one can infer the
existence of only 5 veterinarians who are delayed
in returning previous credits and with salary no more than 15000 \euro). To block the inference channel between $p_{ns}$ and $p_2$, the following privacy additive sanitization is performed:
\begin{equation}
\label{eq:addsantipinstance}
supp(p_{ns})+k, \forall p \subseteq p_{ns}
\end{equation}
After this transformation, the new support of pattern $p_{ns}$ is 72.
The new support value of pattern $p_{ns}$ changes the satisfaction of
Condition 1 in Definition \ref{def:dinstance} in the following way:
$\frac{supp(p_n)}{supp(p_{ns})+k}= \frac{37}{64+8}=0.513$ is less than $0.9 \cdot \frac{supp(p_1)}{p_s}=0.9 \cdot \frac{20}{34}=0.53$. Then, in this scenario, making the collection of patterns in Table \ref{tab:54} $k$-anonymous makes $d$-explainable pattern $p_1$ $d$-unexplainable.
}
\end{scenario}
\begin{table}[ht]
\caption{Scenario \ref{scenario4}: Examples of frequent patterns extracted from Table \ref{tab:51}}
\label{tab:55}
\centering
\begin{tabular}{l c}
\hline
{\em Patterns} & {\em Support} \\
\hline
{\bf $p_s$}:\texttt{\{female, veterinarian\}}  & 30  \\
$p_2:$\texttt{\{female, veterinarian, salary $>$ 15000\}} & 29 \\
$p_1:$\texttt{\{female, veterinarian, No\}} & 29 \\
$p_n:$\texttt{\{paid-delay, veterinarian, No\}} & 23 \\
$p_{ns}:$\texttt{\{paid-delay, veterinarian\}} & 27 \\
$p_d:$\texttt{\{female, veterinarian, paid-delay\}} & 30 \\
\hline
\end{tabular}
\end{table}
\begin{scenario}
\label{scenario4}
{\em
Table \ref{tab:55} illustrates an example of frequent patterns
that come from
the data set in Table \ref{tab:51} with $\sigma=15$.
Let $DI_b:$ \emph{\{Sex = female\}}, $DI_e:$\emph{\{Credit\_history = paid-delay\}}, $f=slift$, $\alpha=1.25$, $k=2$ and $p=0.9$. Suppose the PD pattern $p_1$ in Table \ref{tab:55} is 1.25-discriminatory ({\em i.e.} there is inferred discrimination against veterinarian women applicants). Pattern $p_1$ is $d$-unexplainable pattern w.r.t. pattern $p_n$ because Condition 1 of Definition \ref{def:dinstance} does not satisfy ($\frac{supp(p_n)}{supp(p_{ns})}= \frac{23}{27}=0.85$ is less than $0.9 \cdot \frac{supp(p_1)}{supp(p_s)}=0.87$)  while Condition 2 of Definition \ref{def:dinstance} is satisfied ({\em i.e.} $\frac{supp(p_d)}{supp(p_s)}=\frac{30}{30}=1$ is higher than 0.9). Then, $p_1$ is a $d$-unexplainable pattern ({\em i.e.} the inferred discrimination against veterinarian women applicants is not explainable by their delay in returning previous credits). On the other hand,
although the support of each pattern in the collection is higher than $k$,
there is an inference channel between patterns $p_s$ and $p_2$;
note that $supp(p_s)-supp(p_2)=30-29=1$ is smaller than 2. To block the inference channel between $p_s$ and $p_2$, privacy pattern sanitization is performed according to Expression (\ref{eq:addsanti}). After this transformation, the new support value of pattern $p_s$ is 32.
The new support value of pattern $p_{ns}$ satisfies Condition 1 of Definition \ref{def:dinstance} ($\frac{supp(p_n)}{supp(p_{ns})}= \frac{23}{27}=0.85$ is higher than $0.9 \cdot \frac{supp(p_1)}{supp(p_s)+k}=0.815$) while it does not change the satisfaction of Condition 2 of Definition \ref{def:dinstance} ($\frac{supp(p_d)}{supp(p_s)+k}=\frac{30}{32}=0.93$ is higher than 0.9). Thus, in this scenario, making the collection of patterns in Table \ref{tab:54} $k$-anonymous
turns $d$-unexplainable pattern $p_1$ into a $d$-explainable pattern.
}
\end{scenario}
To summarize, using a privacy pattern sanitization method for making $\mathcal{F}(D, \sigma)$ $k$-anonymous can make $\mathcal{F}(D, \sigma)$ more or less $d$-explainable $\alpha$-protective w.r.t. $slift$. We also observe a similar
behavior for alternative discrimination measures.
Hence, what makes sense is first to obtain a
$k$-anonymous version of $\mathcal{F}(D, \sigma)$
and then look for
$d$-unexplainable patterns in it.
After that, we use anti-discrimination pattern sanitization methods proposed in Section \ref{sec:nDG} for making $\mathcal{F}(D, \sigma)$ $d$-explainable $\alpha$-protective.
According to Theorem \ref{theoprivacy},
these methods cannot make $\mathcal{F}(D, \sigma)$ non-$k$-anonymous as a result of their transformation.
The above procedure (first dealing with $k$-anonymity and
then with $d$-explainability and  $\alpha$-protectiveness)
is encoded in
Algorithm \ref{alg:UnExplainDiscAnonyPattSanit}.
Given an original pattern set $\mathcal{F}(\mathcal{D}, \sigma)$, denoted
by $\mathcal{FP}$ for short, a discriminatory threshold $\alpha$, an
anonymity threshold $k$, a discrimination measure $f$, an explainable discrimination threshold $d$, protected groups $DI_b$, legally-grounded groups $DI_e$, and a class item $C$ which denies some benefit, Algorithm \ref{alg:UnExplainDiscAnonyPattSanit} starts by obtaining $\mathcal{FP}'$, which is
a $k$-anonymous version of $\mathcal{FP}$ (Step 3). It calls Algorithm 3 in
\cite{AtzoriBGP08} to do this.
Then, the algorithm uses the function \textsc{DetUnExplainPatt} in Algorithm \ref{alg:DetUnExplainPatt} to determine the subset $\mathcal{D}_{bad}$ which contains $d$-unexplainable patterns in $\mathcal{FP}'$ (Line 4). Then, the algorithm sorts $\mathcal{D}_{bad}$ by descending order of $impact$ to transform first those patterns in $\mathcal{D}_{bad}$ with maximum impact on making other patterns $\alpha$-protective (Lines 4-8). Finally, the algorithm uses the function \textsc{UnExplainAntiDiscPattSanit} in Algorithm \ref{alg:UnExplainAntiDiscPattSanit} to
make each $d$-unexplainable pattern in $\mathcal{FP}'$ $\alpha$-protective using anti-discrimination pattern sanitization w.r.t. $f$ (Line 9).
\begin{algorithm}[h]
\caption{{\sc Unexplainable discrimination protected and anonymous pattern sanitization}}
\label{alg:UnExplainDiscAnonyPattSanit}
{\small \begin{algorithmic}[1]
\STATE Inputs: $\mathcal{FP}:=\mathcal{F}(D, \sigma)$, $k$,
$DI_b$, $DI_e$,
explainable discrimination threshold $d$, discrimination measure $f$, $\alpha$, $C=${class item with negative decision value}
\STATE Output: $\mathcal{FP}''$: $d$-explainable $\alpha$-protective $k$-anonymous frequent pattern set
\STATE $\mathcal{FP}' = PrivacyAdditiveSanitization(FP,k)$ \hspace{2mm}  //Algorithm 3 in \cite{AtzoriBGP08}
\STATE $\mathcal{D}_{bad} \rightarrow $ {\bf Function} \textsc{DetUnExplainPatt}($\mathcal{FP}'$, $DI_e$, $DI_b$, $f$, $\alpha$, $C$)
\FORALL{$p \in \mathcal{D}_{bad}$}
\STATE Compute $impact(p) = |\{p': A', B', C\} \in \mathcal{D}_{bad}$ s.t. $p' \subset p|$
\ENDFOR
\STATE Sort $\mathcal{D}_{bad}$ by descending $impact$
\STATE $\mathcal{FP}'' \leftarrow $ {\bf Function} \textsc{UnExplainAntiDiscPattSanit}($\mathcal{FP}'$, $\mathcal{D}_{bad}$, $DI_b$, $f$, $\alpha$, $C$)
\STATE Output: $\mathcal{FP}''$
\end{algorithmic}}
\end{algorithm}
       
\section{Experiments}  

\label{sec:experiments}
This section presents the experimental evaluation of the approaches we proposed in this chapter. First, we describe the utility measures and then the empirical results.
In the sequel, $\mathcal{FP}$ denotes the set of frequent patterns extracted from database $\mathcal{D}$ by the Apriori algorithm \cite{Apriori}; $\mathcal{FP}'$ denotes the $k$-anonymous version of $\mathcal{FP}$ obtained by the privacy pattern sanitization method; $\mathcal{FP}''$ denotes the $\alpha$-protective $k$-anonymous version of $\mathcal{FP}$ obtained by Algorithm \ref{alg:AntiDiscAnonyPattSanit}, and $\mathcal{FP}^*$ denotes the $\alpha$-protective version of $\mathcal{FP}$ obtained by Algorithm \ref{alg:AntiDiscPattSanit}. We also denote by $\mathcal{TP}$ any transformed pattern set, {\em i.e.}, either $\mathcal{FP}'$, $\mathcal{FP}''$ or $\mathcal{FP}^*$. We used the Adult and German credit datasets introduced in Section \ref{sec:datasets}.
We used the ``train'' part of Adult dataset to obtain $\mathcal{FP}$ and any transformed pattern set $\mathcal{TP}$
For our experiments with the Adult dataset, we set $DI_b:$ \emph{\{Sex = female, Age = young\}} (cut-off for Age = young: 30 years old).
For our experiments with German credit dataset, we set
$DI_b:$ \emph{\{Age = old, Foreign worker = yes\}} (cut-off for Age = old: 50 years old).
\subsection{Utility Measures}
\label{sec:utilitymeasures}
To assess the information loss incurred
to achieve privacy and anti-discrimination in frequent pattern discovery,
we use the following measures.
\begin{itemize}

\item \textbf{Patterns with changed support}. Fraction of original frequent patterns in $\mathcal{FP}$ which have their
support changed in any transformed pattern set $\mathcal{TP}$:
$$\frac{|\{\langle I, supp(I) \rangle \in \mathcal{FP} : supp_{\mathcal{FP}}(I) \neq supp_{\mathcal{TP}}(I)\}|}{|\mathcal{FP}|}$$
\item \textbf{Pattern distortion error}. Average distortion w.r.t. the original support of frequent patterns:
$$\frac{1}{|\mathcal{FP}|} \cdot \sum_{I \in \mathcal{FP}} (\frac{supp_{\mathcal{TP}}(I) - supp_{\mathcal{FP}}(I)}{supp_{\mathcal{FP}}(I)})$$
\end{itemize}
The purpose of these measures is assessing the distortion introduced when
making $\mathcal{FP}$ $\alpha$-protective $k$-anonymous, in comparison with the distortion introduced by either making $\mathcal{FP}$ $\alpha$-protective or making $\mathcal{FP}$ $k$-anonymous, separately. The above measures are evaluated considering  $\mathcal{TP}=\mathcal{FP}''$, $\mathcal{TP}=\mathcal{FP}'$ and $\mathcal{TP}=\mathcal{FP}^*$.
In addition, we measure the impact of our pattern sanitization methods for making $\mathcal{FP}$ $k$-anonymous, $\alpha$-protective and $\alpha$-protective $k$-anonymous on the accuracy of a classifier using the CMAR ({\em i.e.} classification based on multiple association rules) approach \cite{CMAR}. Below, we describe the process of our evaluation:
\begin{enumerate}
\item The original data are first divided into training and testing sets, $\mathcal{D}_{train}$ and $\mathcal{D}_{test}$.
\item The original frequent patterns $\mathcal{FP}$ are extracted from the training set $\mathcal{D}_{train}$ by the Apriori algorithm \cite{Apriori}.
\item
Patterns in $\mathcal{TP}$ which contain the class item are selected as a candidate patterns for classification. They are donated by $\mathcal{TP}_s$. Note that, $\mathcal{TP}$ can be $\mathcal{FP}$, $\mathcal{FP}'$, $\mathcal{FP}^*$ or $\mathcal{FP}''$.
\item To classify each new object (record) in $\mathcal{D}_{test}$, the subset of patterns matching the new record in $\mathcal{TP}_s$ is found.
\item If all the patterns matching the new object have the same class item, then that class value is assigned to the new record. If the patterns are not consistent in terms of class items, the patterns are divided into groups according to class item values ({\em e.g.} denying credit and accepting credit). Then, the effects of the groups should be compared to yield the strongest group.
A strongest group is composed of a set of patterns highly positively correlated and that have good support. To determine this, for each pattern $p:\{X, C\}$, the value of $\max \chi^2$ is computed as follows:
\begin{equation}
\label{eq:maxchi}
\max\chi^2=(\min\{supp(X), supp(C)\}-\frac{supp(X) \times supp(C)}{|\mathcal{D}_{train}|})^2 \times|\mathcal{D}_{train}|\times e
\end{equation}
where
\begin{equation}
\begin{split}
\label{eq:e}
e=\frac{1}{supp(X) \times supp(C)} + \frac{1}{supp(X) \times (|\mathcal{D}_{train}|-supp(C))}
+ \\ \frac{1}{(|\mathcal{D}_{train}|-supp(X)) \times supp(C)}+ {\frac{1}{(|\mathcal{D}_{train}|-supp(X))(|\mathcal{D}_{train}|-supp(C))}}
\end{split}
\end{equation}
Then, for each group of patterns, the $weighted \chi^2$  measure of the group is computed\footnote{For computing $\chi^2$ see http://www.csc.liv.ac.uk/~frans/Notes/chiTesting.html} as $\sum{\frac{\chi^2 \times \chi^2}{max\chi^2}}$.
The class item of the group with maximum $weighted \chi^2$ is assigned to the new record.
%
\item After obtaining the predicted class item of each record in $\mathcal{D}_{test}$, the accuracy of the classifier can be simply computed w.r.t. observed and predicted class items (\em{i.e.} contingency table).
\end{enumerate}
To measure the accuracy of a classifier based on a collection of frequent patterns, we perform the above process considering $\mathcal{TP}=\mathcal{FP}$, $\mathcal{TP}=\mathcal{FP}''$, $\mathcal{TP}=\mathcal{FP}'$ and $\mathcal{TP}=\mathcal{FP}^*$.
Finally, to measure the impact of making $\mathcal{FP}$ $k$-anonymous on discrimination, we use the following measures, which
are evaluated considering $\mathcal{TP}=\mathcal{FP}'$.
\textbf{Discrimination prevention degree (DPD)}. Percentage of $\alpha$-discriminatory patterns obtained from $\mathcal{FP}$ that are no longer $\alpha$-discriminatory in the transformed patterns ($\mathcal{TP}$).
\textbf{New (ghost) discrimination degree (NDD)}. Percentage of $\alpha$-discriminatory patterns obtained from transformed patterns ($\mathcal{TP}$) that were $\alpha$-protective in $\mathcal{FP}$.
%
%
\subsection{Empirical Evaluation}
We now discuss the experiments conducted on the Adult and German credit datasets, for different values of $\alpha$, $k$ and $\sigma$, in terms of the defined utility measures.
Note that all the empirical results presented in the sequel are related to the
worst-case scenario in which the original patterns are made protected against both explainable and unexplainable discrimination. In other words, we generate $\alpha$-protective or $\alpha$-protective $k$-anonymous versions of
the original pattern set.
\subsubsection{Pattern Distortion}
Fig.~\ref{fig:experiprivacy1} and Fig.~\ref{fig:experiprivacy2} show pattern distortion scores observed after making $\mathcal{FP}$ $k$-anonymous in Adult
and German credit, respectively. We show the results for
varying values of $k$ and the
support $\sigma$. It can be seen that the
percentage of patterns whose support has changed (left charts of Fig.~\ref{fig:experiprivacy1} and \ref{fig:experiprivacy2}) and the average distortion introduced (right charts of Fig.~\ref{fig:experiprivacy1} and \ref{fig:experiprivacy2})
increase with larger $k$ and with smaller support $\sigma$, due to the increasing number of inference channels.
Comparing the two datasets, pattern distortion scores in German credit are higher than those in Adult, even taking the same values of $k$ and $\sigma$. This is mainly due to the substantial difference in the number of inference channels detected in the two datasets: the maximum number of inference channels detected in Adult is 500, while in German credit it is 2164.
\begin{figure*}
\centering
\begin{tikzpicture}[thick, scale=1]
\def \plotdist {7.8 cm} 
\def \xsc {0.68} 
\def \ysc {0.63} 
\begin{scope}[xscale=\xsc , yscale=\ysc]
\begin{axis}[
        xlabel={\bf Support} (\%),
        ylabel={\bf Patterns w. Changed Support} (\%),
        ]
    \addplot[mark=square,blue,mark size=3pt,very thick] plot coordinates {
        (10,39.48)
        (15,36.63)
        (20,29.83)
        (25,30.35)
    };
    \addlegendentry{$k=5$}
    \addplot[mark=triangle,green,mark size=3pt,very thick]
        plot coordinates {
        (10,42.19)
        (15,40.1)
        (20,32.45)
        (25,33.92)
        };
    \addlegendentry{$k=10$}
    \addplot[mark=diamond,gray,mark size=3pt,very thick] plot coordinates {
        (10,48.06)
        (15,44.96)
        (20,34.75)
        (25,35.11)
    };
    \addlegendentry{$k=25$}
 \addplot[mark=oplus,red,mark size=3pt,very thick]
        plot coordinates {
        (10,53.7)
        (15,48.78)
        (20,38.36)
        (25,37.5)
        };
    \addlegendentry{$k=50$}
    \addplot[mark=star,black,mark size=3pt,very thick] plot coordinates {
        (10,60.04)
        (15,55.55)
        (20,41.63)
        (25,37.5)
    };
    \addlegendentry{$k=100$}
    \end{axis}
    \end{scope}
\begin{scope}[xshift=\plotdist, xscale=\xsc , yscale=\ysc]
\begin{axis}[
        xlabel={\bf Support (\%)},
        ylabel={\bf Pattern Distortion Error (\%)},
        ]
    \addplot[mark=square,blue,mark size=3pt,very thick] plot coordinates {
        (10,0.15)
        (15,0.04)
        (20,0.034)
        (25,0.15)
    };
    \addlegendentry{$k=5$}
  \addplot[mark=triangle,green,mark size=3pt,very thick] plot coordinates {
        (10,0.36)
        (15,0.11)
        (20,0.08)
        (25,0.045)
      };
    \addlegendentry{$k=10$}
    \addplot[mark=diamond,gray,mark size=3pt,very thick]
        plot coordinates {
        (10,1.19)
        (15,0.44)
        (20,0.26)
        (25,0.12)
        };
    \addlegendentry{$k=25$}
  \addplot[mark=oplus,red,mark size=3pt,very thick]
        plot coordinates {
        (10,2.8)
        (15,0.99)
        (20,0.64)
        (25,0.35)
        };
    \addlegendentry{$k=50$}
 \addplot[mark=star,black,mark size=3pt,very thick]
        plot coordinates {
        (10,6.4)
        (15,2.31)
        (20,1.37)
        (25,0.35)
        };
    \addlegendentry{$k=100$}
    \end{axis}
    \end{scope}
\end{tikzpicture}
\caption{Pattern distortion scores to make the Adult dataset $k$-anonymous}
\label{fig:experiprivacy1}
\end{figure*}
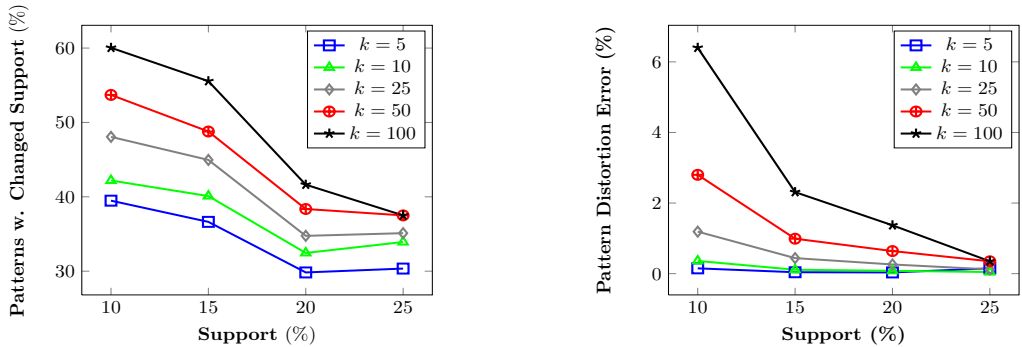
\begin{figure*}
\centering
\begin{tikzpicture}[thick, scale=1]
\def \plotdist {7.8 cm} 
\def \xsc {0.68} 
\def \ysc {0.63} 
\begin{scope}[xscale=\xsc , yscale=\ysc]
\begin{axis}[
        xlabel={\bf Support (\%)},
        ylabel={\bf Pattern w. Changed Support (\%)},
        ]
    \addplot[mark=square,blue,mark size=3pt,very thick] plot coordinates {
        (10,52.03)
        (15,48.66)
        (20,45.25)
        (25,38.84)
    };
    \addlegendentry{$k=3$}
    \addplot[mark=triangle,green,mark size=3pt,very thick]
        plot coordinates {
        (10,60.1)
        (15,57.68)
        (20,52.76)
        (25,46.28)
        };
    \addlegendentry{$k=5$}
    \addplot[mark=diamond,gray,mark size=3pt,very thick] plot coordinates {
        (10,76.2)
        (15,69.44)
        (20,64.48)
        (25,57.58)
    };
    \addlegendentry{$k=10$}
 \addplot[mark=oplus,red,mark size=3pt,very thick]
        plot coordinates {
        (10,80)
        (15,73.78)
        (20,69.03)
        (25,65.23)
        };
    \addlegendentry{$k=15$}
    \end{axis}
    \end{scope}
\begin{scope}[xshift=\plotdist, xscale=\xsc , yscale=\ysc]
\begin{axis}[
        xlabel={\bf Support (\%)},
        ylabel={\bf Pattern Distortion Error (\%)},
        ]
    \addplot[mark=square,blue,mark size=3pt,very thick] plot coordinates {
        (10,4.9)
        (15,3.22)
        (20,1.89)
        (25,1.1)
    };
    \addlegendentry{$k=3$}
  \addplot[mark=triangle,green,mark size=3pt,very thick] plot coordinates {
        (10,11.79)
        (15,7.8)
        (20,4.5)
        (25,2.57)
      };
    \addlegendentry{$k=5$}
    \addplot[mark=diamond,gray,mark size=3pt,very thick]
        plot coordinates {
         (10,34.23)
        (15,23.99)
        (20,14.34)
        (25,8.16)
        };
    \addlegendentry{$k=10$}
  \addplot[mark=oplus,red,mark size=3pt,very thick]
        plot coordinates {
       (10,70.45)
        (15,41.18)
        (20,24.7)
        (25,16.1)
        };
    \addlegendentry{$k=15$}
    \end{axis}
    \end{scope}
\end{tikzpicture}
\caption{Pattern distortion scores to make the German credit dataset $k$-anonymous}
\label{fig:experiprivacy2}
\end{figure*}
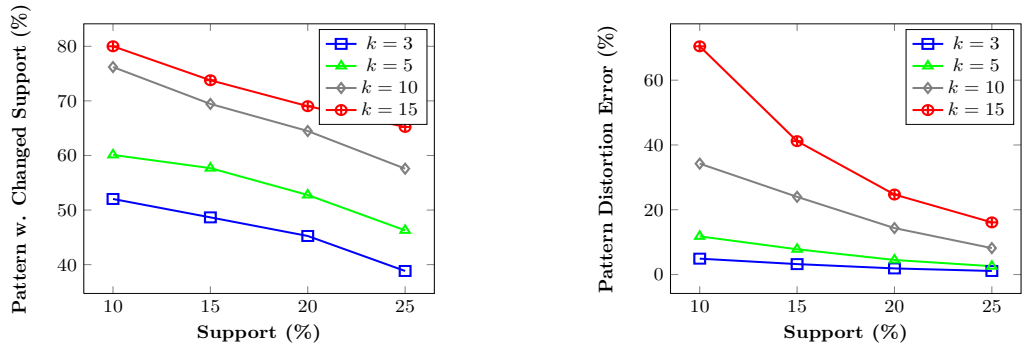
Fig.~\ref{fig:experidiscrmination1} and Fig.~\ref{fig:experidiscrmination2} show pattern distortion scores observed after making $\mathcal{FP}$ $\alpha$-protective in Adult and German credit, respectively. We take $f=slift$
and we show the results for varying values of $\alpha$ and support $\sigma$.
It can be seen that distortion scores increase with smaller $\alpha$ and smaller $\sigma$, because
the number of $\alpha$-discriminatory patterns increases.
 Also in this case the values of the distortion scores for Adult are
 less than for German credit. We performed the same experiments for
 discrimination measures other than $slift$ and we observed a similar behavior.
When comparing Figs.~\ref{fig:experiprivacy1}-\ref{fig:experiprivacy2} and Figs.~\ref{fig:experidiscrmination1}-\ref{fig:experidiscrmination2}, we observe that the percentage of patterns with changed support and the average distortion introduced are higher after the application of privacy pattern sanitization in both datasets. In other words, we observe that
guaranteeing privacy produces more distortion than guaranteeing anti-discrimination.
This is due to the number of inference channels detected in our experiment (for different values of $k$), which is higher than the number of $\alpha$-discriminatory patterns detected (for different values of $\alpha$.)
%
Fig.~\ref{fig:experiprivacydiscrmination1} and Fig.~\ref{fig:experiprivacydiscrmination2} show the pattern distortion scores observed after making $\mathcal{FP}$ $\alpha$-protective $k$-anonymous in the Adult and German credit datasets,
respectively. We take $f=slift$ and we show the results for varying values of $k$, $\alpha$ and $\sigma$.
Since the number of inference channels increases with $k$,
the number of $\alpha$-discriminatory patterns
increases as $\alpha$ becomes smaller, and both numbers increase
as $\sigma$ becomes smaller, the percentage of patterns with modified support (left charts of Fig.~\ref{fig:experiprivacydiscrmination1} and Fig.~\ref{fig:experiprivacydiscrmination2}) and the average distortion introduced (right charts of Fig.~\ref{fig:experiprivacydiscrmination1} and Fig.~\ref{fig:experiprivacydiscrmination2})
have
the same dependencies w.r.t. $k$, $\alpha$ and $\sigma$.
We performed the same experiments for other discrimination measures and we observed a similar behavior.
Comparing the two datasets, here we also observe that the values of
the distortion scores for the Adult dataset are less than for the German credit dataset.
\begin{figure*}
\centering
\begin{tikzpicture}[thick, scale=1]
\def \plotdist {7.1 cm} 
\def \xsc {0.68} 
\def \ysc {0.63} 
\begin{scope}[xscale=\xsc , yscale=\ysc]
\begin{axis}[
legend style={
at={(1.28,0.98)},
},
        xlabel={\bf Support (\%)},
        ylabel={\bf Patterns w. Changed Support (\%)},
        ]
    \addplot[mark=square,blue,mark size=3pt,very thick] plot coordinates {
        (10,5.02)
        (15,5.03)
        (20,4.59)
        (25,2.97)
    };
    \addlegendentry{$\alpha=1.2$}
    \addplot[mark=triangle,green,mark size=3pt,very thick]
        plot coordinates {
        (10,4.55)
        (15,4.34)
        (20,4.26)
        (25,1.78)
        };
    \addlegendentry{$\alpha=1.3$}
    \addplot[mark=diamond,gray,mark size=3pt,very thick] plot coordinates {
       (10,3.94)
        (15,3.99)
        (20,2.29)
        (25,1.78)
    };
    \addlegendentry{$\alpha=1.4$}
 \addplot[mark=oplus,red,mark size=3pt,very thick]
        plot coordinates {
        (10,2.7)
        (15,0.52)
        (20,0)
        (25,0)
        };
    \addlegendentry{$\alpha=1.5$}
    \addplot[mark=star,black,mark size=3pt,very thick] plot coordinates {
       (10,0.54)
        (15,0)
        (20,0)
        (25,0)
    };
    \addlegendentry{$\alpha=1.6$}
    \end{axis}
    \end{scope}
\begin{scope}[xshift=\plotdist, xscale=\xsc , yscale=\ysc]
\begin{axis}[
legend style={
at={(1.28,0.98)},
},
        xlabel={\bf Support (\%)},
        ylabel={\bf Pattern Distortion Error (\%)},
        ]
    \addplot[mark=square,blue,mark size=3pt,very thick] plot coordinates {
        (10,0.88)
        (15,0.66)
        (20,0.56)
        (25,0.33)
    };
    \addlegendentry{$\alpha=1.2$}
  \addplot[mark=triangle,green,mark size=3pt,very thick] plot coordinates {
        (10,0.44)
        (15,0.27)
        (20,0.19)
        (25,0.1)
      };
    \addlegendentry{$\alpha=1.3$}
    \addplot[mark=diamond,gray,mark size=3pt,very thick]
        plot coordinates {
        (10,0.18)
        (15,0.07)
        (20,0.04)
        (25,0.006)
        };
    \addlegendentry{$\alpha=1.4$}
  \addplot[mark=oplus,red,mark size=3pt,very thick]
        plot coordinates {
       (10,0.07)
        (15,0.006)
        (20,0)
        (25,0)
        };
    \addlegendentry{$\alpha=1.5$}
 \addplot[mark=star,black,mark size=3pt,very thick]
        plot coordinates {
        (10,0.001)
        (15,0)
        (20,0)
        (25,0)
        };
    \addlegendentry{$\alpha=1.6$}
    \end{axis}
    \end{scope}
\end{tikzpicture}
\caption{Pattern distortion scores to make the
Adult dataset $\alpha$-protective}
\label{fig:experidiscrmination1}
\end{figure*}
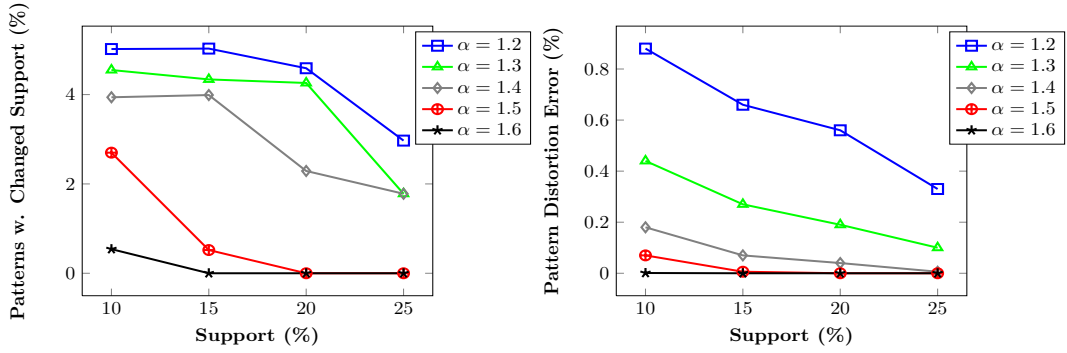
\begin{figure*}
\centering
\begin{tikzpicture}[thick, scale=1]
\def \plotdist {7.8 cm} 
\def \xsc {0.68} 
\def \ysc {0.63} 
\begin{scope}[xscale=\xsc , yscale=\ysc]
\begin{axis}[
        xlabel={\bf Support} (\%),
        ylabel={\bf Patterns w. Changed Support} (\%),
        ]
     \addplot[mark=square,blue,mark size=3pt,very thick] plot coordinates {
        (10,2.6)
        (15,1.58)
        (20,0.84)
        (25,0.5)
    };
    \addlegendentry{$\alpha=1.2$}
  \addplot[mark=triangle,green,mark size=3pt,very thick] plot coordinates {
       (10,2.48)
        (15,1.48)
        (20,0.84)
        (25,0.5)
      };
    \addlegendentry{$\alpha=1.4$}
    \addplot[mark=diamond,gray,mark size=3pt,very thick]
        plot coordinates {
        (10,2.32)
        (15,1.45)
        (20,0.84)
        (25,0.5)
        };
    \addlegendentry{$\alpha=1.6$}
  \addplot[mark=oplus,red,mark size=3pt,very thick]
        plot coordinates {
       (10,2.2)
        (15,1.45)
        (20,0.76)
        (25,0.5)
        };
    \addlegendentry{$\alpha=1.8$}
 \addplot[mark=star,black,mark size=3pt,very thick]
        plot coordinates {
         (10,2)
        (15,1.2)
        (20,0.54)
        (25,0.35)
        };
    \addlegendentry{$\alpha=2$}
    \end{axis}
    \end{scope}
\begin{scope}[xshift=\plotdist, xscale=\xsc , yscale=\ysc]
\begin{axis}[
        xlabel={\bf Support (\%)},
        ylabel={\bf Pattern Distortion Error (\%)},
        ]
      \addplot[mark=square,blue,mark size=3pt,very thick] plot coordinates {
        (10,5.2)
        (15,3.3)
        (20,1.5)
        (25,0.65)
    };
    \addlegendentry{$\alpha=1.2$}
  \addplot[mark=triangle,green,mark size=3pt,very thick] plot coordinates {
        (10,4.75)
        (15,2.48)
        (20,1.5)
        (25,0.65)
      };
    \addlegendentry{$\alpha=1.4$}
    \addplot[mark=diamond,gray,mark size=3pt,very thick]
        plot coordinates {
        (10,4.2)
        (15,1.91)
        (20,1.5)
        (25,0.65)
        };
    \addlegendentry{$\alpha=1.6$}
  \addplot[mark=oplus,red,mark size=3pt,very thick]
        plot coordinates {
      (10,3.8)
        (15,1.46)
        (20,0.63)
        (25,0.65)
        };
    \addlegendentry{$\alpha=1.8$}
 \addplot[mark=star,black,mark size=3pt,very thick]
        plot coordinates {
        (10,3.4)
        (15,1.12)
        (20,0.47)
        (25,0.15)
        };
    \addlegendentry{$\alpha=2$}
    \end{axis}
    \end{scope}
\end{tikzpicture}
\caption{Pattern distortion scores to make the German dataset $\alpha$-protective}
\label{fig:experidiscrmination2}
\end{figure*}
%

\begin{figure*}
\centering
\begin{tikzpicture}[thick, scale=1]
\def \plotdist {7.1 cm} 
\def \xsc {0.68} 
\def \ysc {0.63} 
\begin{scope}[xscale=\xsc , yscale=\ysc]
\begin{axis}[
legend style={
at={(1.28,0.98)},
},
        xlabel={\bf Support (\%)},
        ylabel={\bf Patterns w. Changed Support (\%)},
        ]
     \addplot[mark=square,blue,mark size=3pt,very thick] plot coordinates {
        (5,46.48)
        (10,42.11)
        (15,39.93)
        (20,33.11)
    };
    \addlegendentry{$\alpha=1.2$, $k=5$}
  \addplot[mark=triangle,green,mark size=3pt,very thick] plot coordinates {
       (5,52.35)
        (10,44.66)
        (15,42.88)
        (20,35.73)
      };
    \addlegendentry{$\alpha=1.2$, $k=10$}
    \addplot[mark=diamond,gray,mark size=3pt,very thick]
        plot coordinates {
       (5,64.4)
        (10,55.25)
        (15,50.34)
        (20,40.65)
        };
    \addlegendentry{$\alpha=1.2$, $k=50$}
  \addplot[mark=oplus,red,mark size=3pt,very thick]
        plot coordinates {
     (5,43.68)
        (10,40.57)
        (15,36.97)
        (20,29.83)
        };
    \addlegendentry{$\alpha=1.5$, $k=5$}
 \addplot[mark=star,black,mark size=3pt,very thick]
        plot coordinates {
        (5,50.42)
        (10,43.19)
        (15,40.27)
        (20,32.45)
        };
    \addlegendentry{$\alpha=1.5$, $k=10$}
  \addplot[mark=asterisk,brown,mark size=3pt,very thick]
        plot coordinates {
        (5,63.67)
        (10,54.01)
        (15,48.95)
        (20,38.36)
        };
    \addlegendentry{$\alpha=1.5$, $k=50$}
    \end{axis}
    \end{scope}
\begin{scope}[xshift=\plotdist, xscale=\xsc , yscale=\ysc]
\begin{axis}[
        xlabel={\bf Support (\%)},
        ylabel={\bf Pattern Distortion Error (\%)},
        ]
    \addplot[mark=square,blue,mark size=3pt,very thick] plot coordinates {
        (5,1.96)
        (10,1.04)
        (15,1.14)
        (20,0.85)
    };
    \addlegendentry{$\alpha=1.2$, $k=5$}
  \addplot[mark=triangle,green,mark size=3pt,very thick] plot coordinates {
        (5,3.08)
        (10,1.27)
        (15,1.21)
        (20,0.92)
      };
    \addlegendentry{$\alpha=1.2$, $k=10$}
    \addplot[mark=diamond,gray,mark size=3pt,very thick]
        plot coordinates {
        (5,13.17)
        (10,3.81)
        (15,2.2)
        (20,1.6)
        };
    \addlegendentry{$\alpha=1.2$, $k=50$}
  \addplot[mark=oplus,red,mark size=3pt,very thick]
        plot coordinates {
     (5,0.62)
        (10,0.23)
        (15,0.05)
        (20,0.034)
        };
    \addlegendentry{$\alpha=1.5$, $k=5$}
 \addplot[mark=star,black,mark size=3pt,very thick]
        plot coordinates {
        (5,1.57)
        (10,0.44)
        (15,0.12)
        (20,0.08)
        };
    \addlegendentry{$\alpha=1.5$, $k=10$}
  \addplot[mark=asterisk,brown,mark size=3pt,very thick]
        plot coordinates {
       (5,11.02)
        (10,3)
        (15,1.01)
        (20,0.64)
        };
    \addlegendentry{$\alpha=1.5$, $k=50$}
    \end{axis}
    \end{scope}
\end{tikzpicture}
\caption{Pattern distortion scores to make the Adult dataset $\alpha$-protective $k$-anonymous}
\label{fig:experiprivacydiscrmination1}
\end{figure*}
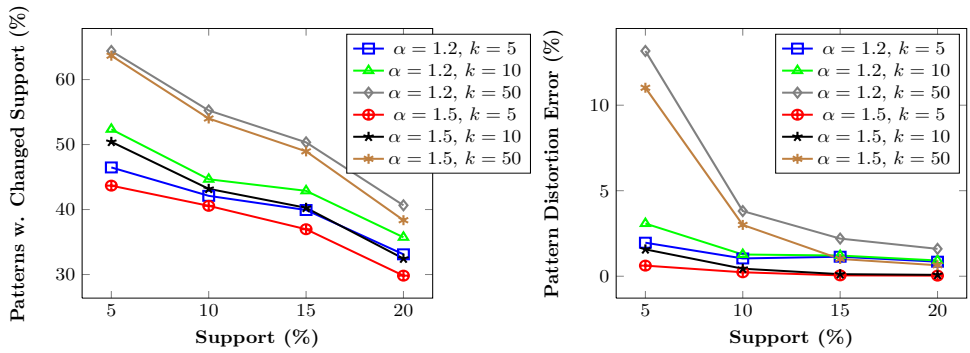
\begin{figure*}
\centering
\begin{tikzpicture}[thick, scale=1]
\def \plotdist {7.4 cm} 
\def \xsc {0.64} 
\def \ysc {0.60} 
\begin{scope}[xscale=\xsc , yscale=\ysc]
\begin{axis}[
legend style={
at={(1.45,0.98)},
},
        xlabel={\bf Support (\%)},
        ylabel={\bf Patterns w. Changed Support (\%)},
        ]
     \addplot[mark=square,blue,mark size=3pt,very thick] plot coordinates {
        (5,55.48)
        (10,52.11)
        (15,49.69)
        (20,45.9)
    };
    \addlegendentry{$\alpha=1.2$, $k=3$}
  \addplot[mark=triangle,green,mark size=3pt,very thick] plot coordinates {
       (5,69.35)
        (10,64.66)
        (15,59.69)
        (20,53.79)
      };
    \addlegendentry{$\alpha=1.2$, $k=5$}
    \addplot[mark=diamond,gray,mark size=3pt,very thick]
        plot coordinates {
       (5,79.4)
        (10,73.25)
        (15,69.44)
        (20,64.68)
        };
    \addlegendentry{$\alpha=1.2$, $k=10$}
  \addplot[mark=oplus,red,mark size=3pt,very thick]
        plot coordinates {
     (5,54.68)
        (10,50.57)
        (15,48.69)
        (20,45.26)
        };
    \addlegendentry{$\alpha=1.6$, $k=3$}
 \addplot[mark=star,black,mark size=3pt,very thick]
        plot coordinates {
        (5,64.42)
        (10,61.19)
        (15,57.69)
        (20,52.76)
        };
    \addlegendentry{$\alpha=1.6$, $k=5$}
  \addplot[mark=asterisk,brown,mark size=3pt,very thick]
        plot coordinates {
        (5,76.67)
        (10,73.01)
        (15,69.44)
        (20,64.68)
        };
    \addlegendentry{$\alpha=1.6$, $k=10$}
    \end{axis}
    \end{scope}
\begin{scope}[xshift=\plotdist, xscale=\xsc , yscale=\ysc]
\begin{axis}[
legend style={
at={(1.28,0.98)},
},
        xlabel={\bf Support (\%)},
        ylabel={\bf Pattern Distortion Error (\%)},
        ]
    \addplot[mark=square,blue,mark size=3pt,very thick] plot coordinates {
        (5,17.16)
        (10,12.04)
        (15,8.28)
        (20,3.65)
    };
    \addlegendentry{$\alpha=1.2$, $k=3$}
  \addplot[mark=triangle,green,mark size=3pt,very thick] plot coordinates {
        (5,23.08)
        (10,18.27)
        (15,12.26)
        (20,6.1)
      };
    \addlegendentry{$\alpha=1.2$, $k=5$}
    \addplot[mark=diamond,gray,mark size=3pt,very thick]
        plot coordinates {
        (5,41.17)
        (10,33.81)
        (15,26.57)
        (20,15.57)
        };
    \addlegendentry{$\alpha=1.2$, $k=10$}
  \addplot[mark=oplus,red,mark size=3pt,very thick]
        plot coordinates {
     (5,11.62)
        (10,8.23)
        (15,5.94)
        (20,2.63)
        };
    \addlegendentry{$\alpha=1.6$, $k=3$}
 \addplot[mark=star,black,mark size=3pt,very thick]
        plot coordinates {
        (5,18.57)
        (10,14.44)
        (15,9.42)
        (20,5.14)
        };
    \addlegendentry{$\alpha=1.6$, $k=5$}
  \addplot[mark=asterisk,brown,mark size=3pt,very thick]
        plot coordinates {
       (5,40.02)
        (10,32.03)
        (15,24.79)
        (20,14.7)
        };
    \addlegendentry{$\alpha=1.6$, $k=10$}
    \end{axis}
    \end{scope}
\end{tikzpicture}
\caption{Pattern distortion scores to make the German dataset $\alpha$-protective $k$-anonymous}
\label{fig:experiprivacydiscrmination2}
\end{figure*}
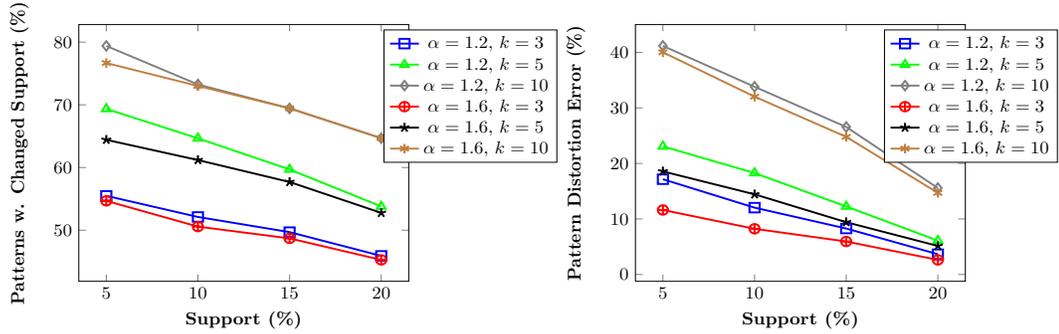

If we compare Figs.~\ref{fig:experiprivacy1}-\ref{fig:experiprivacy2} and Fig.~\ref{fig:experiprivacydiscrmination1}-\ref{fig:experiprivacydiscrmination2}, we observe only a marginal difference between the distortion introduced when making $\mathcal{FP}$ $\alpha$-protective $k$-anonymous, and the
distortion introduced when making it only $k$-anonymous.
For instance, in the experiment where we make Adult $\alpha$-protective $k$-anonymous (left chart of Fig.~\ref{fig:experiprivacydiscrmination1}) with minimum support 10\% and $k=5$, the percentage of patterns with changed support is 42.1\% (in the worst case $\alpha=1.2$), while when making the pattern
set $k$-anonymous (left chart of Fig.~\ref{fig:experiprivacy1}) we get a value of 39.48\%. In addition, the average distortion introduced (right chart of Fig.~\ref{fig:experiprivacydiscrmination1}) is 1.04\%  (in the worst case $\alpha=1.2$) to obtain an $\alpha$-protective $k$-anonymous version of original patterns, while it is 0.15\% to obtain a $k$-anonymous version of it (right chart of Fig.~\ref{fig:experiprivacy1}). As a consequence, we can (empirically) conclude that we provide protection against both the privacy and discrimination threats
with a marginally higher distortion w.r.t. providing protection against the privacy threat only.
\subsubsection{Preservation of the Classification Task}
Tables~\ref{tab:accuracyadult} and~\ref{tab:accuracygerman} show the accuracy of classifiers obtained from $\mathcal{FP}$, $\mathcal{FP}'$, $\mathcal{FP}''$ and $\mathcal{FP}^*$ in the Adult and German credit datasets
for $f=slift$, $\sigma=10\%$ and different values of $\alpha$ and $k$. We do not observe a significant difference between the accuracy of the classifier obtained from an $\alpha$-protective $k$-anonymous version of the original pattern set and the accuracy of
the classifier obtained from either a $k$-anonymous
or an $\alpha$-protective version.
In addition, the accuracy of the classifier decreases with larger $k$ and with smaller $\alpha$.
When comparing the two datasets,
we observe less accuracy for the German credit dataset; this is
consistent with the higher distortion observed above for this dataset.
Note that the low values of accuracy in Tables \ref{tab:accuracyadult} and \ref{tab:accuracygerman} are related to worst-case scenario ({\em i.e.} maximum value of $k$ and minimum value of $\alpha$).
\begin{table}[ht]
\caption{Adult dataset: accuracy of classifiers}
\label{tab:accuracyadult}
\centering
\begin{tabular}{c c c c c c}
\hline
{\bf $k$} & {\bf $\alpha$} & {\bf $\mathcal{FP}$} & {\bf $\mathcal{FP}'$} & {\bf $\mathcal{FP}''$} & {\bf $\mathcal{FP}^*$}\\
\hline
5 & 1.2 & 0.744 & 0.763 & 0.724 & 0.691  \\
5 & 1.5 & 0.744 & 0.763 & 0.752 & 0.739 \\
50 & 1.2 & 0.744 & 0.751 & 0.682& 0.691  \\
50 & 1.5 & 0.744 & 0.751  & 0.746&  0.739  \\
\hline
\end{tabular}
\end{table}
\begin{table}[ht]
\caption{German dataset: accuracy of classifiers}
\label{tab:accuracygerman}
\centering
\begin{tabular}{c c c c c c}
\hline
{\bf $k$} & {\bf $\alpha$} & {\bf $\mathcal{FP}$} & {\bf $\mathcal{FP}'$} & {\bf $\mathcal{FP}''$} & {\bf $\mathcal{FP}^*$}\\
\hline
3 & 1.2 & 0.7  & 0.645  & 0.582 & 0.572  \\
3 & 1.8 & 0.7 &  0.645 & 0.624  & 0.615  \\
10 &  1.2 & 0.7 & 0.583 & 0.561& 0.572 \\
10 & 1.8 &  0.7  & 0.583   & 0.605 &  0.615 \\
\hline
\end{tabular}
\end{table}
\subsubsection{Degree of Discrimination}
Finally, Tables~\ref{fig:AdultDiscrmination} show the discrimination degree measures (DPD and
NDD) obtained after applying privacy pattern sanitization to frequent patterns extracted from Adult and German credit, respectively, for
$f=slift$ and different values of $\alpha$, $k$ and $\sigma$.
 As stated in Section \ref{sec:PrivDisc}, applying privacy pattern sanitization can eliminate or create discrimination. Tables~\ref{fig:AdultDiscrmination} clearly highlight this fact: the values of DPD and NDP increase with $k$. This is because more inference channels are detected for larger values of $k$ and our method perturbs the support of more patterns. As a consequence, the impact on anti-discrimination may increase.
\begin{table*}
	\centering
\caption{Discrimination utility measures after privacy
pattern sanitization: Adult (top); German credit (bottom)}
\label{fig:AdultDiscrmination}
\includegraphics[width=\linewidth]{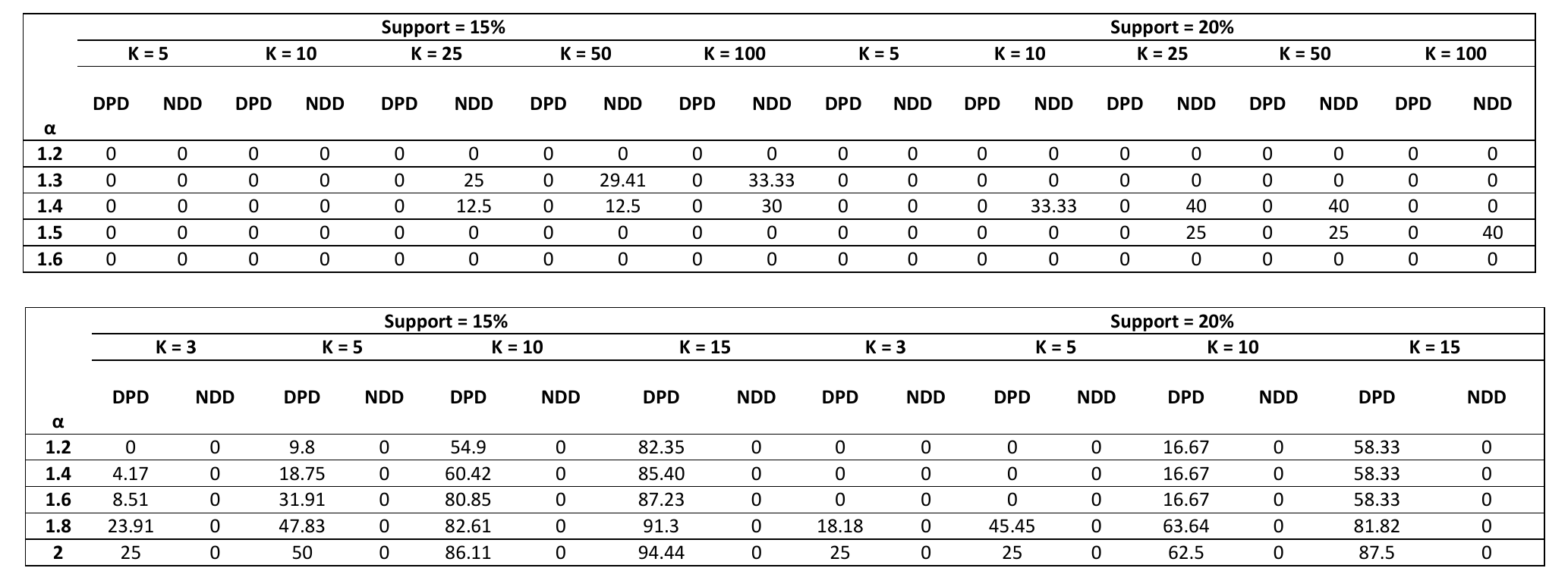}
\end{table*}
Comparing the results obtained in the two datasets, we observe that in Adult usually the value of NDD is higher than DPD, while in German credit it is the other way round. This shows that in the Adult dataset the privacy pattern sanitization tends to make $\mathcal{FP}$ less $\alpha$-protective; in German credit the situation is reversed: the privacy pattern sanitization tends to make
$\mathcal{FP}$ more $\alpha$-protective (NDD = 0\%).
All empirical results presented above are related to
the worst-case scenario, {\em i.e.}, we compared the
utility of the original pattern set with the utility
of the patterns protected against both explainable and unexplainable discrimination (that is, the $\alpha$-protective or $\alpha$-protective $k$-anonymous versions of the original pattern set).
However, protecting the original pattern set against only unexplainable discrimination  (that is, generating $d$-explainable $\alpha$-protective or $d$-explainable $\alpha$-protective $k$-anonymous versions) can lead to less or equal information loss and pattern distortion. This is because pattern sanitization methods transform a smaller number of patterns if the number of $d$-explainable patternsis greater than zero. Table~\ref{fig:explainbleDiscrmination} clearly highlights this fact. It shows the percentage of $d$-explainable patterns among $\alpha$-discriminatory ones obtained from the original pattern set and a $k$-anonymous version of it, for the Adult and German credit datasets,
$\alpha=1.2$, $\sigma=10\%$, $f=slift$ and several values of $d$ and $k$.
In this experiment, we assume that all the itemsets excluding the ones
belonging to PD attributes are legally grounded.
\begin{table*}[h]
	\centering
\caption{Percentage of $d$-explainable patterns detected in $\mathcal{FP}$ and $\mathcal{FP}'$}
\label{fig:explainbleDiscrmination}
\includegraphics[width=\linewidth]{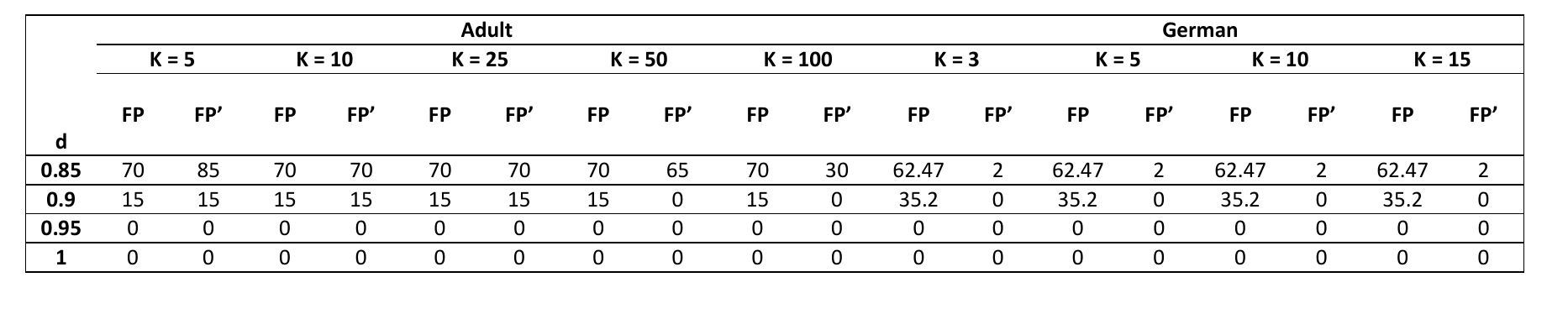}
\end{table*}
We can observe that the percentage of $d$-explainable patterns decreases with larger $d$. In addition, as stated in Section \ref{sec:explainable}, applying privacy pattern transformation to make $\mathcal{FP}$ $k$-anonymous can make $\mathcal{FP}$ more or less $d$-explainable $\alpha$-protective.

\section{An Extension Based on Differential Privacy}
     
\label{sec:DPdiscussion}
In this section, we present how simultaneous anti-discrimination and privacy can be achieved in frequent pattern discovery while satisfying differential privacy (instead of $k$-anonymity) and $\alpha$-protection.
Differential privacy (see Definition~\ref{def:diffpriv}) is composable according to the following lemma.
\begin{lemma}[Sequential composition, \cite{Dwork2006}]
\label{lem:composition}
If there are $M$ randomized algorithms $\mathcal{A}_1,\cdots,\mathcal{A}_M$,
whose privacy guarantees are $\epsilon_1, \cdots, \epsilon_M$-differential
privacy, respectively, then any function $g$ of
them, $g(\mathcal{A}_1,\cdots, \mathcal{A}_M)$, is
$\sum_{i=1}^M$-differentially private.
\end{lemma}
We refer
to $\epsilon$ as the privacy budget of a privacy-aware data analysis
algorithm. When an algorithm involves multiple steps, each step
uses a portion of $\epsilon$ so that the sum of these portions is no
more than $\epsilon$.
\subsection{Differentially Private Frequent Pattern Set}
As mentioned in Section \ref{sec:difffreq}, the authors in \cite{Li2012,Bhaskar2010} propose  algorithms for publishing frequent patterns while satisfying differential privacy. The notion of $\epsilon$-differentially private
frequent pattern set can be defined as follows.
\begin{definition}[$\epsilon$-differentially private frequent pattern set]
\label{def:diffFP}
Given a collection of frequent patterns $\mathcal{F}(\mathcal{D}, \sigma)$ and a differential privacy budget $\epsilon$, $\mathcal{F}(\mathcal{D}, \sigma$) is $\epsilon$-differentially private
if it can be obtained using a randomized algorithm $\mathcal{ALG}$ satisfying
$\epsilon$-differential privacy.
\end{definition}
Bhaskar {\em et al.} in \cite{Bhaskar2010} propose two
algorithms for publishing differentially private frequent patterns.
Their approach
falls in the post-processing category and it
has two steps. In the first step, one
selects the $K$ patterns to be released. In the second step, one releases the support values of these patterns after adding noise to them. The privacy budget $\epsilon$ is evenly divided between the two steps.
As explained in detail in \cite{Li2012}, this approach works reasonably
well for small values of $K$; however, for larger values of $K$, the
accuracy is poor. In the sequel, we use the approach proposed in \cite{Li2012}, called PrivBasis, which greatly outperforms the proposal of Bhaskar {\em et al.}
\subsection{Achieving an Differentially Private Frequent Pattern Set}
\label{sec:privbasis}
Li {\em et al.} in \cite{Li2012} recently have proposed the PrivBasis
algorithm for publishing a differentially private
version the top $K$ frequent patterns. If one desires to
publish all patterns with support above a given threshold $\sigma$,
{\em i.e.} $\mathcal{F}(\mathcal{D}, \sigma)$, one can first compute the value $K$ such that the $K$-th most frequent pattern
has a support value $\geq \sigma$ and the $K+1$-th pattern has support $<\sigma$, and then use PrivBasis with parameter $K$.
The approach falls in the in-processing category and it
uses a novel notion of basis sets. A $\sigma$-basis set $B=\{B_1,B_2,...,B_w\}$, where each $B_i$ is a set
of items, has the property that any pattern with support value
higher than $\sigma$ is a subset of some basis $B_i$. The algorithm
constructs a basis set while satisfying differential privacy and uses
this set to find the most frequent patterns.
The PrivBasis algorithm consists of the following steps:
\begin{enumerate}
\item Obtain $\lambda$, the number of unique items that are involved in the $K$ most frequent patterns.
\item Obtain $F$, the $\lambda$ most frequent items among
the set $I$ of all items in $\mathcal{D}$.
\item Obtain $P$, a set of the most frequent pairs of items
among $F$.
\item Construct $\sigma$-basis set $B=\{B_1,B_2,...,B_w\}$, using $F$ and $P$.
\item Obtain noisy support values of patterns in candidate set $C(B)=\bigcup_{i=1}^w\{p|p \subseteq B_i\}$; one can then select the top $K$ patterns from $C(B)$.
\end{enumerate}
The privacy budget $\epsilon$ is divided among Steps 1, 2, 3, 5
above. Step 4 does not access the dataset $\mathcal{D}$, and only uses the outputs of earlier steps. Step 1 uses  the exponential mechanism to sample $j$ from $\{1, 2,\cdots, K\}$ with the utility function
$u(\mathcal{D}, j) = (1-|f_K-fitem_j |)|\mathcal{D}|$
where $f_K=\sigma$ is the support value of $K$-th most frequent pattern and $fitem_j$ is the support value of the $j$-th most frequent item.
That is, Step 1 chooses $j$ such that the $j$-th most frequent
item has frequency closest to that of the $K$-th most frequent
pattern. The sensitivity of the above utility function is 1, because
adding or removing a record can affect $f_K$ by
at most $1/|\mathcal{D}|$ and $fitem_j$ by at most $1/|\mathcal{D}|$.
Step 2 differentially privately selects the $\lambda$ most frequent items among
the $|I|$ items and Step 3
differentially privately selects a set of the most frequent pairs of items among $\lambda^2$ patterns. Both steps use repeated sampling without replacement,
where each sampling step uses the exponential
mechanism with the support value of each pattern as its utility.
Step 4 constructs $B$ that covers all maximal cliques in $(F, P)$ (see \cite{Li2012} for details). Step 5 computes the noisy support values of all patterns in $C(B)$ as follows. Each basis $B_i$ algorithm divides all possible records
into $2^{|B_i|}$ mutually disjoint bins, one corresponding to each
subset of $B_i$. For each pattern $p \subseteq B_i$, the bin corresponding to $p$
consists of all records that contain all items in $p$, but
no item in $B_i \backslash p$. Given a basis set $B$, adding noise $Lap(w/\epsilon)$ to each bin count and outputting these noisy counts satisfies $\epsilon$-differential privacy. The array element $b[i][p]$ stores the noisy count of the bin corresponding to pattern $p$ and basis $B_i$. For each basis $B_i$, adding or removing a single
record can affect the count of exactly one bin by exactly
1. Hence the sensitivity of publishing all bin counts for one
basis is 1; and the sensitivity of publishing counts for all
bases is $w$. From these noisy bin counts, one can recover the noisy support values of all patterns in $C(B)$ by summing up the respective noisy bin counts in $b[i][p]$.

We note that due to the possibility of drawing a negative noise from the Laplace distribution, PrivBasis can obtain noisy bin counts in $b[i][p]$ which are negative. This can lead to two problems:
1) some patterns could have negative support values; and 2) we can obtain a collection of frequent patterns with contradictions among them. More specifically, a pattern could have a noisy support value which is smaller than the noisy support values of its superset patterns. In order to avoid the contradictions among the released patterns published by PrivBasis, it is enough to post-process the noisy bin counts in $b[i][p]$ by rounding each count to nearest non-negative integer. This should be done after Line 11 of Algorithm 1 in \cite{Li2012}. Note that, as proven by Hay {\em et al.} \cite{hay2010}, a post-processing of differentially private results does non change the privacy guarantee. Hence, the algorithm PrivBasis will remain differentially private by the above changes. In the following we use PrivBasis with the above update. The problem of achieving consistency constraints among the noisy count values has been also addressed in \cite{hay2010}.

\subsection{Achieving a Discrimination Protected and Differential Private Frequent Pattern Set}
\begin{definition}[$\alpha$-protective $\epsilon$-differentially private frequent pattern set]
\label{def:diffFP_prot}
Given a collection of frequent patterns $\mathcal{F}(\mathcal{D}, \sigma)$, a differential privacy budget $\epsilon$, a discriminatory threshold $\alpha$, protected groups $DI_b$, and a discrimination measure $f$, $\mathcal{F}(\mathcal{D}, \sigma)$ is $\alpha$-protective $\epsilon$-differentially private if it is both $\epsilon$-differentially private and $\alpha$-protective
w.r.t. $f$ and $DI_b$.
\end{definition}
To simultaneously achieve anti-discrimination and differential privacy in frequent pattern discovery, we need to generate an $\alpha$-protective $\epsilon$-differentially private version of $\mathcal{F}(\mathcal{D}, \sigma)$. To do so, first we need to examine how making $\mathcal{F}(\mathcal{D}, \sigma)$ $\epsilon$-differentially private impacts the $\alpha$-protectiveness of $\mathcal{F}(\mathcal{D}, \sigma)$ w.r.t. $f$, where $f$ is one of the measures from Definitions \ref{def:elift}-\ref{def:dlift}.
\begin{theorem}
\label{theo:diffdisc}
The PrivBasis approach for making $\mathcal{F}(\mathcal{D}, \sigma)$ $\epsilon$-differential private can make $\mathcal{F}(\mathcal{D}, \sigma)$ more or less $\alpha$-protective w.r.t. $f$ and $DI_b$.
\end{theorem}
\begin{proof}
The $\epsilon$-differentially private version of $\mathcal{F}(\mathcal{D}, \sigma)$, denoted  by $\mathcal{FP}_d$ for short, generated by PrivBasis is an approximation of $\mathcal{F}(\mathcal{D}, \sigma)$. As a side effect of this transformation due to Laplacian or exponential mechanisms, $\mathcal{FP}_d$ might contain patterns that are not in $\mathcal{F}(\mathcal{D}, \sigma)$ ({\em i.e.} ghost patterns) and might not contain patterns that are in $\mathcal{F}(\mathcal{D}, \sigma)$ ({\em i.e.} missing patterns). Moreover, for patterns that are in both $\mathcal{F}(\mathcal{D}, \sigma)$ and $\mathcal{FP}_d$, $\mathcal{FP}_d$ contains the noisy new support values of original patterns in $\mathcal{F}(\mathcal{D}, \sigma)$. Hence, making $\mathcal{F}(\mathcal{D}, \sigma)$ $\epsilon$-differentially private can lead to different situations regarding the $\alpha$-protectiveness of $\mathcal{FP}_d$ w.r.t. $f$ and $DI_b$. We list such situations below:
\begin{itemize}
\item There are $\alpha$-discriminatory patterns in $\mathcal{F}(\mathcal{D}, \sigma)$ w.r.t. $f$ and $DI_b$ which are not in $\mathcal{FP}_d$ ({\em i.e.} missing patterns). In this situation, making $\mathcal{F}(\mathcal{D}, \sigma)$ $\epsilon$-differentially private makes $\mathcal{F}(\mathcal{D}, \sigma)$ more $\alpha$-protective.
\item There are $\alpha$-discriminatory patterns in $\mathcal{FP}_d$ w.r.t. $f$ and $DI_b$ which are not in $\mathcal{F}(\mathcal{D}, \sigma)$ ({\em i.e.} ghost patterns). In this situation, making $\mathcal{F}(\mathcal{D}, \sigma)$ $\epsilon$-differentially private makes $\mathcal{F}(\mathcal{D}, \sigma)$ less $\alpha$-protective.
\item There are PD patterns ({\em e.g.} $p:\{A,B,C\}$) in both $\mathcal{F}(\mathcal{D}, \sigma)$ and $\mathcal{FP}_d$ that are $\alpha$-protective (resp. $\alpha$-discriminatory) w.r.t. $f$ and $DI_b$ in $\mathcal{F}(\mathcal{D}, \sigma)$ and $\alpha$-discriminatory (resp. $\alpha$-protective) in $\mathcal{FP}_d$. It is because the new noisy support values of patterns in $\mathcal{FP}_d$ can increase (resp. decrease) the values of $f(A,B \rightarrow C)$. In this situation, making $\mathcal{F}(\mathcal{D}, \sigma)$ $\epsilon$-differentially private makes $\mathcal{F}(\mathcal{D}, \sigma)$ less (resp. more) $\alpha$-protective.
\end{itemize}
Hence, the theorem holds. 
\end{proof}
Thus, similar to $k$-anonymity, achieving differential privacy in frequent pattern discovery can achieve anti-discrimination or work against
anti-discrimination.
Hence, what makes sense is first to obtain a $\varepsilon$-differentially private version of $\mathcal{F}(D, \sigma)$
and then deal with discrimination.
An interesting point is that, regardless of whether $k$-anonymity
or $\varepsilon$-differential privacy is used,
achieving privacy impacts anti-discrimination similarly.
Based on this observation, we present Algorithm \ref{alg:AntiDiscDiffPrivPattSanit} to generate an $\alpha$-protective $\epsilon$-differentially private version of an original pattern set w.r.t. $f$ and $DI_b$.
\begin{algorithm}[h]
\caption{{\sc Anti-discrimination differentially private pattern sanitization}}
\label{alg:AntiDiscDiffPrivPattSanit}
{\small \begin{algorithmic}[1]
\STATE Inputs: Database $\mathcal{D}$, $K$, items $I$, differential privacy budget $\epsilon$,
$DI_b$, discrimination measure $f$, $\alpha$, $C=${class item with negative decision value}
\STATE Output: $\mathcal{FP}''$: $\alpha$-protective $\epsilon$-differential private frequent pattern set
\STATE $(\mathcal{FP}_d, b[i][p]) \leftarrow$ PrivBasis($\mathcal{D}$,$I$,$K$,$\epsilon$) \hspace{2mm}  //Algorithm 3 in \cite{Li2012}
\STATE $\mathcal{D}_{D} \leftarrow$ {\bf Function} \textsc{DetDiscPatt}($\mathcal{FP}_d$, $b[i][p]$, $DI_b$, $f$, $\alpha$, $C$)
\STATE Lines 5-8 of Algorithm \ref{alg:AntiDiscAnonyPattSanit} to sort $\mathcal{D}_{D}$
\STATE $\mathcal{FP}'' \leftarrow$ {\bf Function} \textsc{AntiDiscPattSanit}($\mathcal{FP}_d$, $b[i][p]$, $\mathcal{D}_{D}$, $DI_b$, $f$, $\alpha$, $C$)
\STATE Output: $\mathcal{FP}''$
%
%
\end{algorithmic}}
\end{algorithm}
\begin{theorem}
\label{theo:discdiff}
Algorithm \ref{alg:AntiDiscDiffPrivPattSanit} is $\epsilon$-differentially private.
\end{theorem}
\begin{proof}
The only part in Algorithm \ref{alg:AntiDiscDiffPrivPattSanit} that depends on
the dataset is Step 3, which is $\epsilon$-differentially private
because it uses PrivBasis. Starting
from Step 4, the algorithm only performs post-processing,
and does not access $\mathcal{D}$ again. Indeed, in Step 4 the value of $f$ w.r.t. each PD pattern in $\mathcal{FP}_d$; in Step 6 the value of $\Delta_f$ can be computed using the noisy support values of patterns in $\mathcal{FP}_d$ and
the noisy bin counts in array $b[i][p]$ (array $b[i][p]$ replaces parameter $\mathcal{D}$ in functions \textsc{DetDiscPatt} and \textsc{AntiDiscPattSanit}). Adding $\Delta_f$ to the noisy support values of respective patterns in Step 6 is post-processing of differentially private results which remain private as proven in \cite{hay2010}.
\end{proof}
Thus, $\alpha$-protection in $\mathcal{F}(\mathcal{D}, \sigma)$ can be achieved by anti-discrimination pattern sanitization methods proposed in Section \ref{sec:nDG} without violating differential privacy,
just as we could do it without violating $k$-anonymity.
\subsection{Discussion}
In the previous sections, we showed that we can achieve simultaneous discrimination and privacy protection in frequent pattern discovery satisfying differential privacy instead of $k$-anonymity. Indeed, it turns out that
$k$-anonymity and differential privacy are similarly related to
$\alpha$-protection.
In terms of privacy, using differential privacy seems preferable because it
provides a worst-case privacy guarantee. In terms of data utility,
the situation is different. As we discussed in Section \ref{sec:privbasis},
if the approach proposed in \cite{Li2012} does not control the
generation of negative bin counts, it may result in negative support values
and collections of frequent patterns with contradictions among them.
These two problems, which are avoided in the approach based on $k$-anonymity presented in Section \ref{sec:problemsol}, clearly have a negative impact on the utility of the published patterns.
In Section \ref{sec:privbasis} we propose a solution to avoid these problems.
Even if this solution avoids contradiction among the released patterns,
it does not eliminate the information loss. In fact, we observe that
forcing negative bin counts to zero could lead to suppressing
some frequent patterns that without any sanitization would belong to the mining result. As a consequence, if we compare the approach based on $k$-anonymity to the approach based on differential privacy in terms of {\em Misses Cost (MC)},
({\em i.e.}, the fraction of original frequent patterns which do not appear in the published result), and {\em Ghost Cost (GC)} ({\em i.e.}, the fraction of frequent patterns appearing in the published result that are not in the original pattern set), we can argue that with $k$-anonymity the values of MC and GC are zero, while with differential privacy they could be more than zero.
We can compare also the two methods in terms of pattern distortion (Section \ref{sec:utilitymeasures}). Our preliminary experiments
seem to indicate that the value of pattern distortion scores
for the approach based on $k$-anonymity is less than for
the approach based on differential privacy; this is not
surprising, because
with the former approach only the support values of subsets of original patterns are changed while with the latter approach, the support values of all the patterns are changed.
From the above preliminary considerations obtained by analyzing the behavior of the two approaches and their properties, we can conclude that,
as far as privacy is concerned, the differentially private approach is better than the approach based on $k$-anonymity. On the other side, the method based on differential privacy seems to preserve less data utility. Clearly, more
extensive empirical work is needed that takes into account
other data utility measures ({\em i.e.} the accuracy of a classifier).
     \section{Conclusions}
\label{sec:conclusionfpd}
In this chapter, we have investigated the problem of discrimination
and privacy
aware frequent pattern discovery, {\em i.e.} the sanitization of the collection of patterns mined from a transaction database in such a way that neither privacy-violating nor discriminatory inferences can be inferred on the released patterns.
In particular, for each measure of discrimination used in the legal literature we proposed a solution for obtaining a discrimination-free collection of patterns. We also proposed an algorithm to take into account the legal concept of genuine occupational requirement for making an original pattern set protected only against unexplainable discrimination. We also found that our discrimination preventing transformations do not interfere with a privacy preserving sanitization based on $k$-anonymity, thus accomplishing the task of combining the two and achieving a robust (and formal) notion of fairness in the resulting pattern collection.
Further, we have presented extensive empirical results on
the utility of the protected data. Specifically, we evaluate
the distortion introduced by our methods and its effects on classification.
It turns out that
the utility loss caused by simultaneous anti-discrimination
and privacy protection is only marginally higher than the loss caused
by each of those protections separately. This result
supports the practical deployment of our methods.
Finally, we have discussed the possibility of using our proposed framework
while replacing $k$-anonymity with differential privacy.
Although our discrimination prevention method can be combined with the
differentially private transformations, we have argued that doing so
can lead to more information loss in comparison with the $k$-anonymity
based approach.


\chapter[A Study on the Impact of Data Anonymization on Anti-discrimination]{A Study on the Impact of Data Anonymization on Anti-discrimination}
\chaptermark{Impact of Anonymization on Anti-discrimination}
\label{chap:study}


In the previous chapter, we studied the relation of PPDM and DPDM in the context of knowledge publishing (post-processing approach). In this chapter, we study the relation between data anonymization and anti-discrimination (pre-processing approach).
We analyze how different data anonymization techniques ({\em e.g.}, generalization) have an impact on anti-discrimination ({\em e.g.}, discrimination prevention). When we anonymize the original data to achieve the requirement of the privacy model ({\em e.g.}, $k$-anonymity), what will happen to the discriminatory bias contained in the original data?
Our main motivation to do this study is finding an answer for three important questions. First, can providing protection against privacy attacks also achieve anti-discrimination in data publishing? Second, can we adapt and use some of the data anonymization techniques ({\em e.g.}, generalization) for discrimination prevention? Third, can we design methods based on data anonymization to make the original data protected against both privacy and discrimination risks?

\section{Non-discrimination Model}
\label{sec:NDDT}

As mentioned in Chapter \ref{chap:discrelatedwork}, civil rights laws \cite{AustralianLeg,Europeandisc,uspay} explicitly identify the attributes to be protected against discrimination. For instance, U.S. federal laws \cite{uspay} prohibit discrimination on the basis of race, color, religion, nationality, sex, marital status, age and pregnancy. In our context, we consider these attributes as potentially discriminatory (PD) attributes.
Let $DA$ be a set of PD attributes in data table $\mathcal{D} (A_1,\cdots, A_n)$ specified by law.
Comparing privacy
legislation~\cite{eurpprivacy} and
anti-discrimination legislation~\cite{Europeandisc,uspay},
PD attributes can overlap with QI attributes ({\em e.g.} \emph{Sex}, \emph{Age}, \emph{Marital\_status}) and/or sensitive attributes ({\em e.g.} \emph{Religion} in some applications).
A domain $D_{A_i}$ 
is associated with each attribute $A_i$ to indicate the set of values that the attribute can assume.
In previous works on anti-discrimination \cite{peder2008,pederruggi2009,ruggi2010,KamiKAIS,hajian2011LNC,hajianTKDE,CaldersICDM,Calders2011}, as we consider in previous chapters, the
authors propose discrimination discovery and prevention techniques w.r.t.
specific protected groups, {\em e.g.} black and/or female poeple. However, this assumption fails to capture the various nuances of discrimination since minority or disadvantaged groups can be different in different contexts.
For instance, in a neighborhood with almost all black people,
whites are a minority and may be discriminated. Then we consider $A_i=q$ to be a PD item, for every $q\in D_{A_i}$, where $A_i \in DA$, {\em e.g.} $Race=q$ is a PD item for any race $q$, where $DA=\{Race\}$. This definition is also compatible with
the law. For instance, the U.S. Equal Pay Act \cite{uspay} states that: ``a selection rate for {\bf any} race, sex, or ethnic
group which is less than four-fifths of the rate for the
group with the highest rate will generally be regarded
as evidence of adverse impact''.
An item $A_i=q$ with $q\in D_{A_i}$ is a PND item if $A_i \notin DA$, {\em e.g.} $Hours=35$ where $DA=\{Race\}$.
   
Building on Definition \ref{label:disc} and considering the above fact, we introduce the notion of $\alpha$-protection for a data table.

\begin{definition} [$\alpha$-protective data table]
\label{def:protection}
Let $\mathcal{D}(A_1,\cdots, A_n)$ be a data table, $DA$ a set of PD attributes associated with it, and $f$ be one of the measures from  Fig. \ref{fig:Measures}. $\mathcal{D}$ is said to satisfy $\alpha$-protection
or to be $\alpha$-protective w.r.t. $DA$ and $f$ if each PD frequent classification rule $c: A, B \rightarrow C$ extracted from $\mathcal{D}$ is $\alpha$-protective, where $A$ is a PD itemset and $B$ is a PND itemset.

\end{definition}

Note that $\alpha$-protection in $\mathcal{D}$ not only can prevent discrimination against the main protected groups w.r.t. $DA$ ({\em e.g.}, women) but also against any subsets of protected groups w.r.t. $\mathcal{A}\backslash DA$ ({\em e.g.}, women who have a medium salary and/or work 36 hours per week).
Releasing an $\alpha$-protective (unbiased) version of an original data table is desirable to prevent discrimination with respect to $DA$. If the original data table
is biased w.r.t. $DA$, it must be modified before being published
({\em i.e.} pre-processed).
The existing pre-processing discrimination prevention methods are based on data perturbation, either by modifying class attribute values \cite{KamiKAIS,hajian2011LNC,hajianTKDE,Ruggieri2011} or by modifying PD attribute values \cite{hajian2011LNC,hajianTKDE} of the training data.
One of the drawbacks of these techniques is that they cannot be applied (are not preferred)
in countries where data perturbation is not legally accepted (preferred), while generalization is allowed;
{\em e.g.} this is the case in Sweden and other Nordic countries
(see p. 24 of \cite{sweden}). Hence, we focus here
on generalization and suppression.

\section{Data Anonymization Techniques and Anti-discrimination}
\label{sec:privacydisc}

In this section, we study how different generalization and suppression schemes have impact on anti-discrimination. In other words, when we anonymize $\mathcal{D}$ to achieve the requirement of the privacy model, {\em e.g.} $k$-anonymity, w.r.t. QI, what will happen to $\alpha$-protection of $\mathcal{D}$ w.r.t. $DA$? The problem could be investigated with respect to different possible relations between PD attributes and other attributes ({\em i.e.} QI and sensitive attributes) in $\mathcal{D}$.
 In this context, we consider the general case where all attributes are QI expect for the class/decision attribute. Then, each QI attribute can be PD or not. In summary, the following relations
are assumed: (1) $QI \cap C = \emptyset$, (2) $DA \subseteq QI$. As mentioned in Section \ref{sec:NDDT}, PD attributes can overlap with QI and/or sensitive and/or non-sensitive attributes. Considering all attributes as QI such that $DA \subseteq QI$ can cover all the above cases.

\begin{example}{\em
\label{exp:61}

Table~\ref{tab:61} presents raw customer credit data, where each record represents a customer's specific information. {\em Sex}, {\em Job}, and {\em Age} can be taken as QI attributes.
The class attribute has two values, {\em Yes} and {\em No}, to indicate
whether the customer has received credit or not. Suppose the privacy model is $k$-anonymity and $k=2$. Table~\ref{tab:61} does not satisfy 2-anonymity w.r.t. $QI=\{Sex, Job, Age\}$.
%
}

\end{example}

\begin{example}{\em
Continuing Example \ref{exp:61}, suppose $DA=\{Sex\}$, $\alpha=1.2$ and $f=slift$. Table ~\ref{tab:61} does not satisfy 1.2-protection w.r.t. $f$ and $DA$ since for frequent PD rule $c$ equal to
$\{Sex=female\} \rightarrow\ Credit\_approved=no$ we have $slift(c)=\frac{2/3}{1/4}=2.66$. Then Table ~\ref{tab:61} is biased w.r.t. women.
}
\end{example}

\begin{table}[t]
\caption{Private data table with biased decision records}
\label{tab:61}
\centering
\begin{tabular}{|c|c|c|c|c|}
\hline
{\bf ID} & {\bf Sex} & {\bf Job } & {\bf Age} & {\bf Credit\_approved }\\

\hline
1 & Male & Engineer & 35 & Yes \\
\hline
2 & Male & Engineer & 38 & Yes \\
\hline
3 & Male & Lawyer & 38 & No \\
\hline
4 & Female & Writer & 30 & No \\
\hline
5 & Male & Writer & 30 & Yes \\
\hline
6 & Female & Dancer & 31 &  No \\
\hline
7 & Female & Dancer & 32 & Yes \\
\hline

\end{tabular}

\end{table}


\subsection{Global Recoding Generalizations and Anti-discrimination}

In this section, by presenting different scenarios, we will show that using global recoding generalizations ({\em i.e.} full-domain generalization, subtree generalization and sibling generalization) to achieve $k$-anonymity w.r.t. QI in $\mathcal{DB}$ can lead to different situations regarding the $\alpha$-protection of $\mathcal{D}$ w.r.t. $DA$.


{\em Global recoding generalizations not offering $\alpha$-protection}. It can happen in different scenarios. First, consider a data table $\mathcal{D}$ with the same attributes as the one in Table \ref{tab:61}, but many more records, and let $DA=\{Job\}$ and $QI=\{Sex, Job, Age\}$. Suppose $\mathcal{D}$ is biased with respect to dancers or a subgroup of dancers, {\em e.g.} dancers who are women ({\em i.e.} $\mathcal{D}$ does not satisfy $\alpha$-protection w.r.t. $DA=\{Job\}$). Generalizing all instances of 30, 31 and 32 values to the the same generalized value $[30, 35)$ to achieve $k$-anonymity w.r.t. QI in $\mathcal{D}$ using full-domain generalization, subtree or sibling generalization cannot achieve $\alpha$-protection w.r.t. $DA=\{Job\}$, based on Definition~\ref{def:protection}.
Second, consider a data table $\mathcal{D}$ with the same attributes as the one in Table \ref{tab:61}, but many more records, and let $DA=\{Job\}$ and $QI=\{Sex, Job, Age\}$. Suppose $\mathcal{DB}$ is biased with respect to dancers. Generalizing all instances of {\em Dancer} and {\em Writer} values to the same generalized value {\em Artist} to achieve $k$-anonymity in $\mathcal{D}$ w.r.t. QI using full-domain generalization or subtree generalization, might cause the
{\em Artist} node to inherit the biased nature of {\em Dancer}. Then, this generalization cannot achieve $\alpha$-protection w.r.t. $DA=\{Job\}$. Third, consider a data table $\mathcal{D}$ with the same attributes as the one in Table \ref{tab:61}, but many more records, and let $DA=\{Age\}$ and $QI=\{Sex, Job, Age\}$. Suppose $\mathcal{D}$ is not biased ({\em i.e.} $\mathcal{D}$ is $\alpha$-protective) with respect to $DA$. It means that all PD frequent rules w.r.t. $DA$ extracted from it are not
$\alpha$-discriminatory. However, $\mathcal{D}$ might
contain PD rules which are $\alpha$-discriminatory and not frequent,
{\em e.g.} $\{Age=30, Sex=Male\} \rightarrow Credit\_approved=no$, $\{Age=31, Sex=Male\} \rightarrow Credit\_approved=no$, $\{ Age=32, Sex=Male\} \rightarrow Credit\_approved=no$. Generalizing all instances of 30, 31 and 32 values to the same generalized value $[30, 35)$ to achieve $k$-anonymity w.r.t. QI in $\mathcal{D}$ using full-domain generalization, subtree or sibling generalization, can cause new frequent PD rules to appear, which might be $\alpha$-discriminatory and discrimination will show up after generalization, {\em e.g.} $\{ Age=[30-35), Sex=Male\} \rightarrow Credit\_approved=no$.

{\em Global recoding generalizations offering $\alpha$-protection}. Consider Table \ref{tab:61} and let $DA=\{Sex\}$ and $QI=\{Sex, Job, Age\}$. Suppose that Table~\ref{tab:61} is biased with respect to women or any subgroup of women, {\em e.g.} women who are 30 years old and/or who are dancers ({\em i.e.} Table \ref{tab:61} does not satisfy $\alpha$-protection w.r.t. $DA=\{Sex\}$). Generalizing all instances of  {\em Female} values to the same generalized value {\em Any-sex} to achieve $k$-anonymity w.r.t QI in Table \ref{tab:61} can also achieve $\alpha$-protection w.r.t. $DA=\{Sex\}$, based on Definition \ref{def:protection}.

To Summarize, using global recoding generalizations to achieve the requirement of the privacy model ({\em i.e.} $k$-anonymity), depending on the generalization, can make original data less more or less protected against discrimination.

\subsection{Local Recoding Generalizations and Anti-discrimination}
\label{sec:cell}

In this section, by analyzing different scenarios, we will show how using local recoding generalization, {\em i.e.} cell generalization, to achieve $k$-anonymity w.r.t. QI in $\mathcal{D}$, has an impact on $\alpha$-protection of $\mathcal{D}$ w.r.t. $DA$.
As mentioned in Section \ref{sec:privacy}, in contrast to global recoding generalizations, in cell generalization some instances of a value may remain ungeneralized while other instances are generalized.

Consider Table \ref{tab:61} and let $DA =\{Sex\}$ and $\alpha=1.2$. Table \ref{tab:61} does not satisfy 1.2-protection w.r.t. $f=slift$ and $DA$,  since for frequent PD rule $c$ equal to
$\{Sex=female\} \rightarrow credit\_approved=no$, by using the definitions of confidence and $slift$ (Expressions (\ref{eq:conf}) and (\ref{eq:slift}), resp.), we have $slift(c)=\frac{supp(A, C)/supp(A)}{supp(\neg A, C)/supp(\neg A)}=\frac{2/3}{1/4}=2.66$. Table \ref{tab:61} neither satisfies 1.2-protection w.r.t. $f=elift$ and $DA$, since for PD rule $c$, by using the definitions of confidence and $elift$ \footnote{when $B$ (PND itemset) in PD rule is empty, $elift$ reduces to the standard lift \cite{peder2008}
} (Expressions (\ref{eq:conf}) and (\ref{eq:elift}), resp.), we have $elift(c)=\frac{supp(A, C)/supp(A)}{supp(C)/|\mathcal{DB}|}=\frac{2/3}{3/7}=1.55$.

\begin{table}[t]
\caption{Different types of cell generalization}
\label{tab:62}
\centering
\begin{tabular}{|c|c|c|c|c|}
\hline
{\bf ID} & {\bf Sex} & {\bf Job } & {\bf Age} & {\bf Credit\_approved }\\

\hline
 1 & {\bf (1) Male $\Rightarrow$ any-sex} & Engineer & 35 & {\bf Yes} \\
\hline
2 & Male & Engineer & 38 & Yes \\
\hline
3 & {\bf (2) Male $\Rightarrow$ any-sex} & Lawyer & 38 & {\bf No} \\
\hline
4 & Female & Writer & 30 & No \\
\hline
5 & Male & Writer & 30 & Yes \\
\hline
6 & {\bf (3) Female $\Rightarrow$ any-sex} & Dancer & 31 &  {\bf No} \\
\hline
 7 & {\bf (4) Female $\Rightarrow$ any-sex} & Dancer & 32 & {\bf Yes} \\
\hline

\end{tabular}

\end{table}

Generalizing some instances of {\em Male} and/or {\em Female} values to the same generalized value {\em Any-sex} to achieve $k$-anonymity w.r.t. $QI=\{Job,Sex,Age\}$ in Table \ref{tab:61} using cell generalization can lead to different impacts on 1.2-protection of Table \ref{tab:61} w.r.t. $DA=\{Sex\}$. The impact depends on the value of class attribute ({\em e.g.} {\em Yes} or {\em No}) of each record in which the value of PD attribute  ({\em e.g.} {\em Female} or {\em Male}) is generalized.
Table \ref{tab:62} shows four types of cell generalization that can happen to achieve $k$-anonymity in Table \ref{tab:61} with numbers (1), (2), (3) and (4). Below, we analyze the impact of each type on 1.2-protection of Table \ref{tab:61} w.r.t. $DA$.

\begin{itemize}
\item Type (1). Generalizing an instance of {\em Male} value to the generalized value {\em Any-sex} while the value  of {\em Credit\_approved} attribute in the record is {\em Yes} cannot make Table \ref{tab:61} more or less 1.2-protective w.r.t. $f=elift$ but it can make Table \ref{tab:61} more 1.2-protective w.r.t. $f=slift$. It is because this type of cell generalization cannot change the value of $elift(c)$ but it can  decease the value of $slift(c)$, which is in this example $slift(c)=\frac{2/3}{1/3}=2$. This type of cell generalization increases the denominator of equation $\frac{supp(A, C)/supp(A)}{supp(\neg A, C)/supp(\neg A)}$ while keeping the numerator unaltered.

\item Type (2). Generalizing an instance of {\em Male} value to the generalized value {\em Any-sex} while the value of {\em Credit\_approved} attribute in the record is {\em No} cannot make Table \ref{tab:61} more or less 1.2-protective w.r.t. $f=elift$ but it can make Table \ref{tab:61} less 1.2-protective w.r.t. $f=slift$. It is because this type of cell generalization cannot change the value of $elift(c)$ but it can increase the value of $slift(c)$. This type of cell generalization decreases the denominator of equation $\frac{supp(A, C)/supp(A)}{supp(\neg A, C)/supp(\neg A)}$ while keeping the numerator unaltered.

\item Type (3). Generalizing an instance of {\em Female} value to the generalized value {\em Any-sex} while the value of {\em Credit\_approved} attribute for the record is {\em No} can make Table \ref{tab:61} more 1.2-protective w.r.t. $f=elift$ since it can decrease the value of $elift(c)$, which is in this example $elift(c)=\frac{1/2}{3/7}=1.16$. This type of cell generalization decreases the numerator of equation $\frac{supp(A, C)/supp(A)}{supp(C)/|\mathcal{DB}|}$ while keeping the denominator unaltered. In addition, this generalization can also make Table \ref{tab:61} more 1.2-protective w.r.t. $f=slift$ since it can decrease the value of $slift(c)$, which is in this example $slift(c)=\frac{1/2}{1/4}=2$. This type of cell generalization decreases the numerator of equation $\frac{supp(A, C)/supp(A)}{supp(\neg A, C)/supp(\neg A)}$ while keeping the denominator unaltered.
\item Type (4). Generalizing an instance of {\em Female} value to the generalized value {\em Any-sex} while the value of {\em Credit\_approved} attribute for the record is {\em Yes} can make Table \ref{tab:61} less 1.2-protective w.r.t. both $f=elift$ and $f=slift$ since it can increase the values of $elift(c)$ and $slift(c)$, which are in this example $elift(c)=\frac{2/2}{3/7}=2.33$ and $slift(c)=\frac{2/2}{1/4}=4$, respectively. This type of cell generalization increases the numerator of equations $\frac{supp(A, C)/supp(A)}{supp(C)/|\mathcal{DB}|}$ and $\frac{supp(A, C)/supp(A)}{supp(\neg A, C)/supp(\neg A)}$, respectively, while keeping the denominators unaltered.

\end{itemize}

Summarizing, using cell generalization to achieve the requirement of privacy model ({\em e.g.} $k$-anonymity), depend on how many records in each above types modified, can make original data table less or more protected against discrimination. In addition, only the generalization of type (3) can make the original data table $\alpha$-protective w.r.t. both $f=elift$ and $f=slift$ if enough number of records are modified. We can conclude that although
cell generalization leads to less data distortion than global
recoding generalizations, it can have less positive impact on
discrimination removal than global recoding generalizations.

\subsection{Multidimensional Generalizations and Anti-discrimination}

By presenting different scenarios, we also study the impact of using multidimensional generalizations to achieve $k$-anonymity w.r.t. QI in
$\mathcal{D}$ on $\alpha$-protection of $\mathcal{D}$ w.r.t. $DA$ and we observe the similar trend as cell generalization. For the sake of
brevity and due to similarity with Section \ref{sec:cell},
we do not recall the details here.

\subsection{Suppression and Anti-discrimination}

In this section, by presenting different scenarios, we will show that using suppression techniques ({\em i.e.} record suppression, value suppression and cell suppression) to achieve $k$-anonymity w.r.t. QI in
$\mathcal{D}$ can lead to different situations regarding the  $\alpha$-protection of $\mathcal{D}$ w.r.t. $DA$.
As shown in Section \ref{sec:cell}, Table \ref{tab:61} does not satisfy 1.2-protection w.r.t. $DA =\{Sex\}$ and both $f=slift$ and $f=elift$, since
for PD rule $c$ equal to
$\{Sex=female\} \rightarrow credit\_approved=no$ we
have $slift(c)=2.66$ and $elift(c)=1.55$.

Suppressing an entire record to achieve $k$-anonymity in Table \ref{tab:61} w.r.t. $QI=\{Job,Sex,Age\}$ using record suppression can lead to different impacts on the 1.2-protection of Table \ref{tab:61} w.r.t. $DA=\{Sex\}$. The impact depends on the value of PD attribute ({\em e.g.} {\em Female} or {\em Male}) and the value of class attribute ({\em e.g.} {\em Yes} or {\em No}) in the suppressed record.
Table \ref{tab:63} shows four types of record suppression which can happen to achieve $k$-anonymity w.r.t. QI in Table \ref{tab:61} with numbers (1), (2), (3) and (4). Below, we analyze the impact of each type on $\alpha$-protection of Table \ref{tab:61} w.r.t. $DA$.

\begin{itemize}
\item Type (1). Suppressing an entire record with the value of {\em Male} in {\em Sex} attribute and the value of {\em Yes} in {\em Credit\_approved} attribute can make Table \ref{tab:61} more 1.2-protective w.r.t. $f=elift$ since it can decrease the value of $elift(c)$, which is in this example $elift(c)=\frac{2/3}{3/6}=1.33$. This type of record suppression increases the denominator of equation $\frac{supp(A, C)/supp(A)}{supp(C)/|\mathcal{D}|}$ while keeping the numerator unaltered. In addition, this suppression can also make Table \ref{tab:61} more 1.2-protective w.r.t. $f=slift$ since it can decease the value of $slift(c)$, which is in this example $slift(c)=\frac{2/3}{1/3}=2$. This type of record suppression increases the denominator of equation $\frac{supp(A, C)/supp(A)}{supp(\neg A, C)/supp(\neg A)}$ while keeping the numerator unaltered.


\item Type (2). Suppressing an entire record with the value of {\em Male} in {\em Sex} attribute and the value of {\em No} in {\em Credit\_approved} attribute can make Table \ref{tab:61} less 1.2-protective w.r.t. both $f=elift$ and $f=slift$ since it can increase the values of $elift(c)$ and $slift(c)$. This type of record suppression decreases the denominator of equations $\frac{supp(A, C)/supp(A)}{supp(C)/|\mathcal{D}|}$ and $\frac{supp(A, C)/supp(A)}{supp(\neg A, C)/supp(\neg A)}$, respectively, while keeping the numerators unaltered.

\item Type (3). Suppressing an entire record with the value of {\em Female} in {\em Sex} attribute and the value of {\em No} in {\em Credit\_approved} attribute cannot make Table \ref{tab:61} more or less 1.2-protective w.r.t. $f=elift$ since it cannot change the value of $elift(c)$ substantially, which is in this example $elift(c)=\frac{1/2}{2/6}=1.5$. This happen because this type of record suppression decreases the numerator of equation $\frac{supp(A, C)/supp(A)}{supp(C)/|\mathcal{D}|}$ while also decreasing its denominator. However, this type of record suppression can make Table \ref{tab:61} more 1.2-protective w.r.t. $f=slift$ since it can decrease the value of $slift(c)$, which is in this example $slift(c)=\frac{1/2}{1/4}=2$. This suppression decreases the numerator of $\frac{supp(A, C)/supp(A)}{supp(\neg A, C)/supp(\neg A)}$ while keeping the denominator unaltered.


\item Type (4). Suppressing an entire record with the value of {\em Female} in {\em Sex} attribute and the value of {\em Yes} in {\em Credit\_approved} attribute can make Table \ref{tab:61} less 1.2-protective w.r.t. both $f=elift$ and $f=slift$ since it can increase the value of $elift(c)$ and $slift(c)$, which are in this example $elift(c)=\frac{2/2}{3/6}=2$ and $slift(c)=\frac{2/2}{1/4}=4$, respectively.

\end{itemize}

\begin{table}[t]
\caption{Different types of record suppression}
\label{tab:63}
\centering
\begin{tabular}{|c|c|c|c|c|}
\hline
{\bf ID} & {\bf Sex} & {\bf Job } & {\bf Age} & {\bf Credit\_approved }\\

\hline
 1 & {\bf (1) Male} & Engineer & 35 & {\bf Yes} \\
\hline
2 & Male & Engineer & 38 & Yes \\
\hline
3 & {\bf (2) Male} & Lawyer & 38 & {\bf No} \\
\hline
4 & Female & Writer & 30 & No \\
\hline
5 & Male & Writer & 30 & Yes \\
\hline
 6 & {\bf (3) Female} & Dancer & 31 &  {\bf No} \\
\hline
7 & {\bf (4) Female} & Dancer & 32 & {\bf Yes} \\
\hline

\end{tabular}

\end{table}

To summarize, using record suppression depending on how many records in each of the above types suppressed can make original data table more or less protected against discrimination after achieving privacy protection. In addition, only record suppression of type (1) can make original data table  $\alpha$-protective w.r.t. both $f=elift$ and $f=slift$ if a sufficient number of records are suppressed.

As mentioned in Section \ref{sec:privacy}, value suppression refers to suppressing every instance of a given value in a data table. Then, depending on which attribute values are suppressed after achieving privacy protection, value suppression can offer $\alpha$-protection or not. Cell suppression refers to suppressing some instances of a given value in a data table. Then, similarly to cell generalization,
depending on the values of suppressed cells and the class values of respective records, cell suppression can make original data more or less protected against discrimination. Thus, similarly to cell generalization, suppression techniques have less positive impact on discrimination removal than global recoding generalizations.
Finally, Table \ref{tab:64} summarizes the results we obtained in this chapter.

\begin{table*}[t]\footnotesize
\caption{Summary of results}
\label{tab:64}
\centering
\scalebox{0.9}{
\begin{tabular}{|c|c|c|c|}
\hline
{\bf Data Anonymization techniques} & {\bf Achieve $\alpha$-protection} & {\bf Against $\alpha$-protection} & {\bf No impact } \\

\hline
Global recoding generalizations  & \checkmark & \checkmark & \checkmark  \\
\hline
Cell generalization/Cell suppression Type (1) & \checkmark &  & \checkmark  \\
\hline
Cell generalization/Cell suppression Type (2) &  & \checkmark & \checkmark \\
\hline
Cell generalization/Cell suppression Type (3) & \checkmark &  &   \\
\hline
Cell generalization/Cell suppression Type (4) &  & \checkmark &  \\
\hline
Multidimensional generalization & \checkmark & \checkmark & \checkmark  \\
\hline
Record suppression Type (1) & \checkmark &  &  \\
\hline
Record suppression Type (2) & & \checkmark &   \\
\hline
Record suppression Type (3) & \checkmark & & \checkmark \\
\hline
Record suppression Type (4) & & \checkmark & \\
\hline
Value suppression & \checkmark & & \checkmark \\
\hline

\end{tabular}
}
\end{table*}

\section{Conclusion}
\label{sec:conclusionstudy}

In this chapter, we have investigated the relation between data anonymization techniques and anti-discrimination to answer an important question: 
how privacy protection via data anonymization impacts the discriminatory bias contained in the original data. By presenting and analyzing different scenarios, we learn that we cannot protect original data against privacy attacks without taking into account anti-discrimination requirements ({\em i.e.} $\alpha$-protection). This happens because data anonymization techniques can work against anti-discrimination.
In addition, we exploit the fact that some data anonymization techniques ({\em e.g.} specific full-domain generalization) can also protect data against discrimination. Thus, we can adapt and use some of these techniques for discrimination prevention. Moreover, by considering anti-discrimination requirements during anonymization, we can present solutions to generate privacy- and discrimination-protected datasets. We also find that global recoding generalizations have a more positive impact on discrimination removal
than other data anonymization techniques.


\chapter[Generalization-based Privacy Preservation and
Discrimination Prevention in Data Publishing and Mining]{Generalization-based Privacy Preservation and
Discrimination Prevention in Data Publishing and Mining}
\chaptermark{Generalization-based PPDM and DPDM}
\label{chap:GBPPDPDM}


We observe that published data must be {\em both} privacy-preserving
and unbiased regarding discrimination. We present the first
generalization-based approach to simultaneously
offer privacy preservation and discrimination prevention. We formally define the problem, give an optimal algorithm to tackle it and evaluate the algorithm in terms of both general and specific data analysis metrics. It turns out
that the impact of our transformation on the quality of data is the
same or only slightly higher than the impact
of achieving just privacy preservation.
In addition, we show how to extend our approach
to different privacy models and anti-discrimination legal concepts.

\section{Introduction}

Although PPDM and DPDM have different goals, they have
some technical similarities. Necessary steps of PPDM are:
i) define the privacy model ({\em e.g.} $k$-anonymity);
ii) apply a proper anonymization technique ({\em e.g.} generalization) to
satisfy the requirements of the privacy model; iii) measure data
quality loss as a side effect of data distortion
(the measure can be general or tailored to specific data mining tasks).
Similarly, necessary steps for DPDM include: i) define the non-discrimination model according to the respective legal concept ({\em i.e.} {\em $\alpha$-protection} according to the legal concept of  direct discrimination); ii) apply
a suitable data
distortion method to satisfy the requirements of the
non-discrimination model; iii) measure data quality loss
as in the case of DPDM.
\subsection{Motivating Example}
\label{motiv}
Table~\ref{tab:71} presents raw customer credit data, where each record represents a customer's specific information. {\em Sex}, {\em Race}, and working hours named {\em Hours} can be taken as QI attributes.
The class attribute has two values, {\em Yes} and {\em No}, to indicate
whether the customer has received credit. Assume that {\em Salary} is
a sensitive/private attribute
and groups of {\em Sex} and {\em Race} attributes are {\em protected}.
The credit giver wants to publish
a privacy-preserving and non-discriminating version of Table~\ref{tab:71}.
To do that, she needs to eliminate two types of threats against her customers:
\begin{itemize}
\item {\em Privacy threat, e.g., record linkage}: If a record in the table is so specific that only
a few customers match it, releasing the data may
allow determining the customer's identity (record linkage attack) and
hence the salary of that identified customer.
Suppose that the adversary knows that the target identified customer is white
and his working hours are 40. In Table~\ref{tab:71}, record $ID=1$ is the only
one matching that customer, so the customer's salary becomes known.
\begin{table}[t]
\caption{Private data set with biased decision records}
\label{tab:71}
\centering
\begin{tabular}{|c|c|c|c|c|c|}
\hline
{\bf ID} & {\bf Sex} & {\bf Race} & {\bf Hours} & {\bf Salary} & {\bf  Credit\_ approved}\\
\hline
1 & Male & White & 40 & High & Yes \\
\hline
2 & Male & Asian-Pac & 50 & Medium & Yes \\
\hline
3 & Male & Black & 35 &  Medium & No \\
\hline
4 & Female & Black & 35 &  Medium & No \\
\hline
5 & Male & White & 37 & Medium & Yes \\
\hline
6 & Female & Amer-Indian & 37 & Medium & Yes \\
\hline
7 & Female & White & 35 & Medium & No \\
\hline
8 & Male & Black & 35 & High  & Yes \\
\hline
9 & Female & White & 35 & Low  & No\\
\hline
10 & Male & White & 50 & High & Yes\\
\hline
\end{tabular}
\end{table}
\item {\em Discrimination threat}: If credit has been denied to most female customers,
releasing the data may lead to making biased decisions against them when
these data are used for extracting decision patterns/rules as part
of the automated decision making. Suppose that the {\em  minimum support ($ms$)}
required to
extract a classification rule from the data set in Table~\ref{tab:71} is that
the rule be satisfied by at least 30\% of the records. This
would allow extracting the classification rule $r: Sex=female \rightarrow Credit\_approved=no$ from these data. Clearly, using such a rule for credit scoring is discriminatory against female customers.
\end{itemize}
\subsection{Contributions}
We argue that both threats above must be addressed at
the same time, since providing protection against only one of them
might not guarantee protection against the other.
An important question is how we can provide protection against both privacy and discrimination risks without one type of protection working
against the other and with minimum impact on data quality.
In this chapter, we investigate for the first time
the problem of discrimination- and privacy-aware {\em data} publishing,
{\em i.e.} transforming the data, instead of patterns, in order to simultaneously fulfill privacy preservation and discrimination prevention.
Our approach falls into the pre-processing category: it sanitizes
the data {\em before} they are used in data mining tasks
rather than sanitizing the knowledge patterns
extracted by data mining tasks (post-processing).
Very often, knowledge publishing (publishing
the sanitized patterns)
is not enough for the users or researchers, who want to be able to mine the data themselves. This gives researchers greater flexibility in performing the required data analyses. We introduce an
anti-discrimination model that
can cover every possible nuance of discrimination w.r.t. multiple attributes, not only for specific protected groups within one attribute. We show that generalization can be used for discrimination prevention. Moreover, generalization not only can make the original data privacy protected but can also simultaneously make the original data both discrimination and privacy protected. We present an optimal algorithm that can cover different {\em legally grounded} measures of discrimination to obtain all full-domain generalizations whereby the data is discrimination and privacy protected.
The ``minimal'' generalization ({\em i.e.}, the one incurring
the least information loss according to some criterion) can then be chosen.
In addition, we
evaluate the performance of the proposed approach and the data quality loss incurred as a side effect of the data generalization needed
to achieve both discrimination and privacy protection.
We compare this quality loss with the one incurred
to achieve privacy protection only. Finally, we present how our approach can be extended to satisfy different privacy models and anti-discrimination legal concepts.


\section{Privacy Model}
\label{sec:privacymodel}
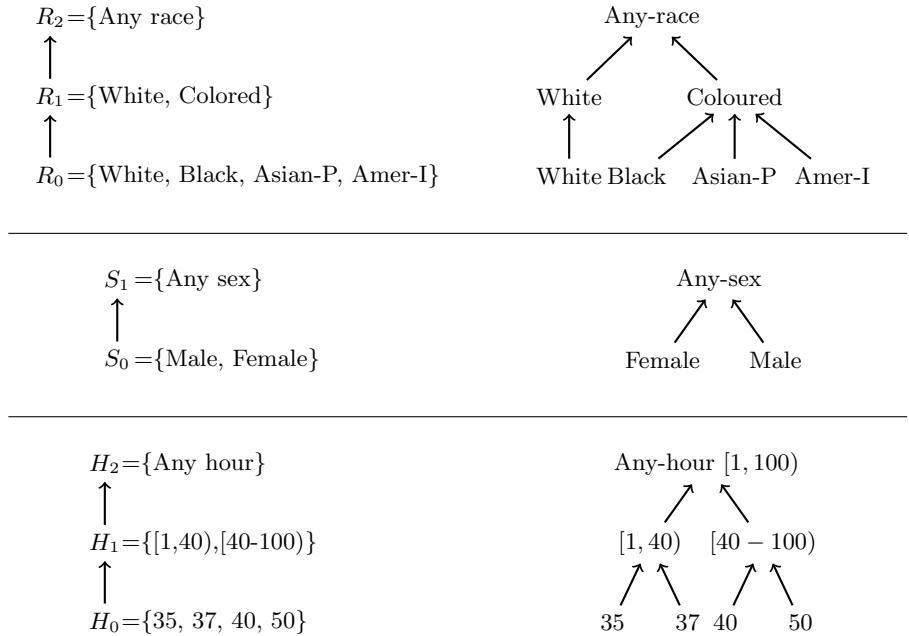
\begin{figure}[t]
\centering
\begin{tikzpicture}
\footnotesize
\tikzstyle{level 1}=[sibling distance=22mm]
\tikzstyle{level 2}=[sibling distance=13mm]
\tikzset{level distance=30pt}
\node (10) at (0,0) {$R_2$}
		child { node (11)	{$R_1$} edge from parent [thick,<-]
		child { node (12)	{$R_0$} edge from parent [thick,<-]
		}
	};
\draw (10) node [right] {\,\,=\{Any race\}};
\draw (11) node [right] {\,\,=\{White, Colored\}};
\draw (12) node[right] {\,\,=\{White, Black, Asian-P, Amer-I\}};
\node (1ar) at (8,0) {Any-race}
	child { node (1w) {White} edge from parent [thick,<-]
		child { node (1ww) {White} edge from parent [thick,<-]}}
	child { node (1c) {Coloured} edge from parent [thick,<-]
		child { node (1b) {Black} edge from parent [thick,<-]}
		child { node (1a) {Asian-P} edge from parent [thick,<-]}
		child { node (1ai) {Amer-I} edge from parent [thick,<-]}
		};
\end{tikzpicture}
\vspace{-0.3truecm}
\begin{center}
\line(1,0){340}
\end{center}
\begin{tikzpicture}
\footnotesize
\tikzset{level distance=30pt}
\node (20) at (0,0) {$S_1$}
		child { node (21)	{$S_0$} edge from parent [thick,<-]
		};
\draw (20) node [right] {\,\,=\{Any sex\}};
\draw (21) node [right] {\,\,=\{Male, Female\}};
\node (2as) at (8,0) {Any-sex}
	child { node (1w) {Female} edge from parent [thick,<-]}
	child { node (1c) {Male} edge from parent [thick,<-]
		};
\end{tikzpicture}
\vspace{-0.3truecm}
\begin{center}
\line(1,0){340}
\end{center}
\begin{tikzpicture}
\footnotesize
\tikzset{level distance=30pt}
\node (30) at (0,0) {$H_2$}
		child { node (31)	{$H_1$} edge from parent [thick,<-]
		child { node (32)	{$H_0$} edge from parent [thick,<-]
		}
	};
\draw (30) node [right] {\,\,=\{Any hour\}};
\draw (31) node [right] {\,\,=\{[1,40),[40-100)\}};
\draw (32) node [right] {\,\,=\{35, 37, 40, 50\}};
\tikzstyle{level 2}=[sibling distance=10mm]
\node (3ah) at (8,0) {Any-hour $[1,100)$}
	child { node (3a) {$[1,40)$} edge from parent [thick,<-]
		child { node (3e){$35$} edge from parent [thick,<-]}
		child { node (3f) {$37$} edge from parent [thick,<-]}
		}
	child { node (3b) {$[40-100)$} edge from parent [thick,<-]
		child { node (3c) {$40$} edge from parent [thick,<-]}
		child { node (3d) {$50$} edge from parent [thick,<-]}
		};
\end{tikzpicture}
\caption{An example of domain (left) and value (right) generalization hierarchies of Race, Sex and Hours attributes}
\label{fig:DomainValue}
\end{figure}

As described in Chapter \ref{sec:privmeasures}, to prevent record linkage attacks through quasi-identifiers,
Samarati and
Sweeney \cite{Samarati1998,SWEENEY1998} proposed the
notion of $k$-anonymity.
A data table satisfying this requirement is called $k$-anonymous. 
The total number of tuples in $\mathcal{D}$ for each sequence of values in $\mathcal{D}[QI]$ is called {\em frequency set}.
Let $D_i$ and $D_j$ be two domains. If the values of $D_j$ are the generalization of the values in domain $D_i$, we denote $D_i \leq_D D_j$.
A many-to-one {\em value generalization function} $\gamma: D_i \rightarrow D_j$ is associated with every $D_i$, $D_j$ with $D_i \leq_D D_j$. Generalization is based on a {\em domain generalization hierarchy} and a corresponding {\em value generalization hierarchy} on the values in the domains.
A domain generalization hierarchy is defined to be a set of domains that is totally ordered by the relationship $\leq_D$. We can consider the hierarchy as a chain of nodes, and if there is an edge from $D_i$ to $D_j$, it means that $D_j$ is the {\em direct generalization} of $D_i$. Let $Dom_i$ be a set of domains
in a domain generalization hierarchy of a quasi-identifier attribute $Q_i \in QI$. For every
$D_i, D_j, D_k \in Dom_i$ if $D_i \leq_D D_j$ and $D_j \leq_D D_k$, then $D_i \leq_D D_k$. In this case, domain $D_k$ is an {\em implied generalization} of $D_i$. The maximal element of $Dom_i$ is a singleton, which means that all values in each domain can be eventually generalized to a single value. Figure~\ref{fig:DomainValue} left shows possible domain generalization hierarchies for the \emph{Race}, \emph{Sex}
and \emph{Hours} attributes in Table \ref{tab:71}.
Value generalization functions associated with the domain generalization hierarchy induce a corresponding value-level tree, in which edges are denoted by $\gamma$, {\em i.e} direct value generalization, and paths are denoted by $\gamma^+$, {\em i.e.} implied value generalization. Figure~\ref{fig:DomainValue} right shows a value generalization hierarchy with each value in the \emph{Race}, \emph{Sex} and \emph{Hours} domains, {\em e.g.} Colored = $\gamma$(black) and
Any-race $\in \gamma^+$(black).
\begin{figure}[t]
\begin{center}
\begin{tikzpicture}
\footnotesize
\label{fig:lattice}
\node (s0r0) at (0,0) {$\langle S_0, R_0 \rangle$};
\node (s1r0) at (-2,0.6) {$\langle S_1, R_0 \rangle$};
\node (s0r1) at (2,0.6) {$\langle S_0, R_1 \rangle$};
\node (s1r1) at (-2,1.5) {$\langle S_1, R_1 \rangle$};
\node (s0r2) at (2,1.5) {$\langle S_0, R_2 \rangle$};
\node (s1r2) at (0,2.1) {$\langle S_1, R_2 \rangle$};
\draw[thick,->] (s0r0)--(s0r1);
\draw[thick,->] (s0r0)--(s1r0);
\draw[thick,->] (s1r0)--(s1r1);
\draw[thick,->] (s0r1)--(s0r2);
\draw[thick,->] (s0r1)--(s1r1);
\draw[thick,->] (s1r1)--(s1r2);
\draw[thick,->] (s0r2)--(s1r2);
\end{tikzpicture}
\caption{Generalization lattice for the Race and Sex attributes}
\label{fig:lattice}
\end{center}
\end{figure}
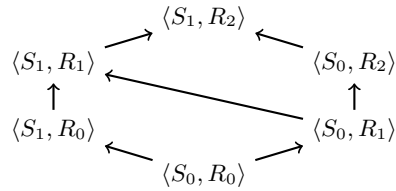
For a $QI=\{Q_1,\cdots, Q_n\}$ consisting of multiple attributes, each
with its own domain, the domain generalization hierarchies
of the individual attributes $Dom_1,\cdots,$\allowbreak$Dom_n$ can be combined to form a multi-attribute {\em generalization lattice}. Each vertex of a lattice is a domain tuple $DT=\langle N_1,\cdots,\allowbreak N_n\rangle$ such that $N_i \in Dom_i$, for $i=1,\cdots,n$, representing a multi-attribute domain generalization. An example for \emph{Sex} and \emph{Race} attributes is presented in Figure~\ref{fig:lattice}.
\begin{definition}[Full-domain generalization] Let $\mathcal{D}$ be a data table having a quasi-identifier $QI=\{Q_1,\cdots, Q_n\}$ with corresponding domain generalization hierarchies $Dom_1,\cdots,$\allowbreak$ Dom_n$. A full-domain generalization can be defined by a domain tuple $DT=\langle N_1,\cdots,N_n\rangle$ with $N_i \in Dom_i$, for every $i=1,\cdots,n$. A full-domain generalization with respect to $DT$ maps each value $q \in D_{Q_i}$  to some $a \in D_{N_i}$ such that
$a=q$, $a = \gamma(q)$  or $a \in \gamma^+(q)$.
\end{definition}
For example, consider Figure~\ref{fig:DomainValue} right
and assume that
values 40 and 50 of {\em Hours} are generalized to
$[40-100)$; then 35 and 37 must be generalized to $[1, 40)$.
 A full-domain generalization w.r.t. domain tuple $DT$ is $k$-anonymous if it yields $k$-anonymity for $\mathcal{D}$ with respect to $QI$.
As introduced in Section \ref{sec:PPDRelatedwork}, in the literature, different generalization-based algorithms have been proposed to $k$-anonymize a data table. They are optimal \cite{Samarati,Lefevre2005,BAYARDO2005} or minimal \cite{IYENGAR2002,Fung2005,Wang2004}. Although minimal algorithms are in general more efficient than optimal ones,
we choose an optimal algorithm ({\em i.e.} Incognito~\cite{Lefevre2005})
because, in this first work about combining
privacy preservation and discrimination prevention in data publishing,
it allows us to study the worst-case toll on efficiency of achieving
both properties.
Incognito is a well-known suite of optimal bottom-up generalization algorithms to generate all possible $k$-anonymous full-domain
generalizations
. 
In comparison with other optimal algorithms, Incognito is more scalable and practical for larger data sets and more suitable for categorical attributes.
%
Incognito is based
on two main properties satisfied for $k$-anonymity:
\begin{itemize}
\item {\bf Subset property}. If $\mathcal{D}$ is $k$-anonymous
with respect to $QI$, then it is $k$-anonymous with respect to any
subset of attributes in $QI$
(the converse property does not hold in general).
\item {\bf Generalization property}. Let $P$ and $Q$ be nodes
in the generalization lattice of $\mathcal{D}$ such that
$D_P \leq_D D_Q$. If $\mathcal{DB}$ is $k$-anonymous with respect to $P$,
then $\mathcal{DB}$ is also $k$-anonymous with respect to $Q$
(monotonicity of generalization).
\end{itemize}
\begin{example}
\label{exp:71}
Continuing the motivating
example (Section~\ref{motiv}), consider Table~\ref{tab:71} and suppose $QI =\{Race, Sex\}$ and $k=3$. Consider the generalization lattice over $QI$ attributes in Fig. \ref{fig:lattice}. Incognito finds that Table~\ref{tab:71} is 3-anonymous with respect to domain tuples $\langle S_1, R_1\rangle$, $\langle S_0, R_2\rangle$ and $\langle S_1, R_2\rangle$.
\end{example}
\section{Non-discrimination Model}
\label{discriminationmodel}

We use the measure presented in Section \ref{sec:NDDT} (see Definition \ref{def:protection}). 


\section{Simultaneous Privacy Preservation and Discrimination Prevention}
\label{sec:problemdefinition}
We want to obtain anonymized data tables that are protected against record linkage and also free from discrimination, more specifically $\alpha$-protective $k$-anonymous data tables defined
as follows.
\begin{definition}[$\alpha$-protective $k$-anonymous data table]
\label{def:fair}
Let $\mathcal{D}(A_1,\cdots, A_n)$ be a data table, $QI=\{Q_1,\cdots,Q_m\}$ a quasi-identifier, $DA$ a set of PD attributes, $k$ an anonymity threshold, and $\alpha$ a discrimination threshold. $\mathcal{D}$ is $\alpha$-protective $k$-anonymous if it is both $k$-anonymous and $\alpha$-protective with respect to $QI$ and $DA$, respectively.
\end{definition}
We focus on the problem of producing a version of $\mathcal{D}$ that is $\alpha$-protective $k$-anonymous with respect to $QI$ and $DA$.
The problem could be investigated with respect to different possible relations between categories of attributes in $\mathcal{D}$.
$k$-Anonymity uses quasi-identifiers for re-identification, so
we take the worst case for privacy in which
all attributes are QI except the class/decision attribute.
On the other hand, each QI attribute can be PD or not.
In summary, the following relations
are assumed: (1) $QI \cap C = \emptyset$, (2) $DA \subseteq QI$.
Taking the largest possible QI makes sense indeed.
The more attributes are included in QI, the more protection $k$-anonymity
provides and, in general, the more information loss it causes.
Thus, we test our proposal in the worst-case privacy scenario.
On the discrimination side,
as explained in Section \ref{sec:algorithm}, the more attributes are included in QI, the more protection is provided by
$\alpha$-protection.
\subsection{The Generalization-based Approach}
\label{sec:approach}
To design a method, we need to consider the impact of data generalization on discrimination.
\begin{definition}
\label{def:PDiscnode}
Let $\mathcal{D}$ be a data table having a quasi-identifier
$QI=\{Q_1,\cdots, Q_n\}$ with corresponding domain generalization hierarchies $Dom_1,\cdots,Dom_n$. Let $DA$ be a set of PD attributes associated with $\mathcal{D}$. Each node $N$ in $Dom_i$ is said to be PD if $Dom_i$ corresponds to one of the attributes in $DA$ and $N$ is not the singleton of $Dom_i$. Otherwise node $N$ is said to be PND. 
\end{definition}
Definition~\ref{def:PDiscnode} states that not only
ungeneralized
nodes of PD attributes are PD but also the generalized nodes of these domains
are PD.
For example, in South Africa, about 80\% of the population is Black,
9\% White, 9\% Colored and 2\% Asian.
Generalizing the {\em Race} attribute
in a census of the South African population to {\em \{White, Non-White\} }
causes the {\em Non-White} node to
inherit the PD nature of {\em Black, Colored and Asian}.
We consider the singleton nodes as PND because generalizing all instances of all values of a domain to single value is PND, {\em e.g.} generalizing all instances of male and female values to {\em any-sex} is PND.
\begin{example}
\label{exp:72}
Continuing the motivating
example, consider $DA=\{Race, Sex\}$ and Figure~\ref{fig:DomainValue} left. Based on Definition~\ref{def:PDiscnode}, in the domain generalization hierarchy of {\em Race}, {\em Sex} and {\em Hours},  $R_0$, $R_1$, $S_0$ are PD nodes, whereas $R_2$, $S_1$, $H_0$, $H_1$ and $H_2$ are PND nodes.
\end{example}
When we generalize data ({\em i.e.} full-domain generalization) can we achieve $\alpha$-protection? By presenting two main scenarios we show that the answer can be yes or no depending on the generalization:
\begin{itemize}
\item When the original data table $\mathcal{D}$ is biased
versus some protected groups  w.r.t. $DA$ and $f$ ({\em i.e.}, there is at least a frequent rule $c: A \rightarrow C$ such that $f(c) \geq \alpha$, where $A$ is PD itemset w.r.t. $DA$), a full-domain generalization can make $\mathcal{D}$ $\alpha$-protective if it includes the generalization of the respective protected groups ({\em i.e.} $A$).
\item When the original data table $\mathcal{D}$ is biased versus a subset of the protected groups w.r.t. $DA$ ({\em i.e.}, there is at least a frequent rule $c: A, B \rightarrow C$ such that $f(c) \geq \alpha$, where $A$ is a PD itemset and $B$ is PND itemset w.r.t. $DA$), a full-domain generalization can make $\mathcal{D}$ $\alpha$-protective
if any of the following holds: (1) it includes the generalization of the respective protected groups ({\em i.e.} $A$); (2) it includes the generalization of
the attributes
which define the respective subsets of the protected groups ({\em i.e.} $B$); (3) it includes both (1) and (2).
\end{itemize}
Then, given the generalization lattice of $\mathcal{D}$ over QI, where $DA \subseteq QI$, there are some candidate nodes for which $\mathcal{D}$ is $\alpha$-protective ({\em i.e.}, $\alpha$-protective full-domain generalizations).
Thus, we observe the following.
\begin{observation}
$k$-Anonymity and $\alpha$-protection can be achieved simultaneously in $\mathcal{D}$ by means of full-domain generalization.
\end{observation}
\begin{example}
Continuing Example~\ref{exp:71}, suppose $f=elift$ and consider the generalization lattice over $QI$ attributes in Fig. \ref{fig:lattice}. Among three 3-anonymous full-domain generalizations, only $\langle S_1, R_1\rangle$ and $\langle S_1, R_2\rangle$ are also 1.2-protective with respect to $DA = \{Sex\}$.
\end{example}
%
Our task is to obtain $\alpha$-protective $k$-anonymous full-domain generalizations.
The naive approach is the sequential way: first, obtain $k$-anonymous full-domain generalizations and then restrict to the subset of these that are $\alpha$-protective. Although this would solve the problem, it is a very expensive
solution: discrimination should be measured
for each $k$-anonymous full-domain generalization to determine
whether it is $\alpha$-protective. In the next section we present a more efficient algorithm that takes advantage of the common properties of $\alpha$-protection and $k$-anonymity.
\subsection{The Algorithm}
\label{sec:algorithm}
In this section, we present an optimal algorithm for obtaining all possible full-domain generalizations with which $\mathcal{D}$ is $\alpha$-protective $k$-anonymous.
\subsubsection{Foundations}
\begin{observation}[Subset property of $\alpha$-protection]
From Definition \ref{def:protection}, observe that if $\mathcal{D}$ is $\alpha$-protective with respect to $DA$, it is $\alpha$-protective w.r.t. any subset of attributes in $DA$. The
converse property does not hold in general.
\end{observation}
For example, if Table \ref{tab:71} is 1.2-protective w.r.t $DA=\{Sex, Race\}$, Table \ref{tab:71} must also be 1.2-protective w.r.t. $DA=\{Sex\}$ and $DA=\{Race\}$.
Otherwise put, if Table \ref{tab:71} is not 1.2-protective w.r.t. $DA=\{Sex\}$
 or it is not 1.2 protective w.r.t. $DA=\{Race\}$,
it cannot be 1.2-protective w.r.t. $DA=\{Sex, Race\}$. This is in correspondence with the subset property of $k$-anonymity. Thus, $\alpha$-protection w.r.t.
all strict subsets of $DA$ is a necessary (but not sufficient) condition for $\alpha$-protection w.r.t. $DA$. Then, given generalization hierarchies over QI, the generalizations that are not $\alpha$-protective w.r.t. a subset $DA'$ of $DA$ can be discarded along with all their descendants in the hierarchy.
To prove the generalization property of $\alpha$-protection, we need a
preliminary well-known
mathematical result, stated in the following lemma.
\begin{lemma}
\label{l:minmax}
Let $x_1,\cdots,x_n, y_1,\cdots,y_n$ be positive integers and let
$x=x_1+\cdots+x_n$ and $y=y_1+\cdots+y_n$.
Then
\[\min_{1\leq i \leq n}\left\{\frac{x_i}{y_i}\right\}\leq \frac{x}{y}\leq\max_{1\leq i \leq n}\left\{\frac{x_i}{y_i}\right\}.\]
\end{lemma}
\begin{proof}
Without loss of generality, suppose that $\frac{x_1}{y_1}\leq \cdots \leq \frac{x_n}{y_n}$.
Then
\[\frac{x}{y}=\frac{y_1}{y}\frac{x_1}{y_1}+\cdots+\frac{y_n}{y}\frac{x_n}{y_n}
\leq \left (\frac{y_1}{y}+\cdots+\frac{y_n}{y}\right)\frac{x_n}{y_n}\leq \frac{x_n}{y_n}.\]
The other inequality is proven analogously.
\end{proof}
\begin{proposition}[Generalization property of $\alpha$-protection]
\label{generalization}
Let  $\mathcal{D}$ be a data table and $P$ and $Q$ be nodes in
the generalization lattice of
$DA$ with $D_P \leq_D D_Q$.
If $\mathcal{D}$ is $\alpha$-protective w.r.t. to $P$ considering
minimum support $ms=1$ and discrimination measure $elift$ or $clift$, then
$\mathcal{D}$ is also $\alpha$-protective w.r.t. to $Q$.
\end{proposition}
\begin{proof}
Let $A^1,\ldots,A^n$ and $A$ be itemsets in $P$ and $Q$, respectively,
such that $\{A^1,\ldots,A^n\}=\gamma^{-1}(A)$. That is, $A$ is the
generalization of $\{A^1,\ldots,A^n\}$.
Let $B$ be an itemset from attributes in $QI\setminus DA$,
and $C$ a decision item. For simplicity, assume that
$supp(A^i,B)> 0$ for $i=1,\ldots,n$.
According to Section~\ref{sec:NDDT}, for the PD rule
$c: A,B\rightarrow C$,
\[elift(c)= \frac{\frac{supp(A,B,C)}{supp (A,B)}}{\frac{supp(B,C)}{supp (B)}}\,\, \text{ and }\,\, clift(c)=
\frac{\frac{supp(A,B,C)}{supp (A,B)}}{\frac{supp(X,B,C)}{supp (X,B)}},\]
where $X$ is the most favored itemset in Q with
respect to $B$ and $C$.
Since $supp(A,B)=\sum_{i} supp(A^i,B)$, and $supp(A,B,C)=\sum_{i} supp(A^i,B,C)$, by Lemma~\ref{l:minmax} we obtain that
\[\frac{supp(A,B,C)}{supp(A,B)}\leq \max_{i} \frac{supp(A^i,B,C)}{supp(A^i,B)}.\]
Hence if none of the rules
$A^i,B\rightarrow C$ are
$\alpha$-discriminatory with respect to the measure $elift$,
then the rule $A,B\rightarrow C$ is not $\alpha$-discriminatory.
Now we consider the measure $clift$.
Let $Y$ be the most favored itemset in $P$ with respect to
the itemsets $B$ and the item $C$. By following an analogous
argument, we obtain that
\[
\frac{supp(X,B,C)}{supp (X,B)}
\geq
\frac{supp(Y,B,C)}{supp (Y,B)}.
\]
Therefore if none of the rules
$A^i,B\rightarrow C$ are
$\alpha$-discriminatory with respect to the measure $clift$,
then $c$ is not $\alpha$-discriminatory.
\end{proof}
%
For example, considering $DA=\{Race\}$ and $f=elift$ or
$f=clift$,
 based on the generalization property of $k$-anonymity, if Table \ref{tab:71} is 3-anonymous w.r.t. $\langle R_0,H_0\rangle$, it must be also 3-anonymous w.r.t. $\langle R_1,H_0\rangle$ and $\langle R_0,H_1\rangle$. However, based on the generalization property of $\alpha$-protection, if Table \ref{tab:71} is 1.2-protective w.r.t. $\langle R_0,H_0\rangle$,
it must be also 1.2-protective w.r.t. $\langle R_1,H_0\rangle$,
which contains the generalization of the attributes  in $DA$,
but not necessarily w.r.t. $\langle R_0,H_1\rangle$
(the latter generalization is for an attribute not in $DA$).
Thus, we notice that the generalization property of $\alpha$-protection is weaker than the generalization property of $k$-anonymity, because
the former is only
guaranteed for generalizations of attributes in $DA \subseteq QI$,
whereas the latter holds for generalizations of any attribute in $QI$.
Moreover, the generalization property has a limitation. Based on Definition \ref{def:protection}, a data table is $\alpha$-protective w.r.t. $DA$
if all PD frequent rules extracted from the data table are not
$\alpha$-discriminatory w.r.t. $DA$. Hence, a data table might
contain PD rules which are not $\alpha$-protective and not frequent, {\em e.g.} {\em Age=25, City=NYC $\rightarrow$ Credit=no}, {\em Age=27, City=NYC $\rightarrow$ Credit=no}, {\em Age=28, City=NYC $\rightarrow$ Credit=no}, where $DA=\{Age\}$. However, after generalization, frequent PD rules can appear which might be $\alpha$-discriminatory and discrimination will show up, {\em e.g.} {\em Age=[25-30], City=NYC $\rightarrow$ Credit=no}.
This is why the generalization property of $\alpha$-protection
requires that $\alpha$-protection w.r.t. $P$ hold for all PD rules,
frequent and infrequent
(this explains the condition $ms=1$ in Proposition~\ref{generalization}).
The next property allows improving the efficiency of the algorithm for obtaining $\alpha$-protective $k$-anonymous data tables by means of full-domain generalizations. Its proof is straightforward.
\begin{proposition}[Roll-up property of $\alpha$-protection]
Let  $\mathcal{D}$ be a data table
with records in a domain tuple $DT$,
let $DT'$ be a domain tuple with $DT\leq_D DT'$, and let $\gamma:DT\to DT'$ be
the associated generalization function. The support of an itemset $X$ in $DT'$
is the sum of the supports of the itemsets in $\gamma^{-1}(X)$.
\end{proposition}
\subsubsection{Overview}
We take Incognito as an optimal anonymization algorithm based on
the above properties and extend it to generate the set of all possible $\alpha$-protective $k$-anonymous full-domain generalizations of $\mathcal{D}$.
Based on the subset
property for $\alpha$-protection and $k$-anonymity,
the algorithm, named $\alpha$-protective Incognito, begins by checking single-attribute
subsets of QI, and then iterates by checking
$k$-anonymity and $\alpha$-protection with respect to increasingly larger subsets, in a
manner reminiscent of
\cite{Apriori}.
Consider a graph of candidate multi-attribute generalizations (nodes) constructed from a
subset of QI of size $i$. Denote this subset by $C_i$.
The set of direct multi-attribute generalization relationships (edges) connecting these nodes is denoted by $E_i$.
The $i$-th iteration of $\alpha$-protective Incognito performs a search that
determines first the $k$-anonymity status and second the $\alpha$-protection status of table $\mathcal{D}$ with respect to each candidate generalization in $C_i$. This is accomplished
using a modified bottom-up breadth-first search, beginning
at each node in the graph that is not the direct
generalization of some other node.
A modified breadth-first
search over the graph yields the set of multi-attribute
generalizations of size $i$ with respect to which $\mathcal{D}$ is $\alpha$-protective $k$-anonymous  (denoted by $S_i$).
After obtaining the entire $S_i$, the algorithm constructs the set
of candidate nodes of size $i+1$
($C_{i+1}$),
and the edges
connecting them
($E_{i+1}$)
using the subset property.
\subsubsection{Description}
Algorithm~\ref{alg71} describes $\alpha$-protective Incognito.
In the $i$-th iteration, the algorithm determines the $k$-anonymity status of $\mathcal{D}$ with respect to each node in $C_i$ by computing the frequency set in one of the following ways: if the node is root, the frequency set is computed using $\mathcal{D}$. Otherwise, for non-root nodes, the frequency set is computed using all parents' frequency sets. This is based on the roll-up property for $k$-anonymity.
If $\mathcal{D}$ is $k$-anonymous with respect to the attributes of the node,
the algorithm performs
two actions. First, it marks all direct generalizations of the node as $k$-anonymous. This is based on the generalization property for $k$-anonymity:
these generalizations need not
be checked anymore for $k$-anonymity in the subsequent search
iterations. Second, if the node contains at least one PD attribute and $i \leq \tau$ (where $\tau$ is the discrimination granularity level, see definition
further below), the algorithm determines the $\alpha$-protection status of $\mathcal{D}$ by computing function {\em Check $\alpha$-protection}($i$, $node$) (see Algorithm \ref{alg72}).
If $\mathcal{D}$ is $\alpha$-protective w.r.t. the attributes of
the node, the algorithm marks
as $\alpha$-protective $k$-anonymous
all direct generalizations of the node which are $\alpha$-protective according
to the generalization property of $\alpha$-protection.
The algorithm will not check them anymore for $\alpha$-protection in the subsequent search iterations. Finally, the algorithm constructs $C_{i+1}$ and $E_{i+1}$ by considering only nodes in $C_i$ that are marked as $\alpha$-protective $k$-anonymous.

\begin{algorithm}[h]
\caption{{\sc $\alpha$-protective Incognito}}
\label{alg71}
{\footnotesize \begin{algorithmic}[1]
\REQUIRE Original data table $\mathcal{D}$, a set $QI=\{Q_1,\cdots, Q_n\}$
of quasi-identifier attributes, a set of domain generalization hierarchies $Dom_1,\cdots, Dom_n$, a set of PD attributes $DA$, $\alpha$, $f$, $k$, C=\{Class item\}, $ms$=minimum support, $\tau \leq |QI|$
\ENSURE The set of $\alpha$-protective $k$-anonymous full-domain generalizations
\STATE $C_1$=\{Nodes in the domain generalization hierarchies of attributes in $QI$\}
\STATE $C_{PD}=\{\forall C \in C_1$ s.t. $C$ is PD\}
\STATE $E_1$=\{Edges in the domain generalization hierarchies of attributes in $QI$\}
\STATE $queue$=an empty queue
\FOR{$i=1$ to $n$}
\STATE //$C_i$ and $E_i$ define a graph of generalizations
\STATE $S_i$=copy of $C_i$
\STATE \{roots\}=\{all nodes $\in C_i$ with no edge $\in E_i$ directed to them\}
\STATE Insert \{roots\} into $queue$, keeping $queue$ sorted by height
\WHILE{$queue$ is not empty}
\STATE $node$ = Remove first item from $queue$
\IF{$node$ is not marked as $k$-anonymous or $\alpha$-protective $k$-anonymous}
\IF{$node$ is a root}
\STATE $frequencySet$= Compute the frequency set of $\mathcal{D}$ w.r.t. attributes of $node$ using $\mathcal{D}$.
\ELSE
\STATE $frequencySet$= Compute the frequency set of $\mathcal{D}$ w.r.t. attributes of $node$ using the parents' frequency sets.
\ENDIF
\STATE Use $frequencySet$ to check $k$-anonymity w.r.t. attributes of $node$
\IF{$\mathcal{D}$ is $k$-anonymous w.r.t. attributes of $node$}
\STATE Mark all direct generalizations of $node$ as $k$-anonymous
\IF{$\exists N \in C_{PD}$ s.t. $N \subseteq node$ and $i \leq \tau$}
\IF{$node$ is a root}
\STATE $MR$=  {\sc Check $\alpha$-protection}($i$, $node$) of $\mathcal{D}$ w.r.t. attributes of $node$ using $\mathcal{D}$.
\ELSE
\STATE $MR$= {\sc Check $\alpha$-protection}($i$, $node$) of $\mathcal{D}$ w.r.t. attributes of $node$ using parents' $I_l$ and $I_{l+1}$
\ENDIF
\STATE Use $MR$ to check $\alpha$-protection w.r.t. attributes of $node$
\IF{$MR=case_3$}
\STATE Mark all direct generalizations of $node$ that contain the generalization of $N$ as $k$-anonymous $\alpha$-protective
\ELSIF{$MR=case_1$}
\STATE Delete $node$ from $S_i$
\STATE Insert direct generalizations of $node$ into $queue$, keeping $queue$ ordered by height
\ENDIF
\ENDIF
\ELSE
\STATE Steps 31-32
\ENDIF
\ELSIF{$node$ is marked as $k$-anonymous}
\STATE Steps 21-36
\ENDIF
\ENDWHILE
\STATE $C_{i+1}, E_{i+1}=GraphGeneration(S_i, E_i)$
\ENDFOR
\STATE Return projection of attributes of $S_n$ onto $\mathcal{D}$ and $Dom_1,..., Dom_n$
\end{algorithmic}}
\end{algorithm}
The discrimination granularity level $\tau \leq |QI|$ is one of the inputs of $\alpha$-protective Incognito. The larger $\tau$, the more protection regarding discrimination will be achieved. The reason is that, if the algorithm can check the status of $\alpha$-protection in $\mathcal{D}$ w.r.t. nodes which contain more attributes ({\em i.e.}, finer-grained subsets of protected groups), then more possible local niches of discrimination in $\mathcal{D}$ are discovered. 
However, a greater $\tau$ leads to more computation by $\alpha$-protective Incognito, because
$\alpha$-protection of $\mathcal{D}$ should be checked in more
iterations.
In fact, by setting $\tau < |QI|$, we can provide a trade-off between efficiency and discrimination protection.
As mentioned above, Algorithm \ref{alg72} implements the {\em Check $\alpha$-protection}($i$, $node$) function to check the $\alpha$-protection of $\mathcal{D}$ with respect to the attributes of the node. To do it in an efficient way,
first
the algorithm generates the set of $l$-itemsets of attributes of $node$ with their support values, denoted by $I_l$, and the set of $(l+1)$-itemsets of attributes of $node$ and class attribute, with their support values, denoted by $I_{l+1}$,
where $l=i$ is the number of items in the itemset. In SQL language, $I_l$ and $I_{l+1}$ is obtained from $\mathcal{D}$ by issuing a suitable query.
This computation is only necessary for root nodes in each iteration; for non-root nodes,
$I_l$ and $I_{l+1}$ are obtained from $I_l$ and $I_{l+1}$ of parent nodes based on the roll-up property of $\alpha$-protection. Then, PD classification rules ({\em i.e.} $PD_{groups}$) with the required values to compute each $f$ in Figure~\ref{fig:Measures} ({\em i.e.} $n_1$, $a_1$, $n_2$ and $a_2$) are obtained by scanning $I_{l+1}$.
\begin{algorithm}[h]
\caption{{\sc Check $\alpha$-protection ($i$, $node$)}}
\label{alg72}
{\footnotesize \begin{algorithmic}[1]
\STATE $l=i$
\STATE $I_l$=\{$l$-itemsets containing attributes of $node$\}
\STATE $I_{l+1}$=\{$(l+1)$-itemsets containing attributes of $node$ and class item $C$\}
\FOR{each $R \in I_{l+1}$}
\STATE $X=R \backslash C$
\STATE $a_1=supp(R)$
\STATE $n_1=supp(X)$ \hspace{3 mm} // $X$ found in $I_l$
\STATE $A$=largest subset of $X$ containing protected groups w.r.t. $DA$
\STATE $T=R\backslash A$
\STATE $Z=\neg A \cup T$
\STATE $a_2=supp(Z)$ \hspace{3 mm}  // Obtained from $I_{l+1}$
\STATE $n_2=supp(Z \backslash C)$ \hspace{3 mm} // Obtained from $I_{l}$
\STATE Add $R: A, B \rightarrow C$ to $PD_{groups}$ with values $a_1$, $n_1$, $a_2$ and $n_2$
\ENDFOR
\STATE Return $MR$={\sc Measure\_disc}($\alpha$, $ms$, $f$)
\end{algorithmic}}
\end{algorithm}
During the scan of $I_{l+1}$, PD classification rules $A, B \rightarrow C$ ({\em i.e.} $PD_{groups}$) are obtained with the respective values $a_1=supp(A, B, C)$, $n_1=supp(A, B)$
(note that $supp(A, B)$ is in $I_l$), $a_2=supp(\neg A, B, C)$
(obtained from $I_{l+1}$), and $n_2=supp(\neg A, B)$ 
(obtained from $I_l$).
By relaxing $\tau$ we can limit the maximum number of itemsets in $I_l$ and $I_{l+1}$ that are generated during the execution of $\alpha$-protective Incognito.
After obtaining $PD_{groups}$ with the values $a_1$, $a_2$, $n_1$ and $n_2$, Algorithm \ref{alg72} computes the function {\em Measure\_disc} ($\alpha$, $ms$, $f$) (see Algorithm \ref{alg73}).
\begin{algorithm}[h]
\caption{{\sc Measure\_disc($\alpha$, $ms$, $f$)}}
\label{alg73}
{\footnotesize \begin{algorithmic}[1]
\IF{$f=slift$ or $olift$}
\IF{$\exists$ a group ($A, B \rightarrow C$) in $PD group$ which is frequent
w.r.t. $ms$ and $\alpha$-discriminatory w.r.t. $f$}
\STATE Return $MR=Case_1$ \hspace{3 mm} // $\mathcal{D}$ is not $\alpha$-protective w.r.t. attributes of $node$
\ELSE
\STATE Return $MR=Case_2$ \hspace{3 mm} // $\mathcal{D}$ is $\alpha$-protective w.r.t. attributes of $node$
\ENDIF
\ENDIF
\IF{$f=elift$ or $clift$}
\IF{$\exists$ a group ($A, B \rightarrow C$) in $PD group$ which is frequent w.r.t. $ms$ and $\alpha$-discriminatory w.r.t. $f$}
\STATE Return $MR=Case_1$ \hspace{3 mm} // $\mathcal{D}$ is not $\alpha$-protective w.r.t. attributes of $node$
\ELSIF{$\exists$ a group ($A, B \rightarrow C$) in $PD group$ which is infrequent w.r.t. $ms$ and $\alpha$-discriminatory w.r.t. $f$}
\STATE Return $MR=Case_2$ \hspace{3 mm} // $\mathcal{D}$ is $\alpha$-protective w.r.t. attributes of $node$
\ELSIF{$f=clift$ and $\exists$ a group ($A, B \rightarrow C$) in $PD group$ which is infrequent w.r.t. $ms$ whose confidence is lower than the confidence of
the most favored item considered in the computation of $clift$}
\STATE Return $MR=Case_2$ \hspace{3 mm} // $\mathcal{D}$ is $\alpha$-protective w.r.t. attributes of $node$
\ELSE
\STATE Return $MR=Case_3$ \hspace{3 mm} // $\mathcal{D}$ is $\alpha$-protective w.r.t. attributes of $node$ and subsets of its generalizations
\ENDIF
\ENDIF
%
\end{algorithmic}}
\end{algorithm}
This function
takes $f$ as a parameter and is based on the
generalization property of $\alpha$-protection. If $f=slift$ or $f=olift$ and if there exists at least one frequent group $A, B \rightarrow C$ in $PD_{groups}$ with $slift(A, B \rightarrow C) \geq \alpha$, then $MR=case_1$ ({\em i.e.} $\mathcal{D}$ is not $\alpha$-protective w.r.t. attributes of $node$). Otherwise, $MR=case_2$ ({\em i.e.} $\mathcal{D}$ is $\alpha$-protective w.r.t. attributes of $node$). If $f=elift$ or $f=clift$, the generalization property of $\alpha$-protection is satisfied, so if there exists at least one frequent group $A, B \rightarrow C$ in $PD_{groups}$ with $elift(A, B \rightarrow C) \geq \alpha$,
then $MR=case_1$. Otherwise if there exists at least one infrequent group $A, B \rightarrow C$ in $PD_{groups}$ with $elift(A, B \rightarrow C) \geq \alpha$, then $MR=case_2$. Otherwise if all groups in $PD_{groups}$, frequent and infrequent, have $elift(A, B \rightarrow C) < \alpha$, $MR=case_3$.
It is worth mentioning that in the $i$-th iteration of $\alpha$-protective Incognito, for each node in $C_i$, first $k$-anonymity will be checked and then $\alpha$-protection. This is because
the algorithm only checks $\alpha$-protection for the nodes that
contain at least one PD attribute, while $k$-anonymity is checked for all nodes. Moreover, in some iterations, the algorithm does not check $\alpha$-protection if $\tau < |QI|$.
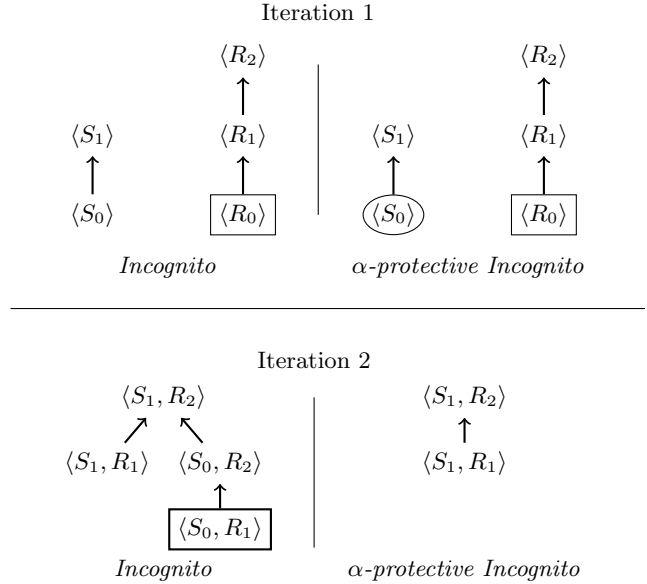
\begin{figure}[t]
\centering
\begin{tikzpicture}[grow'=up]
\footnotesize
\tikzset{level distance=30pt}
\node at (3,2.7) {Iteration 1};	
\node at (0,0) {$\langle S_0\rangle $}
		child { node {$\langle S_1 \rangle$} edge from parent [thick,->]
	};
\node[rectangle,draw] at (2,0) {$\langle R_0\rangle $}
		child { node {$\langle R_1 \rangle$} edge from parent [thick,->]
			child { node {$\langle R_2 \rangle$} edge from parent [thick,->]}
	};
\node at (1,-0.7) {\emph{Incognito}};	
\draw (3,0) -- (3,2);	
\node[ellipse,draw,scale=0.7] (100) at (4,0) {\phantom{$\langle S_0\rangle $}}
		child { node (101)	{$\langle S_1 \rangle$} edge from parent [thick,->]
	};
\node at (100) {$\langle S_0\rangle $};
\node[rectangle,draw] (110) at (6,0) {$\langle R_0\rangle $}
		child { node(111)	{$\langle R_1 \rangle$} edge from parent [thick,->]
			child { node(112)	{$\langle R_2 \rangle$} edge from parent [thick,->]}
	};
\node at (5,-0.7) {\emph{$\alpha$-protective Incognito}};	
\end{tikzpicture}

\vspace{-0.6truecm}

\begin{center}
\line(1,0){240}
\end{center}

\begin{tikzpicture}
\footnotesize
\tikzset{level distance=25pt}
\node at (2,0.5) {Iteration 2};	
\node (000) at (0,0) {$\langle S_1,R_2\rangle $}
		child { node (001)	{$\langle S_1,R_1 \rangle$} edge from parent [thick,<-]}
		child { node (010)	{$\langle S_0,R_2 \rangle$} edge from parent [thick,<-]
			child { node[rectangle,draw] (020)	{$\langle S_0,R_1 \rangle$}
			edge from parent [thick,<-]}
	};
\draw (2,0) -- (2,-2);
\node (010) at (4,0) {$\langle S_1,R_2\rangle $}
		child { node (011)	{$\langle S_1,R_1 \rangle$} edge from parent [thick,<-]
			};
\node at (0,-2.3) {\emph{Incognito}};	
\node at (4,-2.3) {\emph{$\alpha$-protective Incognito}};	
\end{tikzpicture}
\caption{The candidate 1- and 2-attribute generalization of Table \ref{tab:71} by Incognito(left) and $\alpha$-protective Incognito (right)}
\label{fig:exp}
\end{figure}
\begin{example}
Consider Table \ref{tab:71} and suppose that $QI=\{Sex, Race\}$ and $Credit\_approved$ are the only attributes in the data table, and $DA=\{Gender\}$, $k=3$ and $\alpha=1.2$. The first iteration of Incognito (Figure \ref{fig:exp} up-left) finds that Table \ref{tab:71} is not 3-anonymous w.r.t. $\langle R_0 \rangle$. The second iteration of Incognito performs a breadth-first search to determine the $k$-anonymity status of Table \ref{tab:71} w.r.t. 2-attribute generalizations of $\langle Sex,Race \rangle$ by considering all the 1-attribute generalizations
of the first iteration except $\langle R_0 \rangle$. Incognito (Figure \ref{fig:exp} down-left) finds that Table \ref{tab:71} is not 3-anonymous w.r.t. $\langle S_0, R_1 \rangle$, while other generalizations are 3-anonymous. The first iteration of $\alpha$-protective Incognito (Figure \ref{fig:exp} up-right) finds that Table \ref{tab:71} is not 3-anonymous w.r.t. $\langle R_0 \rangle$ and  Table \ref{tab:71} is not 1.2-protective w.r.t. $\langle S_0 \rangle$. The second iteration of $\alpha$-protective Incognito performs a breadth-first search to determine the $k$-anonymity and $\alpha$-protection status of Table \ref{tab:71} w.r.t. 2-attribute generalizations
of $\langle Sex,Race \rangle$ by considering all the 1-attribute generalizations of the first iteration except $\langle R_0 \rangle$ and $\langle S_0 \rangle$. $\alpha$-protective Incognito (Figure \ref{fig:exp} down-right) finds that Table \ref{tab:71} is  3-anonymous and 1.2-protective w.r.t. both $\langle S_1, R_1 \rangle$ and $\langle S_1, R_2 \rangle$.

 \end{example}
\section{Experiments}
\label{sec:expri}
Our first objective is to evaluate the performance of $\alpha$-protective Incognito (Algorithm~\ref{alg71}) and compare it with Incognito. Our second objective is to evaluate the quality of unbiased anonymous data
output by $\alpha$-protective Incognito, compared to that of the anonymous data output by plain Incognito, using both general and specific data analysis metrics.
We implemented all algorithms using Java and IBM DB2.
All experiments were performed
on an Intel Core i5 CPU with 4 GB of RAM. The software included windows 7 Home Edition and DB2 Express Version 9.7.
We considered
different values of $f$, $DA$, $k$, $\alpha$ and $\tau$ in our experiments.
\subsection{Dataset}
In our experiments, we used the Adult data set introduced in Section \ref{sec:datasets}. We ran experiments on the training set of the Adult dataset.
We used the same 8 categorical attributes
used in \cite{Fung2005}, shown in Table~\ref{tab:72}, and obtained their generalization hierarchies
from the authors of \cite{Fung2005}. For our experiments, we set $ms=5\%$ and 8 attributes in Table~\ref{tab:72} as QI, and $DA_1=\{Race, Gender, Mari$\allowbreak$tal\_status\}$, $DA_2=\{Race, Marital\_status\}$ and $DA_3=\{Race, Gender\}$. The smaller $ms$, the more computation and
the more discrimination discovery. In this way,
we considered
a very demanding scenario in terms of
privacy (all 8 attributes were QI) and anti-discrimination (small $ms$).
\begin{table}[ht]
\caption{Description of the Adult data set}
\label{tab:72}
\centering
\begin{tabular}{|c|c|c|}
\hline
{\bf Attribute} & {\bf \#Distinct values} & {\bf \#Levels of hierarchies} \\
\hline
\hline
Education & 16 & 5  \\
\hline
Marital\_status & 7 & 4 \\
\hline
Native\_country & 40 & 5  \\
\hline
Occupation & 14 & 3  \\
\hline
Race & 5 & 3   \\
\hline
Relationship & 6 & 3 \\
\hline
Sex & 2 & 2 \\
\hline
Work-class & 8 & 5 \\
\hline
\end{tabular}
\end{table}
%
%
\subsection{Performance}
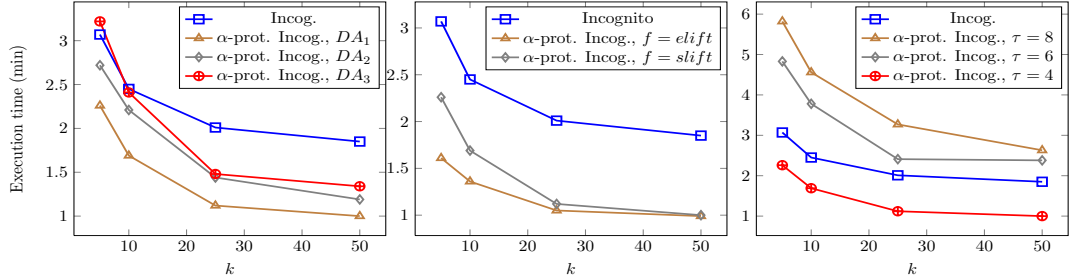
\begin{figure*}
\centering
\begin{tikzpicture}[thick, scale=0.84]
\def \plotdist {5.4cm} 
\def \xsc {0.72} 
\def \ysc {0.65} 
\begin{scope}[xscale=\xsc , yscale=\ysc]
\begin{axis}[
        xlabel=$k$,
        ylabel=Execution time (min),
        ]
 \addplot[mark=square,mark size=3pt,blue,very thick] plot coordinates {
        (5,3.07)
        (10,2.45)
        (25,2.01)
        (50,1.85)
    };
    \addlegendentry{Incog.}
    \addplot[mark=triangle,brown,mark size=3pt,very thick]
        plot coordinates {
          (5,2.26)
        (10,1.69)
        (25,1.12)
        (50,1)
        };
    \addlegendentry{$\alpha$-prot. Incog., $DA_1$}
    \addplot[mark=diamond,gray,mark size=3pt,very thick] plot coordinates {
        (5,2.72)
        (10,2.21)
        (25,1.44)
        (50,1.19)
    };
    \addlegendentry{$\alpha$-prot. Incog., $DA_2$}
\addplot[mark=oplus,red,mark size=3pt,very thick] plot coordinates {
        (5,3.22)
        (10,2.41)
        (25,1.48)
        (50,1.34)
    };
    \addlegendentry{$\alpha$-prot. Incog., $DA_3$}
    \end{axis}
    \end{scope}
\begin{scope}[xshift=\plotdist, xscale=\xsc , yscale=\ysc]
\begin{axis}[
        xlabel=$k$,
        ]
 \addplot[mark=square,blue,mark size=3pt,very thick] plot coordinates {
        (5,3.07)
        (10,2.45)
        (25,2.01)
        (50,1.85)
    };
    \addlegendentry{Incognito}
    \addplot[mark=triangle,brown,mark size=3pt,very thick]
        plot coordinates {
        (5,1.61)
        (10,1.36)
        (25,1.05)
        (50,0.99)
        };
    \addlegendentry{$\alpha$-prot. Incog., $f=elift$}
    \addplot[mark=diamond,gray,mark size=3pt,very thick] plot coordinates {
         (5,2.26)
        (10,1.69)
        (25,1.12)
        (50,1)
    };
    \addlegendentry{$\alpha$-prot. Incog., $f=slift$}
    \end{axis}
    \end{scope}
\begin{scope}[xshift=2*\plotdist, xscale=\xsc , yscale=\ysc]
\begin{axis}[
        xlabel=$k$,
        ]
 \addplot[mark=square,blue,mark size=3pt,very thick] plot coordinates {
        (5,3.07)
        (10,2.45)
        (25,2.01)
        (50,1.85)
    };
    \addlegendentry{Incog.}
    \addplot[mark=triangle,brown,mark size=3pt,very thick]
        plot coordinates {
        (5,5.82)
        (10,4.56)
        (25,3.27)
        (50,2.63)
        };
    \addlegendentry{$\alpha$-prot. Incog., $\tau=8$}
    \addplot[mark=diamond,gray,mark size=3pt,very thick] plot coordinates {
        (5,4.83)
        (10,3.78)
        (25,2.41)
        (50,2.38)
    };
    \addlegendentry{$\alpha$-prot. Incog., $\tau=6$}
      \addplot[mark=oplus,red,mark size=3pt,very thick] plot coordinates {
        (5,2.26)
        (10,1.69)
        (25,1.12)
        (50,1)
    };
    \addlegendentry{$\alpha$-prot. Incog., $\tau=4$}
    \end{axis}
\end{scope}
%
\end{tikzpicture}
\caption{Performance of Incognito and $\alpha$-protective Incognito for several values of $k$, $\tau$, $f$ and $DA$. Unless otherwise specified, $f=slift$
, $DA=DA_1$ and $\tau=4$.}
\label{fig:executions2}
\end{figure*}

Figure \ref{fig:executions2} reports the execution time of $\alpha$-protective Incognito, for different values of $\tau$, $DA$, $f$, $k$ in comparison with Incognito.
We observe that for both algorithms, as the size of $k$ increases, the performance of algorithms improves. This is mainly because, as the size of $k$ increases, more generalizations are pruned as part of smaller subsets, and less
execution time is needed. Moreover, we observe that Incognito is faster than $\alpha$-protective Incognito only if the value of $\tau$ is very high ({\em i.e.} $\tau=6$ or $\tau=8$). By decreasing the value of $\tau$, $\alpha$-protective Incognito runs even faster than Incognito. The explanation is that, with $\alpha$-protective Incognito, more generalizations are pruned as part of smaller subsets by checking both $k$-anonymity and $\alpha$-protection, and less execution
time is needed. The difference between the performance of the two algorithms gets smaller when $k$ increases. In addition, because of the generalization property of $\alpha$-protection with respect to $elift$, $\alpha$-protective Incognito is faster for $f=elift$ than for $f=slift$. However, this difference is not substantial since, as we mentioned in Section \ref{sec:algorithm}, $\alpha$-protection should consider all frequent and infrequent PD rules. In summary, since $\alpha$-protective Incognito provides extra protection ({\em i.e.} against discrimination) in comparison with Incognito, the cost is
sometimes a longer execution time, specifically
when the value of $\tau$ is very high, near to $|QI|$. When $\tau$ is small, $\alpha$-protective Incognito can even be faster than Incognito. Hence, given
that discrimination discovery is an intrinsically expensive task, our solution is optimal and offers reasonable performance for off-line application.
%
\subsection{Data Quality}
\begin{figure*}
\centering
\begin{tikzpicture}[thick]
\def \plotdist {7.3 cm} 
\def \xsc {0.67} 
\def \ysc {0.63} 
\begin{scope}[xscale=\xsc , yscale=\ysc]
\begin{axis}[
legend style={
at={(0.1,0.9)},
anchor=north west
},
width=10cm,
height=7cm,
  ybar,
  ytick pos=left,
  bar width = 5pt,
  xlabel=$k$,
  ylabel=Minimum GH,
  ytick={14,15,16,17},
  xtick={1,...,5},
  xticklabels={%
        5,
        10,
        25,
        50,
        100
        },
        ]
         \addplot 
         plot coordinates {
        (1,14)
        (2,15)
        (3,15)
        (4,15)
        (5,15)
    };
    \addlegendentry{I}
    \addplot 
    plot coordinates {
         (1,14)
        (2,15)
        (3,15)
        (4,15)
        (5,16)
    };
    \addlegendentry{II}
    \addplot 
     plot coordinates {
      (1,14)
        (2,15)
        (3,15)
        (4,15)
        (5,15)
    };
    \addlegendentry{III}
      \addplot 
        plot coordinates {
         (1,15)
        (2,15)
        (3,16)
        (4,16)
        (5,17)
        };
    \addlegendentry{IV}
      \addplot 
      plot coordinates {
        (1,15)
        (2,15)
        (3,16)
        (4,16)
        (5,16)
    };
    \addlegendentry{V}
    \end{axis}
    \end{scope}
\begin{scope}[xshift=\plotdist, xscale=\xsc , yscale=\ysc]
\begin{axis}[
  ybar,
  legend style={
  at={(0.1,0.9)},
anchor=north west
},
  bar width = 5pt,
    width=10cm,
    height=7cm,
 xlabel=$k$,
        ylabel=Minimum DR,
    xtick={1,...,5},
    xticklabels={%
        5,
        10,
        25,
        50,
        100
        },
        ]
         \addplot 
         plot coordinates {
        (1,0.104)
        (2,0.127)
        (3,0.139)
        (4,0.139)
        (5,0.182)
    };
    \addlegendentry{I}
 \addplot 
    plot coordinates {
        (1,0.104)
        (2,0.127)
        (3,0.149)
        (4,0.149)
        (5,0.195)
    };
    \addlegendentry{II}
 \addplot 
     plot coordinates {
        (1,0.104)
        (2,0.127)
        (3,0.149)
        (4,0.149)
        (5,0.195)
    };
    \addlegendentry{III}
      \addplot 
        plot coordinates {
        (1,0.120)
        (2,0.149)
        (3,0.149)
        (4,0.149)
        (5,0.248)
        };
    \addlegendentry{IV}
      \addplot 
      plot coordinates {
        (1,0.120)
        (2,0.149)
        (3,0.149)
        (4,0.149)
        (5,0.248)
    };
    \addlegendentry{V}
    \end{axis}
    \end{scope}
\end{tikzpicture}
\caption{General data quality metrics. Left, generalization height (GH).
Right, discernibility ratio (DR). Results are given for
$k$-anonymity (I); and $\alpha$-protection $k$-anonymity with
$DA_3$, $\alpha=1.2$ (II); $DA_3$, $\alpha=1.6$ (III); $DA_1$, $\alpha=1.2$ (IV);  $DA_1$, $\alpha=1.6$ (V).
In all cases $f=slift$.}
\label{fig:dataquality1}
\end{figure*}
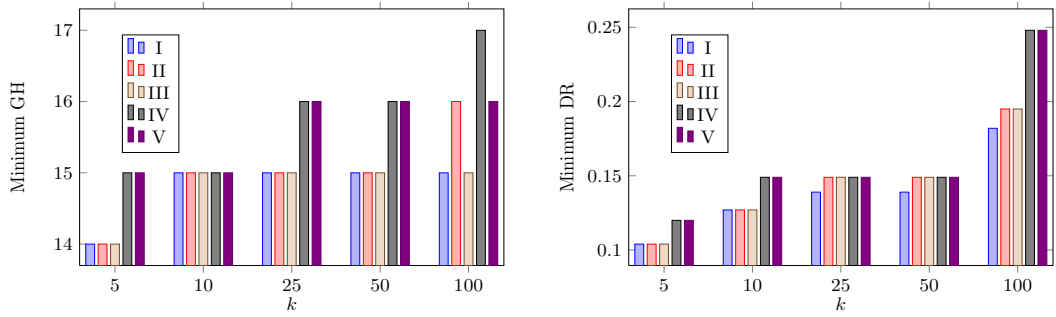

Privacy preservation and discrimination prevention are one side of
the problem we tackle. The other side is retaining information
so that the published data remain practically useful. Data quality can be measured in general or with respect to a specific data analysis task ({\em e.g.} classification).
First, we evaluate the data quality of the
protected data obtained by $\alpha$-protective Incognito and Incognito using
as general metrics the generalization height~\cite{Lefevre2005,Samarati} and
discernibility~\cite{BAYARDO2005}. The generalization height (GH) is the height of an anonymized data table in the generalization lattice. Intuitively, it
corresponds to the number of generalization steps that were performed. The
discernibility metric charges a penalty to each record for being indistinguishable from other records. For each record in equivalence QI class $qid$, the penalty is $|\mathcal{DB}[qid]|$. Thus, the discernibility cost is equivalent to the sum of the $|\mathcal{DB}[qid]|^2$. We define the discernibility ratio (DR)
as DR$=\frac{\sum_{qid}|\mathcal{DB}[qid]|^2}{|\mathcal{DB}|^2}$. Note that:
i) $0 \leq$ DR $\leq 1$;
ii) lower DR and GH mean higher data quality. From the list of full-domain generalizations
obtained from Incognito and $\alpha$-protective Incognito, respectively,
we compute the
minimal full-domain generalization w.r.t. both GH and DR
for each algorithm and compare them.
Second, we measure the data quality of the anonymous data obtained by $\alpha$-protective Incognito and Incognito for a classification task using the classification metric CM from \cite{IYENGAR2002}. CM charges a penalty for each record generalized to a $qid$ group in which the record's class is not the majority class. Lower CM means higher data quality. From the list of full-domain generalizations
obtained from Incognito and $\alpha$-protective Incognito, respectively,
we compute the
minimal full-domain generalization w.r.t. CM
for each algorithm and we compare them. In addition, to evaluate the impact of our transformations on the accuracy of a classification task, we first
obtain the minimal full-domain generalization w.r.t. CM to anonymize the training set. Then, the same generalization is applied
to the testing set to produce a generalized testing set.
Next, we build a classifier on the anonymized training set
and measure the classification accuracy (CA) on the generalized
records of the testing set. For classification models we
use the well-known decision tree classifier J48 from the Weka software package \cite{weka}. We also measure the classification accuracy
on the original data without anonymization. The difference represents the cost in terms of classification accuracy
for achieving either both privacy preservation and discrimination prevention
or privacy preservation only.
Fig. \ref{fig:dataquality1} summarizes the data quality results using general metrics for different values of $k$, $DA$ and $\alpha$, where $f=slift$.
We found that the data quality of $k$-anonymous tables ({\em i.e.} in terms
of GH and DR) without $\alpha$-protection
is equal or slightly better than the quality of
$k$-anonymous tables with $\alpha$-protection. This is because
the $\alpha$-protection $k$-anonymity requirement provides extra protection
({\em i.e.}, against discrimination) at the cost of some
data quality loss  when $DA$ and $k$ are large and
$\alpha$ is small. As $k$ increases, more generalizations
are needed to achieve $k$-anonymity,
which increases GH and DR. We performed the same
experiment for other values of $f$, and we observed a similar trend (details
omitted due to lack of space).

\begin{figure*}
\centering
\begin{tikzpicture}[thick]
\def \plotdist {7.3 cm} 
\def \xsc {0.67} 
\def \ysc {0.63} 
\begin{scope}[xscale=\xsc , yscale=\ysc]
\begin{axis}[
legend style={
at={(0.3,0.9)},
},
        xlabel=$k$,
        width=10cm,
        height=7cm,
        ylabel=Minimum CM,
        ytick={0.27,0.28,0.288},
               ]
    \addplot[mark=square,blue,mark size=3pt,very thick] plot coordinates {
        (5,0.271)
        (10,0.271)
        (25,0.271)
        (50,0.271)
        (100,0.272)
    };
    \addlegendentry{I}

 \addplot[mark=diamond,gray,mark size=3pt,very thick] plot coordinates {
         (5,0.271)
        (10,0.271)
        (25,0.271)
        (50,0.271)
        (100,0.272)
    };
    \addlegendentry{II}

  \addplot[mark=triangle,brown,mark size=3pt,very thick] plot coordinates {
      (5,0.271)
        (10,0.271)
        (25,0.271)
        (50,0.271)
        (100,0.272)
    };
    \addlegendentry{III}

    \addplot[mark=triangle,green,mark size=3pt,very thick]
        plot coordinates {
        (5,0.271)
        (10,0.271)
        (25,0.271)
        (50,0.271)
        (100,0.288)
        };
    \addlegendentry{IV}
   
      \addplot[mark=oplus,red,mark size=3pt,very thick] plot coordinates {
        (5,0.271)
        (10,0.271)
        (25,0.271)
        (50,0.271)
        (100,0.281)
    };
    \addlegendentry{V}
   
    \end{axis}
    \end{scope}
\begin{scope}[xshift=\plotdist, xscale=\xsc , yscale=\ysc]
\begin{axis}[
legend style={
at={(0.3,0.6)},
},
		width=10cm,
        height=7cm,
        xlabel=$k$,
        ylabel= CA (\%),
        ytick={80.76,81,81.5,82,82.5},
               ]
   \addplot[mark=dash,black,mark size=3pt,very thick] plot coordinates {
        (5,82.36)
        (10,82.36)
        (25,82.36)
        (50,82.36)
        (100,82.36)
    };
    \addlegendentry{0}
    \addplot[mark=square,blue,mark size=3pt,very thick] plot coordinates {
        (5,82.01)
        (10,82.01)
        (25,82.01)
        (50,82.01)
        (100,82)
    };
    \addlegendentry{I}
    \addplot[mark=diamond,gray,mark size=3pt,very thick] plot coordinates {
         (5,81.95)
        (10,81.95)
        (25,81.95)
        (50,81.95)
        (100,81.85)
    };
    \addlegendentry{II}
  \addplot[mark=triangle,brown,mark size=3pt,very thick] plot coordinates {

        (5,81.98)
        (10,81.98)
        (25,81.98)
        (50,81.98)
        (100,81.88)
      
    };
    \addlegendentry{III}
    \addplot[mark=triangle,green,mark size=3pt,very thick]
        plot coordinates {
         (5,81.9)
        (10,81.9)
        (25,81.9)
        (50,81.9)
        (100,80.76)
        };
    \addlegendentry{IV}

      \addplot[mark=oplus,red,mark size=3pt,very thick] plot coordinates {
        (5,81.93)
        (10,81.93)
        (25,81.93)
        (50,81.93)
        (100,80.86)
    };
    \addlegendentry{V}
   
    \end{axis}
    \end{scope}
\end{tikzpicture}
\caption{Data quality for classification analysis
Left, classification metric (CM).
Right, classification accuracy, in percentage (CA). Results are given for the original data (0); $k$-anonymity (I); and $\alpha$-protection $k$-anonymity with
$DA_3$, $\alpha=1.2$ (II); $DA_3$, $\alpha=1.6$ (III); $DA_1$, $\alpha=1.2$ (IV);  $DA_1$, $\alpha=1.6$ (V).
In all cases $f=slift$.
}
\label{fig:dataquality2}
\end{figure*}
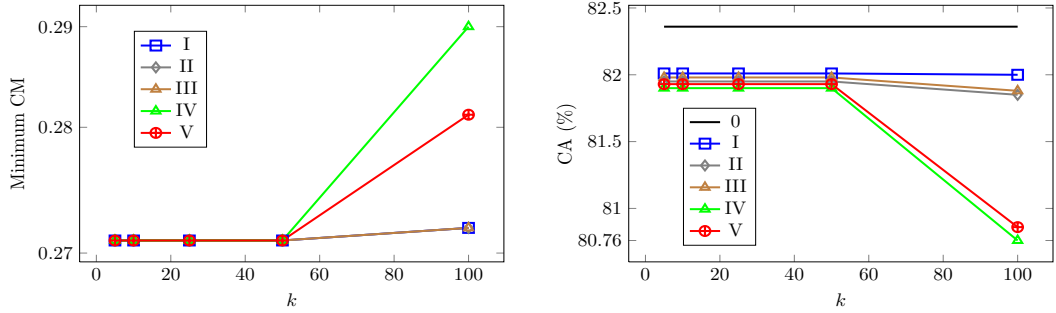

The left chart of Fig. \ref{fig:dataquality2} summarizes
the data quality results using the
classification metric (CM) for different values
of $k$, $DA$ and $\alpha$, where $f=slift$.
We found that the data quality of $k$-anonymous tables ({\em i.e.} in terms of CM) without $\alpha$-protection
is equal or slightly better than the quality of $k$-anonymous tables with $\alpha$-protection. This is because the $\alpha$-protection $k$-anonymity requirement provides extra protection ({\em i.e.}, against discrimination)
at the cost of some data quality loss when $DA$ and $k$ are large.
The right chart of Fig. \ref{fig:dataquality2} summarizes the impact of
achieving $k$-anonymity or $\alpha$-protection $k$-anonymity on the
percentage classification accuracy (CA) of J48 for different values of $k$, $DA$ and $\alpha$, where $f=slift$. We observe a similar trend as for CM. The accuracies of J48 using $k$-anonymous tables without $\alpha$-protection are equal or slightly better than the accuracies of J48 using $k$-anonymous tables with $\alpha$-protection.
%
%
%
%
%
\section{Extensions}
\label{sec:alternative}
We consider here alternative privacy models and anti-discrimination
requirements.
\subsection{Alternative Privacy Models}
\subsubsection{Attribute Disclosure}
In contrast to $k$-anonymity, the privacy models in attribute linkage assume the existence of sensitive attributes in $\mathcal{D}$
such that $QI \cap S = \emptyset$.
As shown in \cite{Machanavajjhala2006,Li2007}, by using full-domain generalizations over QI, we can obtain data tables protected against attribute disclosure.
Considering attribute disclosure risks, we focus on the problem of producing an anonymized version of $\mathcal{D}$ which is protected against attribute disclosure and free from discrimination ({\em e.g.}, $\alpha$-protective $l$-diverse data table). We study this problem considering the following possible relations between $QI$, $DA$ and $S$:
\begin{itemize}
\item $DA \subseteq QI$: It is possible that the original data are biased in the subsets of the protected groups which
are defined by sensitive attributes ({\em e.g.} women who have medium salary). In this case, only full-domain generalizations which
include the generalization of protected groups values can make $\mathcal{D}$ $\alpha$-protective. This is because the generalization is only performed over QI attributes.
\item $DA \subseteq S$: A full-domain generalization over QI can make
the original data $\alpha$-protective only if $\mathcal{D}$ is biased in the subsets of protected groups which
are defined by QI attributes. In other scenarios, {\em i.e.}, when data is biased versus some protected groups or subsets of protected groups which
are defined by sensitive attributes,
full-domain generalizations over QI cannot make $\mathcal{D}$ $\alpha$-protective. One possible solution
is to generalize attributes which are both sensitive and PD ({\em e.g.}, {\em Religion} in some applications),
even if they are not in QI.
\end{itemize}
\begin{observation}
If $DA \subseteq QI$, $l$-diversity/$t$-closeness
 and $\alpha$-protection can be achieved simultaneously in $\mathcal{D}$ by means of full-domain generalization.
\end{observation}
%
Since the subset and generalization properties are also satisfied
for $l$-diversity and $t$-closeness, to obtain all full-domain
generalizations with which data is $\alpha$-protective and protected against attribute disclosure, we take $\alpha$-protective Incognito and make the following changes: 1) every time a data table is tested for $k$-anonymity, it
is also tested
for $l$-diversity or $t$-closeness; 2) every time a data table is tested for $\alpha$-protection, it is tested w.r.t. attributes of $node$ and sensitive attributes. This can be done by simply updating the {\em Check $\alpha$-protection} function. Just as the data quality of $k$-anonymous data tables without $l$-diversity or $t$-closeness is slightly better than the quality of
$k$-anonymous data tables with $l$-diversity or $t$-closeness,
we expect
a similar slight quality loss when adding $l$-diversity or $t$-closeness
to $k$-anonymity $\alpha$-protection.
\subsubsection{Differential Privacy}
We define a differential private data table as an anonymized data table generated by a function (algorithm) which is differentially private.
The general structure of differentially private data release approaches is to first build a contingency table of the original raw data over the database domain. After that, noise is added to each
frequency count in the contingency table to satisfy differential
privacy. However, as mentioned in \cite{FungDiff}, these approaches are not suitable for high-dimensional data with a large domain because when the added noise is relatively large compared to the count, the utility of the data is significantly decreased. In~\cite{FungDiff}, a generalization-based algorithm
for differentially private data release is presented.
It first probabilistically generates a generalized contingency table and then adds noise to the counts. Thanks to generalization,
the count of each partition is typically much larger than the added noise. In this way, generalization helps to achieve a differential private version of  $\mathcal{D}$ with higher data utility.
Considering the differential privacy model, we focus on the problem of producing an anonymized version of  $\mathcal{D}$ which is differentially private and free from discrimination with respect to $DA$. Since the differentially
private version of $\mathcal{D}$ is an approximation of $\mathcal{D}$ generated at random, we have the following observation.
\begin{observation}
Making original data table $\mathcal{D}$ differentially private using
Laplace noise addition can make $\mathcal{D}$ more or less $\alpha$-protective w.r.t. $DA$ and $f$.
\end{observation}
Given the above observation and the fact that generalization can help to achieve differential privacy with higher data quality, we propose to obtain a noisy generalized contingency table of $\mathcal{B}$ which is also $\alpha$-protective. To do this, one solution is to add uncertainty to an algorithm that generates all possible full-domain generalizations with which $\mathcal{D}$ is $\alpha$-protective.
As shown in \cite{FungDiff}, for higher values of the privacy budget, the quality of differentially private data tables is higher than the quality of $k$-anonymous data tables, while for smaller value of the privacy budget it is the other way round.
Therefore, we expect that
differential privacy plus discrimination prevention
will compare similarly to
the $k$-anonymity plus discrimination prevention presented in
the previous sections of this chapter.
\subsection{Alternative Anti-discrimination Legal Concepts}
Unlike privacy legislation, anti-discrimination legislation is very
sparse and includes different legal concepts, {\em e.g.} direct and indirect discrimination and the so-called genuine occupational requirement.
\subsubsection{Indirect Discrimination}
Indirect discrimination occurs when the input does not contain PD attributes, but discriminatory decisions against protected groups might be indirectly made because of the availability of some background knowledge;
for example, discrimination
against black people might occur if the input data contain \emph{Zipcode} as attribute (but not {\emph Race}) and one knows that
the specific zipcode is mostly inhabited by black people \footnote{\small http://eur-lex.europa.eu/LexUriServ/LexUriServ.do?uri=CELEX:62011CJ0385:EN:Not} ({\em i.e.}, there is high correlation between \emph{Zipcode} and \emph{Race} attributes).
Then, if the protected groups do not exist in the original data table or have been removed from it due to privacy or anti-discrimination constraints, indirect discrimination still remains
possible.
Given $DA$, we define background knowledge as the correlation between $DA$ and PND attributes which are in $\mathcal{D}$:
\[ \mathcal{BK}=\{A_i \rightarrow A_x | A_i \in \mathcal{A}, A_i \mbox{ is PND and } A_x \in DA\} \]
Given $\mathcal{BK}$, we define $IA$ as a set of PND attributes in $\mathcal{D}$ which are
highly correlated to $DA$, determined according to
$\mathcal{BK}$. Building on Definition \ref{label:disc}, we introduce the notion of non-redlining $\alpha$-protection for a data table.
\begin{definition} [Non-redlining $\alpha$-protected data table]
\label{def:indirectprotection}
Given $\mathcal{D}(A_1,\cdots, A_n)$, $DA$, $f$ and $\mathcal{BK}$, $\mathcal{D}$ is said to satisfy non-redlining $\alpha$-protection or to be non-redlining $\alpha$-protective w.r.t. $DA$ and $f$ if each PND frequent classification rule $c: D, B \rightarrow C$ extracted from $\mathcal{D}$ is $\alpha$-protective, where $D$ is a PND itemset of $IA$ attributes and $B$ is a PND itemset of $\mathcal{A} \backslash IA$ attributes.
\end{definition}
%
%
%
Given $DA$ and $\mathcal{BK}$, releasing a non-redlining  $\alpha$-protective version of an original table is desirable to prevent indirect discrimination against protected groups w.r.t. $DA$.
Since indirect discrimination against protected groups originates from the correlation between $DA$ and $IA$ attributes, a natural
countermeasure is to diminish this correlation.
Then, an anonymized version of original data table protected against indirect discrimination ({\em i.e.} non-redlining $\alpha$-protective) can be generated by generalizing $IA$ attributes. As an example, generalizing all instances of 47677, 47602 and 47678 zipcode values to the same generalized value 467** can prevent indirect discrimination against black people living in the 47602 neighborhood.
\begin{observation}
If $IA \subseteq QI$, non-redlining $\alpha$-protection can be achieved in $\mathcal{D}$ by means of full-domain generalization.
\end{observation}
Consequently, non-redlining $\alpha$-protection can be achieved with each of the above mentioned privacy models based on full-domain generalization
of $\mathcal{D}$ ({\em e.g.} $k$-anonymity), as long as $IA \subseteq QI$.
Fortunately, the subset and generalization properties satisfied
by $\alpha$-protection are also satisfied by non-redlining $\alpha$-protection.
Hence, in order to obtain all possible full-domain generalizations with
which $\mathcal{D}$ is indirect discrimination and privacy protected, we take $\alpha$-protective Incognito and make the following changes: 1) add $\mathcal{BK}$ as the input of the algorithm and determine $IA$ w.r.t. $\mathcal{BK}$, where PD attributes are removed from $\mathcal{D}$; 2) every time a data table is tested for $\alpha$-protection, test it for non-redlining $\alpha$-protection instead.
Considering the above changes, when combining indirect
discrimination prevention and privacy protection, we expect similar
data quality and algorithm performance as we had when combining
direct discrimination prevention and privacy protection.
\subsubsection{Genuine occupational requirement}
The legal concept of genuine occupational requirement refers to detecting that part of the discrimination which may be explained by other attributes \cite{Calders2011}, named legally grounded attributes; {\em e.g.},
denying credit to women may be explainable if most of them
have low salary or delay in returning previous credits. Whether low salary or delay in returning previous credits is an acceptable legitimate argument to deny credit is for the law to determine.
Given a set $LA$ of legally grounded attributes in $\mathcal{D}$, there
are some works which attempt to cater technically
to them in the anti-discrimination protection~\cite{Ruggieri2011,Calders2011,Dwork2012}. The general idea is to prevent only unexplainable (bad) discrimination. Loung et al. \cite{Ruggieri2011} propose a variant of the $k$-nearest neighbor (k-NN) classification which labels each record in a data table as discriminated or not. A record $t$ is discriminated
if: i) it has a negative decision value in its class attribute; and ii) the difference between the proportions of $k$-nearest neighbors of $t$ w.r.t. $LA$ whose decision value is the same of $t$ and
belong to the same protected-by-law groups as $t$ and the ones that
do not belong to the same protected groups as $t$ is
greater than the discrimination threshold.
This implies that the negative decision for $t$ is not explainable on the basis of the legally grounded attributes, but it is biased by group membership. We say that a data table is protected only against unexplainable discrimination w.r.t. $DA$ and $LA$ if the number of records labeled as discriminated is zero (or near zero). An anonymized version of an original data table which is protected  against unexplainable discrimination can be generated by generalizing $LA$ and/or $DA$ attributes. Given a discriminated record, generalizing $LA$ and/or $DA$ attributes can decrease the difference between the two above mentioned proportions. Hence, an anonymized version of original data table which is protected against unexplainable discrimination and  privacy protected can be obtained using full-domain generalization over QI attributes as long as $DA \subseteq QI$ and $LA \subseteq QI$.
\section{Conclusions}
\label{sec:conclusiongen}
We have investigated the problem of discrimination- and privacy-aware
data publishing and analysis,
{\em i.e.}, distorting an original data set in such a way that neither privacy-violating nor discriminatory inferences can be made on the released data sets. To study the impact of data generalization ({\em i.e.} full-domain generalization) on discrimination prevention, we applied generalization not only for making the original data privacy-protected but also for making them protected against discrimination. We found that a subset of $k$-anonymous full-domain generalizations with the same
or slightly higher data distortion than the rest
(in terms of general and specific data analysis metrics)
are also $\alpha$-protective. Hence, $k$-anonymity and $\alpha$-protection
can be combined to attain privacy protection and
discrimination prevention in the published data set.
We have adapted to $\alpha$-protection two well-known
properties of $k$-anonymity, namely the subset and the generalization
properties. This has allowed us to propose an $\alpha$-protective
version of Incognito, which can take as parameters several
 legally grounded measures of discrimination and generate
 privacy- and discrimination-protected full-domain generalizations.
 We have evaluated the quality of data output by this algorithm,
 as well as its execution time. Both turn out to be nearly as good
 as for the case of the plain Incognito, so the toll paid
 to obtain $\alpha$-protection is very reasonable. Finally,
we have sketched how our approach can be extended to
satisfy alternative privacy guarantees or anti-discrimination legal constraints.

\chapter{Conclusions}
\label{conclusions}


In the information society, massive and automated data collection occurs as a consequence of the ubiquitous digital traces we all generate in our daily life. The availability of such wealth of data makes its publication and analysis highly desirable for a variety of purposes, including policy making, planning, marketing, research, etc. Yet, the real and obvious benefits of
data analysis and publishing have a dual, darker side. There are at least two potential threats for individuals whose information is published: privacy invasion and potential discrimination. Privacy
invasion occurs when the values of published sensitive attributes can be linked to specific individuals (or companies). Discrimination is unfair or unequal treatment of people based
on membership to a category, group or minority, without regard to individual characteristics.

On the legal side, parallel to the development of privacy legislation, anti-discrimination legislation has undergone a remarkable expansion, and it now prohibits discrimination
against protected groups on the grounds of race, color, religion, nationality, sex, marital status, age and pregnancy, and in a number of settings, like credit and insurance, personnel selection and wages, and access to public services. On the technology side, efforts
at guaranteeing privacy have led to developing privacy preserving data mining (PPDM) and efforts at fighting discrimination have led to
developing anti-discrimination techniques in data mining. Some proposals are oriented to the discovery and measurement of discrimination, while others deal with preventing data mining (DPDM) from becoming itself a source of discrimination, due to automated decision making based on discriminatory models extracted from inherently biased datasets. 

However, up to now, PPDM and DPDM have been studied in isolation. Is it sufficient to focus on one and ignore the other? Is it
sufficient to guarantee data privacy while allowing automated
discovery of discriminatory profiles/models? In this thesis, we argue
that the answer is no. If there is a chance to create a trustworthy
technology for knowledge discovery and deployment, it is with
a holistic approach which faces both privacy and discrimination
threats (risks). We explore the relationship between PPDM and
DPDM in both contexts of data and knowledge publishing. We design holistic approaches capable of addressing {\emph both} threats together in significant data mining processes. 
This thesis is the first work inscribing simultaneously privacy and anti-discrimination with a  by design approach in data publishing and mining. It is a first example of a more comprehensive set of ethical values that is inscribed into the analytical process. 

\section{Contributions}

In more detail, our contributions are:

\begin{enumerate}

\item We develop a new pre-processing
discrimination prevention methodology including different data
transformation methods that can prevent direct discrimination,
indirect discrimination or both of them at the same time.
To attain this objective, the first step is to measure
discrimination and identify categories and groups of individuals that
have been directly and/or indirectly discriminated in the decision-making
processes; the second step is to transform data in the proper way to
remove all those discriminatory biases. Finally, discrimination-free
data models can be produced from the transformed dataset without seriously
damaging data quality. The experimental results
reported demonstrate that the proposed techniques are quite successful
in both goals of removing discrimination and preserving data quality. 

\item We have investigated the problem of discrimination
and privacy
aware frequent pattern discovery, {\em i.e.} the sanitization of the collection of patterns mined from a transaction database in such a way that neither privacy-violating nor discriminatory inferences can be inferred on the released patterns.
We found that our discrimination preventing transformations do not interfere with a privacy preserving sanitization based on $k$-anonymity, thus accomplishing the task of combining the two and achieving a robust (and formal) notion of fairness in the resulting pattern collection.
Further, we have presented extensive empirical results on
the utility of the protected data. Specifically, we evaluate
the distortion introduced by our methods and its effects on classification.
It turns out that
the utility loss caused by simultaneous anti-discrimination
and privacy protection is only marginally higher than the loss caused
by each of those protections separately. This result
supports the practical deployment of our methods. Moreover, we have discussed the possibility of using our proposed framework
while replacing $k$-anonymity with differential privacy.

\item We have investigated the relation between data anonymization techniques and anti-discrimination to answer an important question: 
how privacy protection via data anonymization impacts the discriminatory bias contained in the original data. By presenting and analyzing different scenarios, we learn that we cannot protect original data against privacy attacks without taking into account anti-discrimination requirements ({\em i.e.} $\alpha$-protection). This happens because data anonymization techniques can work against anti-discrimination.
In addition, we exploit the fact that some data anonymization techniques ({\em e.g.} specific full-domain generalization) can also protect data against discrimination. Thus, we can adapt and use some of these techniques for discrimination prevention. Moreover, by considering anti-discrimination requirements during anonymization, we can present solutions to generate privacy- and discrimination-protected datasets.
We also find that global recoding generalizations have a more positive impact on discrimination removal than other data anonymization techniques.

\item We have investigated the problem of discrimination- and privacy-aware
data publishing and mining,
{\em i.e.}, distorting an original data set in such a way that neither privacy-violating nor discriminatory inferences can be made on the released data sets. To study the impact of data generalization ({\em i.e.} full-domain generalization) on discrimination prevention, we applied generalization not only for making the original data privacy-protected but also for making them protected against discrimination. We found that a subset of $k$-anonymous full-domain generalizations with the same
or slightly higher data distortion than the rest
(in terms of general and specific data analysis metrics)
are also $\alpha$-protective. Hence, $k$-anonymity and $\alpha$-protection
can be combined to attain privacy protection and
discrimination prevention in the published data set.
We have adapted to $\alpha$-protection two well-known
properties of $k$-anonymity, namely the subset and the generalization
properties. This has allowed us to propose an $\alpha$-protective
version of Incognito, which can take as parameters several
 legally grounded measures of discrimination and generate
 privacy- and discrimination-protected full-domain generalizations.
 We have evaluated the quality of data output by the proposed algorithm,
 as well as its execution time. Both turn out to be nearly as good
 as for the case of the plain Incognito, so the toll paid
 to obtain $\alpha$-protection is very reasonable. 

\end{enumerate}

\section{Publications}

The main publications supporting the content of this thesis are the following:

\begin{itemize}

\item Sara Hajian, Josep Domingo-Ferrer and Antoni Mart\'{\i}nez-Ballest\'e. 	Discrimination prevention in data mining for intrusion and crime detection. In {\em IEEE Symposium on Computational Intelligence in Cyber Security-CICS 2011}, pp. 47-54, 2011.
\item Sara Hajian, Josep Domingo-Ferrer and Antoni Mart\'{\i}nez-Ballest\'e. Rule protection for indirect discrimination prevention in data mining. In 	{\em Modeling
Decisions for Artificial Intelligence-MDAI 2011}, LNCS 6820, pp. 211-222. Springer, 2011.
\item Sara Hajian and Josep Domingo-Ferrer. Anti-discrimination and privacy protection in released data sets. In {\em Joint Eurostat/UNECE Work Session on Statistical Data Confidentiality 2011}, Tarragona, Catalonia, Oct. 26-28, 2011.
\item Sara Hajian and Josep Domingo-Ferrer.  A methodology for direct and indirect discrimination prevention in data mining. {\em IEEE Transactions
on Knowledge and Data Engineering}, to appear (available
on-line from 22 March 2012).
\item Sara Hajian and Josep Domingo-Ferrer and Antoni Mart\'{\i}nez-Ballest\'e. Antidiscriminaci\'on en la detecci\'on de delitos e intrusiones. {\em Actas de la XII Reuni\'on Espa\~{n}ola sobre Criptolog\'{\i}a y
Seguridad de la Informaci\'on RECSI-2012}.
\item Sara Hajian, Anna Monreale, Dino Pedreschi, Josep Domingo-Ferrer and Fosca Giannotti. Injecting discrimination and privacy awareness into pattern discovery. In {\em 2012 IEEE 12th International Conference on Data Mining Workshops}, pp. 360-369. IEEE Computer Society, 2012.
\item Sara Hajian and Josep Domingo-Ferrer. A study on the impact of data anonymization on anti-discrimination. In {\em 2012 IEEE 12th International Conference on Data Mining Workshops}, pp. 352-359. IEEE Computer Society, 2012.
\item Sara Hajian and Josep Domingo-Ferrer. Direct and Indirect Discrimination Prevention Methods. In {\em Discrimination and Privacy in the Information Society 2013}. pp. 241-254.

\end{itemize}

Submitted articles that are still pending acceptance:

\begin{itemize}

\item Sara Hajian, Josep Domingo-Ferrer, Anna Monreale, Dino Pedreschi and Fosca Giannotti. Discrimination- and Privacy-aware Frequent Pattern Discovery. 

\item Sara Hajian, Josep Domingo-Ferrer and Oriol Farr\`as. Generalization-based Privacy Preservation and Discrimination Prevention in Data Publishing and Mining.

\end{itemize}

\section{Future Work}
 
This thesis is the first one addressing in depth the problem of providing protection against both privacy and discrimination threats in data mining, and it opens several future research avenues:


\begin{itemize}


\item Explore the relationship between discrimination prevention and privacy preservation in data mining considering alternative data anonymization techniques, other than those studied in this thesis.

\item Propose new measures and techniques for measuring and preventing indirect discrimination. This will require to further
study the legal literature on discrimination.

\item Improve the existing algorithms to achieve better utility and performance.

\item Empirically compare the proposed approaches based on differential privacy with the proposed approaches based on $k$-anonymity in terms of different data utility measures. 

\item Extend the existing approaches and algorithms to other data mining tasks.

\item Make the algorithms and analyses applicable to a variety of input data.

\item Present real case studies in the context of discrimination discovery and prevention in data mining.

\item Study the problem of on-line discrimination prevention. In the on-line case, the protection should be performed at the very time of a service request.

\item Extend concepts and methods to the analysis of
discrimination in social network data.

\end{itemize}

\bibliographystyle{plain}

\end{document}